# How on Earth:
## *Flourishing in a Not-for-Profit World by 2050*

***An initiative of the Post Growth Institute – www.postgrowth.org***

***Jennifer Hinton***

***Donnie Maclurcan***

***Please note that this is a draft from August 2016. An updated version is currently being edited.***

This book presents both a critique of the current economic system and a vision for a more sustainable economy; one that serves people and planet. It is important to know that this is written for a general, international audience and has not been peer-reviewed in a formal manner. Although the core ideas in this book will not change, it is still being edited for optimal writing tone and framing, as well as some further theory development and refinement.

We are releasing this working draft because we feel the core ideas it presents are so important that they need to reach the minds of people who will work with them and develop them.

If you would like to work with or build on the ideas put forth in this draft, please cite it as:

Hinton, Jennifer & Maclurcan, Donnie (2016) *How on Earth: Flourishing in a Not-for-Profit World by 2050* (working draft). Ashland, OR: Post Growth Publishing.

Please note that many of the references and notes are incomplete and inconsistent in their referencing style. We have included links wherever possible.

We would greatly appreciate feedback from peers and can be contacted via email at:

Jennifer Hinton: jen(at)postgrowth.org

Donnie Maclurcan: donmaclurcan(at)gmail.com

# 1. When Profit is a Means to an End

## *Profit can be more generative than you might imagine*

In a healthy economy, money is constantly circulating. This simple premise forms the basis of the entire economic model outlined in this book. Beneath ideology and political persuasion, this wisdom also sits within the hearts of each and every one of us. This can be seen in a simple experiment we have been conducting with diverse audiences since 2010.[1]

In the experiment, we ask audience members to draw a horizontal line across a page. Above the line, they are to draw a symbol, shape or simple image that represents a healthy, sustainable economy. Below the line, they replicate the process, but this time drawing something that represents our present economy.

A remarkable pattern always emerges.

For a sustainable economy, most people visualize something cyclical and balanced. Circles, the infinity symbol, and wave patterns are the three most common images.

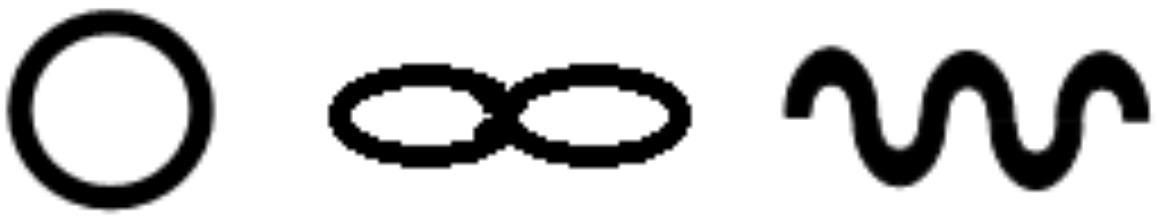

**Images representing a sustainable, healthy economy**

For our present economy, most people visualize something hierarchical or unbalanced. Triangles, straight lines, and jagged patterns are the three most common images.

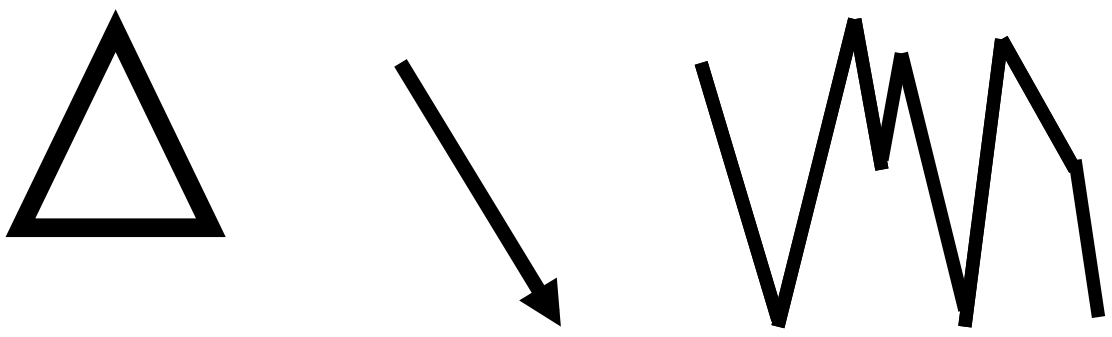

**Images representing our present economy**

Most people carry the inherent wisdom and intuition that a healthy, sustainable economy must be balanced and it must use money and resources in a circular way. **This *principle of circulation* must be built into the very way our economy functions;** not as a side-effect, not as an afterthought, but as the way that economic activity naturally flows.

What we have now, in contrast, is a linear economy, focused on throughput, with massive amounts of resources going to waste, and money accumulating in the hands of a few. Keeping resources and money circulating in this economy is an afterthought, not an inherent feature of the system. In fact, many of us who are trying to work for a more sustainable economy often feel like we're swimming upstream or fighting a doomed battle against the very rules of the economic system, itself.

---

[1] Our experiment has involved audiences that include students, educators, politicians, investors, CEOs, environmentalists, and social justice activists.

Luckily, it doesn't have to be that way. The kind of business with which most of us are familiar is not the only kind of business. An entirely different model is sitting right in front of us and has gone mostly unnoticed, despite its increasingly large presence. And it points the way to the kind of healthy, balanced economy that so many of us desire.

The economic model that we're introducing in this book has the circulation of money written into its DNA. In our vision of the economy, all players are looked after through fair wages, and any surplus is fed back into the system. After all, when it comes to money, what more does anyone need than a good enough wage to cover today's expenses and to put something aside for the future?

This approach is aligned with the principles of ecosystems, in which each animal and each species takes only what it needs and any surplus is passed on or left behind, to be used by other members of the ecosystem. Ecosystems don't have any equivalents of bloated savings accounts or offshore tax havens. The hoarding of wealth would be destructive to the system as a whole.

Can you imagine lions in the savannah hunting all day to accumulate as much of a surplus of meat as possible, hiding it in a cave and every once in a while giving some to the vultures and hyenas? It's a laughable scenario. In reality, the lions hunt for as much food as they need, then spend the rest of the day resting and playing. They leave some of their kill behind for other species in the ecosystem to eat. If the lions accumulated more than they need, it would disrupt the entire food chain. They would decimate the species that they prey on. The vultures and hyenas would not have enough to eat and would starting dying off. And the numbers of vultures and hyenas in the savannah would affect the numbers of many other animals through their predatory patterns, which in turn would affect other species. The lions' greed would not only decimate some species and deprive others of vital nutrition, but they might very well find themselves starving in the end, as the whole system breaks down. This is what is happening in our current system. Due to the structure and rules of our economy, a small minority has accumulated most of the money and this concentration is weakening the whole system. The key lies in what happens to the *surplus* in the economy.

The bold hypothesis we present in this book is that **businesses that see profit as a means to achieving deeper goals will increasingly outperform those that prioritize profit as a goal in itself.** Our main aim in this first chapter is to shed a new light on profit, re-formulate its place and potential in the world, and offer a glimpse of a much healthier economy based on this new perspective.

## The Potential of Profit

What do you think of when you hear the word 'profit'? What sort of thoughts and feelings arise?

Many of us think of a wide range of associated words all the way from 'greed', 'excess', and 'selfishness', to 'business', 'reward', and 'investment'. We all have some sort of feelings about profit because it plays such a central role in society. Indeed, profit seems to make the world go round.

We've found that many people feel profit is a very positive thing; it motivates economic activity and keeps the 'invisible hand'[2] of the market going. It is the just reward for working hard and taking the risks of going into business. On the flip side, others feel that profit is inherently evil. It motivates all kinds of destructive behavior and encourages people to act selfishly and hoard wealth. And some of us have very mixed feelings about profit, acknowledging that it has both pros and cons.

Whatever feelings you have about profit, no one can deny its relevance. So it's important for us to have a clear understanding of exactly what profit is and what it can be. Profit is simply the financial surplus generated by business activity, 'the difference between the amount earned and the amount spent in buying, operating, or producing something.'[3] Thus, profit is neither inherently good nor bad. But, as the surplus of economic activity, it is important. And what happens to the surplus in our economy is central to whether we have a healthy or destructive economy.

We argue that, **in order to determine whether business is generative (contributing to the health of the whole economy) or extractive (taking away from the health of the whole economy), one must ask the question, "*Who* profits from the profit?"**

## The Company that Changed Our Lives

Many people know about the destructive activities that are carried out every day for the sake of maximizing profit. We constantly hear about how companies are profiting from the exploitation of workers in sweatshops, increasing their revenue by overfishing or cutting costs by cutting corners. But relatively few people are aware of the businesses and entrepreneurs all over the world that approach profit in a very different way, in a generative way. One such company sparked the idea for this book, and changed our lives forever.

In 2009, a man named Colin Saltmere gave a presentation at a conference in Brisbane, Australia. He described his civil engineering company, Myuma, as having 50 employees and an annual turnover of 17 million Australian dollars. He presented their work, which included the construction of roads in north-west Queensland. Then, Saltmere said five words that changed our lives, "And we're not-for-profit".

This was a paradigm-shifting moment for us. How could this company, which does so much work and generates so much revenue, be *not*-for-profit? Furthermore, if an engineering company could be not-for-profit, then might there be not-for-profit companies in other unlikely sectors of the economy, too? If so, would it be possible to have an entire economy based on not-for-profit business? Could that help us move beyond our current crises? The potential felt enormous. We had to know more.

In a follow-up conversation, Saltmere told us that, as an Aboriginal group of engineers, they wanted to do work that would help the community. So they started their engineering firm as a not-for-profit (NFP). This means that Myuma is based on a social mission and that, after it has paid all its expenses, including wages, it must use 100% of its profits to fulfill its social mission.

---

2 The Invisible Hand is a concept first described by political philosopher Adam Smith in the late 1700s. A central concept in capitalism, it describes how the decisions of self-interested economic actors lead to the fair allocation of resources for all, through the mechanism of supply and demand, almost as if an invisible hand was moving the economy. (The Concise Encyclopedia of Economics: Adam Smith (1723- 1790) , http://www.econlib.org/library/Enc/bios/Smith.html - lfHendersonCEE2BIO-084_footnote_nt467)

3 "Profit," Oxford Dictionaries, http://www.oxforddictionaries.com/definition/english/profit

Thus the seed was planted that business not only can, but *should*, be generative for the wider community. And, in fact, the entire economy should be based on generative businesses. With NFPs constantly cycling their surplus back into the real economy, the difference between for-profit and not-for-profit business could be the difference between having a linear, extractive economy and having a circular, generative economy. An economy based on NFP enterprise might be just what we need to move society in a more sustainable direction. It would take us beyond the current, stale debate about whether the market should be more heavily regulated or if it should be allowed to operate more freely, because the functioning of an NFP market economy would be so different from the for-profit market economy. As far as we know, this was a completely new vision of what the market could be. **This could offer a realistic alternative to both capitalism and state socialism.**

Of course, many people had understood the generative nature of more socially-oriented businesses long before we started working on these ideas. People like Nobel Prize recipient, Muhammad Yunus, had been talking about 'social business' for years. However, such visions focused on adding social purpose to business, regardless of the profit-orientation of a company. They hadn't seen the importance of the for-profit/NFP distinction in terms of expressing whether businesses see profit as a means to an end or as an end in itself. To our knowledge, no one had ever gone so far as to imagine that NFP business could stand alone, as the core of the economy.

Although we had both been working in the areas of environmentalism, social justice, sustainability and nonprofits for almost a decade, learning about Myuma gave us a new angle for investigation. Inspired by their story and excited about the potential of Myuma's business model to transform the economy, we spent the next five years learning as much as possible about the landscape of NFP enterprise. What we found further fueled our enthusiasm. A significant amount of data had already been collected by the Center for Civil Society Studies at Johns Hopkins University.[4] We were able to see from their work that the foundation for an NFP global economy is already being laid. Not-for-profit businesses are emerging all over the world and in every sector.

## Introducing Not-for-Profit Business

In order to understand NFP business, we must first understand the most common form of business: for-profit business. It's quite simple. Most businesses have owners, shareholders and investors who expect to make money from the business. They expect what's called a 'return on investment'. This means that they invest in the business expecting to get more money out than they put in. In most cases, they want to gain as much wealth from the company as possible, to 'maximize' their gain. To this end, the majority of businesses are set up in order to maximize profits for their owners, shareholders and investors. This is business, as most of us know it. It seems natural that a business should have shareholders and investors, or at least owners, who receive a portion of the profit. In legal terms, though, there is a specific name for this type of business. It is aptly called *for-profit* business.

[4] Salamon, L., et al., (2013), *The State of Global Civil Society and Volunteering: Latest Findings from the Implementation of the UN Nonprofit Handbook,* Baltimore, MD.

The legal interpretation of a business being for-profit is almost universal. While the way for-profit businesses handle their financial surplus varies, and some of them may never make any profit, all for-profit companies have the *ability* to privately distribute profits. A company that can distribute profits to individuals (such as owners, shareholders, investors, partners, workers, managers and board directors) is legally for-profit. In such businesses, profit is either the main goal or one of the main goals of doing business. For-profit companies see profit not as a means to an end, but rather as an end itself.

But what if there were a business model oriented towards a deeper purpose and used profit as a means to achieving that purpose? What if there were a way of doing business that didn't have any private owners and had to use all of its profit – the surplus left over after paying for all business costs, including wages - for the good of the community rather than distributing it to private owners? Well, that's exactly what not-for-profit business is.

Most of us have learned that businesses and nonprofits are two very different creatures. And traditionally, nonprofits and businesses have inhabited quite disparate realms. Businesses generate their own revenue and maximize profits, while nonprofits rely largely on charity and grants in order to do mission-driven work. In fact, the nonprofit world has long been dependent on the business world for financial support, in both direct and indirect ways. Historically, businesses have been the moneymakers, while nonprofits have been the caretakers.

However, nonprofits are increasingly going into business to generate their own revenue and the status quo is rapidly changing. These trends are converging to create an entirely new sphere in the economy: the realm of NFP business. Not-for-profit businesses could be thought of as money-making caretakers.

Indeed, NFP companies such as Bupa, an international healthcare group, and Mozilla, the world-renowned software developer, have made profits in recent years, cycling 100% of those profits back into their missions.[5,6] As one executive we spoke with put it, "We're not for-profit, but we're not for loss either.[7] This is because profit helps them better meet their mission and deliver their services.

The NFP ethos is very different from for-profit companies, which see the provision of goods and services as a way to earn profits, rather than the other way around. The for-profit mindset has put the cart before the horse. Not-for-profit business puts profit back into its proper place in the economy.

Thus, we can distinguish between profit-ability (the ability to generate a financial surplus, thereby enhancing a business's sustainability) and profit-maximization (the primary focus on maximizing profits). Companies like Myuma, Bupa and Mozilla illustrate that for-profit is not the only way to run a business.

---

[5] Bupa (2015), "Longer, Healthier, Happier Lives: Annual Report 2014," http://www.bupa.com/annualreport/pdf/bupa_full_report.pdf.

[6] Mozilla (2014), "Mozilla Foundation and Subsidiaries - 2013 Audited Financial Statements," https://static.mozilla.com/moco/en-US/pdf/Mozilla_Audited_Financials_2013.pdf.

[7] Personal correspondance with author, SE Housing Coop.

Three points universally distinguish a 'not-for-profit' entity from a 'for-profit' entity. All NFPs have:

1. A social and/or environmental mission and hold it as their top priority;

2. No ability to privately distribute profit or assets; and

3. No individual owners or shareholders.

What qualifies as a social or environmental mission? In many countries, the NFP definition allows for a very broad range of purposes. It includes missions that benefit the public at large, like protecting a forest or taking care of a community park. It also includes missions that benefit a marginalized group of people, like impoverished families or people struggling with homelessness. It can also include missions that benefit a select group of people who aren't vulnerable. NFP community centers and credit unions, for instance, exist for the benefit of their members, regardless of whether those members are struggling or not. An NFP's mission can even be as simple as 'providing high quality products'. This gives us a very wide range of groups that can be NFP, which is why companies like Myuma, Bupa and Mozilla can be NFP.

However, this can also leave the door open for some organizations to register as NFPs that are actually in business to maximize wealth for private individuals. Some NFPs, such as mutual funds in the U.S., have the mission of providing high quality services to their customers, but their service is to maximize capital gains, which is the antithesis of the NFP legal distinction. As such, we exclude from our interpretation of 'not-for-profit' any organization whose primary service is to maximize capital gains for individuals.

Therefore, we could add as a fourth provision that NFPs:

4. *Cannot* have as their mission the maximization of capital gains for individuals or for-profit companies.

Another feature that distinguishes many NFPs is one that further prevents private gain from an NFP: a windup clause (or asset lock) that prohibits any individual gain from company assets in the event that the entity closes down or is dissolved.

The asset lock is very important, because an NFP distributing its assets is very similar to an NFP distributing its profits. Legislation that allows NFPs to distribute their assets to individuals blurs the for-profit/NFP distinction. For instance, an NFP golf club should not be able to distribute its assets to individual members when the company dissolves. Instead, all of those assets should be passed on to other NFPs (this is legally required of all 501c(3)s in the U.S. and all Community Interest Companies in the U.K.). Otherwise, individuals can make quite a significant capital gain from the golf club's assets. Distribution of an NFP's assets can also create a litigation nightmare, in terms of who gets what, especially if people have donated to the NFP.

On their own, each of the elements above could not transform the economy. But together, the rules that differentiate NFP businesses from for-profit businesses have the power to change the economic game as we know it.

This is why 'NFP' is an invaluable standard by which to gauge the intent and potential impact of a company. It's a vital factor in discerning whether a business is generating value for the wider community[8] or just for a few individuals.

For some people, 'not-for-profit business' sounds like an oxymoron. Few people appear to know that not-for-profits can do business, or that businesses can be not-for-profit. In fact, 'not-for-profit' is often interpreted as meaning 'no profit' – with the assumption- that 'not-for-profit' organizations cannot or do not make any financial surplus. Yet, around the world, NFPs can make as much profit[9] as they want. In fact, without any financial surplus, they might not be able to invest in stronger operations or withstand dips in income.

Not-for-profit means just what it says: making a profit is not the primary purpose of these organizations. Such companies do not exist *for* profit; they exist for a deeper purpose. Rather than focusing on profit as a goal itself, profit is a tool to help them achieve their mission.

This misunderstanding that NFPs cannot make a profit leads people to think of NFPs as distinct from social enterprises, sustainable businesses or cooperatives. Many entrepreneurs feel they must choose between starting a social enterprise *or* an NFP. Or they feel that running a sustainable business means that it must be for-profit. In fact, the term 'for-profit' is hardly ever used in discussions about business, because it's taken for granted that all businesses are for-profit, by default. This perception is finally shifting.

Social enterprises, sustainable businesses, cooperatives, and even multinational corporations can all be NFP. The key factor that differentiates NFPs from for-profit business is that they must put all of their profits back into their social mission and cannot distribute any portion of profits to individuals (such as workers, management and board directors). This is an age-old distinction, but never before has it been so important because never before have so many NFPs been doing so much business.

It's important to remember that profit refers to the surplus that remains *after* wages are paid, in both for-profit and NFP companies. So, the fact that NFP organizations cannot distribute profit doesn't have any detrimental effect on wages or salaries, as those are always considered operating costs – part of running a healthy business is paying your staff.

Some readers might wonder if there is a difference between an NFP using financial surplus to increase salaries the next year and a for-profit company distributing its financial surplus to owners at the end of the year. The difference is profound. Not-for-profit employees are paid to do valuable work without which the NFP could not survive. In other words, they earn their salaries by creating value. Furthermore, when the manager of an NFP submits a budget to the NFP's board for review, the board considers this budget in light of both the company's financial position and the efficient allocation of resources needed to achieve its mission. Salaries are generally only raised if doing so will help the NFP better fulfill its purpose. This is in stark contrast with for-profit companies, which are expected to maximize financial surplus in order to deliver dividends to owners, who don't have to work to earn that money. In the for-profit context, purposes other than profit are generally taken into account only if doing so will help the company deliver higher profits. The focus on profit-maximization encourages short-term thinking, speculative behavior, and cutting costs (often reducing wages and lowering the quality of products and services). Good NFP salaries, on the other hand, promote longer-term thinking and further value creation for the company and the community it serves, because profit and wages are in service of a deeper purpose.

---

8 By wider community, we refer to different levels of community, including neighborhoods, religious or other social communities, towns, cities, nations, bioregions or the entire planet. Different NFPs aim at meeting the needs of different communities, but the important aspect is that they are not set up to enrich a few individuals; there is always some sort of community at the heart of NFPs

9 In many places, nonprofit organizations refer to excess funds at the end of any given financial year as 'surplus' rather than profit. Within NFP enterprise, we use the terms surplus and profit interchangeably irrespective of the level to which income is self-generated.

It's not just the way that NFP enterprises are transforming the business world that changes the game; it's also how they're transforming the world of traditional nonprofits. As more and more NFPs move into business, it also becomes important to distinguish between traditional, charity-dependent nonprofits and NFP businesses. Often the terms 'nonprofit' and 'not-for-profit' are used interchangeably, but we make a distinction in this book. We believe that the mindset and economic role of NFP enterprise is so different from that of traditional nonprofits that it is worth recognizing it as a totally new category, as a new segment of the economy.

We have chosen to use the terms 'NFP business', 'NFP enterprise', or simply 'an NFP' to differentiate this new breed of purpose-based businesses from traditional nonprofits, which depend on charity, volunteers and grants. We decided to use these terms because 'nonprofit' tends to evoke more of the 'no-profit' misunderstanding, while 'Not-for-Profit' is more straightforward in describing the mission of the company, or rather, what its mission is not. It's not for-profit.[10]

Why define something by what it is not? If all NFP businesses are driven by purpose, why not speak of 'for-benefit' or 'for-purpose' businesses? We intentionally avoid such language, because the words 'benefit' and 'purpose' are too vague to show what happens with a company's surplus. They leave the questions of "who benefits?" and "for what purpose?" unanswered. This is clearly illustrated by the fact that the benefit corporation in the U.S. is actually a for-profit business model and the B Corp certification (B as in benefit) is only available to for-profit companies.[11]

Although in most countries there is no legal distinction between traditional nonprofits and NFP enterprises, for the purposes for this book, when we refer to an NFP business or enterprise we include organizations that generate at least 50%[12] of their funding from the sale of goods and services. This means that NFP enterprises can receive ongoing philanthropy and grants, but cannot depend on these for the majority of their revenue.

Not-for-profit enterprises are able to use strategic business approaches to generate revenue that go far beyond raising money via grants, donations and other forms of philanthropy. And that's what makes NFPs different from traditional nonprofits: they seek to be financially independent and they generate a significant portion of their revenue through the sale of goods and services. They also generally have different approaches to management, finance, productivity, innovation, service delivery, and worker participation.

[10] This is further reinforced by the Oxford English Dictionary's definitions of these terms. Not-for-profit (n.): An organization, corporation, etc., which does not operate for the purpose of making a profit. Nonprofit (n.): A nonprofit-making organization; spec. a charity. Additionally, 'not-for- profit enterprise' is already a very familiar term in Australia, where the inspiration for this book came from.

[11] See: http://benefitcorp.net/faq

[12] The UK Cabinet Office considers a charity as being a 'very good fit' for the social enterprise model when it earns over 25% of its revenue from the sale of goods and services (REF: U.K. Cabinet Office). We've decided to aim a bit higher.

We will delve further into these distinctions and their nuances in greater detail in the following chapters. For now, the take-away message is that NFP business is mission-oriented and uses all resources, including profit, to achieve social or environmental goals, whereas for-profit business is profit-oriented and can distribute profit to private individuals.

We are explaining these distinctions so thoroughly because the NFP framework is so widely misunderstood, and yet it is a necessary ingredient in any recipe for a healthier economy. With NFP businesses largely geared towards the greater good, profit tends to be constructive and generative. This is because NFPs generate surplus with the intention of being better able to achieve their social or environmental goals. With for-profit businesses largely geared towards self-interest, profit tends to be destructive and degenerative. This is because surplus is principally created with the intention of accumulating private wealth and power. That's what the profit motive is all about.[13]

Not-for-profit business is not perfect. Such companies face many of the same challenges that for-profit companies experience, such as achieving sustainable streams of revenue. They also face many of the same challenges that traditional nonprofits encounter, such as drifting from their mission.

However, an economy built on the NFP framework and ethic is healthier for people and the planet than an economy founded on the for-profit framework and ethic. The for-profit ethic entails an extractive mentality, seeking to pull value out of business into the hands of a few owners and investors; whereas the NFP ethic entails a generative mentality, seeking to use profit to generate value for the wider community.

[13] The profit motive is a central concept in capitalism, which refers to economic actors being primarily motivated by financial gain.

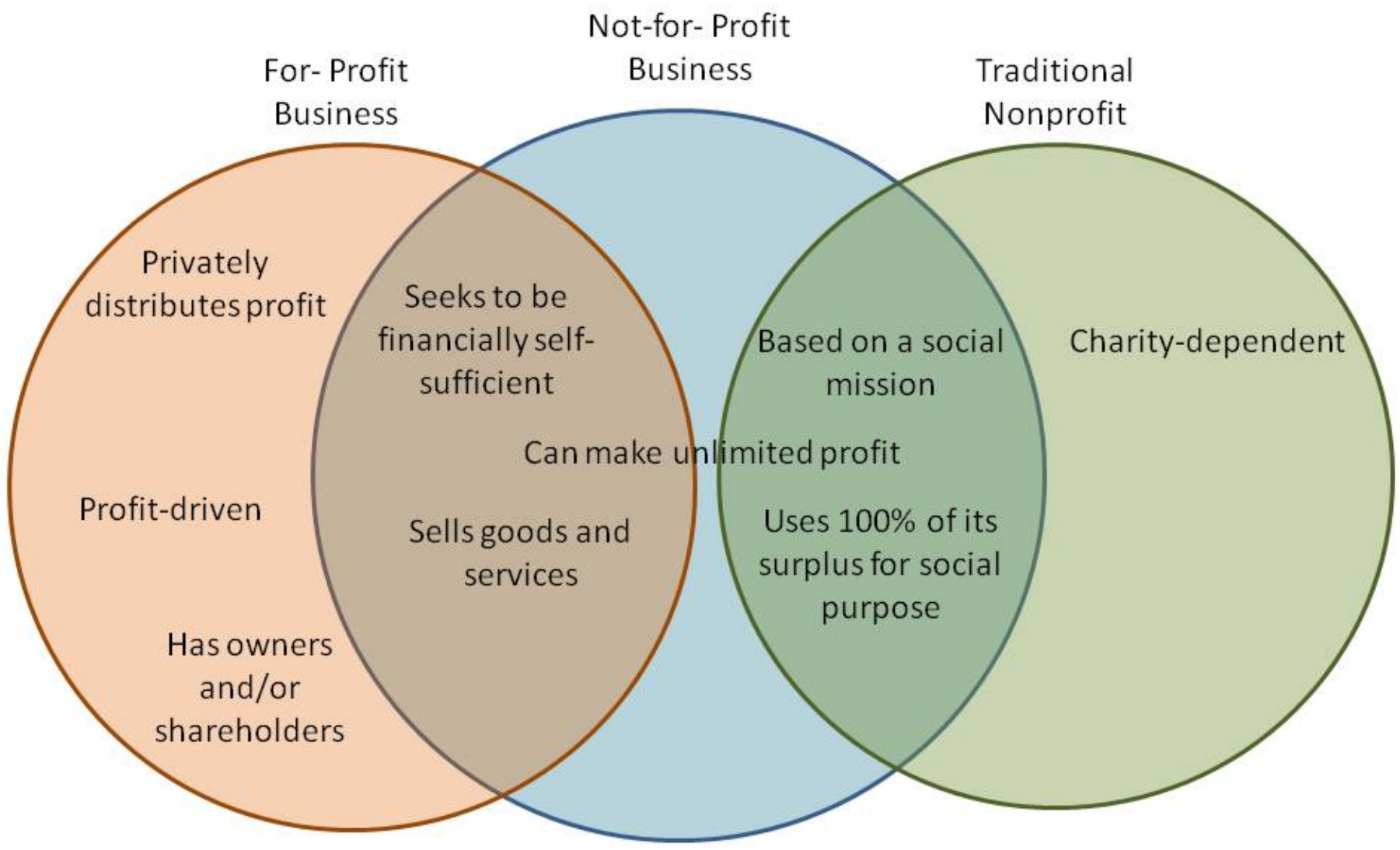

### NFP Business: The ideal hybrid

**Not-for-profit enterprise is the ideal hybrid between the innovative, efficient aspects of for-profit businesses and the mission-driven nature of traditional nonprofits.** It is a business model built for purpose - the purpose of 'meeting needs'. These enterprises don't face the pressure to sacrifice social and environmental wellbeing in the name of profit, which is something that most for-profit businesses confront. Without obligations to shareholders and with minimal dependency on donors, NFP enterprises have the freedom and ability to get their work done in a way that maximizes social, environmental and economic outcomes.

While the traditional nonprofit model may be diminishing in prominence, it's no surprise that the NFP enterprise model is on the rise, [14] with activity across sectors as diverse as telecommunications, retail, manufacturing, software development, construction, healthcare and the food industry.

## The Business Spectrum

Although we're drawing a line in the sand between for-profit and NFP business, it's important to acknowledge that not all for-profit businesses are created with the same single-minded drive towards profit.

Many small, locally-owned businesses do not maintain a singular focus on profit and are often happy to forgo revenue for a cause that helps their community, such as donating goods and services to homeless shelters. In addition to companies that voluntarily act in less profit-oriented ways, we also see the emergence of new legal business models that seek to push for-profit business in a more purpose-driven direction.

The benefit corporation, for example, is a relatively new for-profit business model introduced in some U.S. states, whose legal structure was designed to enable the profit-orientation of the traditional corporation to exist alongside a social mission. Benefit corporations subscribe to the 'triple bottom line' philosophy; seeking to balance people, planet and profit. Note that profit is still a goal in itself, rather than just a means to help meet the needs of people and planet. Benefit corporations are therefore predisposed to the for-profit ethic.

[14] Salamon, L., et al., 2013.

Many people have also heard of social enterprises, which are often either assumed to be all for-profit, or all NFP. In most regions, social enterprises can be either for-profit or NFP. For-profit social enterprises fall into the triple bottom line category – seeking to balance profit as a goal alongside social or environmental goals.

Worker-owned cooperatives are another example of less profit-oriented businesses that are legally for-profit. Owned and managed by the workers themselves, profits of a worker co-op can be distributed to individuals, making worker co-ops 'for- profit'. And, in addition to their wages, workers usually look forward to their dividend from profits. The worker-owned co-op model is still based on the profit motive and the for-profit ethic, but to a much lesser extent than typical for-profit corporations. And, of course, worker co-ops distribute their profit more equitably than companies that only give profit to a narrow group of shareholders and investors. In fact, sometimes worker co-ops even choose to limit the maximum amount that can be paid out in dividends.

Benefit corporations, for-profit social enterprises and worker co-ops all represent important steps away from the strictly profit-oriented mentality of corporations traded on the stock market, which prioritize delivering higher quarterly profits for shareholders. They are a move in the NFP direction and play an important part in a larger trend that we're observing.

In order to better understand these differences, we can conceive of businesses being situated along two spectrums; one relating to their ownership structure and the other relating to the extent to which they are profit-driven. The graph below illustrates these spectrums with a few examples of the many business models that exist worldwide.

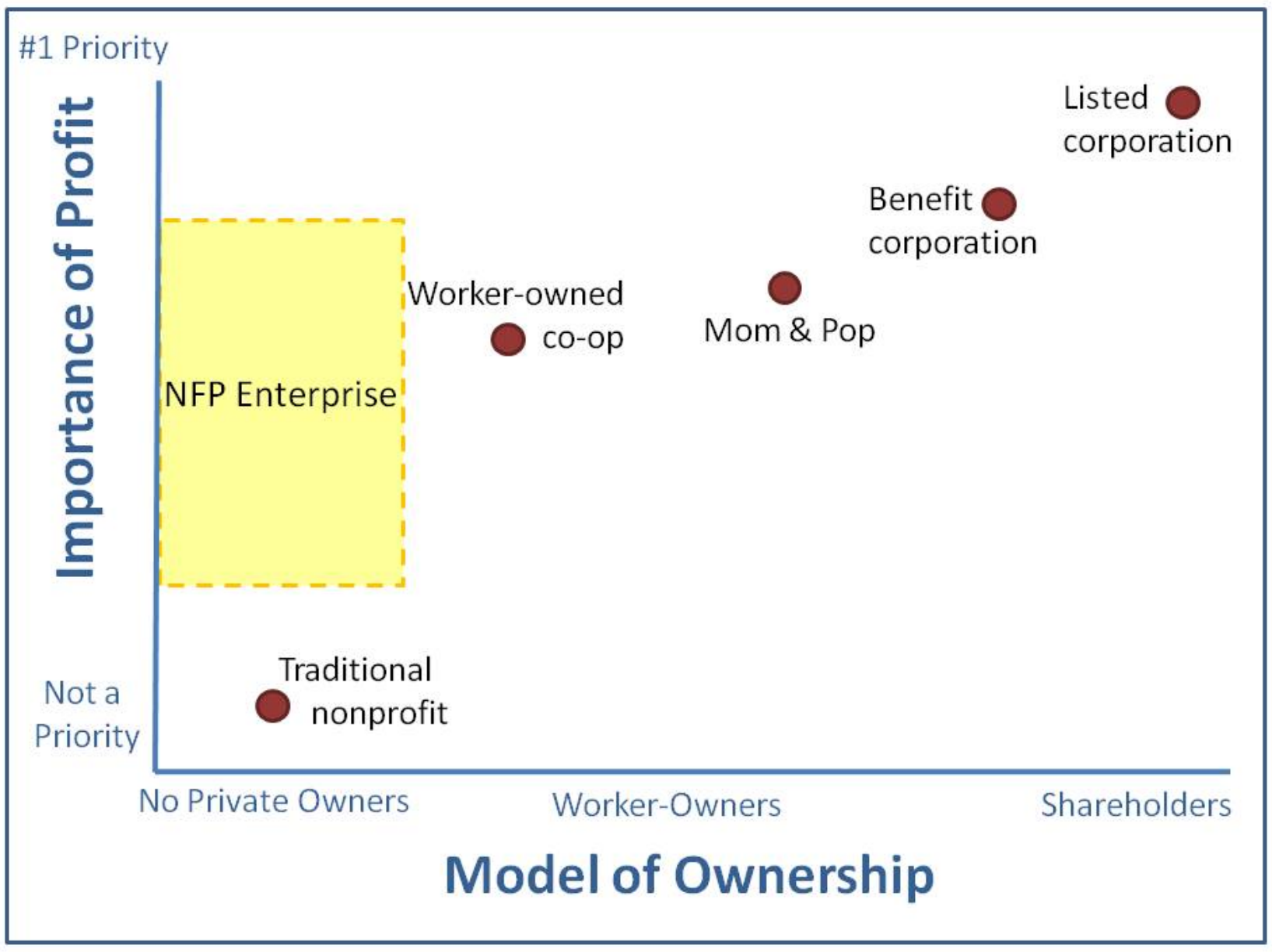

### The NFP- For-Profit Business Spectrum

On the far-right end of the ownership spectrum is the publicly listed corporation (i.e.- businesses that are traded on the stock market), owned by shareholders. You can see from the vertical position on the graph that profit is the number one priority for such companies; they seek to maximize profit. Benefit corporations have private owners and shareholders who don't work for the business, but they have a social purpose in addition to their desire to maximize profit. Many small family-owned companies, also known as 'mom and pop shops', would be near the center, because the owners of these companies usually do a certain amount of work for the business and profit is rarely the singular priority. Towards the middle of the horizontal axis, the members of worker-owned co-ops are the owners and profit is still a priority, but can co-exist with other priorities, including worker wellbeing and other ethical practices. On the left-side of the spectrum, in the ideal 'zone of NFP enterprise', the there is no private ownership, and these businesses can be profitable but profit is never a top priority. Traditional nonprofits are outside of the zone of NFP enterprise, as they don't see profit as much of a priority at all. The vision we're putting forth in this book is one in which most of the economy exists within the zone of NFP enterprise, a Not-for-Profit World.

The NFP World economy still has a great amount of diversity. Taking a closer look at what exists in the zone of NFP enterprise, we discover many different forms including: incorporated associations; companies limited by guarantee; non-distributing cooperatives (including some producer co-ops and all consumer co-ops); social enterprises (in many European countries this model is NFP by default); and Muhammad Yunus' Type 1 Social Businesses (the Type 2 Social Business model is for-profit).

Various government-owned enterprises can also be considered NFP enterprises, such as motor vehicle departments (which charge fees for the service of providing license plates and driver's licenses), as well as most public transportation and municipal utility companies.

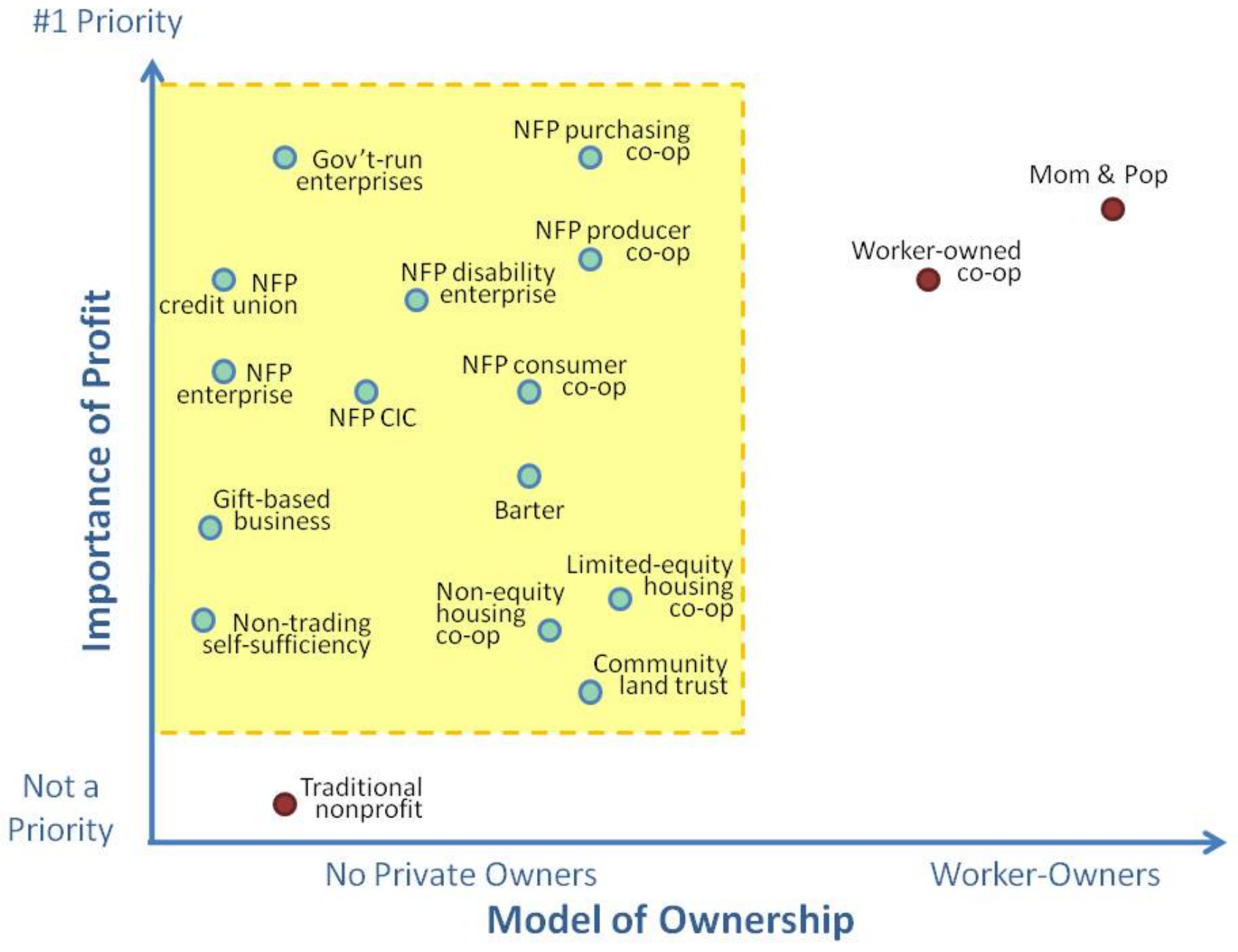

### Diversity of the NFP Economy

The common theme, and what makes an NFP economy so much healthier than the current for-profit economy, is that there are no private owners, which means that the surplus benefits a wider community, not just a few individuals. In coming chapters we will see why keeping the financial surplus of business in the real economy fundamentally changes the entire system. We will also take a closer look at the diversity that exists within the not-for-profit economy. But first, let's look at how we ended up with a for-profit economy.

# Profit in the Bigger Picture

We presently live in a for-profit world. The vast majority of the businesses in our current economic system are for-profit companies. This means an incredible amount of the economy's surplus goes into the pockets of individual investors and business owners.

This might seem well and good, an accepted norm of how things are, but the fact that all of these business models allow for the private distribution of profit and assets does something very damaging to the dynamics of the economy.

## The Pump and the Siphon

**In a healthy economy, wealth keeps circulating throughout the system, so that we all have enough to meet our needs.** We call this the 'Wealth Circulation Pump'. Indeed, this is what happens in our economy. Much like a water circulation pump keeps hot water flowing through the pipes of a house, we take money in the form of wages and send money back out into the flow of the economy by spending it. If the wealth circulation pump is functioning properly, then money keeps flowing naturally around the economy to where it's needed.

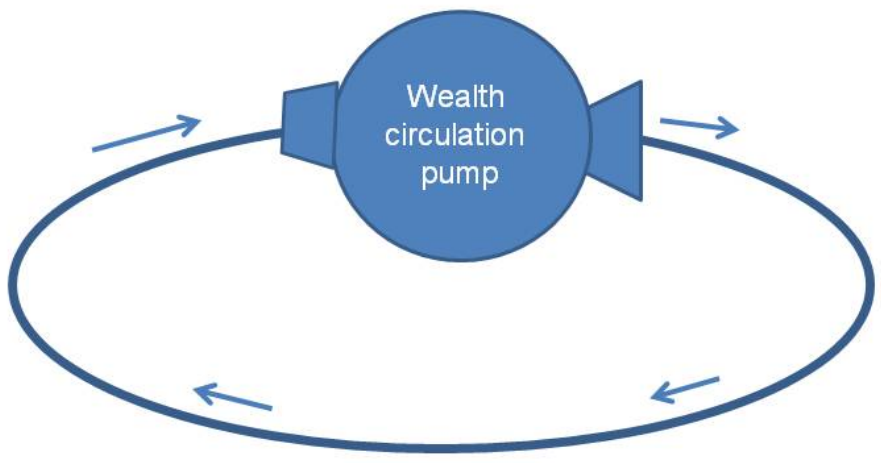

The Wealth Circulation Pump

Theories of capitalism[i] say that the wealth circulation pump is primed mainly by capitalists investing in businesses. The idea is that the innate urge of investors and business owners to make more money compels them to reinvest their wealth in more business activities and this reinvestment benefits the economy, at large. It creates more jobs, which allows more people to participate in more economic activities. This has been called the Invisible Hand of the market. It is supposed to keep money constantly pumping through the economy

It is clear that this is not what's happening in today's economy. So far, the 21st century has been an age of extreme wealth inequality.[15, 16] We can all sense that something is interfering with the Wealth Circulation Pump. What is it? We call it the 'Wealth Extraction Siphon'.

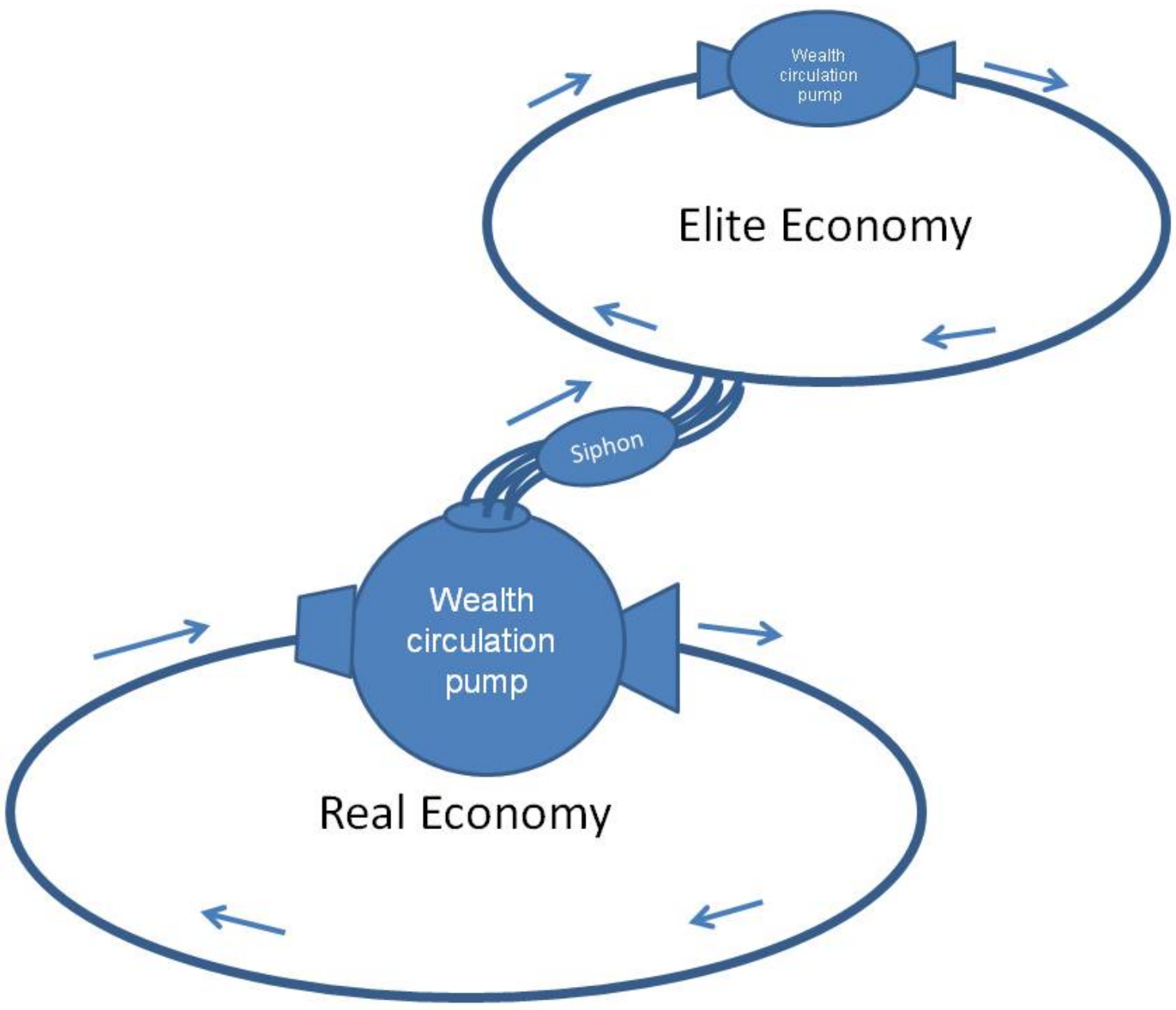

**Wealth Extraction Siphon**

This is how it works. For-profit businesses distribute surplus to owners, investors and shareholders. They then use their wealth primarily to invest in more business ownership (equity) to receive more profits and capital gains. They receive more wealth from the returns on their investments and use that new wealth to invest in more equity, and so it continues in a feedback loop. The people who own the most equity in businesses are the wealthiest people in society and their wealth allows them to buy even more equity in businesses (they don't need to go into debt to do this), so they are able to multiply their wealth like no one else. This does contribute to the wealth circulation pump to an extent, as the capitalists' businesses pay wages, allowing their employees to continue to buy the things they need. But their aim is to take out more than they put in and they do that very well. Thus, their activities extract more wealth than they circulate, which is why capitalism has an overall trend towards greater inequality.[17] The little amount of wealth that trickles down from the rich simply cannot compensate for the speed at which wealth is being extracted from the real economy.

Recent studies show that the super-rich receive a large portion of their income from the real economy of goods and services, but rather than putting that wealth back into the real economy, they put most of it into the elite economy of luxury goods and services and financial assets.[18, 19, 20, 21] We call this the elite economy because it is a speculative market that only those who have a certain amount of extra money can afford to participate in.

---

[15] Hardoon, D. (2015), "Wealth: Having It All and Wanting More," Oxfam, http://hdl.handle.net/10546/338125.

[16] Wilkinson, R. G. and Pickett, K., (2009), *The Spirit Level: Why Greater Equality Makes Societies Stronger,* Bloomsbury Press, New York.

[17] Piketty, T. and Goldhammer, A., (2014), *Capital in the Twenty-First Century,* The Belknap Press of Harvard University Press, Cambridge Massachusetts.

As the incomes of the rich continue to rise while the average household income continues to drop, the elite economy becomes even less accessible to the average person. A recent Oxfam report found that the richest 1% of the world's population now own more than the other 99% combined.[22]

Although this level of inequality rightly angers many people, it's very important to understand that wealthy people are simply doing what is considered rational in an economy that revolves around profit and the private accumulation of wealth. This is not a phenomenon of a few bad apples acting greedily and intentionally creating their own elite economy, despite economic rules that encourage contributing to the wider community. Instead, it is a natural outcome of a system that holds greed as the primary source of motivation. It's a fundamental feature of the for-profit economy. **For-profit business acts as a siphon, sucking the surplus out of the real economy into the elite economy.**

Some people might ask, "But the wealthy are investing in the stock market. Isn't that part of the pump?"

Most of us believe that the stock market serves primarily as a mechanism of investment - that when people buy shares in a company, they are investing in that company and providing it with the capital it needs to grow. However, this is only the case when a company offers shares on what's called the *primary* market for the first time, through an Initial Public Offering, or via a further stock issuance. The vast majority of stock market transactions happen in the *secondary* market (like the New York Stock Exchange), and do not give companies any capital or investment funds.[23] Rather this is the domain of pure speculation. When shares are traded after their initial sale, money just bounces between traders, never touching the company's bank account (or the real economy).

The for-profit system encourages speculation and short-term thinking; two ingredients that have proven to be extremely destructive on a planet that we must all share and look after for future generations. When you buy stock in a company on Wall Street, you are more or less placing a bet, speculating that your chosen company will generate more profit than last year or, at the very least, it will not lose money. Just as when you bet on a horse or a dog at the racetrack, if the company gains money, so do you, as a speculator. If the company loses money, so do you. It is speculative. It is gambling. Big traders often make gains from the

18 We will describe how this happens in more detail in chapter 3.
19 Capgemini and RBC Wealth Management (2015), "World Wealth Report 2015," https://www.worldwealthreport.com/sites/all/themes/wwr/images/WWR2015-GeneralInfographic.jpg.
20 The RSA (2014), "David Harvey on the Contradictions of Capitalism," https://www.youtube.com/watch?v=f9dLcGJ5NI0.
21 Frank, R. (2015),"Where the Very Rich Make Their Income," CNBC, http://www.cnbc.com/2015/04/09/where-the-rich-make-their-income.html.

22 Hardoon, D., et al. (2016), "An Economy for the 1%: How Privilege and Power in the Economy Drive Extreme Inequality and How This Can Be Stopped," Oxfam, http://hdl.handle.net/10546/592643.

23 McMenamin, J., (1999), *Financial Management: An Introduction,* Routledge, New York, NY, 75.

losses of smaller traders, who have less experience and knowledge about the stock market.[24] Because most of the big traders on the stock market are the global elite, the money just sloshes around the elite economy after it's extracted from the real economy.[25] Ironically, the wealth circulation pump seems to be working just fine in the elite economy.

Even in the case of companies that aren't traded on the stock exchange, investors and owners are looking to extract as much wealth from the company as possible, with few exceptions.

When we need our economy to support both people and planet for the long-term, is it really appropriate for our primary business model to be centered on speculative investment?

## Our For-Profit World

In a world with for-profit business at its heart, profits are the primary objective. Thus, in our current global economy, profits are generally created in the most cost-efficient ways, regardless of social and ecological consequences. A very clear example of this is how profit-maximization has been driving companies to outsource labor, often moving factories to countries where wages are lower and environmental, health and safety regulations are lax or nonexistent. This is largely done in order to increase shareholder value, deliver higher profits to owners, and stay competitive against other companies that are also trying to maximize profits.

Planned obsolescence is also symptomatic of this tendency to maximize profit at all costs. Products are often designed to break down or become obsolete as soon as possible so that consumers will buy more items more frequently. It feels like every year the things we buy are less and less durable and need to be replaced more frequently. Most of us have experienced this when a household item like a printer or vacuum cleaner stops working and we take it to a shop to be repaired, only to be told that it would be cheaper to simply buy a newer version of the product. In part, this is because companies are using cheaper materials and cheaper production methods to maximize profits.[26] But even if products don't physically break down, companies find ways to keep their profit margins high, such as selling software that becomes incompatible with newer applications after a year or two.

On the quantity versus quality of products, legendary industrialist Henry Ford said, "There is one rule for the industrialist and that is: Make the best quality of goods possible at the lowest cost possible, paying the highest wages possible."[27] We have strayed far from that wisdom. The rule for most businesses today could be articulated as, "Make the most profits possible at the lowest cost possible, including paying the lowest wages possible."

---

24 Long, H., "Who's Getting Rich Off the Stock Market?," *The Buzz*, CNN Money, September 24, 2014, http://money.cnn.com/2014/09/18/investing/stock-market-investors-get-rich/.

25 Austin, L. & Williams, R. (2015) 'Composition of Income Reported on Tax Returns in 2012', Tax Notes, Tax Policy Center: https://www.taxpolicycenter.org/publications/composition-income-reported-tax-returns-2012/full

26 Midler, P., (2011), *Poorly Made in China: An Insider's Account of the Tactics Behind China's Production Game,* Wiley, Hoboken, N.J.

27 Andersen, E., "21 Quotes from Henry Ford on Business, Leadership and Life," Forbes, May 31, 2013, http://www.forbes.com/sites/erikaandersen/2013/05/31/21-quotes-from-henry-ford-on-business-leadership-and-life/.

A growing segment of the population criticizes such destructive business practices, but fails to see that profit-maximization is what most of these companies are set up to do. It's their purpose; their raison d'être. It's in the name: *for-profit* business.

What's more, the social stratification that results from economic inequality in a for-profit system tears communities apart and generates status envy, exacerbating the conditions for a rampant consumer culture that further alienates people and destroys ecosystems.[28] It is a vicious, for-profit cycle.

Again, this is not because these business owners, investors and shareholders are awful people. They are all just playing by the rules of the for-profit game. We cannot blame them for winning a game that incentivizes and favors unsavory kinds of behavior. Instead, we must promote different priorities, rules and social norms.

## A Brief History of Profit

To believe that a different set of rules and standards is even possible, it helps to know a little more about how dramatically the approaches to profit and the wider economy have changed over time.

In the 1970s, philosopher of science Thomas Kuhn described how paradigm shifts occur in science. A paradigm is simply the most common way of understanding the world around us, a frame of thinking that includes both assumptions and the way information is interpreted. Kuhn claimed that scientific paradigms are constantly developing and evolving but that from time to time, a paradigm shift happens when the old way of doing science no longer seems valid or effective and new ways of seeing the world emerge to explain the phenomena that the old paradigm couldn't. Kuhn noted as an example the famous paradigm shift that happened in the field of astronomy when Galileo was able to use a telescope to show how the Earth revolves around the sun. Before that, most astronomers based all of their calculations and theories on the assumption that the Earth was the center of the universe and the sun, as well as all other planets, revolved around Earth. However, there were things that the geo-centric view couldn't explain. As a result, the existing paradigm faced a crisis and a new paradigm, the helio-centric view, eventually replaced it.

Just as there have been paradigm shifts in the way we understand the universe, so have there been shifts in the way we see the economy. The concept of profit is central to these economic paradigm shifts. **As our social, political and cultural needs change, so does our idea of what should be done with economic surplus.** This means that the way we organize society and politics, as well as the cultural context in which businesses are embedded, influences our understanding of profit and vice versa. These aspects are all constantly co-evolving. The same is true for concepts like 'business', 'ownership', and 'markets'. We inevitably interpret these ideas through our cultural lens or filter. Societies throughout the world and throughout time have treated profit in different ways. From a historical perspective, the profit motive is relatively new and what feels normal to us is just a snap-shot in time.

---

[28] Wilkinson, R. G. and Pickett, K., (2009), *The Spirit Level: Why Greater Equality Makes Societies Stronger,* Bloomsbury Press, New York.

For the first 100,000 years of human existence, the notion that people should strive to produce and store surplus for private gain was largely irrelevant. Not only was food storage unnecessary and impractical for foragers who moved seasonally through the landscape, but the social organization and cultural values of these kin-based societies also had a different focus entirely. Foragers lived communally, underpinned by the values and practices of kinship, reciprocity and subsistence.[29] Deep connections with the other members of the group and an understanding of ecological interconnections with other species in the region were essential to survival. These foraging bands focused on satisfying human needs,[30] which they saw as limited.[31, 32] The forager worldview clearly entails an ethic of enough.

While forager economies did not involve surplus production, the peasant communities which were common in ancient civilizations did. These societies typically responded to surplus by institutionalizing wealth-leveling practices that periodically redistributed any marked difference in individual material wealth. Their customs ensured that inevitable differences in productivity and fortune did not disrupt the social stability of the community.

Indeed, redistributive practices such as this can still be found across many indigenous cultures.[33] They are powerful reminders that other ways of treating surplus beyond private accumulation have long existed and that **the present model of wealth accumulation that dominates our lives is a relatively new development in human history.** In socio-economic systems where redistributive practices are institutionalized, people are motivated by cultural goals which can only be achieved by putting the surplus back into social institutions.[34] Surplus becomes simply the vehicle for achieving social respect, and does not lead to differences in access to resources.

Such customs are supported by cultural assumptions and worldviews that typically stigmatize greed, over-consumption and hoarding. For example, in the Native American Cree language ‘wetiko’[35] refers to a greedy person, a cannibal spirit, or a man-eating monster. Found in many Native American cultures, this concept is used to describe someone who has collected more than they need and is thus considered to have a social disease.[36] For these cultures, greed equates with social disruption because it breaks the bonds needed to sustain society.

---

29 Sahlins, M., (1974), *Stone Age Economics,* Aldine Transaction.
30 Bodley, J. H., (2012), *Anthropology and Contemporary Human Problems,* AltaMira Press, Lanham, MD.
31 Foster, G. M., "Peasant Society and the Image of Limited Good," *American Anthropologist* 67(2) (1965): 293-315.

32 Service, E. R., (1971), *Primitive Social Organization: An Evolutionary Perspective,* Random House, New York,.
33 See, for instance, the potlatch of indigenous communities of the Pacific Northwest, in which individuals of high social standing hold feasts or ceremonies for the entire community, (more information at: Encyclopedia Britannica ‘Potlatch’).

34 Wolf, E. R., (1982), *Europe and the People without History,* University of California Press, Berkeley.

35 Sometimes spelled ‘whitiko’, ‘windigo’ or ‘wendigo’.
36 Levy, P., (2013), *Dispelling Wetiko: Breaking the Curse of Evil,* North Atlantic Books, Berkeley, Calif.

This period of time, when most societies were forager bands or peasant communities and there was little orientation to surplus production, can be seen as the **Pre-Profit Era** of human history. Today's focus on profit was non-existent. Instead, there was a common understanding of the importance of sharing any surplus with the group for the sake of social stability and overall wellbeing, and this understanding led to cultural norms and rituals that enforced standard practices of wealth circulation.

More formal food production systems began to appear at around 11,000 BCE. Intrinsic to this "agricultural revolution" were cultivation practices designed to create the regular surplus needed to support more stratified political institutions. Thus, the Pre-Profit Era gave way to the **For- Profit Era**. In this paradigm, profit became a goal in itself and it was explicitly used to support an elite class.[37]

These early societies had cultural and social institutions that allowed divinely-ordained elites to claim ownership of goods produced by lower-ranking people.[38] Only the elites got to keep surplus and commoners faced accusations of theft if they did so.[39] For most of the For-Profit Era, private accumulation of surplus by commoners remained morally problematic, largely due to cultural narratives that considered the privatization of profit socially unacceptable and, in many cases, damnable under religious teaching.[40] Luca Pacioli, an Italian mathematician that lived in the 15th century, echoed the Bible's teaching that all wealth belongs to God,[41] a statement that exemplifies the most common attitude towards wealth accumulation and private profit at that time. And, of course, with the advent of Islam in the 6th century AD, the majority of banks in the Muslim world were forbidden to charge interest under Sharia law. Elites in most societies were also expected to use their wealth to erect lavish public buildings, to fund annual community festivities and to perform ritual functions.[42] European nobility, for example, used their family wealth to erect the palaces, churches and fortresses. Power could only be demonstrated by culturally-appropriate displays of wealth, and noblesse oblige was expected.

Starting with the Protestant Reformation (16th century), during the Enlightenment (17th- 18th century) and on into the Industrial Revolution (19th- 20th century), the economy gradually became more focused on profit. This was part of a larger shift towards mercantilism and imperialism, wherein governments controlled trade in order to expand their empires and maximize national gain. Although much of the imperialism and colonialism of those times was justified by spreading European Christianity, there was a corresponding secularization throughout Europe. Life increasingly revolved around economic and political institutions rather than religious ones. This transition away from religious authority facilitated the development of a new worldview that allowed for more self-determination and encouraged the creation of personal wealth and the development of a globalized market economy. But it

---

37 Heilbroner, R. L., (1999), *The Worldly Philosophers: The Lives, Times, and Ideas of the Great Economic Thinkers,* Simon and Schuster, New York.
38 Wolf, E. R., (1982), *Europe and the People without History,* University of California Press, Berkeley.
39 Ibid
40 Heilbroner, R. L., (1999), *The Worldly Philosophers: The Lives, Times, and Ideas of the Great Economic Thinkers,* Simon and Schuster, New York.
41 Anielski, M., (2007), *The Economics of Happiness: Building Genuine Wealth,* New Society Publishers, 56.
42 Heilbroner, R. L., (1999), *The Worldly Philosophers: The Lives, Times, and Ideas of the Great Economic Thinkers,* Simon and Schuster, New York.

was Adam Smith's notion of a market economy in the late 1700s, that ultimately cemented the social acceptability of private profit and marked the emergence of capitalism.[43, 44]

Smith put forth the then radical notion that society would be better off if people were allowed to interact and negotiate in the market as they wanted, letting supply and demand balance out in a way that would govern the nature and quantity of what was traded. Known as the 'Invisible Hand' of the market economy, Smith's most influential idea[45] was that if each individual acts out of their own self-interest in the marketplace, it will add up to social benefit for all, because everyone will find the best way to meet their own needs. **As capitalism took hold through the Industrial Revolution, personal gain acted as the main motivation for conducting and expanding business activity. The profit motive was born.** For the first time in history, it became widely accepted that a common person, not related to nobility, royalty or the church, could hope to financially gain from business.[46] Little wonder it became so intoxicating.

The co-evolution of the English language with the profit motive clearly illustrates the corresponding shift in cultural values. For instance, the word 'wealth' emerged in the English language in the 14th century and originally meant 'wellbeing, related to health' or bodily well-being. Only later did the term come to be synonymous with material possessions and money.[47] The same goes for 'prosperity', which originally meant 'happiness.'[48]

As economies shifted from mercantilism to capitalism, corporations and limited liability companies became popular vehicles for doing business. Before 1819, corporate charters in the U.S. were only issued if it was in the public's interest and they were tightly regulated by the government.[49] Forming a corporation back then usually required a legislative act. All investors generally had an equal say in corporate governance and corporations had to align their activities with the purposes expressed in their charters, which had to be in the public's interest. A corporation's charter could be revoked by the state if the corporation was not acting for the common good and many charters were dissolved as soon as their projects came to an end.[50], [51] In essence, until 1819, corporations in the U.S. had to be socially-oriented,

43 Ibid
44 Skidelsky, R., "Life after Capitalism," *Al Jazeera*, July 7, 2011, http://www.aljazeera.com/indepth/opinion/2011/07/201176105512730267.html.

45 The Invisible Hand is such a significant part of Smith's legacy not because he put so much emphasis on it, but rather because political philosophers and economists who came after him praised the idea. Many scholars believe that the concept is commonly misunderstood (see Merropol for instance) (Ref: Meeropol). We refer to the Invisible Hand in terms of how it is most commonly understood and, thus, how it impacts economic policy and behavior.
46 Max Weber's concept of the "Protestant Ethic" also explains how the rise of Protestantism in Christianity brought about a shift in values which allowed for and even encouraged personal rewards for hard work (Ref: Max Weber, Protestant Ethic and the Spirit of Capitalism).
47 See http://www.etymonline.com/index.php?term=wealth&allowed_in_frame=0

48 See http://www.etymonline.com/index.php?allowed_in_frame=0&search=prosperity

49 Horowitz, M., (1977), *The Transformation of American Law, 1780-1860,* Harvard University Press, Cambridge, MA, 112.
50 Ibid
51 Mitchell, L. E., (2007), *The Speculation Economy: How Finance Triumphed over Industry,* Berrett-Koehler Publishers, San Francisco, CA.

mission-based and were held accountable for delivering social goods by law. In their earliest forms, corporations were very much like NFP businesses are today.

## The Emerging Not-for-Profit Era

Nowadays, the for-profit system is driving inequality and ecological destruction and it's no surprise that human wellbeing in most countries of the world has stagnated or declined over the last decade.[52] The crises of the for-profit system increasingly appear to be unresolvable from within the for-profit paradigm. Hence, we find ourselves in the transition period between two eras. More effective ways of viewing the economy, business and profit are evolving and ushering in a new age.

Although for-profit business has taken center stage since the beginning of the Industrial Revolution, it has not been the only model in the business ecosystem. There were also many formative not- for-profit developments in the 19th century. This is when cooperatives were first formed in order to unite for the benefit of all members through collective ownership and democratic management of the business, a step in the NFP direction. This is also when the first formal nonprofits emerged as legal vehicles for charity work and philanthropy. Early nonprofit organizations found ways to raise money for their cause, in addition to relying on donations. Some did even raise funds through the sale of goods and services, such as the Goodwill thrift shops started in 1895 by Reverand Edgar J. Helms.[53] These were the first enterprising nonprofits.

After the Second World War, we saw the development of welfare states throughout the West, as a manifestation of a new sense of responsibility for taking care of all of the citizens of a nation. This new sense of responsibility coincided with the emergence of not-for-profit enterprise in more or less the same form we see it today, and the momentum behind it has been building up over the past six decades to where it is now. An increasing number of NFP enterprises have emerged all over the world in just the last couple of decades, in almost every sector of the economy. This growing trend is also reflective of a cultural shift in business and in the economy. These businesses have chosen to incorporate as NFPs because it's the ideal framework to ensure they remain 100% dedicated to their social mission and it is the best way of showing their customers and communities that they'll never be tempted to compromise in the name of speculation and greed.

The founders of the Mozilla Corporation, for instance, created a nonprofit foundation as its sole shareholder to prevent it from becoming a privately-owned company.[54]

---

[52] Helliwell, J., et al. (2015), "World Happiness Report 2015," Sustainable Development Solutions Network, http://worldhappiness.report/wp-content/uploads/sites/2/2015/04/WHR15.pdf.

[53] Institute for Social Entrepreneurs (2008), "Evolution of the Social Enterprise Industry: A Chronology of Key Events."

[54] Lee, J.-A., (2009), "The Neglected Role of Nonprofit Organizations in the Intellectual Commons Environment" (Doctoral dissertation), Stanford University, http://law.stanford.edu/wp-content/uploads/2015/03/JyhAnLee-dft2009.pdf.

'Not-for-profit' is not just a legal status for organizations. The term also expresses an underlying ethic and value system in which one's own individual benefit comes from the benefit of the whole community. It is a wisdom that acknowledges that we are unconditionally interdependent with the communities in which we live. Each individual's level of wellbeing is an indication of the whole community's wellbeing and vice versa. The NFP ethic is an ethic of enough; that is, I will take just enough, so that others in my community can also have enough and we can flourish together. Think again of the lions in the savannah who take just enough. Fortunately, this ethic of sufficiency is currently experiencing a revival.

These NFP enterprises also represent a shift in the narratives and beliefs that guide business and economic activity. The profit motive and the image of the selfish, ever-calculating, rational 'economic man' are increasingly questioned in business and academia. There are new understandings which show that the drive for individual gain is not nearly as central to human behavior as previously thought.[55, 56, 57, 58] Recent research in the fields of economics, psychology and cognitive science sheds more light on the complexities of human nature and behavior. Researcher Dan Pink's investigation of drive and motivation, for instance, shows that most of our behavior is motivated by factors other than profit, such as autonomy, mastering new skills, creativity, socializing and contributing to the greater whole.[59] Researchers in the relatively new fields of behavioral economics and social psychology are also finding that people do not act nearly as rationally as most economists have long assumed.[60, 61, 62] Our brains are not hard-wired to make us act out of narrow self-interest, but are very flexible and are constantly being shaped by the environment in which we find ourselves because we are such a social and empathic species.[63, 64, 65]

---

55 Bowles, S. and Gintis, H., (2013), *A Cooperative Species: Human Reciprocity and Its Evolution,* Princeton University Press, Princeton.
56 Pink, D. H., (2009), *Drive: The Surprising Truth About What Motivates Us,* Riverhead Books, New York, NY.
57 Rifkin, J., (2009), *The Empathic Civilization: The Race to Global Consciousness in a World in Crisis,* TarcherPerigee
58 Ariely, D., (2008), *Predictably Irrational: The Hidden Forces That Shape Our Decisions,* Harper, New York, NY.

59 Pink, D. H., (2009), *Drive: The Surprising Truth About What Motivates Us,* Riverhead Books, New York, NY.

60 Ariely, D., (2008), *Predictably Irrational: The Hidden Forces That Shape Our Decisions,* Harper, New York, NY.

61 Henrich, J., et al., "The Weirdest People in the World?," *Behavioral and Brain Sciences* 33 (2010): 61-135.
62 Plous, S., (1993), *The Psychology of Judgment and Decision Making,* McGraw-Hill, New York.
63 Bowles, S. and Gintis, H., (2013), *A Cooperative Species: Human Reciprocity and Its Evolution,* Princeton University Press, Princeton.

64 Rifkin, J., (2009), *The Empathic Civilization: The Race to Global Consciousness in a World in Crisis,* TarcherPerigee.
65 Zak, P. J., (2013), *The Moral Molecule: The Source of Love and Prosperity,* Dutton, New York.

Work done in the field of evolutionary biology confirms that cooperation plays just as important a role in evolution as competition.[66, 67, 68] If this seems hard to believe, think of your body, which is made up of over 30 trillion cells, 90% of which are bacteria.[69, 70] Are all of these different cells and bacteria fighting each other and competing for resources? On the contrary, our bodies are evidence of incredible levels of cooperation in evolution. This can also be seen in ecosystems throughout the world. The complexity of the ecosystems we see today on our planet would never have been able to emerge without massive levels of cooperation.

All of this new evidence points to something that many of us have intuited for a very long time: **human nature is malleable**. It is not this or that; it's this *and* that. Human nature is far from being black and white and is better thought of as existing along gradated spectrums ranging from generous to greedy, from thrifty to gluttonous, from cooperative to competitive, and everything in between. To paraphrase the great Russian novelist Aleksandr Solzhenitsyn: the line separating good from evil runs not between people, but through every human heart.

We display different aspects of human nature, depending on social norms as well as the extent to which our fundamental needs are met. Perhaps the best word to describe human nature, then, is 'complex'. If our needs are not being adequately met or if we feel resources are scarce, we tend to express the competitive, greedy aspects of human nature.[71] If our needs are being met in healthy ways, we express the more compassionate, cooperative aspects of human nature.[72]

What are the needs that have such a big impact on whether we act from a place of altruism or greed? You might be familiar with Maslow's hierarchy of needs.[73] We find the work of one theorist even more valuable. Based on decades of cross-cultural work, Chilean economist Manfred Max-Neef hypothesized that every human being has nine fundamental needs: subsistence, protection, affection, understanding, participation, leisure, creation, identity and freedom.[74] This framework shows us that the needs of individuals are inexorably tied to the wellbeing of the community, because a person cannot satisfy their need for protection, affection, participation, leisure and identity, without the support of a community.

---

66 Bowles, S. and Gintis, H., (2013), *A Cooperative Species: Human Reciprocity and Its Evolution,* Princeton University Press, Princeton.
67 Sahtouris, E., (2000), *Earthdance: Living Systems in Evolution,* Metalog Books, Santa Barbara, CA.

68 Capra, F. and Luisi, P. L., (2014), *The Systems View of Life: A Unifying Vision,* Cambridge University Press, Cambridge.
69 Eveleth, R., "There Are 37.2 Trillion Cells in Your Body," *SmartNews*, Smithsonian Magazine, October 24, 2013, http://www.smithsonianmag.com/smart-news/there-are-372-trillion-cells-in-your-body-4941473/.

70 TED (2009), "Bonnie Bassler: How Bacteria "Talk"," http://www.ted.com/talks/bonnie_bassler_on_how_bacteria_communicate?language=en.

71 Research shows that even very wealthy people feel a sense of financial insecurity (See, for instance: Wood, G., (2011), "Secret Fears of the Super-Rich", The Atlantic, http://www.theatlantic.com/magazine/archive/2011/04/secret-fears-of-the-super-rich/308419/).
72 Maté, G., (2010), *In the Realm of Hungry Ghosts: Close Encounters with Addiction,* North Atlantic Books, Berkeley, Calif.
73 Huitt, W., "Maslow's Hierarchy of Needs, Educational Psychology Interactive," Valdosta State University.
74 Max-Neef, M., (1991), *Human Scale Development,* Apex Press.

The not-for-profit ethic is based on fulfilling human needs in ways that support individuals, as well as the communities in which individuals are embedded. From a deep sense of purpose, our contributions fulfill the wider community's needs in harmony with our own. Recall the cells in our bodies. Each cell fulfills its own needs and simultaneously contributes to the health of the organ of which it is a part, as well as the whole body.[75] Without contributing to the whole, individual cells would perish. Without fulfilling their own needs, they wouldn't be able to contribute to the whole. This is interdependence.

This understanding of interdependence was embedded in many ancient cultures and remains an integral part of many indigenous societies today. Fortunately, this wisdom is re-emerging in different forms all over the world. Movements such as **the sharing economy, the solidarity economy, open source innovation, peer-to-peer networks and a growing focus on taking care of the commons all show the eagerness of people to work for the greater good.**

In fact, a significant portion of the world's trade has always occurred, and continues to occur, without the profit motive.[76] In the country of Mali, for example, 17% of all trade, including education and healthcare delivery, happens within what's often called the 'gift economy.'[77] Goods and services are shared without the expectation of any direct reciprocation.

Even in fully industrialized economies, a great amount of economic activity is not profit-motivated. In 1996, economist Duncan Ironmonger estimated that the value of goods and services produced by unpaid workers in Australian households was almost the same as the value of the goods and services that paid workers produced for the Australian market.[78] This is the caring economy; parents, family members, friends and neighbors looking after each other in ways that aren't accounted for by most economists. The caring economy covers the basic needs of billions of people, forming the foundation for all other economic activity. None of us ever got a bill from our parents when we turned 18 years old, tallying up how much it cost to raise us. Instead, we just take for granted that parents do the hard work of raising kids without getting paid because they have a deeper source of motivation. Thus, the purpose motive makes the world go round.

While NFP organizations and businesses rely on the purpose motive to achieve their missions, most for-profit companies still rely on the profit motive to maximize financial surplus. The NFP approach is fundamentally different in terms of depth, scope and goals, and it is quickly gaining advantages in the market.

In light of the more complex view of human nature, needs and motivation, it only makes sense that purpose-based business is on the rise. What is really exciting is that this healthier mode of doing business is just part of a much larger transformation underway. Consumer behavior is moving in a more ethical direction. People are craving more purpose in their work. The Internet is making it easier to start businesses. People all over the world increasingly favor access to products over ownership.[79] And technology has made connecting, sharing, swapping, and trading unbelievably easy.

---

[75] Sahtouris, E., (2000), *Earthdance: Living Systems in Evolution,* Metalog Books, Santa Barbara, CA.

[76] Heilbroner, R. L., (1999), *The Worldly Philosophers: The Lives, Times, and Ideas of the Great Economic Thinkers,* Simon and Schuster, New York.

[77] Ibid

[78] Folbre, N., "Measuring Care: Gender, Empowerment and the Care Economy," *Journal of Human Development* 7(2) (2006): 183-199.

All of these transformations in society mean that NFP businesses might soon be outcompeting their for-profit peers due to the serious advantages they hold in the changing marketplace. In a world with rising demand for ethical products and services, organizations that focus on fulfilling human and ecological needs are ahead of the game. Not-for-profit businesses don't have dividends to worry about and can often offer lower prices, primarily because they are NFP. They often receive tax exemptions and the ability to receive tax deductible donations. They more easily draw on the support of passionate volunteers and attract the growing segment of the population that wants their paid work to contribute positively to society. Not-for-profit enterprises' propensity for flatter organizational structures enables exciting prospects for productivity and innovation, as well. Moreover, NFP businesses are freer to truly innovate, due to the absence of owners and shareholders who so often restrict creative energy to only the areas they deem will reap a good financial return.

We are collectively stepping out of the profit motive and into the purpose motive and are likely witnessing the beginning of a whole new economic paradigm: the **Not-for-Profit Era**. It's rooted in a very simple notion: **there is enough for us all if we keep the surplus circulating in an economy that promotes value creation, rather than value appropriation, and shared interest, rather than self-interest**. And this new economic era has the potential to offer a truly sustainable economy that works for both people and planet.

We are not at the end of history. We are somewhere in the larger evolution of human civilization, and the Not-for-Profit Era could very well be the next phase in the evolution of our economy.

Our hope is that this book is only the beginning; that it will spark a much larger field of interest, debate, and research into NFP business and galvanize a social movement to realize the potential of a NFP World economy.

# A Glimpse into the Not-for-Profit World

Imagine waking up and feeling good about going to work, no matter what the nature of your job is. You feel positive and motivated, knowing that your work provides you with a livelihood that also contributes to the wellbeing of the wider community.

How might a world look in which every person woke up feeling this way? What would it be like if every business were an NFP? A Not-for-Profit World would still involve a thriving market. Government, banks, money, loans and interest would remain. It is just that within a NFP framework, these things would have vastly different purposes and consequences.

For instance, when banks can't privately distribute profits and they have no shareholders or owners that they need to keep happy with dividends, they have no reason to exist other than to provide high-quality financial services to their customers, and they have very little to distract them from this mission. They are built to be more transparent and more efficient. Rather than siphoning wealth away from people and communities who take out loans, all profits are allocated according to the NFP bank's social mission, enabling the generation of real community wealth. Now imagine the *entire* banking sector being NFP. Imagine the entire retail sector being NFP. Imagine all manufacturing being NFP. Envision a world in which all energy, food, transportation, housing and all other goods and services are provided by NFP companies.

[79] Botsman, R. and Rogers, R., (2010), *What's Mine Is Yours : The Rise of Collaborative Consumption,* Harper Business, New York.

In a world with NFP business at its heart, profit is a *means* by which social and environmental wellbeing is achieved; it is not a goal in itself. With changes in the nature of incentive and ownership in business, the NFP model enables companies to make truly sustainable decisions, in turn promoting a less consumerist society and doing less harm to the natural environment. In fact, many NFP enterprises have a mission to help regenerate ecosystems.

The NFP World fosters a more equitable economy because it naturally leads to a more balanced distribution of wealth. The requirement that NFP businesses reinvest all their profit into their mission translates into the constant circulation of wealth throughout the whole economy. Money and other resources go to where they're needed most, as the market is made up of mission-driven businesses. Thus, wealth, power and other benefits are distributed more widely and this is particularly important in light of our current inequality crisis. Not-for-profit business restores the functionality of the wealth circulation pump of the market.

In transitioning to an NFP World economy, most people will experience relative increases in wages and salaries, as huge dividend payouts to the world's richest people disappear and companies are able to better value the work their employees do. This would contribute to a lot more equality, on the local, regional and global levels, therein increasing the quality of life for everyone.[80]

The NFP World would also be better for the environment due to what we call the Paradox of Enough**.** In an economy in which wealth concentrates, we consume more but have less, whereas in an economy in which wealth circulates, we consume less but have more. In the for-profit world, we have less social connection, less free time, and less relative wealth. We consume more goods and services in order to compensate for not having enough of the stuff that makes life truly worthwhile. We have too much material wealth and not enough immaterial wealth. But in the NFP World, we have more free time, more cooperation, more dignity, tighter social connections and more equality. Rather than having too much of one kind of wealth and not enough of the other, we have enough of both. And when we have enough, we are better able to meet our needs without the pressure to over-consume.

**For the first time in modern history we have the structures, capabilities and impetus to evolve to a Not-for-Profit World, in which the drivers of good business are harnessed for our collective flourishing.** Not-for-profit enterprise is the keystone that allows us to bridge to a healthier economy, and it's been hiding in plain sight, waiting for us to have the wisdom to see and pursue it.

That is not so say that a transition from the current for-profit economic paradigm to an NFP economic paradigm would be without difficulties. Potential challenges to the development of an NFP World include inertia as well as active resistance from the for-profit world. The for-profit way of thinking about business and the economy is so deeply embedded in social norms throughout the world that it might be very difficult for many people to imagine an economy that doesn't revolve around private accumulation and the profit motive; so many of us have built our lives around these ideas. This is why a transition to an NFP economic paradigm requires a strong social movement, the foundation of which already exists. The NFP World has the potential to resonate with many diverse social and environmental movements all over the world, as a vision worth working towards collectively.

---

[80] Wilkinson, R. G. and Pickett, K., (2009), *The Spirit Level: Why Greater Equality Makes Societies Stronger,* Bloomsbury Press, New York.

This book is about society taking an evolutionary step in the direction of a set of ethics and practices that are healthier for everyone; the planet's ecosystems included. It is not about being against for-profit business models, or governments dictating that business move in a not-for-profit direction. It is about businesses, citizens, consumers, educators, and policy-makers working together to build a brighter future. We have glimpsed a not-for-profit world, and it is beautiful!

# 2. The Power of Not-for-Profit Enterprise

## *Not-for-profit enterprise offers the sustainable business model we've been seeking*

In the early 1970s, just after Bangladesh gained its independence from Pakistan, the Bangladesh Rural Advancement Committee (BRAC) was started by Sir Fazle Hasan Abed and two dozen volunteers. The country had been ravaged by war and genocide and many people were struggling to meet their basic needs. Abed was an executive at Shell Oil at the time, but he quit his job to help build houses, rehabilitate farmland and open healthcare clinics through the organization he founded as BRAC. Over time, Bangladesh recovered and BRAC's mission transformed from disaster relief to long-term development.[81]

The organization now has 115,000 employees, making it the largest NFP business in the world in terms of employment.[82] BRAC's social mission is to meet the needs of financially disadvantaged people in Bangladesh. Specifically, they run educational programs and provide healthcare services to rural populations. Rather than depending on donations and philanthropy to fund the good work they do, BRAC runs businesses. They fund their social efforts through many different revenue streams, operating banks, food processing plants, professional print and copy shops, and department stores that sell products made by rural artisans. In 2011, BRAC's revenue was $422,139,409[ii], 80 percent of which was generated through business activities. It is estimated that BRAC positively impacts the lives of 135 million people.[83]

BRAC is a shining example of how generative NFP enterprise can be: using strategic business approaches to support a mission that benefits the wider community, seeing profit merely as a tool to make the world a better place, not a goal in itself.

## Transformational Times

The story of BRAC gives us a glimpse into a much larger global trend. Something amazing is happening in the economy. An undercurrent that has gone mostly unnoticed until now has the potential to change our entire global economy for the better.

What comes to mind when you hear the words 'not-for-profit'?

Many people think of a mixed bag of words like: charity, foundation, bureaucracy, difficult, struggling, inefficient, passionate, relationships, corporate-funded, values-based mission, and volunteers. In our experience, the words 'not-for-profit' and 'nonprofit' conjure up a wide range of feelings in people. We used to have those same mixed feelings about 'not-for-profit' and 'nonprofit' ourselves. So it came as a surprise to us that one of the most exciting transformations happening in the economy right now is taking place in the nonprofit sector.

Take these revealing numbers, for example: between the years 2000 and 2010, the for-profit sector in the U.S. experienced a 6% decrease in employment, while the nonprofit sector experienced a 17% *increase* in employment. During that same time period, wages in the for-profit sector fell by 1%, while wages in the nonprofit sector increased by 29%.[84]

---

81 Smillie, I. (2009) *Freedom From Want: The Remarkable Success Story of BRAC, the Global Organization That's Winning the Fight Against Poverty*. Sterling, VA: Kumarian Press.

82 Ibid

83 BRAC World (2013) 'Thinking Big with BRAC' video. (https://www.youtube.com/watch?v=HKwBxqD9sRg)

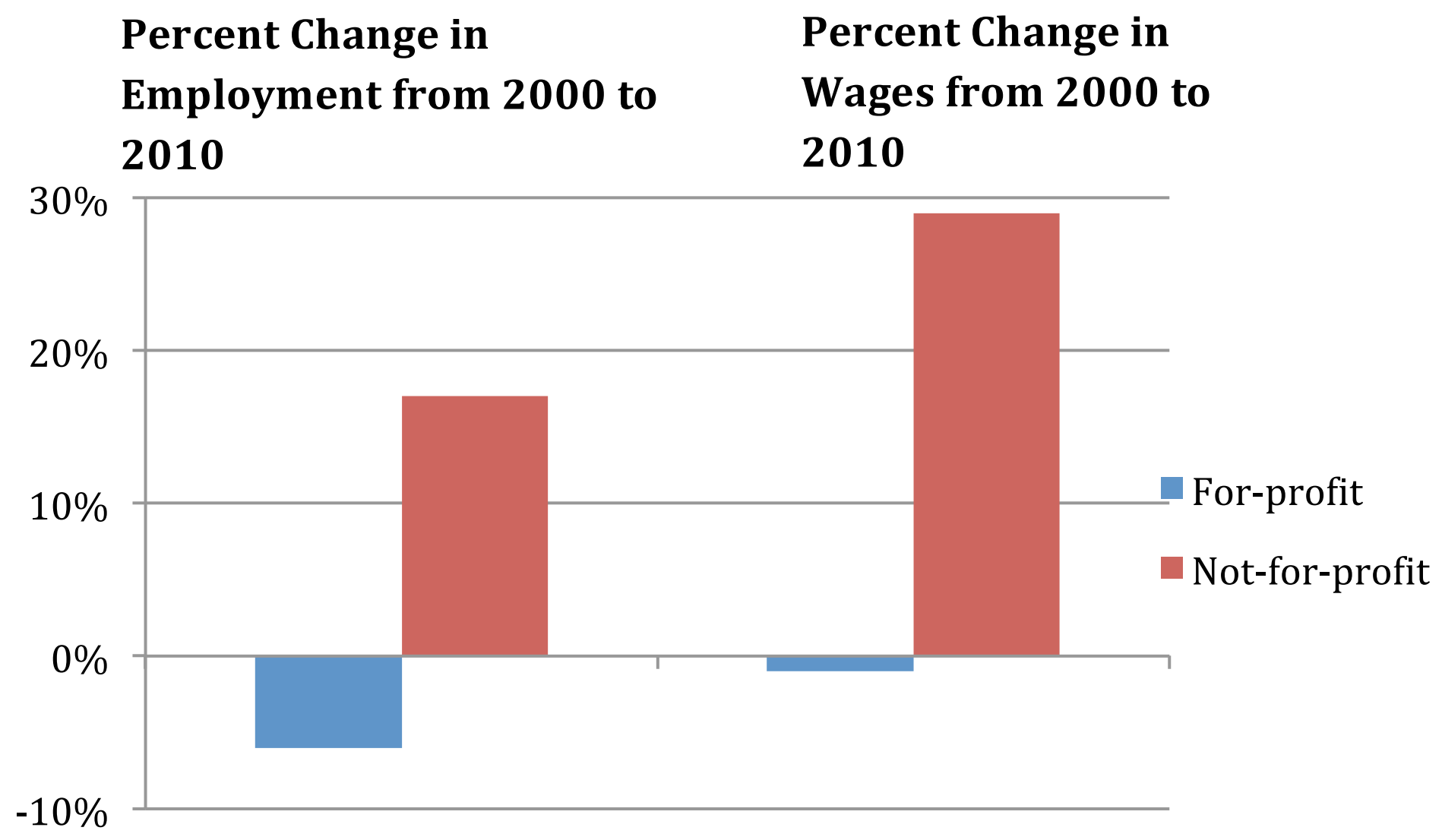

**2.1 - Chart of FP versus NFP Employment and Wages, 2000- 2010 (U.S.)**[85][86]

These statistics run counter to the story that we've all been told – that the profit motive makes the world go around and rich capitalists keep employment and wages high. These numbers are even more counter-intuitive if you think of what happened during this period of time in the U.S. – the Great Financial Crisis began. The government had less money, consumption dropped, corporate profits fell, and people tended to hoard rather than spend or donate their extra money. This is clearly reflected in the for-profit numbers above, but shouldn't wages and employment have fallen in all sectors of the economy, across the board? What could possibly explain the fact that wages and employment in nonprofits actually rose, while they fell in the for-profit sector?

These numbers tell a story about the direction our economy might be heading. The economic crisis of the last eight years has had very different impacts on the for-profit and nonprofit sectors of the economy. As the financial industry experienced big hits that sent shockwaves throughout the global economy, slowing economic growth in most of the 'developed' world, for-profit companies experienced major losses. As a result, they laid off employees and decreased wages in order to cut costs, as the statistics above show. They also had to decrease their sponsorship and philanthropic giving to nonprofits. This meant sink or swim for much of the nonprofit sector. Instead of falling victim to the decline in grants, donations and philanthropy, many nonprofits recruited volunteers, board members and managers with a background in business and quickly learned how to become more financially self-sufficient through generating their own income; essentially they started going into business. And those that did so successfully have flourished.[87] In the latter half of the 20th century, nonprofits typically responded to declines in grants and philanthropy by cutting costs, cutting services and trying harder to get donations. Now there is a different approach that many are adopting: shifting to an enterprise model.

---

84 Roeger, K., A. Blackwood, and S. Pettijohn (2012) *Nonprofit Almanac 2012*. Washington, D.C.: Urban Institute Press.
85 This is our graph based on the numbers provided in the report. The report refers to the nonprofit sector and the private sector, rather than for-profit. And it does not distinguish between traditional nonprofits and NFP businesses.
86 Ibid

As a result, nonprofits are steadily moving into the market. These NFP companies are selling goods, offering services on a sliding-scale fee basis, and even leasing their property in strategic ways. Data from all over the world shows that nonprofits are increasingly developing business plans and revenue models that enable financial independence.[88][89][90] This trend has been strengthening in the last few years and the nonprofit sectors of 27 countries are increasingly generating their own income through various forms of self-managed business.[91]

In 2013, nonprofits in the U.S. reported over $1.74 trillion in revenue, 72 percent of which came from program fees and contract work.[92] Nonprofit revenues in the U.S. grew at a rate of 41 percent from 2000 to 2010.[93]

Forty-five percent of registered charities in the U.K. currently identify themselves as social enterprises and 92% want to increase their earned income.[94] In 2013, earned revenue constituted more than half of the income of civil society organizations in the U.K. (over 75% in some regions).[95] And a recent study in Australia concluded that the country's nonprofit sector is in a state of transition towards a social enterprise model.[96]

Whether or not a nonprofit can effectively do business mostly boils down to the organization's culture and mindset. Many organizations still operate from the mindset of being the 'nonprofit sidekick' of the for-profit economy. They see themselves as being the people who require philanthropy and donations to do the 'good work' that business is too busy making money to do. This dependence-mindset is part of these organizations' internal culture. They maintain this dependence by using a great deal of their time and resources to apply for grants and ask for donations and philanthropy.

Their train of thought is that they can't charge for the good they're doing for society because the people who benefit from their goods and services can't afford to pay for them, and someone must subsidize this good work. Who? Philanthropists, donors and the government. There's almost the feeling that charging people for goods and services that benefit the community will cheapen their work or twist their ethics. There's a prevailing idea, not just among nonprofits, but throughout society, that business and charity are two very different things that should not be mixed.

---

87 Salamon, L., et al., 2013

88 Kam-Tong Chan, Yu-Yuan Kuan, Shu-Twu Wang, (2011) "Similarities and divergences: comparison of social enterprises in Hong Kong and Taiwan", *Social Enterprise Journal*, 7(1): 33-49.

89 Haugh, H. (2005). 'A research agenda for social entrepreneurship'. *Social Enterprise Journal*, 1: 1-12.

90 Dees, J. G., Emerson, J., Economy, P. (2001). *Enterprising nonprofits: A toolkit for social entrepreneurs*. New York, NY: John Wiley & Sons.

91 Salamon, L., et al., (2013)

92 National Center for Charitable Statistics website (2013) 'Quick Facts About Nonprofits'. (http://nccs.urban.org/data-statistics/quick-facts-about-nonprofits)

93 Rifkin J. (2014) 'The Rise of Anti-Capitalism', *The New York Times*, March 15. (https://www.nytimes.com/2014/03/16/opinion/sunday/the-rise-of-anti-capitalism.html)

94 Jervis, J. (2013) Overwhelming Majority of Charities are Eager to Start Trading, *The Guardian*, May 28. (http://www.theguardian.com/social-enterprise-network/2013/may/28/charities-social-enterprise-finance)

95 National Council for Voluntary Organisations. (2013) 'What are the sector's different sources and types of income?', *UK Civil Society Almanac 2013*. (http://data.ncvo.org.uk/a/almanac13/what-are-the-sectors-different-sources-and-types-of-income-2/)

96 Logue, D. & Zappala, G. (2014) *The Emergence of the 'Social Economy': the Australian not-for-profit sector in transition*. Sydney, Australia: University of Technology Sydney.

But **shouldn't all businesses be set up to benefit society?** Should companies provide goods and services that *don't* benefit society? Or goods and services that actually *harm* society, just in order to make money, and then donate a bit of that money to try to compensate for the harm they've caused or the gap that they haven't filled, because it wouldn't make them as much money?

Traditional nonprofits that maintain the dependence-mindset aren't asking these important questions. And they aren't doing themselves or society any favors by staying in a position of dependence.

Dependence on philanthropy and grants often means that nonprofits must prioritize the funders' goals over their own. This can be detrimental to the nonprofit's mission. Funders usually require goal-setting, monitoring and evaluation to be done within their own financial time horizons, driving short-term thinking in many nonprofits, at the expense of long-term, systemic pursuits.

An article published in 2005 in the Harvard Business Review, titled 'Should nonprofits seek profits?', warned against nonprofits trying to earn their own income.[97] The authors believed that going into business was just too risky for nonprofits, because so many income-earning endeavors fail in the nonprofit space and society cannot afford for nonprofits to fail, as their services are so important in helping the disadvantaged.

The article describes how one nonprofit's manager came to see their attempt at earning income through a restaurant-café as the best way to attract grants and philanthropy, even though the restaurant was actually *costing* more money than it was generating. The author used this example to illustrate that nonprofits simply don't have the right mentality to run a business, as if this is an inherent flaw of being nonprofit.

The author is right; that specific organization's failure is probably mostly due to the dependence-mindset being so deeply ingrained that they even saw their business as a way of attracting donors, rather than as a way of generating income to be self-sufficient. A manager with this mindset, who is so focused on the idea of fundraising, will probably not be very good at keeping a business going; if for no reason other than the fact that he or she will constantly be distracted from the business by thinking of ways to get more grants and donors. However, NFP businesses all over the world are proving that this is mindset is not an inbuilt aspect of being NFP.

The same amount of time and resources that so many traditional nonprofits spend chasing grants and philanthropy can be better spent creating ways for the organization to become more financially self-sufficient, which also gives them more independence in terms of decision-making. They can become more self-directed by earning their own income and having no obligations to outside funders. Being financially independent can also contribute to more resilience during economic downturns, as you can adjust your business strategies according to the market's needs, rather than faltering when donations run low.

[97] Foster, W. & Bradach, J. L. (2005) 'Should Nonprofits Seek Profits?', *Harvard Business Review*, February Issue.

Such independence gives NFPs more self-confidence to get their mission done as well. This is another problem with the dependence-mentality in nonprofits that often goes unarticulated. Being dependent makes them feel that they can only accomplish so much on their own. Just like an individual who is financially dependent on someone else; they feel bound and limited. Being more self-sufficient, on the other hand, gives them a sense of empowerment, although it sometimes takes more work. There's also a different sense of pride in the work one does when it is self-directed. This is just as true for organizations as it is for individuals. Kathleen Buescher, President and CEO of Provident Counseling, Inc., a $5 million family services NFP in St. Louis, summed it up perfectly:

"It's my theory that nonprofits in the future will have to fund a lot of their mission this way. We're just not going to have sufficient other money to do it. We'll have to earn it ourselves. And the beauty of making a profit, as we've been able to do during the past 15 years, is that you can do a lot with the money, you can do what you want to do. You can do it how you want to do it for as long as you want to do it and you don't have to make anybody happy except your own board and staff. You don't have to meet anybody else's expectations. That's a very freeing idea, and once you feel it, you don't want to go back to the confines of any other type of funding."[98]

An article on the popular website Fast Company tells the story of VolunteerMatch, an American NFP that helps volunteers go to where they're needed most. VolunteerMatch has been reaping the many benefits of doing more business. The organization went from generating 57% of its revenue from service fees in 2006, to generating 87% of its own revenue in 2009. Greg Baldwin, the president of VolunteerMatch, proudly remarked in an interview, "We are almost completely sustainable due to earned revenue." And because of that element of financial viability, Baldwin is optimistic that the company will be able to better serve tens of thousands of nonprofits in coming years, creating even more positive social outcomes.[99]

Likewise, the YHA recreation and accommodation provider in the United Kingdom generates 95% of its revenue through the sale of goods and services. In 2015 they enjoyed their best year ever of trading, with over 8% growth and a profit of £1.6 million, which is not a big surprise in light of the fact that they scored a 90% on their customer satisfaction surveys. About 400,000 young people stay with them annually, thousands of which receive direct funding support from the YHA. And they're expanding. Over twenty-two million pounds have been invested in their network since 2011, allowing them to build new hostels in Brighton, Cardiff and the Cornwall. All this from a not-for-profit.[100][101]

Perhaps the main reason society has traditionally seen charitable work and business as playing separate and distinct roles in society is because we've been wary of people getting rich in the name of social benefit. But while there is a very good chance for socially-oriented for-profit companies to do this, NFP businesses strike just the right balance. This is because no individuals can own them or make a financial gain from their profits and their number one priority is to meet social needs, which they can better achieve by being financially self-sufficient.

---

98 BRAC (2014) 'BRAC at a glance'. [http://www.brac.net/partnership?view=page]

99 Korngold, A. (2010) 'The Nonprofit Financial Model Never Worked; Here's a New Model that Does', *Fast Company*, October 27. (https://www.fastcompany.com/1698097/nonprofit-financial-model-never-worked-heres-new-model-does)

100 YHA. (2015) *Our Impact: A look back at a year of positive impact*. Derbyshire, UK: YHA. (https://www.yha.org.uk/sites/default/files/YHA-Impact-Review-2015.pdf)

101 YHA website. (2018) 'About' page. (http://www.yha.org.uk/about-yha/yha-today)

Shifting from a philanthropically dependent mindset to one with a business plan moves organizations from a deficit-based mentality to an asset-based mentality. In other words, rather than focusing on needs and problems (deficits), NFP businesses are likely more focused on strengths and potentials (assets). This is a hugely empowering shift for the nonprofit sector.

Not only is the percentage of self-generated income rising within the nonprofit sector, but NFP companies are also expanding their reach into new areas of business. **Markets traditionally dominated by for-profit business, such as construction, manufacturing, software development, food catering and retail, are all arenas for not-for-profit business now.**

# Business Built for Purpose

You might be wondering how so many different kinds of businesses can possibly be not-for-profit. Sure, they might not privately distribute profit, but what about their social mission?

In most countries, the definition of 'social benefit' or 'charitable purpose' is flexible enough to include any kind of mission, other than generating profit, that has some sort of positive social outcome. Some countries, like the UK, determine this by evaluating whether a business does more good than harm for society. This allows for a very broad range of businesses, not just the usual suspects like education, healthcare and community development.

For example, in the retail sector, the Eco-Home Centre is a registered NFP business in Cardiff, Wales, which sells home renovation and construction supplies. Its mission is to "champion sustainable building in Wales"[iii].[102]

In fashion and design, there are companies like Bombolulu in Kenya, which employs people with various disabilities to produce jewelry, textiles, as well as wood and leather products that are sold worldwide through Fair Trade Organizations.[103]

Law firms and real estate companies can also have social missions. The ASU Alumni Law Group, in Arizona, is an NFP legal firm whose mission is to help the community gain access to affordable legal services.[104]

Home Ground Real Estate is Australia's first NFP real estate and property management company. They charge regular real estate fees to help people buy and sell homes, but 100% of their profits go to help people who are at risk of homelessness.[105]

And there are a plethora of NFP energy companies, like Som Energia, Catalunya's first NFP energy cooperative, whose mission is to generate clean, local energy for the community.[106]

---

[102] Eco-Home Centre website (2018) 'Home' page. (http://www.ecohomecentre.co.uk/)
[103] Bombolulu website (2018) 'Home' page. (http://www.apdkbombolulu.org)
[104] Arizona State University Alumni Law Group website (2018) 'Home' page. (http://asualumnilawgroup.org/)
[105] Home Ground Real Estate website (2018) 'Home' page. (http://www.homegroundrealestate.com.au)

In banking and finance, there are examples like the not-for-profit GLS Bank in Germany, which allows members to choose to receive a lower interest rate on their savings in order for the bank to provide loans to charitable projects at a lower cost.[107] Credit unions, a type of NFP co-operative bank found throughout the world, also exist solely to serve the financial needs of their communities. Credit unions gained 10 million members worldwide between 2013 and 2014 and in the U.S., more than 100 million people are members of credit unions.[108,109]

There is also a plethora of very successful NFP insurance companies all over the world, including State Farm Insurance in the U.S. and CRISP Insurance and CBHS Health Fund in Australia. All of these insurance companies exist to provide the best, most affordable insurance coverage they can to their customers and all profits are invested back into that mission.

You can also find NFPs in the transportation industry. The Independent Transportation Network in the U.S. is dedicated to providing dignified transportation for senior citizens and it has over 4000 members, all of whom are also NFP businesses.[110]

There are NFP breweries, like Ex Novo in Portland, Oregon, which has the mission of providing the community with good beer and gives 100% of its profits to charities that help disadvantaged communities.[111] This might sound just like a for-profit brewery that simply donates to charity, but this brewery has no private owners as well as a clear mission that is not related to profit. This means that it won't be tempted to sell as much beer at as a high a price and as low a cost as possible, potentially sacrificing the quality of the product and the working conditions of employees in order to turn a profit for owners and investors. This is the extremely important difference between for-profits and NFPs, regardless of what industry they're in.

There's even an NFP dating site. Humanitarian Dating is a website that helps people who are passionate about resolving the world's crises connect with likeminded people. Any profit they receive from paid membership fees goes right back into better serving their members.[112]

**In fact, there isn't any sector in the real economy in which a business cannot be an NFP and this is why we can envision an entirely NFP world economy.** The idea is very simple: every business can and should create positive social and environmental outcomes, and profit should only be seen as way of achieving those outcomes. In the twenty-first century, we can expect that from our economy.

These companies measure their success by looking at how well they're fulfilling their purpose, not how well they're maximizing profit. And they see having healthy revenue streams as an integral part of how they achieve their deeper goals. They keep the horse in front of the cart.

---

106 Som Energia's website (2018) 'Home' page. (https://www.somenergia.coop/eu/welcome-to-som-energia/)
107 GLS Bank website (2018) 'Home' page. (https://www.gls.de/privatkunden/english-portrait/)
108 World Council of Credit Unions (2014) Global Credit Union Statistical Report 2014, Washington DC: WCCU. (http://www.woccu.org/documents/2014_Statistical_Report)
109 Muckian, M. (2015) 'Credit Union Membership Up by 10 Million Globally', *Credit Union Times*, July 10.
110 ITNAmerica website (2018) 'Programs' page. (http://itnamerica.org/what-we-do/#programs)
111 Ex Novo website. (2018) 'Home' page (http://www.exnovobrew.com)
112 Humanitarian Dating website (2015) 'Home' page. (http://www.humanitariandating.com/index.php) (website not working on Feb 18, 2018, but still an interesting example)

BRAC proves that NFP enterprise is not restricted to meeting needs that can be 'cashed in on'. Rather they fund their social missions that aren't easily commercialized, like education for the financially disadvantaged, by thinking creatively and doing business in other sectors.

Similarly, The Endeavor Foundation in Australia seeks to give people with disabilities employment opportunities. They also provide education, home care, accommodation and assisted holidays to their beneficiaries. In order to support this work, Endeavor offers recycling, packaging, and industrial sewing services. Any profit generated from these business activities goes to making the disability services they provide even more affordable and accessible.[113]

Relying more on earned income exposes NFPs to competition in the marketplace, usually resulting in more efficient and effective operations. Increased cash flow means that their beneficiaries benefit even more.

In an era when inequality is skyrocketing, a large portion of our planet's ecosystems are facing collapse, and levels of wellbeing all over the world are declining, it's no longer enough for the business world to try to minimize or mitigate its negative impacts. We need companies that generate positive outcomes for the whole living community, not just a handful of wealthy business owners and investors. We need an economy that is generative, not extractive.

Fortunately, the kind of generative business model that so many of us have been seeking already exists. Not-for-profit business is better for people and the planet. This is not because all NFP businesses are perfectly eco-friendly and ethical in their activities. **Not-for-profit companies are simply much more likely to work for positive social and environmental outcomes than for-profits** because they:

- exist for a social or environmental purpose;
- see profit only as a means of achieving a deeper purpose, not as the goal itself;
- have a board to hold the company accountable to that purpose;
- are held accountable by the larger public for staying mission-oriented;
- have no private owners or shareholders (so no one gains from cutting ethical corners);
- have a greater ability to be more democratically-run; and
- are in a better position to take feedback from the wider community into consideration.

NFP is truly a business model built for purpose. It represents economic activity that focuses on meeting social and environmental needs. Because the NFP business model doesn't require constant growth in order to maximize profits, it allows us to acknowledge the fact that some needs are best met outside of the market. Based on the ethic of enough and the 'purpose motive', NFP business is a lot less likely to pressure people to consume for consumption's sake, which is good for all the inhabitants of this planet.

Unlike in the for-profit world, where business is fundamentally at odds with ecological limits because companies must constantly grow, NFP enterprises are much more likely to take environmental issues seriously, and to constantly seek to reduce their ecological footprints.

REI, a well-known recreational equipment retail chain, has been an NFP consumer co-op since it opened in 1938. In November of 2015, they decided to give all employees a paid day of leave and keep their doors closed on Black Friday[iv] in order to encourage people to spend the day outdoors with family and friends, rather than participating in the nation-wide spending splurge. REI was making a statement: that they were willing to sacrifice some revenue in order promote the values of human connection and ecological stewardship on a day when most other retailers promote unfettered consumerism and material accumulation at the expense of social and ecological wellbeing. [114]

---

[113] Endeavor website (2018) 'Home' page. (https://www.endeavour.com.au)

Not-for-profits are much freer and more able to choose planet over profit. An NFP can much more easily get away with not making a profit year after year in order to be more eco-friendly in its business activities.

Due to the different values and greater levels of accountability in the NFP business world, there is also a deeper consideration for consumers' real needs. CarShare in Colorado makes this very clear on their website, which says, "We're committed to helping our members share more than just cars. We want to promote community through a shared sense of purpose. By sharing the costs of car ownership, more resources will be available for the things that truly matter – like clean air, water, and open space instead of parking lots. Shared vehicles, in combination with urban cycling and reliable transit, give our nonprofit model the leverage to partner with virtually unlimited opportunities and offer our members the lowest rates possible."[115]

NFP enterprises are also more inclusive. Such businesses have a predisposition toward democratic decision-making, transparency, fairness, and equality, which builds trust and inclusivity. If you think about which companies make a point of employing people with disabilities in your community, names of NFPs will most likely come to mind.

Like Food Connect, many innovative NFPs couple social and environmental missions. Blue Star Recyclers in Colorado, whose core mission is "recycling electronics and other materials to create jobs for people with autism and other disAbilities". They are doing excellent work. Since 2009, they have provided 40 local jobs for people with disabilities in four Colorado communities. They have ethically recycled over 12 million pounds (5.4 million kilograms) of electronics.[116] In 2016 they generated the vast majority of their income, about $1.6 million, from recycling and refurbishment sales.[117] This is a business model that can be found all over the U.S. and even all over the world, with Versability Resources in Virginia, Green Vision Inc in New Jersey, Opportunity E-Cycling in Montana, Garten in Oregon, and Abilities Group in New Zealand, just to name a few.

Along with being mission-driven rather than profit-driven, NFP businesses also contribute to greater socio-economic equality because they keep their financial surplus in the system (in addition to many NFP businesses who aim to alleviate poverty directly through their services). This keeps the wealth circulation pump going strong. As Wilkinson and Pickett illustrated in their foundational book, *The Spirit Level*, a more equal society, without extreme poverty and extreme wealth, is better for everyone.[118]

---

114 McGregor, J. (2015) 'One Big Reason REI Can Decide to Skip Black Friday', *The Washington Post*, October 28.
115 CarShare website (2018) 'Membership FAQs' page. (http://carshare.org/membership-faqs/)
116 Blue Star Recyclers website (2017) 'A 100% Solution'. (http://www.bluestarrecyclers.org/solution.htm)
117 Blue Star Recyclers (2017) 2016 Annual Report, Colorado Springs, CO: Blue Star Recyclers. (http://www.bluestarrecyclers.org/_images/mission-jobs-disabilities/Annual%20Report%202016.pdf)

While many of us have become accustomed to for-profit companies undertaking very destructive business practices like running sweatshops and dumping toxic waste into rivers, it would be shocking and totally unacceptable for an NFP to do such things. It's almost as if we've accepted that the profit motive naturally leads to these negative consequences, but that it's the only way to do business, therefore for-profit enterprise is a necessary evil. The NFP business model offers a real alternative. It raises the bar, setting a much higher ethical standard for business.

As such, we argue that moving beyond for-profit business is not only possible, but it is *necessary* for a more sustainable world[v].

# Ownership and Assets

**In light of the extreme inequality of the 21st century, one of the most significant aspects of the NFP model is that it invites us to re-imagine business ownership**. Ownership of businesses, and of the means of production, is so central to any economy that it has the power to shift an entire economic paradigm.

Strictly speaking, no one can own an NFP. Not-for-profit businesses can have managers, boards and CEOs, but no owners or shareholders. Anytime you hear of member-owned NFP co-ops or having a membership share in an NFP, it's purely nominal. Your 'share' is only what you put in; it can't grow and it can't be traded. So, an NFP 'share' is more like a refundable deposit and the so-called 'owners' don't have any equity. Nobody can really own a share of an NFP or hold equity in it, other than another NFP.[119]

This is an extremely important point of distinction between NFP and for-profit businesses. Equity in a for-profit business can be a tremendous source of wealth for its owners (think of J.P. Morgan, Bill Gates, Mark Zuckerberg and the Walton family, for instance), whereas NFP ownership is more like custodianship or stewardship; it's more of a caretaking relationship.

This caretaking approach demonstrates the 'ethic of enough' inherent in the NFP model. The employees of an NFP are compensated through their salaries and there's generally a limit to how much NFP managers and CEOs can receive. In many parts of the world, this relies on a 'reasonability' test, in which an NFP must prove why a high salary is justified. This is usually done by referring to salaries in peer organizations. Extremely high salaries, like those of many for-profit CEOs, are deemed unacceptable in NFPs by tax agencies and courts of law.

Dividends from for-profit companies, on the other hand, have no defined limits. Indeed, most shareholders want their dividends to be as large as possible, and this is a rational and socially acceptable response in light of the prevailing system conditions. Thus, dividends and private equity represent an ethic of 'never enough', as they can (and should) always be bigger from one year to the next.

---

118 Wilkinson, R.G. & Pickett, K., 2009
119 Glaeser, E. (2003) *The Governance of Not-for-Profit Organizations*. Chicago, IL: University of Chicago Press)

Dividends and private equity are not allowed in the NFP realm because they violate the 'non-distribution' legal mandate of NFPs (also known as 'no private inurement').[120] The idea is that no profit or assets should be privately distributed (i.e., distributed to individuals).

This is why NFPs also have what is commonly referred to as an 'asset-lock'. When NFPs are dissolved or shut down, their assets are not distributed to owners, as with for-profit companies, because there are no owners. Rather, most countries' laws require that, upon closing down, an NFP's assets must be used to further the organization's mission or given to other NFPs. Thus, the assets and wealth stay in the NFP system – and in the real economy. Just as nobody can get rich from an NFP's profit, no one can get rich from an NFP's assets either.

The only area of NFP business in which we've found that the asset-lock is somewhat ambiguous is when it comes to cooperatives. We have not found any legislation that governs what happens to the assets of NFP cooperatives when they shut down. However, in practice, it is basically unheard of for anyone to make a financial gain from assets when an NFP cooperative is dissolved, because it is much easier for the co-op members to decide to donate assets to another co-op than to decide how to divide the assets among a large number of members who all deserve an equal share. In some places, it is considered standard practice for co-ops that are closing down to give all assets to other co-ops, in accordance with the cooperative spirit. Italy has a special national cooperative development fund that promotes the creation of other cooperatives, and when a co-op is dissolved, all remaining assets are transferred to this fund.[121]

If an NFP business is dissolving and owes money to a lender, the debt owed can usually be recovered in the form of assets, but that is only to pay off a debt, not for the lender to make a financial gain. An NFP must justify any transfer of assets or money upon winding down and most tax agencies scrutinize this process very thoroughly to ensure that nobody is benefitting, except for other NFPs.

In many places, there are provisions for ensuring that individuals cannot even indirectly make a financial gain from an NFP. For instance, in the U.S., NFPs cannot pay contractors above the market rate (called 'excess benefit transactions' by the Internal Revenue Service). This is basically an inbuilt mechanism that reduces corruption and nepotism in the NFP business sector.

No one can own an NFP or its assets, but an NFP entity itself can own land and assets. An NFP can also own other businesses and hold equity in them, but because nobody owns the NFP, it acts like a steward or caretaker of the land, assets and businesses it owns.

# Subsidiaries

**Because nobody can own a not-for-profit business, each NFP is best thought of as owning itself**. An important part of this not-for-profit ownership is the ability of NFPs to own other companies (including both for-profit and not-for-profit) and the *inability* of for-profit companies to own NFPs. In other words, NFPs are legally allowed to have for-profit and NFP *subsidiaries*, but for-profit companies are not legally allowed to have NFP subsidiaries.

---

[120] Glaeser, 2003

[121] Thompson, D. (2005) 'Building the Future: Change, challenge, capital, and clusters in Italy's market leader', Cooperative Grocer Network Magazine, November – December. (https://www.grocer.coop/articles/building-future)

For example, Jamie Oliver's gourmet restaurant, Fifteen, in London is owned by the Jamie Oliver Foundation, a charity.[122] All of Fifteen's profits go to the NFP foundation, which uses all of its funds to achieve its stated goal: to shape the health and wellbeing of current and future generations and contribute to a healthier world by providing better access to food education for everyone.[123]

All profits go to a social mission, so it doesn't really matter much whether the subsidiary, Fifteen, is registered as a for-profit or not-for-profit. Because it is owned by the foundation, it is in effect an extension of the not-for-private-profit, essentially making it NFP as well. However, if Fifteen, as a for-profit company, was legally able to own the Jamie Oliver Foundation (which it's not legally able to - this is a hypothetical example), it would mean that Fifteen, which *can* distribute profits to individuals, would be able to siphon money out of the foundation into private pockets. This is why an NFP can own a for-profit, but not the other way around.

Similarly, Tait Communications, a global communications company based in New Zealand, is a for-profit subsidiary of an NFP foundation. Tait sells communications devices, such as the portable radio systems used by emergency response teams, and all of Tait's profits go to research and development, educational programs and other charitable purposes through its owner, the foundation, as was the vision of the company's founder, Sir Angus Tait.[124]

Likewise, the well-known American food company, Newman's Own, is wholly-owned by the Newman's Own Foundation, which uses the subsidiary's profits to work with disadvantaged people, children with medical conditions and people who struggle with malnutrition, as was actor Paul Newman's vision when he started the company.[125]

Starting a subsidiary in order to generate revenue is a popular strategy among NFPs whose cause is difficult to commercialize. It's completely natural for the Eco-Home Centre in Wales to sell building products as part of its mission to support sustainable building. Its business activities are undeniably related to its mission. But it's easier to do this for certain missions than others. For instance, what sort of business activities might be related to offering rehabilitation for substance abuse? What sort of business activities might be related to providing shelter to the homeless? This is where subsidiaries can be extremely helpful for NFPs.

An impressive example of this is Housing Works Bookstore Cafe in New York City. It is a for-profit subsidiary that sells books and serves coffee in order to give all of its profits to Housing Works Group, its NFP owner. Housing Works Group's mission is "to end the dual crises of homelessness and AIDS through relentless advocacy, the provision of lifesaving services, and entrepreneurial businesses that sustain our efforts".[126] Housing Works Group also owns many thrift shop subsidiaries that likewise give all profits to the foundation.

---

122 Fifteen's website (2018) 'Welcome' page. (http://www.fifteen.net/)
123 Jamie Oliver Food Foundation website (2018) 'About' page. (http://www.jamiesfoodrevolution.org)
124 Tait Communications website (2018) 'Our Values' page. (https://www.taitradio.com/about-us/our-values)
125 Newman's Own website (2018) 'About Us' page. (http://newmansownfoundation.org/about-us/)

One of the most successful examples of this model in the U.S. is the 35-year social enterprise partnership between FEI Behavioral Health, a for-profit workforce resilience and crisis management company, and its single shareholder (i.e. – sole owner), the Alliance for Strong Families and Communities, an NFP family services organization. Since beginning this social enterprise structure in 1979, FEI has provided more than $30 million of financial support to the Alliance, its members, and the communities they serve.[127]

A Better Way Ministries in Newnan, Georgia, is a drug rehabilitation and gospel outreach program that employs the men in its care in a number of small, profitable subsidiary companies, including an auto detailing business, a bakery and a moving company. The money generated by these satellite companies allows the parent organization to earn income while helping the program recipients earn a living through engagement in the program.

A study by the Yale School of Management and the Goldman Sachs Foundation on enterprising nonprofits, found that having a subsidiary business to fund its social mission has a positive effect on the overall reputation and mindset of the NFP, improving its "reputation, mission, service and program delivery, entrepreneurial culture, self-sufficiency, and its ability to attract and retain donors and staff."[128]

A very early example of this is Greyston Bakery, which is a for-profit subsidiary of the Greyston Foundation, in New York City. Greyston's mission is to employ and train people who are commonly perceived as 'unemployable', including former drug addicts, prisoners, and recovering alcoholics. Greyston is widely known for its open hiring policy; that they will hire 'anyone who walks through the door'. Selling 4 million pounds of baked goods to 2,500 customers annually, they have been helping the 'unemployable' earn a living wage through dignified work since 1982 and are still going strong.[129,130]

For-profit subsidiaries also serve a very practical function in some countries, as they can keep NFPs from losing their tax exemptions for doing 'unrelated business' - business considered to be disconnected from their mission. For instance, under U.S. law, Housing Works might have a difficult time explaining how selling coffee is related to their mission to help people suffering from AIDS and homelessness. To avoid any problems, they started a subsidiary and, as a for-profit, their subsidiary doesn't have to have a social mission at all. The only risk with for-profit subsidiaries is that they can legally be sold to for-profit companies, so they can technically go from being extensions of NFPs to being extensions of for-profit companies overnight.

---

126 Housing Works website (2018) 'Housing Works Bookstore Café' page. (https://www.housingworks.org/locations/bookstore-café)

127 FEI website (2018) 'History of FEI Behavioral Health' page. (http://www.feinet.com/aboutus/history-fei-behavioral-health)

128 Massarsky, C.W. & Beinhacker, S.L. (2002) *Enterprising Nonprofits: Revenue Generation in the Nonprofit Sector*. Englewood Cliffs, NJ: Yale School of Management and The Goldman Sachs Foundation Partnership on Nonprofit Ventures.

129 Greyston website (2018) 'Mission and History' page. (https://greyston.org/about-us/mission-and-history/)

130 Kickul, J. & Lyons, T.S. (2012) *Understanding Social Entrepreneurship: The Relentless Pursuit of Mission*. New York, NY: Routledge.

Many people were outraged when this happened with National Geographic magazine. For all 127 years of the magazine's history, National Geographic Media was a for-profit subsidiary of the National Geographic Society, a well-known nonprofit. In September of 2015, the National Geographic Society sold 73 percent of their ownership of National Geographic Media to 21st Century Fox, one of the largest for-profit media corporations in the world.[131] Readers far and wide took to social media to protest the buy-out, worried about how it would affect the content of the magazine. Perhaps it would have been better if National Geographic Media were registered as an NFP subsidiary from the start, rather than a for-profit subsidiary, to ensure that a for-profit company could never own it. After all, educating people through the magazine's content is a perfectly acceptable social mission for an NFP to have.

As the example of National Geographic illustrates, it is wise for an NFP to start a not-for-profit subsidiary with a different social mission, rather than a for-profit subsidiary, when possible. This enables an NFP to enjoy the advantages that a for-profit subsidiary would offer, without the risk of it one day being privately-owned and operated for private profit.

A great example of an NFP subsidiary is Tender Funerals, in New South Wales, Australia. Tender Funerals itself is a not-for-profit, and it is owned and run by another NFP called Our Community Project.[132] It's wonderful to know that nobody is making a financial gain from Tender's funeral services. Instead the company uses any surplus to provide better services to the community. This begs the question: shouldn't every funeral service provider be not-for-profit?[133]

# Boards

Although an NFP has no owners, it must have a board (also called 'committee' in some countries) that holds the business accountable for using its resources to accomplish its mission.

For-profit companies often have boards, too, but there are some crucial differences between for-profit and NFP boards. First of all, the mission of for-profits is to maximize profit, so boards hold the company accountable for maximizing profit, or in the case of triple bottom line businesses, balancing profit as a goal with social and environmental concerns. In fact, most corporate boards are primarily tasked with protecting the shareholders' interests.[134, 135] An NFP board protects the interests of the community and holds the company accountable for working towards its stated social and environmental goals. It serves as a light house for the organization, helping it steer clear of trouble. Secondly, for-profit board members are usually paid, as part of the for-profit mentality is to reward everything monetarily, while most NFP board members are not paid.[136]

---

131 Sessa-Hawkins, M. (2015) 'Rupert Murdoch's 21st Century Fox buys National Geographic media', *PBS News Hour*, September 9. (https://www.pbs.org/newshour/science/national-geographic-fox-enter-profit-venture)
132 Our Community Project website (2016) Our Community Project Annual Report 2015- 2016. (https://docs.wixstatic.com/ugd/c48554_d64015f82b6e44768c0e1139b3bdab42.pdf)
133 Our Community Project website (2018) 'Tender Funerals' page. (https://www.ourcommunityproject.org.au/tender-funerals-back)
134 Investopedia website (2018) 'Board of Directors – B of D' page. (https://www.investopedia.com/terms/b/boardofdirectors.asp)
135 Epstein, M.J. & McFarlan, F.W. (2011) 'Nonprofit vs. For-profit Boards: critical differences', *Strategic Finance*, March: 28.

The board is a vital part of any NFP, as it helps the organization maintain integrity, accountability and transparency. With the purpose motive driving board participation and decision-making, as opposed to the profit motive, NFP boards are likely to take multiple perspectives into account. These boards can also encourage a more democratic way of managing the business, as they can break the traditional 'owner - manager - worker' hierarchy that has been so prevalent in business structures for the past few hundred years. Not-for-profit boards can work directly with managers *and* with workers, to decide what's best for the organization.

It might sound like the NFP board adds a layer of bureaucracy that could slow a business down, but boards do not typically make many management decisions. The management team of the NFP takes care of most decisions, just like in a for-profit company. A board might only convene a few times per year, just to check in on the NFP's progress and make sure it is sticking to its commitments. Boards have the duty of approving an NFP's budget and any major changes to the budget and they also play a role in making major decisions that will change the structure of an NFP, such as merging with another NFP or starting a subsidiary company.

Although most countries' legislation allows for paid staff members to be on the board, there is also a requirement for a certain percentage of board members to be non-paid and non-staff, in order to maintain a degree of objectivity in decision-making processes. In this regard, there are strict rules about conflicts of interest. Any board member who stands to benefit from a decision must declare their conflict of interest and abstain from voting on the issue at hand. Board members are expected to hold each other accountable for declaring such conflicts of interest. As most NFP board members are volunteering their valuable time for a cause, they tend to take their responsibilities in this position quite seriously. As such, the NFP board means that there's less need for government or outside authorities to act as the conscience of the company or to balance out the self-interest of the owners and managers.

That's not to say that a board is immune to corruption, which we'll address later, but the requirement to have a mission-driven board is a critical difference between the for-profit and NFP business models, as it adds a layer of accountability.

# Raising Capital

Since the advent of capitalism, the main way to start or grow a business has been to raise capital from private investors. Investors typically decide whether to invest in a startup business based on how much money they believe they'll get back from it in the form of equity, interest, and dividends. In the business world, this is known as ROI (Return on Investment). This sort of investment is something that the owners and managers of companies can draw on at various stages of the business's development, depending on their needs for greater financial liquidity or their desire for more personal wealth. If they want to grow, they can raise more capital by approaching more investors with a plan for how they will deliver a high return on investment.

[136] Ibid

NFPs can't raise capital in the ways that for-profit businesses traditionally do, and for-profit subsidiaries that are wholly-owned by NFPs can't raise capital in this way either. They don't have shareholders or owners other than maybe an NFP parent organization, so they can't offer equity, shares or dividends in exchange for investment without the NFP parent selling part of its ownership, putting the whole organization at risk of drifting from the social mission. This is why raising capital investment has been a massive advantage that for-profits have traditionally had over NFPs. That is, until recently.

The landscape of the economy has changed tremendously over the last couple of decades and NFPs no longer face the same barriers they used to in the market. Only 20 years ago, it would have been quite rare to see businesses starting up as NFPs. However, the Age of Technology has changed everything and NFP startups are on the rise.[137] The internet has not only helped level the playing field in terms of capital-raising, but it has also significantly brought down the costs of starting and running a business, which means that NFPs and for-profits alike no longer have to spend huge amounts of money to start and build a successful business.

When planning how to start an NFP, grants, donations, membership fees and volunteers often come to mind. Grants from foundations or government agencies and donations of money, time, and other assets from the community can provide a great financial base on which to build a successful NFP business, as long as there is a plan to generate enough revenue to become financially independent after the startup period. This is similar to a for-profit business that relies on an initial injection of capital, investment and loans, to get started, but seeks to stand on its own two feet as soon as possible.

In addition to using donations, grants, membership fees and volunteers to start up, many new NFP businesses take loans from banks and private investors. This is a perfectly acceptable way of raising capital in the NFP world, as there is a cap on how much a private investor will get back on his or her loan – a set interest rate. The interest is considered compensation for the time it takes the company to pay back the loan (like a fee) as well as compensation for the loss of the money's value over time due to inflation (a dollar declines in real purchasing power over time). Interest is best thought of as the price of the loan, and is not a private gain for the lender. Many places even have usury laws that place restrictions on the rate of interest allowed in lending, so that 'price' is not exploitative. In many U.S. states, for instance, it is not legal to give consumer loans with an interest rate of over 10%.[138]

Loan repayments (including interest) are considered a business expense, since it is a set amount that is calculated and budgeted at the beginning of the year. A dividend cannot be calculated and budgeted at the beginning of the year, as it depends on the amount of profit generated by the company[vi]. That's why a dividend can never be considered a business expense and is instead considered private distribution of profit[vii]. Because the amount of a dividend may change at any time, there is a speculative aspect to it, and inherent in that is an ethic of 'never enough'.

---

[137] Robehmed, N. (2013) 'A New Nonprofit Model: Meet the Charitable Startups', *Forbes*, December 16. (https://www.forbes.com/sites/natalierobehmed/2013/12/16/a-new-nonprofit-model-meet-the-charitable-startups/#7e8ce8787aab)

[138] Lectic Law Library website (2018) 'State Interest Rates and Usury Limits' page. (https://www.lectlaw.com/files/ban02.htm)

In addition to traditional loans, new forms of capital-raising are increasingly available to NFPs. Instead of a large financial return on investment, NFPs promise a social return on investment (SROI). This means that investors will benefit in social ways from their investment. Their investment will benefit under-served segments of the community and/or the environment, and thus will benefit everyone in the community, as well as the investors themselves.

Although much innovation is happening in the sphere of NFP investing, a certain set of principles apply to all NFP investment and capital-raising:

- There can be no private equity (no shares, no stock, no ownership).
- No private distribution of profit or assets is allowed.
- Repayment of investment must be budgeted for as a business expense, not taken from profits.
- The amount an investor receives back is fixed (either by an absolute number or by an interest rate), so it is non-speculative investment.

The Internet plays a very important role in opening up new opportunities for NFP startups to access capital from a larger array of sources. Crowdfunding provides a way for people to draw on global networks of support for their cause. A startup NFP can run a crowdfunding campaign to raise money from hundreds, thousands or even millions of ‘social investors’ (i.e. – those interested in a social return on investment) who will fund the startup, often in return for a reward, also known as a ‘perk’.

Many startups entice crowdfunders by pre-selling their product or service (the perk) – this is in part how we produced this book you are reading. A consultancy startup might offer a professional consultation for a crowdfunding investment of $150, or a startup café might offer a punch card for ten coffees as a perk for a $50 investment. As an NFP startup, offering your goods or services in exchange for crowdfunding investment can be a great way to pre-sell your products and grow your market reputation from the get-go. There is a much larger aspect of crowdfunding than just taking donations - it is about reciprocity.

This is what we experienced with the crowdfunding campaign for the book you’re reading right now. We raised nearly $21,000 to fund the research, writing and design of the book, but we also got the moral support of all our investors (i.e., our crowdfunders). They helped spread the word (advertising the book by word of mouth) for the three years between the time of the crowdfunding campaign and our publishing date. It was also very inspiring and motivating for our research and writing team to feel that we had a whole global network of support for our book, even before the first page was written.

It’s important to note that crowdfunding can be done in a for-profit or an NFP way. There are numerous examples of for-profit startups using crowdfunding to start their company and then the owners reap the profits. This was the case with Oculus Rift virtual reality headsets. Oculus Rift raised $2.4 million of capital investment through a crowdfunding campaign.[139] It rewarded contributors with headsets among other perks. However, , a year and half later, when Facebook bought the company for $2 billion many of Oculus Rift’s crowdfunders felt a bit used, as they didn’t receive a penny.[140] One calculated that he could have gotten more than $43,000 from the sale to Facebook if he had had equity in the company, but instead he was

[139] Jeffries, A. (2014) ‘If you back a Kickstarter project that sells for $2 billion, do you deserve to get rich?’, The Verge, March 28. (http://www.theverge.com/2014/3/28/5557120/what-if-oculus-rift-kickstarter-backers-had-gotten-equity)
[140] Ibid

left with the, by then, obsolete virtual reality headset that was a perk for contributing (Ref: Ibid). This sort of experience has led people to create equity-based crowdfunding, which offers shares or equity in the company in return for contributions. Equity-based crowdfunding is, of course, a form of for-profit investment and does not apply to NFPs.

Non-equity based crowdfunding is an ideal tool for NFP startups, though. Companies that are registered as NFPs are often seen as more trustworthy than for-profits, because they are so clearly mission-driven, so they're more likely to appeal to crowdfunders who are looking to put their money into something that makes a positive contribution and aligns with their ethics. Crowdfunders of NFPs can be sure that all of their investment has gone to a mission they believe in, not to lining the pockets of a for-profit's owners.

A famous recent example of a successful, large-scale crowdfunding campaign to raise capital for a NFP startup is that of Ocean Clean-up. In 2012, a young man from the Netherlands named Boyan Slat gave a TED Talk about the technology he'd been innovating with others to clean up plastic pollution in the oceans, something that people all over the world are experiencing. Millions of people have been emotionally affected by photos and videos on the Internet of sea animals suffering from eating too much plastic or getting caught in plastic nets and bags, so this was is a very popular innovation. Boyan and his colleagues were able to raise $2 million in capital from their crowdfunding campaign to turn their idea into an NFP business. Some of the perks they offered to their crowdfunders included a movie about plastic in the oceans, a reusable shopping bag, and a children's book about saving the ocean.

Ocean Cleanup has a business plan for how to make Ocean Cleanup financially self-sufficient, using revenue gained from selling the plastic they collect from the oceans to fund the good work they're doing. They don't pay the board, and employees are paid a decent, but not extravagant wage. This is a company that truly embodies the spirit of NFP enterprise: a mission that benefits a very large community, an ethic of enough, and the investing of all surplus back into their mission. That's why they were able to raise so much startup capital for their endeavor.[141]

Novel ways of raising capital, like crowdfunding, have allowed a whole new generation of NFP enterprises to emerge, and these young NFPs are in turn coming up with more new ways of raising capital. There is now an incredible amount of creative thinking and social innovation happening in the area of NFP capital-raising.

The Internet is increasing accessibility to loans as well. Various online platforms and websites have popped up that enable startup companies and projects to connect with investors who are looking to lend money to socially-oriented projects at a reasonable interest rate[viii]. This is known as peer-to-peer lending (or P2P lending). Cutting out the 'middle man', like banks and other financial institutions, P2P lending allows loan transactions to happen much more quickly and across much larger geographical distances. An activist in Belgium can go to a P2P lending platform to get funding for a project they want to do, and can connect with an investor halfway around the world, in Sydney. They negotiate the terms of the loan; the interest rate, payment schedule and what should happen in case of late payments or default. And once they reach an agreement, voila, the loan is made, the activist has enough money to start their project and the lender will be receiving loan repayments in a few months with a fair interest rate.

---

[141] Ocean Cleanup website (2018) 'Foundation Details' page.
(https://www.theoceancleanup.com/foundation-details.html)

As NFP businesses grow in number and strength, so do the organizations that are set up to help NFPs access capital. The Nonprofit Finance Fund is a veteran in this field and continues to lead the way. Since 1980, they have served thousands of NFPs throughout the U.S., by providing over $700 million in loans and access to other forms of finance such as grants, tax credits and capital, in support of $2.3 billion of projects for thousands of organizations across the nation.[142]

There is a general trend for NFP businesses to use debt-based capital, whereas for-profits tend to seek equity-based capital. In both forms, the company owes the investor something. In debt-based capital, the business owes the lender the principal amount of the loan or bond until it's paid off, and in equity-based capital, the business actually sells a piece of itself for capital investment. We argue that debt-based capital is healthier, because it allows the company to maintain its focus, values and ethics, without having to take new owners' or shareholders' considerations into account. Debt-based capital is temporary, while equity-based capital is permanent.

The community bond, also known as a social impact bond or social benefit bond, is another powerful debt-based capital-raising instrument for NFPs. Community bonds are a lot like small loans[ix], but a company gets a lot of these small loans from multiple sources in the community with a low interest rate. A company can sell community bonds to many members of the community for relatively small amounts of money. Then, over time, the company will pay back the principal amount (the original amount the bond was bought for) – say $500 - plus interest. Community bonds allow a startup business to tap into the support of the community by asking for smaller sums from a large number of people. This contrasts with traditional ways of capital-raising, typically asking for a large sum of money from just a few sources, such as banks or wealthy investors.

Glas Cymru, a NFP company in Wales, used community bonds to buy Welsh Water in 2001. Over a period of 18 months, Glas Cymru raised £1.9 billion in bonds, allowing it to acquire the Welsh Water company from Western Power Distribution, a for-profit corporation. They paid down the bonds in 6 years, providing a small financial return and a big social return to their community investors.

Some of the successes of the group to date include:

- "some £3 billion invested to improve drinking water quality, environmental protection and customer service – at no cost to the taxpayer
- £150 million returned to customers in the form of 'customer dividends' and some £10 million of support for disadvantaged customer groups via social tariffs and an assistance fund, and
- lower average customer bills in real terms than in 2000, in part due to the best record in the sector in cost reduction and improved efficiency."[143]

It's important to remember that Glas Cymru's investors don't hold equity or shares in the company, as it is an NFP. A bondholder is not the same thing as a shareholder. A shareholder usually does have equity in the company, meaning he or she essentially owns part of the company. A bondholder, like a lender, can be financially compensated for the time they've lent their money away, via interest payments, but the bondholder cannot gain from the company's profits, where shareholders can with shares. The bondholder is paid from the NFP's budget, not from its profits.

---

142 Nonprofit Finance Fund website (2018) 'Financing' page. (http://www.nonprofitfinancefund.org/financing)

143 Glas Cymru website (2015) 'Company information' page. (http://www.dwrcymru.com/en/company-information/glas-cymru.aspx)

The fact that dividends depend on the profit made each year or quarter means there is a speculative aspect to the dividends paid out to shareholders. The same goes for owners with equity in a company, who are hoping that the company will be more valuable every year, so that their personal stake in the company is also worth more. The extreme of this speculative tendency in for-profit capital-raising is venture capital. Venture capital is a form of investment in which an investor or a group invests a sum of money into a company, with the explicit expectation to profit as much from the company as possible in as short amount of time as possible. It can be very risky for both the investor and the investee, as both stand to lose a lot of money.

NFPs have lots of options when it comes to raising capital, including loans, crowdfunding, peer-to-peer lending, and community bonds, in addition to more traditional forms of nonprofit finance, like grants, donations and volunteers[x].

Not-for-profit businesses can use any combination of these different strategies to meet their capital needs. As more social innovation continues to happen in order to make capital-raising easier and more accessible to these new businesses, further advances in this area are expected.[xi]

# Taxation

A business can be NFP in almost any sector, anywhere in the world, but it may or may not get tax exemptions. For instance, only about half of all nonprofits in the U.S. are 501(c)3 certified, the most common kind of tax-exempt status in that country.[144] Tax benefits for NFPs come in many forms. They can be exempt from paying corporation taxes, sales and use taxes, payroll taxes, and property taxes, depending on a region's tax laws.

Tax exemptions for NFPs can be a tricky thing, depending on local legislation. For instance, in the U.S., if an NFP earns more than 20 percent or so of its revenue from business activities that aren't directly related to its stated social purpose, the NFP might lose its tax-exempt status.[145]

In Canada, an NFP risks losing its tax exemptions if it makes a large amount of profit or accumulates a large sum of money, which the Canada Revenue Agency might interpret as being profit-oriented. Instead, Canadian law states that surplus for NFPs is acceptable if it is unintentional and unanticipated.[146]

In most cases, even if an organization loses its tax exemptions, it can maintain its legal NFP status, because tax-exemptions are an 'add-on' to NFP legal status. NFPs that engage primarily in social business activities might be happy to be taxed like for-profit businesses, because gaining tax-exempt status might put too many restrictions on how they operate or might even prevent them from engaging in their business activities.[147]

---

144 National Center for Charitable Statistics (2018) 'National Taxonomy of Exempt Entities'. (http://nccs.urban.org/classification/national-taxonomy-exempt-entities)
145 The Editors (2011) Does My Nonprofit Need to Pay Tax? Understanding Unrelated Business Income Tax, *Nonprofit Quarterly*, December 25. (https://nonprofitquarterly.org/2011/12/25/does-my-nonprofit-need-to-pay-tax-understanding-unrelated-business-income-tax/)
146 Canada Customs and Revenue Agency (2001) IT-496R Income Tax Act: Nonprofit Organizations, August 2.

In the U.K., NFPs are exempt from taxes on income and gains as long as that money is used for charitable purposes,[148] but Canada and Australia have their own unique set of criteria for tax-exemptions.[149][150]

Is a tax exemption an unfair advantage, since NFPs are starting to compete more in the market with for-profits? Why do NFPs get tax benefits that for-profits don't? And should they really be entitled to these benefits?

These questions have a surprisingly simple answer. Not-for-profit companies reduce the need for taxation in the first place, by fulfilling social needs that the government would otherwise have to fund. Blue Star Recyclers, for instance, have calculated that they saved the taxpayers $233,353 in 2015 by employing people with disabilities.[151] Imagine what would happen to the millions of people who are benefitting from the NFPs we mention throughout this chapter - people struggling with addiction, homelessness, illness, disabilities and financial hardship – if there were no NFPs. Either the whole world would be a lot worse off for it, or taxes would have to be a lot higher so governments could address those peoples' needs.

In this light, it only seems fair that NFP businesses should not have to pay the same taxes as for-profit companies, which aim to financially benefit owners and investors through profit-maximization. In contrast, the wider community benefits from the financial surplus of NFPs. Thus, NFPs very much deserve their financial advantage.

NFP tax exemptions are even more justifiable considering the enormous subsidies and tax cuts without which most for-profits would be unprofitable. Often called 'corporate welfare', large companies all over the world in the agricultural, banking, finance, shipping and fossil fuel industries receive incredible amounts of government assistance in the form of subsidies and tax breaks.[152] Not only do they enjoy tax cuts and subsidies, they also have the resources to seek advice on exploiting loopholes to avoid paying their fair share of tax.

This scenario is not restricted to large for-profit companies. Thirty-four percent of U.S. for-profit banks get a tax advantage through their Subchapter S Status for small businesses.[153] The Credit Union National Association estimates that concessions to small for-profit banks resulted in nearly $1 billion of foregone revenue for the U.S. Treasury in 2016, while

---

147 Community Enterprise Law website (2018) Nonprofit organizations; 'Nonprofits without tax exemption' section. (http://communityenterpriselaw.org/forming-community-enterprise/nonprofit-organizations-in-the-sharing-economy/)
148 U.K. government website (2018) 'Charities and Tax' page. (https://www.gov.uk/charities-and-tax/tax-reliefs)
149 Canadian Revenue Agency website (2018) Income Tax Act (R.S.C., 1985, c. 1 (5th supp.)) (http://laws-lois.justice.gc.ca/eng/acts/I-3.3/section-149.html)
150 Australian Tax Office website (2018) Income tax exempt organisations. (https://www.ato.gov.au/non-profit/your-organisation/do-you-have-to-pay-income-tax-/income-tax-exempt-organisations/)
151 Blue Star Recyclers (2016) 2015 Annual Report, Colorado Springs, CO: Blue Star Recyclers.
152 Klein, N. (2014) *This Changes Everything.* Camp Hill, PA: Simon & Schuster.
153 Schenk, M. (2017) 'The $1 billion bank benefit', CUNA News online. (http://news.cuna.org/articles/112361-the-1-billion-bank-benefit)

privately distributing profits and without providing the substantial member benefits credit unions provide.[154]

Not-for-profit tax exemptions can also act as another incentive for NFPs to remain as accessible, effective and transparent as possible. In exchange for these tax exemptions, there is a higher expectation for NFPs to disclose how they use their resources. An NFP owes the public documentation of how it used the income that society forgoes the right to tax.[155]

In addition to tax exemptions, some governments offer NFPs other financial benefits. For example, NFPs in Australia get reductions on the annual business registration fee they have to pay.[156] And in the U.S., some states give nonprofit organizations immunity from certain legal liabilities and other states limit NFP liability by enacting a damage cap, so that NFPs aren't as vulnerable as for-profits with regards to certain legal issues.[157]

# NFP Enterprises Come in Many Different Forms

Creating a distinct category for NFP enterprise might make it seem like a homogenous sector. In reality, there is an incredible amount of variety in the NFP business world. In addition to traditional nonprofits trying to become more self-sufficient, we're seeing more people starting businesses and cooperatives as NFPs. **Old nonprofit models are shifting and new NFP models are emerging, making for a diverse and exciting new space in the economy.**

Not-for-profit businesses around the world meet the needs of hundreds of millions of customers and beneficiaries everyday through the wide variety of goods and services they sell.[158] You might be surprised at how many of these businesses you're already familiar with.

A world-renowned, older not-for-profit enterprise is the international YMCA, which provides local fitness and leisure centers, as well as childcare and teaching services. The YMCA is global, but YMCA Canada is a great example of how the business model works.

"(T)he YMCA provides vital community services that are having a positive impact on some of Canada's most pressing social issues—from chronic disease to unemployment, social isolation, poverty, inequality and more."

In 2014, all 50 member associations of the YMCA in Canada employed over 22,000 people and provided services to over 2.3 million people. And they earned 87% of their income through membership dues, course fees, and government contracts. Only 4% came from contributions and fundraising[xii]. [159]

---

154 Ibid

155 (McLaughlin, p. 123 (ask Donnie))

156 Australian Securities and Investments Commission (2018) 'Special Purpose Companies'. (http://asic.gov.au/for-business/registering-a-company/steps-to-register-a-company/special-purpose-companies/)

157 Legal Information Institute (2018) Non-profit organizations, Cornell Law School website. (https://www.law.cornell.edu/wex/non-profit_organizations)

158 Salamon, L. et al., (1999) *Global Civil Society: Dimensions of the Nonprofit Sector*. Baltimore, MD: The Johns Hopkins Center for Civil Society Studies.

159 YMCA Canada website (2015) Annual Report 2014: Further together. (http://ymca.ca/CWP/media/YMCA-National/Documents/Annual%20Reports/Annual-Report-2014-EN.pdf)

Goodstart Early Learning Center is one of Australia's most well-known childcare providers and it's an NFP (Ref: Goodstart website). In 2014, they operated 644 centers, caring for 72,500 children, and generated a surplus of $40 million Australian dollars (Ref: Goodstart Annual Report 2014, p.8 & 16). Because they're NFP, this surplus allowed them to help 8500 children who are considered vulnerable (financially or otherwise), and keep cost increases below the industry average (Ref: Goodstart Annual Report 2014, p. 8)

In healthcare in the U.S., world-renowned hospitals like the Mayo Clinic and Cedars Sinai are NFP. The Blue Cross Blue Shield insurance franchises in many states are also NFP (including in Michigan, Massachussetts, Arkansas, Maryland, and Minnesota).

The U.S. Public Broadcasting Service (PBS) and National Public Radio (NPR) are both not-for-profit businesses, generating income through charging their member stations fees for the content they create.

As incongruous as it may sound, there are even Fortune 500 NFP businesses. The United Services Automobile Association (commonly known as the USAA) provides insurance, auto, and financial services to its 11.2 million members. Its net worth went from $20 billion in 2011 to $27.8 billion in 2015 and any surplus it generates is used by the USAA to better serve its members.[160]

Bosch, the world-renowned home appliances company (not to mention, a Fortune 500 company) is 92% owned by the Robert Bosch Stiftung GmbH, a nonprofit foundation in Germany, so it can be seen as 92% NFP. (The rest of the ownership rights are held by the Bosch family and the Bosch corporate managers.) In 2013, the company reported a $3 billion profit,[161] the vast majority of which went into better serving its customers and its foundation's social mission of furthering international understanding, welfare, education, the arts and culture, and research and teaching in the humanities, social sciences and natural sciences.[162]

There are also smaller, less well-known, but impactful NFP businesses, such as TROSA (Triangle Residential Options for Substance Abusers) in North Carolina, which operates several service-related businesses, including lawn care, a furniture shop and the largest independent moving company in the region. These subsidiary businesses raise funds for its primary mission, which is operating a two-year residential addiction treatment program. The businesses also provide the backdrop for their rehabilitation programs, such as vocational training and mentoring, by employing program participants.[163]

In Belgium, De Kringwinkel is a chain of more than 100 second-hand shops which serves the triple purpose of selling quality second-hand products at very reasonable prices, keeping perfectly good items out of landfills, and providing employment to people who might not otherwise be able to work.

"Over 80% of the more than 4,500 people working in the organisation are longtime unemployed or have a limited education level. De Kringwinkel invests strongly in its image and communication and evolved from a 'dusty' unpopular image to a hip, young and original image. De Kringwinkel proves that nonprofit can be very cool."[164]

---

160 USAA website (2018) 'Financial strength' page. (https://www.usaa.com/inet/pages/about_usaa_corporate_overview_financial_strength)
161 Fortune 500 list (2015) (http://fortune.com/global500/2013/robert-bosch-gmbh-131/)
162 Bosch website (2018) 'Robert Bosch Stiftung' page. (https://www.bosch.com/our-company/sustainability/society/robert-bosch-stiftung/)
163 TROSA website (2015) 'About us' page. (http://www.trosainc.org/about-us)

Not only can NFPs be found in a very wide variety of sectors, but they can also be found throughout the world, and they are certainly not limited to the global North. Here is a taste of the diversity we found in conducting our research in English (imagine how many more there must be all over the globe, accessible only by other languages!).

Grameen Shakti is an excellent example of an NFP renewable energy company in a 'developing' country. Based in Bangladesh, Grameen Shakti has installed more than 1.6 million solar home systems, with all of their services benefiting approximately 18 million people.[165]

Established in 2001 in Pakistan, Akhuwat is an NFP that provides a range of services, including university-level education, business management training and a cloth bank, but it is best known for its social impact through microfinance. It offers interest-free microloans to people facing economic hardship, giving them guidance and training to build a livelihood for themselves. They charge a small fee for loan applications and they have a program in which former borrowers are encouraged to give back to the organization after they have achieved financial independence. Akhuwat's services have benefitted over 1 million families in Pakistan and the business generated a financial surplus of more than 365 million rupees in 2016, which will allow them to make even more interest-free loans in the future.[166, 167]

## NFP Social Enterprises

In addition to traditional nonprofits generating more of their own income, an increasing number of new companies are starting up as NFPs. This is largely a response to worsening socio-economic and environmental conditions, which has spawned an increasing demand among consumers for ethical products and services as well as a growing desire for purpose-driven work.

People often ask us, "Isn't social enterprise what you're talking about?" It's true that many social enterprises are NFP, but as we've mentioned before, a fair number of social enterprises are for-profit and in some countries the social enterprise landscape is very confusing in terms of differentiating between for-profits and NFPs. It's important to remember that the term 'social enterprise' doesn't tell you if a company sees profit as a goal or if it's allowed to distribute profit to individuals. This distinction is vital when we're talking about whether a business model is truly generative or not. Therefore, we are careful to distinguish NFP social enterprises as being in a class of their own.

In the UK, the Community Interest Company (CIC) limited by guarantee is a not-for-profit social enterprise model. CICs in the United Kingdom can be either limited by guarantee (the form that is NFP) or limited by shares (the form that is for-profit because it has shareholders

---

164 De Kringwinkel's website (2015) 'English' page. (https://www.dekringwinkel.be)

165 Grameen Shakti website (2015) (http://www.gshakti.org/index.php?option=com_content&view=category&layout=blog&id=54&Itemid=78)

166 Akhuwat website (2016) 'Impact' page. (http://www.akhuwat.org.pk/overview/#impact)

167 Deloitte Yousuf Adil Accountants (2016) Akhuwat Financial Statements for the Year Ended June 30, 2016. (http://www.akhuwat.org.pk/wp-content/uploads/2017/10/AuditReportfortheyearendedJune302016.pdf)

and can distribute profit). Almost 80 percent of CICs have chosen to go the not-for-profit route.[168]

Co-Wheels Car Club is one such CIC providing car-sharing services all over England and Wales. It is the largest car club in the UK and its mission is to help its members save money, reduce car ownership and create environmental benefit.[169]

Bread Share Community Bakery in Edinburgh, Scotland, is another not-for-profit CIC. Their mission is to enhance their community by providing high-quality, healthy baked goods, as well as a place for social interaction.[170]

The number of community interest companies on the public register on the 30th of November 2013 was 8666 and the Regulator of CICs has said that the number of community interest companies has been growing exponentially.[171]

All of these companies are purpose-based, reinvest 100% of their profits into their missions, and have an 'asset-lock', so nobody can personally gain from the assets if they dissolve.

## NFP Cooperatives and Mutuals

Many, but not all, cooperative and mutual businesses are NFP. Cooperatives and mutuals exist to benefit their members. Cooperatives have a special focus on democratic management by members, but in most places, there is no significant legal difference between mutual business models and cooperative models. Some mutuals and cooperatives are owned by members and distribute profit to members,[172] which makes them for-profit. Others are non-distributing and have asset-locks, which makes them NFP. In Australia, for instance, about three-quarters of cooperatives are established as NFPs.[173]

Although most other countries don't seem to keep official tabs on whether cooperatives are distributing or non-distributing, we feel comfortable saying that it's likely that most cooperatives in the world are NFP due to the prominence and proliferation of consumer cooperatives. The consumer co-op is an NFP model that is popular all over the world[xiii]. Most commonly found in the realms of retail, banking, electricity provision, child care, education and medicine, these are businesses run by and for the consumer members. The members might get a 'patronage dividend' from the profit at the end of the year, proportional to the amount they spent there, but the amount of the 'dividend' will never be more than what the customer has spent at the store during the year, because the 'dividend' is proportional to the amount they spent. For instance, a customer who bought $800 worth of products over the year will get more back than someone who spent $200, but neither of them will ever get anywhere near the amount they put into the co-op; otherwise the co-op would go out of business, losing more money than it makes.

If this sounds confusing, think of it this way. Where did the profits come from? The customers paying for the company's services all year. So it's not a dividend, in the traditional sense, if the people who receive the dividend are the same people who contributed to the profit as customers in the first place. It's really just a refund or rebate for the products and services they have already bought. In effect, the consumer co-op's profits allow it to make the products that customers bought cheaper, which is the mission of the co-op: to provide affordable products and services. Because the consumers are the ones receiving 'patronage dividends', there is no profit motive or drive to maximize profit, as this would just entail the consumers paying more.[xiv]

---

168 (Ref: CIC Regulator)
169 (Ref: Co-Wheels Car Club website)
170 (Ref: Bread Share website)
171 (Ref: Regulator of Community Interest Companies Annual Report 2011/2012)
172 Ref: Cooperative models – UK Parliament
173 (Ref: Australian Parliament website)

Examples of consumer food co-ops around the world include Bios Coop in Thessaloniki, Greece; the Alfalfa House in Sydney, Australia; Hansalim in South Korea; SaludCoop in Colombia; the Ashland Food Co-op in Oregon; and Coop Danmark, in Denmark.

Not all consumer co-ops are food-related - they have also emerged as a way for distributing electricity in areas where making a profit would be difficult, like rural Wisconsin. The Central Wisconsin Electric Cooperative was established in 1938 by local citizens to bring electricity to rural communities in Wisconsin, and this consumer cooperative puts any and all profit right back into serving those communities. As a cooperative, every member has a say in how the business is managed.[174]

Another business model in the sphere of consumer cooperatives is the credit union[xv]. Credit unions are member-run banks whose mission is to provide financial services to its members. As with all consumer co-ops, 'dividends', rebates and refunds that members might receive from a credit union at the end of the year never exceed the amount of money that the members have put into the credit union. Credit unions have experienced an impressive rise in popularity in recent years, especially in the U.S. where membership growth rates have been higher every year since 2010.[175] In 2015, the U.S. had a total of 104 million credit union members, which is nearly 1/3 of the entire American population, according to the United States Census.[176]

This is probably because customers usually get better deals from credit unions than from mainstream banks, which is especially true for lower income members.[177] Banks collect 2.5 times more in fees from low-balance checking accounts annually ($218 on average) than on high-balance accounts ($90 on average), while credit unions collect an average of $80 on low-balance accounts and $42 on high-balance accounts.[178] This is an embodiment of the official motto of the credit union industry, which captures the NFP business ethic perfectly: *Not for profit, not for charity, but for service.*[179] Since the start of the economic downturn in 2008, most for-profit banks have pulled back from small business lending, while small business loans from credit unions have continued to grow each year.[180]

Hope Federal Credit Union in Jackson, Mississippi, is an excellent example of how a credit union can maximize social outcomes because it is not profit-oriented.

174 (Ref: CWEC website)
175 (Ref: CUNA Mutual report)
176 (Ref: CUNA Mutual report, US Census)
177 (Ref: CUNA powerpoint)
178 (Ref: **CUNA powerpoint**)
179 (Ref: Firefighters First Credit Union website)
180 (REF: CUNA powerpoint slide 13)

"In 2011, Hope made 85% of its business loans to minority and women-owned businesses, loans supported through community and economic development programs, loans to businesses in distressed communities and loans for community facilities. What's more, nearly 75 percent of its loans were made in high-poverty/low-income communities – a rate nearly 30 percentage points higher than the average for regional banks."[181]

Hope has more than 26,000 members and it offers a viable alternative to 'unbanked' and 'underbanked' citizens, many of whom have struggled significantly with debt owed to for-profit institutions.[182]

It's not only in the U.S. that credit unions are doing well. As of 2014, there were 57,000 credit unions in 105 countries with 217 million members and assets totaling $1.8 trillion.[183]

Consumer cooperatives can also take the form of regular retail shops. Mountain Equipment Co-op in Canada is one of the country's largest consumer cooperatives, providing outdoor clothing and equipment. It has about 4.3 million members, 47,523 of whom voted in the 2014 co-op elections.[184] That same year, the co-op generated 336 million Canadian dollars in sales.[185]

Perhaps even more interesting are examples of the consumer co-op model being applied to senior living. Cooperative Services Inc. is a network of senior living cooperatives, with 58 different locations in four different states in the U.S. Resident-members of CSI actively participate in managing their co-op. They never have to wonder where their rent money goes because they write the annual budget themselves, cooperatively.

"By design, CSI co-ops are not-for-profit, so at the end of the year, any money left in the budget stays with the individual co-op. The resident members decide how to use any surplus cash for the betterment of their co-op."[186]

And this trend is not restricted to the English-speaking world. Not-for-profit cooperatives are notably abundant in Japan. In 2007, Japanese consumer co-ops generated a total of $22 billion in revenue and had 17 million members.[187] There is also a growing movement of university, medical and housing co-ops in the island nation.[188]

In the category of NFP mutuals, which are essentially cooperative insurance companies, there is another Fortune 500 company: Nippon Life, which has 10 million policyholders in Japan, South Korea and the U.S.[189] In 2015, it had over $500 billion in assets and profits totalling $5.15 billion.[190]

---

181 (Ref: HOPE's website.)

182 (Ref: hopecu.org, NCBA doc)
183 (Ref: World Council of Credit Unions 2014 Statistical Report)
184 (Ref: MEC annual report 2014)
185 Ref: Ibid
186 (Ref: CSI Support & Development Services)
187 (Ref: Cooperative Grocer Network)
188 (Ref: Ibid)
189 (Ref: Nippon Life website)
190 (Ref: Nippon Annual Report 2015)

Other international examples of mutual insurance companies include: Liberty Mutual and Nationwide Insurance in the U.S., Unimutual in Australia, Argos Mutual in Argentina, Åland Mutual Insurance Company in Finland, and Mutuelle Olivier in Lebanon.[191]

## Community-Owned Enterprises

There is a growing amount of momentum behind community-ownership.[192] When a community collectively owns something, this is a form of NFP ownership, as there are no individual owners or shareholders. The ownership is set up in a way that benefits the entire community, not just a few privileged people. Community ownership models are being innovated in a wide variety of sectors.

Community-owned renewable energy is one of the most exciting of these innovations. Denmark was a pioneer in this field, with many of its numerous community wind projects dating back to the 1980s.[193] The way it works is that a community pools its resources in order to make the capital investment of buying the wind turbines, which will then pay for themselves by providing free energy within 7-10 years after installation. After those 7-10 years, the community just sits back and benefits from clean, renewable, cost-free energy. Some communities even make money from selling their surplus energy into a national or regional grid, with any profit made by the community used to benefit the community.[194]

Community ownership can take many different forms. In the case of wind turbines, for example, the municipality or city can own them, or a new NFP corporation can be formed to represent the community's ownership (like community-based enterprises in the U.K.). Or a NFP cooperative can be formed, to ensure democratic control of the turbines, as is the case with many of Denmark's community wind installations. In any case, the ownership is purpose-driven and does not allow for private financial gain.

Often, community ownership is a response to privatization in a certain sector. When people feel that a basic need is not being met adequately due to private ownership, a natural reaction is to 'communitize' that product or service. Next Century Cities is an initiative dedicated to spreading community-owned broadband throughout the U.S., city by city. In the face of restricted access to the Internet due to the rising prices of major for-profit broadband providers, the people who created Next Century Cities decided enough was enough; that the Internet has become too essential for both work and daily life for people not to have. To date, 128 cities' mayors in the U.S. have joined Next Century Cities, committing to make inaccessibility of the Internet a thing of the past.[195]

When a community owns something, it is more likely to be used in the community's best interest than if it were in private hands. In 2013, a local group of 150 concerned community members in Derbyshire, England, formed the Melbourne Area Transition cooperative, which raised money to buy almost 10 acres (nearly 40,500 m$^2$) to be used for the benefit of the community.[196] They are currently using part of the land to grow fresh, organic food for local

191 (Ref: ICMIF Members)
192 (Ref: NEF report)
193 (Ref: Danish Wind Co-ops Can Show Us the Way)
194 (Ref: Danish Wind Co-ops Can Show Us the Way)
195 (Refs: Next Century Cities website and Shareable article)
196 (Ref: Transition article)

schools and they are preserving the rest as a woodland area where community members can go to play, relax and just enjoy nature.[197]

Groups all over the world are using community ownership to protect land, forests, and rivers for the good of the community. The idea of community forests has taken off in Latin America in recent years and a lot of innovation has taken place there. Some community forests have set up multiple subsidiary enterprises based on the forest, like businesses that provide timber, water, tourism, and non-timber forest products. Some of these enterprises are not profitable (they may even operate at a loss), but communities float profits from one business to others in order to make them all work. It's like a supportive web of businesses. For these communities, the enterprises are important not because they make profits per se, but because they generate local employment, so people from the community do not need to go to the city in search of work. For instance, in a community forest in Ixtlan de Juarez, Mexico, the most profitable business has been in timber and the other businesses, like water and tourism don't generate as much money, but that doesn't matter to the community, as profit isn't their main goal. Similar community forest models have been observed in The Gambia as well as First Nations communities in Quebec.[198]

## Government-Owned Corporations

Government-ownership is basically just community-ownership on a larger scale. And sometimes they can be the same thing, as is the case with municipalities owning wind turbines. From the local through to the national and federal levels, government-owned corporations can also be considered NFP businesses, as all profits must be reinvested into government missions, which are meant to serve society. Profits cannot be privately distributed. In most countries, these corporations encompass things like water works, electricity grids, airports and sea ports.

In the U.S., the Bank of North Dakota is a unique example of a government-owned company, as it is the only state-owned bank in the United States. Because it is focused on returning value to the communities of North Dakota instead of private shareholders, it has stayed away from most of the risky and speculative activity that so many for-profit banks have become tied up in, which kept the bank and the state of North Dakota insulated from the economic downturn in 2008.[199] In fact, it was boasting record profits in 2009, while most private, for-profit banks were in turmoil in the U.S.[200] It is also a great example of how the NFP ethic guides companies to be more cooperative, as the Bank of North Dakota seeks to support, rather than compete with local banks. Following the obvious success of North Dakota's model, 17 other states have introduced bills to operate public banks.[201]

This isn't unique to the U.S. Some analysts, including former Home Minister of India P. Chidambaram, credited the nationalized banks as having helped the Indian economy withstand the global financial crisis of 2007-2009.[202] In fact, 40% of all banks in the world are publicly-owned, most of which are in 'emerging economy' countries that largely escaped the financial crisis.[203]

---

197 (Ref: Melbourne Area Transition website)
198 (Ref: Donnie's interview with CF expert Fernanda Tomaselli)

199 (Ref: Mother Jones interview)
200 (Ref: Mother Jones interview; Ellen Brown's work)
201 (Ref: Ellen Brown)

Although they represent an especially powerful NFP model, public banks are not the only kind of government-owned corporation. The Bulgarian State Railway Corporation is government-owned. The Power and Water Corporation in the Northern Territory of Australia is government-owned. The Suez Canal Authority in Egypt is, too. As are Électricité de France, the New Zealand Post Office, and Telkom Indonesia (a telecommunications provider), just to name a few examples from around the world. And plenty of airlines are also government-owned and run, including Iran Air, ELAL in Israel, Iceland Air, Swissair, TWA in the U.S., SATENA in Colombia, and LADE in Argentina.

The world of government-owned enterprises is expansive, and represents an important part of the NFP business sector and the world economy.

## NFP Business Landscape

A wide variety of businesses can be NFP - and it's important to note that common words used to identify businesses like 'multinational corporation', 'local company', ' small business', 'ethical business', 'social enterprise', 'co-operative', and 'family- business' do not tell us if the firm is profit-oriented or purpose-oriented. These labels do not tell us whether there are private owners or shareholders. They do not tell us whether the profit is being extracted into the elite economy at the end of the year or if it is being cycled back into the real economy. In trying to create a more sustainable economy, **the business language we're all used to is not sufficient, nor is it precise enough to describe whether a business is fundamentally generative or not**, that is, whether it is contributing to power of the Wealth Circulation Pump or that of the Wealth Extraction Siphon.

---

202 (Ref: India Stock Market Law and Handbook)
203 (Ref: other Ellen Brown article)

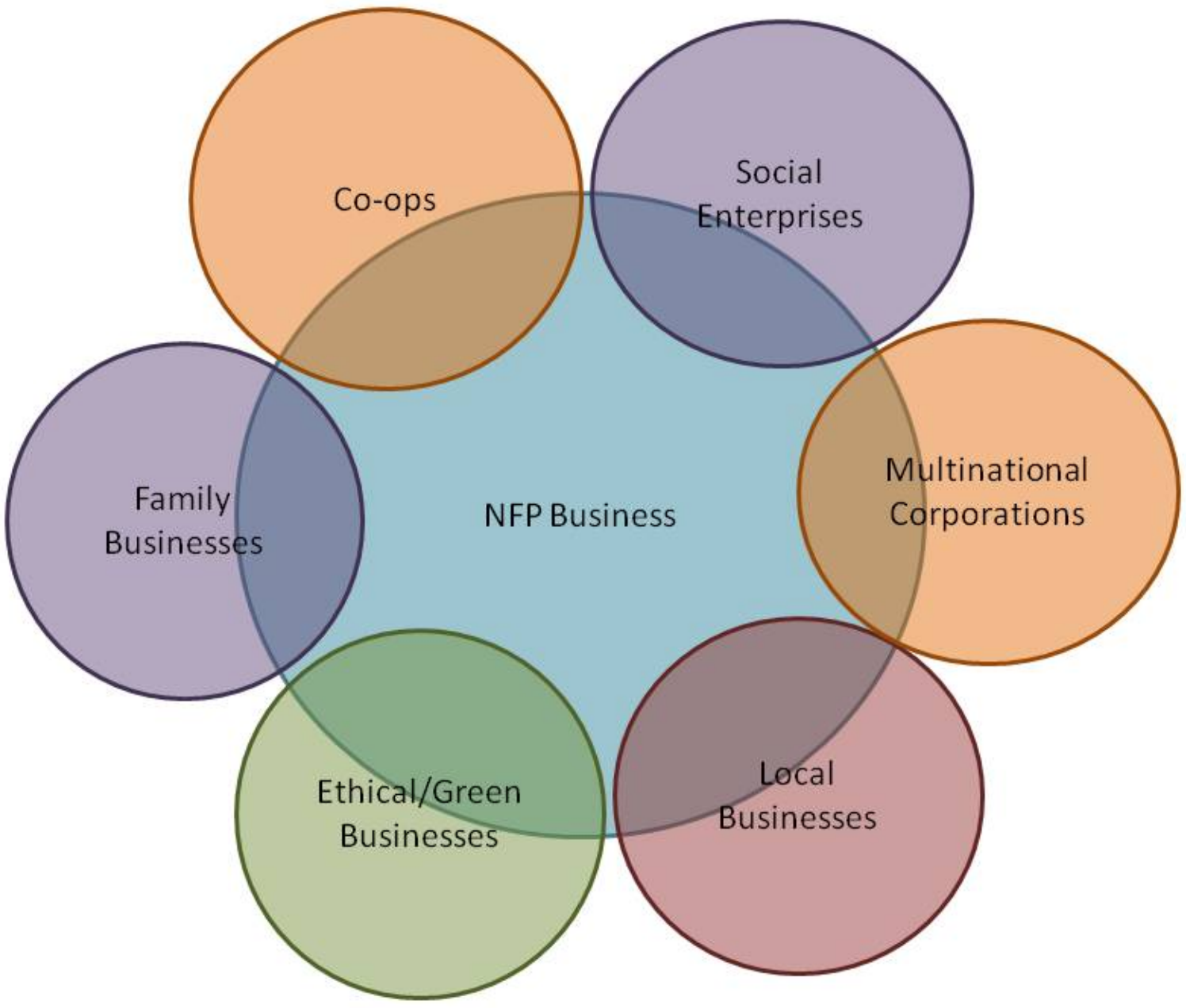

**A Wide Variety of Businesses Can Be NFP**

If we zoom in on the 'Zone of NFP Enterprise' on the business model spectrum[xvi], we can see some of the great diversity in the realm of NFP business and note that there is a lot of potential for even more diverse forms to emerge.

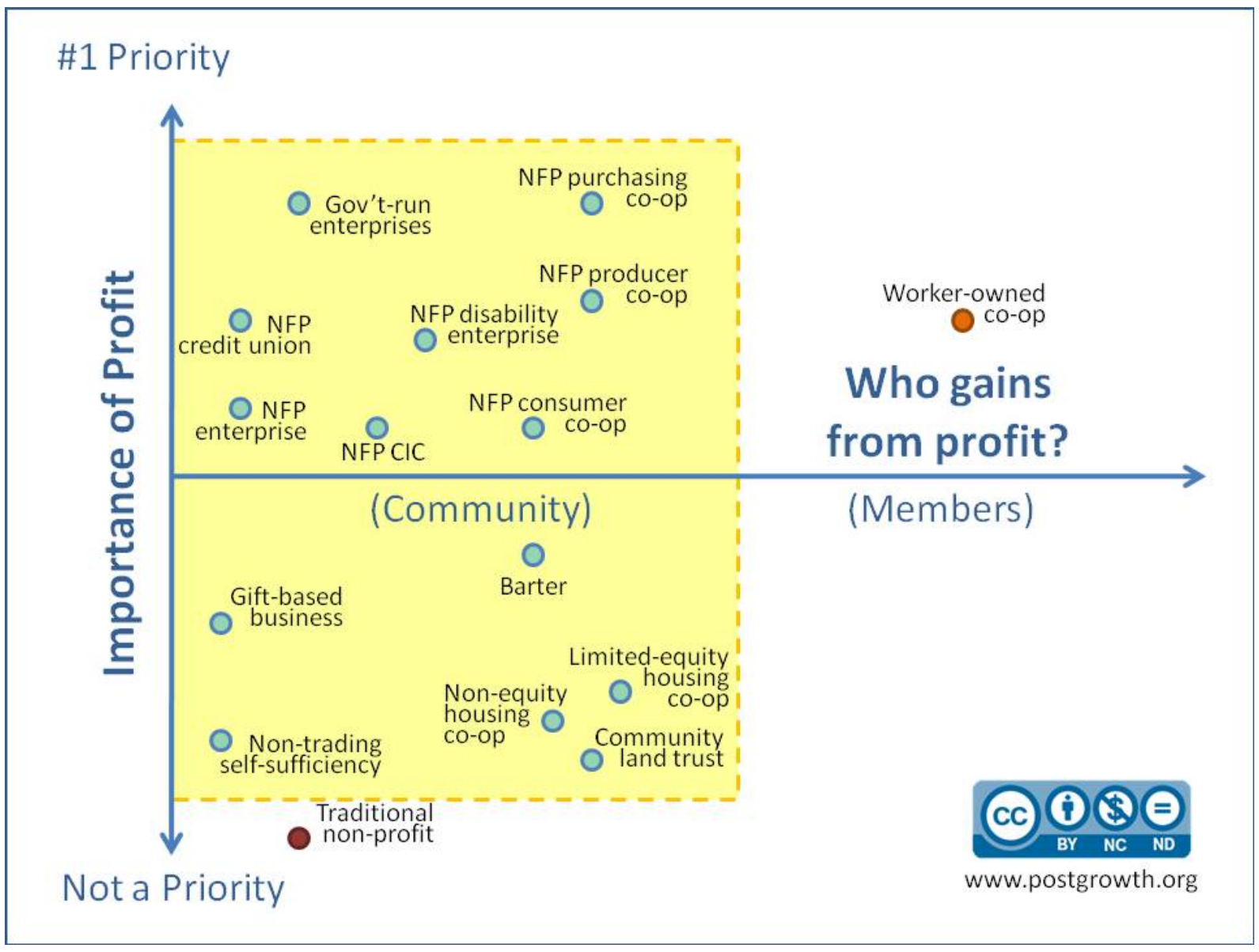

**Some of the Diversity of the Landscape of NFP Enterprise**[xvii]

Traditional nonprofits are outside of the box because they do not focus enough on generating their own revenue, or becoming financially self-sufficient. On the other hand, if an NFP or for-profit subsidiary worries too much about profit, and goes too far up on the vertical axis, it is likely to drift from its mission, which defeats its whole reason for existing.

Not-for-profit business has the potential to be just as diverse as for-profit business, which is featured to the right of the dotted line (see the image below). Everything in the NFP box is mission-driven, embodies an ethic of 'enough' and has no private owners. Our vision of the Not-for-Profit World economy is to have the core of economy in that broad box – thus creating an entire economy truly built for purpose. Now it's time to think *inside* the box!

With the emergence of new for-profit business models that have social missions, profit distribution caps, asset-locks and capped investment strategies, along with the growing number of co-ops and NFP enterprises all over the world, business trends seem to be moving in the NFP direction. In fact, we argue that the business models to the left in the image below are actually becoming more competitive in the market than the business models to the right, due to the fact that they can more adequately meet the needs of society in the 21st century.

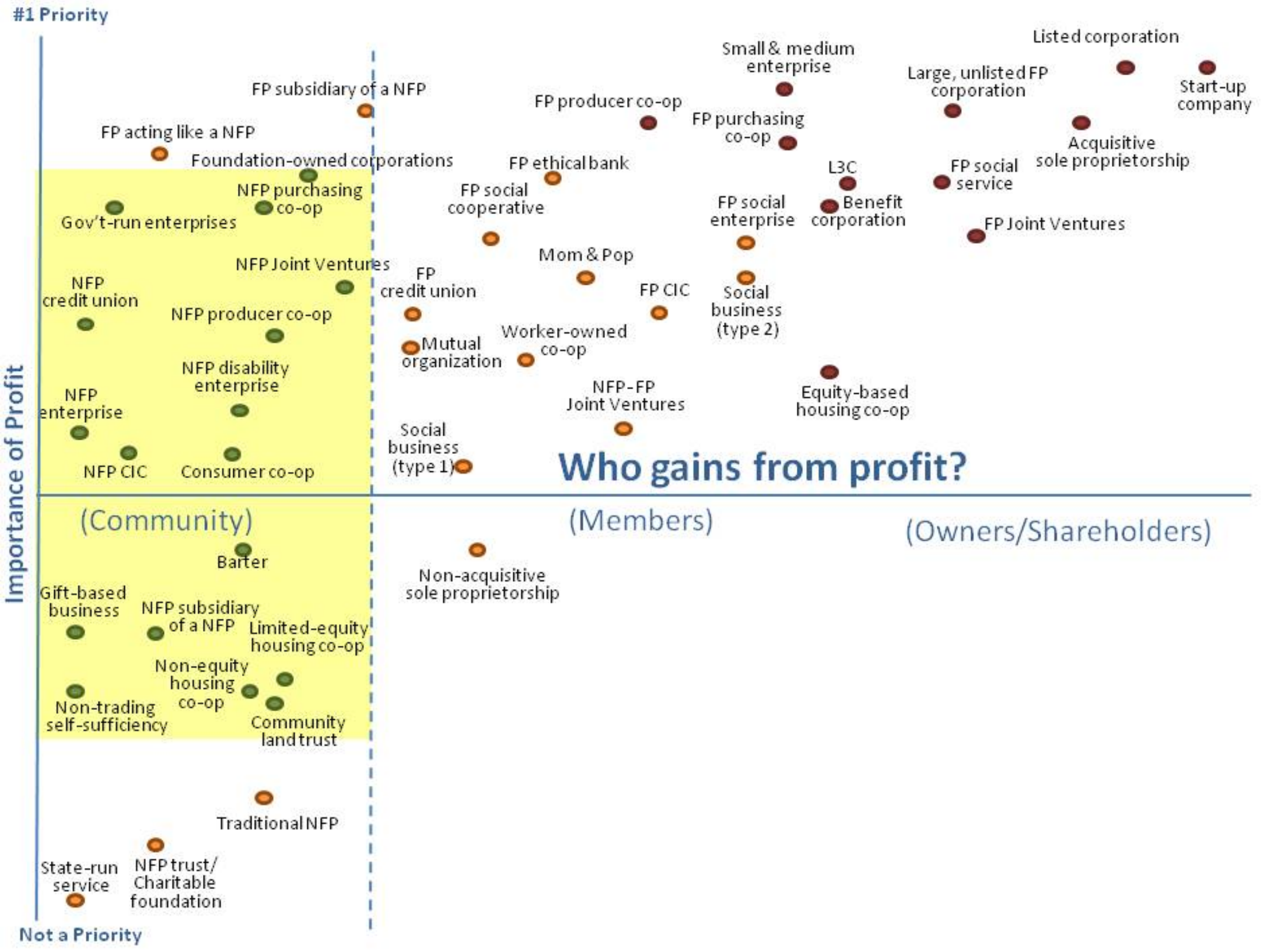

**Landscape of All Business Models**

Companies with common for-profit legal status can also act like NFPs, cycling all of their profits back into social or environmental missions. A for-profit Limited Liability Company can create clauses in its by-laws or statutes that state: a) the social or environmental mission of the company, and b) that the company will never privately distribute its profits or assets, but instead use all surplus to achieve its mission. This means that even in places that don't have legal infrastructure that supports NFP business, for-profit companies can act like NFPs and join the growing momentum behind a purpose-driven economy.

## What's in a Name?

From what we have said so far about NFPs, it might seem easy to distinguish an NFP from a for-profit entity - we thought it was when we started working on this book. However, the dividing line is not always so clear. Some terms can make for-profits seem like NFPs and vice versa, so it's important to be aware of those as we enter an era in which this distinction is becoming ever more important.

Much of this confusion is due to nonprofits moving into the business sphere at a rapid pace. It is such a new phenomenon that our vocabulary about business has not yet adapted. Up to now, all business has been assumed to be for-profit. When we think of business, most of us think of for-profit business by default and that's why our business language was developed to refer specifically to this type of business. As NFPs move into business, things can get kind of blurry.

For instance, Harvard University, which is an NFP business, is often described as a private university. But what does 'private' mean? This word conjures up all kinds of different understandings. Many take it to mean privately-owned by individuals, but as we know, NFPs cannot be privately-owned. Taken at face value, it may seem that Harvard is a for-profit university. But it is not. Are they just trying to make it seem more like a business than a traditional nonprofit? Maybe. But the most likely reason they use the word 'private' is to distinguish Harvard from public universities. In the U.S., universities are usually categorized as being either public or private. Private is taken to mean 'not state-owned'. There is no real third term for independent, not-for-profit universities like Harvard. Yet this growing trend means that NFP enterprises should start collectively carving out their own distinct space in all sectors. In the case of American universities, it would be a lot clearer if we begin to categorize universities as NFP, public, or for-profit; rather than just private or public.

Many NFP co-ops also use for-profit language, perhaps without realizing it. An NFP co-op is not technically 'owned' by the members, as many U.S. credit unions and consumer co-ops claim to be. Many of these same NFP co-ops refer to their members as 'shareholders', which of course is also legally impossible in NFPs. A shareholder owns a portion of a company, and seeks to make financial gains from the part of the company he or she owns in the form of a dividend payout from the profits every year. Since no one can own an NFP and an NFP cannot privately distribute profit, the words 'owner' and 'shareholder' have no place in the description of NFP co-ops.

Likewise, an NFP co-op might refer to a required founding 'share' for a nominal amount, but there is no title, equity, trade-ability or transferability of shares and, the ownership is therefore, 'not-for-profit'. In simple terms, no one can own a share of an NFP, except for another NFP. Using the word 'share' in the context of an NFP co-op is purely symbolic and can actually do more harm than good, making people think the co-op is for-profit. The inaccurate use of these words can be confusing and misleading.

Many NFP co-ops also say that their members receive 'dividends'. We have found this to be the case for quite a few NFP credit unions and energy co-ops, which at first made us doubt whether they were really NFP. Yet these dividends are really just refunds or rebates that the co-ops give back to their customers at the end of the year. No one receiving dividends from an NFP is making a financial gain from them. The same is true for 'patronage dividends' and 'capital credits'. These are just rebates or partial refunds in the NFP arena, but the word 'dividend' makes it sound like shareholders are looking to make a financial gain at the end of the year. It makes the refund sound speculative when it is not. Sometimes interest on loans is even referred to as equity by NFPs. But again, as no one can own an NFP, no one can have equity in it. Equity is another one of these for-profit terms that implies private ownership and even speculation. Interest on loans that a company pays to its lenders is very different from equity-based dividends that only for-profit companies can distribute.

No one can blame NFPs for using for-profit language; it has been natural for them as they move into the business arena to use the business jargon that already exists. But as the NFP business sector starts to realize its full potential, we encourage these enterprises to proudly use a more accurate vocabulary. There are no member owners, just members and that's part of the strength of the model. There are no shares, shareholders, or dividends, and that's why we can feel sure that they're dedicated to a social mission, and not profit-maximization. There's no private equity and that's the *beauty* of it.

Let's embrace these important differences and carve out more space for this burgeoning sector that's transforming the economy for the better. Let's give it more credit.

Until now, most law-makers have not realized the importance of the for-profit/NFP distinction. As a result, some legal structures also blur these lines. For example, Community Interest Companies in the UK are often thought of by the public as NFP entities, but there are in fact both for-profit and not-for-profit CICs. Community Interest Companies limited by share, as their name implies, do sell shares in the company and can privately distribute up to 30% of the profits. However, the vast majority of incorporated CICs are not-for-profit (the limited by guarantee form)[204] because the entrepreneurs behind these companies see it as a model that better aligns with their purpose and vision. Thus, the NFP entrepreneur represents a new livelihood for the 21$^{st}$ century.

Likewise, the benefit corporation model in the U.S., which is only for-profit, sounds like it is NFP. In fact, many people have asked us why we don't use the term 'for-benefit corporation' instead of 'NFP', as it sounds so much more positive, and seemingly defines the company by what it is, rather than what it is not. The misleading nature of the for-profit benefit corporation name is the reason. It goes to show how many different interpretations the words 'benefit' or 'purpose' can have, whereas 'not-for-profit' has a very specific legal interpretation that is nearly universal throughout the world, and has been tried and tested over 200 years. In fact, we strongly feel that what the world needs right now is exactly *not*-for-profit business.

Many people also assume that all social enterprise models and all social cooperative models are NFP. However, we've seen that social enterprise legal models don't require a company to be for-profit or NFP in most countries in the world.

An exception to this is the social co-op in southern Europe. Social cooperative models in Greece and Italy are often referred to as NFP, even in government documents, but they are actually for-profit according to international legal definitions, as these models allow for the private distribution of between 20 and 35 percent of profits.[205]

Although these models are a step in the right direction (a step away from the narrow focus on profit-maximization) they fall short of being able to bring about systemic change. They still privatize the surplus of the system and put the cart before the horse.

---

204 (Ref: CIC regulator)
205 (Ref: Social Cooperatives in Greece; Social Co-operatives in Italy).

We are not the first to recognize the importance of making clear distinctions between a company that privately distributes profit and one that doesn't. There's a reason why the not-for-profit legal status exists and is so heavily defended by courts of law. This wisdom has been around since the birth of for-profit corporations, centuries ago. It's a clear distinction in legal terms. Now we just have to use accurate words to describe it in business terms, and to show people the potential of NFP models.

To our knowledge, this book is one of the first to point out the importance of the for-profit versus NFP distinction in terms of its ramifications for how the *entire economy* works[xviii]. This is a new frontier and clarifying the territory is always part of pioneering work. One of the main messages of this book is that we should embrace NFP enterprise as the generative business model it is.

# Not-for-Profit Enterprise is on the Rise

People all over the world are increasingly recognizing the strengths of the NFP way of doing business and this is why NFP business is on the rise.[206] It is such a new trend that there is not an overwhelming amount of data comparing NFP enterprise to traditional nonprofits and for-profit business. The purpose of this book is to reveal the existence and rise of NFP enterprise to a broad audience (even if we don't know how fast or big that rise is) and, just as importantly, to illustrate the incredible potential NFP enterprise has to fundamentally transform the economy for the better. We're putting forth a vision for a healthier economy that can meet the challenges of this century and we feel like there is enough data about NFP enterprise to show that this vision is worth pursuing. This book is an open invitation to put these ideas to the test.

The Urban Institute, Johns Hopkins University and a handful of independent researchers are tracking this NFP enterprise trend[xix], but it is incredible how little attention NFP business gets when it is right under our noses. Political economist Gar Alperovitz points out that almost 40 percent of the 200 largest enterprises in the twenty largest U.S. cities are NFPs, like universities and hospitals.[207]

More nonprofits are enterprises than you might realize - just think of all the associations whose membership fees pay for staff. Here are some statistics to show just how strong the NFP business sector is, all over the world:

- In Scotland, a nation of just over 5 million people, the NFP sector generated over £4.9 billion a year and employed 138,000 people in over 45,000 organizations in 2012- 13.[208]
- And in that same timeframe, 2012-2013, the NFP sector in Australia contributed 55 billion Australian dollars to the economy, which was twice as large as the economic contribution of Tasmania that year.[209]
- In 2010, the NFP sector in Canada contributed over $100 billion to the nation's GDP (7 percent of the total GDP), four times more than auto manufacturing.[210] And these numbers don't even factor in the economic impact of NFP volunteers.

---

206 Salamon, 2013
207 (Ref: America Beyond Capitalism)
208 (Ref: SCVO report)
209 (Ref: The Not-for-Profit Sector in Australia: Fact Sheet)
210 (Ref: Wellesly Institute report).

- One in three Americans is a member of a credit union and membership is rising exponentially, by 2.25 percent a year.[211]
- There are 4,600 Community Development Corporations in the U.S., focused on setting up low-income housing schemes, all of which are not-for-profit.[212]
- Twenty-five percent of the electricity in the U.S. is provided by NFP co-ops and public utilities.[213]
- In Bangladesh, the Rural Electrification Board has connected 30 million people to the electrical grid through NFP rural electric cooperatives.[214]
- In Santa Cruz, Bolivia, an NFP called SAGUAPAC is the largest urban water cooperative in the world. Its 183,000 water connections serve 1.2 million people - three quarters of the city's population - with "one of the purest water quality measures in Latin America."[215]
- The United Kingdom has 10,600 Community Interest Companies (CICs), 78 percent of which are not-for-profit.[216] And from 2014 to 2015, CIC registration rose by 13 percent.[217]
- It is estimated that there are about 20,000 social enterprises in Australia, most of which are NFP, and that number is also growing.[218]

That's a lot of people benefitting from NFP businesses and, importantly, no owners or shareholders are getting rich from it. All of the money generated by all of these NFPs stays in the real economy. Imagine if all that profit was instead fueling the elite economy.

Trends seem to point towards ongoing growth in the NFP business sector. Between 2001 and 2011, the NFP business sector in the U.S. grew at a rate of 25 percent (compared to the for-profit sector's 0.5 percent growth) and the sector's growth continues to massively outpace the growth of the overall economy.[219] On a global scale, it is estimated that 7.4 percent of the world's workforce is employed by nonprofits – and that probably doesn't even take into account all of the employees of mutual insurance companies and NFP cooperatives.[220]

We have seen a dramatic rise of the use of the words 'social enterprise' and a steady rise in the use of 'not-for-profit' in books since 1980.[221] And NFP enterprise incubators - spaces where NFP startups are intentionally developed and given support - are popping up all over the place.

Green White Space, for instance, is an NFP business that helps other NFP businesses start up, scale and sustain themselves through self-generated revenue. Green White Space is based in Sweden, but has projects and partners all over the world.

211 (Ref: CUNA)
212 (Ref: Community Wealth statistics)
213 (Ref: The Rise of the New Economy Movement)
214 (Ref: ILO and ICA, 2014)
215 (Ref: ILO and ICA, 2014)
216 (Ref: CIC Regulator)
217 (Ref: Regulator of the CIC Annual Report 2014/2015)
218 (Ref: FASES, Probono article)
219 (Ref: Urban Institute; NY Times article)
220 (Ref: The Nonprofit Times)
221 Ref: Google Books Ngram Viewer social enterprise search; Google Book Ngram Viewer NFP search

"Our vision is to prove a new model successful; a model where the not-for-profit enterprise truly competes with more traditional business structures."[222]

The Cascades Hub in Oregon, initiated by Donnie Maclurcan, co-author of this book, is also dedicated to helping entrepreneurs start NFP businesses. It runs workshops, boot camps and classes on how to start and sustain an NFP enterprise.[223]

As more of us come to crave meaning in our work, and as the challenges of the twenty-first century only increase, we can expect the rise of NFP enterprise to gain momentum in response to society's changing needs.

With this new understanding of how generative business can really be when profit is just a means to meeting deeper goals, we are now able to see our current economy with new eyes. It turns out that for-profit business has played a central role in the economic, social and ecological crises we're experiencing.

---

[222] (Ref: Green White Space's website)
[223] (Ref: Cascades Hub website)

# 3. Crises of the For-Profit World

## *Our for-profit world is leading to collapse*

At a party given by a billionaire on Shelter Island, American author Kurt Vonnegut asked his friend and fellow author, Joseph Heller, "Joe, how does it make you feel to know that our host only yesterday may have made more money than your novel 'Catch-22' has earned in its entire history?"

Joe replied, "It doesn't bother me. I've got something he'll never have."

Kurt asked, "What could that possibly be?"

Joe answered, "Enough."

## The Common Roots of the Crises

Before exploring how the Not-for-Profit World model can work, we need to have a deeper understanding of how and why the for-profit world in which we currently live isn't working well.

When we look a bit deeper, we find that the worsening social, economic and environmental conditions of the 21st century are symptoms of a deeper crisis of purpose and perception in Western culture.[224] This crisis manifests in our for-profit economic system and its fundamental assumptions.

### The For-Profit Story

Stories create the world in which we live. The manmade world around us is a manifestation of a story. The world is manifested and shaped according to our views and beliefs. And those views and beliefs are really just the stories we've inherited and tweaked to explain why the world is the way it is.[225]

For instance, the idea that progress means moving forward in time is a story; a belief. Historians and anthropologists have found that many indigenous cultures, including most Native American cultures, view time as cyclical, not linear, because they built their societies according to the natural cycles of the moon, the sun and the seasons.[226] They share a collective story that says time is a cycle that continuously moves around and around. This is very different from the Western civilization's story about time. In the Western world, we assume everything must constantly be developing, progressing, and moving forward over time – we envision time as a ruler by which to measure progress.[227] Think of all the charts and graphs that have time as their horizontal axis (we even have some in this book). It's linear and

224 (Ref: paraphrased from Stephen Sterling's dissertation)
225 (Ref: Harari – *A Brief History of Humankind*; Eisenstein- *Ascent of Humanity*; Eisler – *The Chalice and the Blade*)
226 (Refs: Charles Eisenstein, *Ascent of Humanity*; Donald Fixico, *The American Indian Mind in a Linear World)*
227 (Ref: Ibid)

it's moving in one direction – from left to right. We're not saying that one story of time is better or more correct than the other, but rather just acknowledging that they're both stories.

Remember Kuhn's paradigm shifts? A paradigm can be thought of as the main guiding story of a society. Before Copernicus and Galileo, European society organized around the story that the Earth was the center of the universe. After Galileo's telescopic discovery, society began to organize around the story that the Earth and other planets revolve around the Sun – that we are part of a heliocentric solar system.

Many scholars argue that it is the very ability to share stories that separates us from other species.[228] This ability has allowed us to create complex forms of social organization. The capacity to transmit stories, to describe the world around us, to talk about ideas and visions for the future, has enabled us to organize into very large groups to work towards common goals. Only by having shared goals, and a shared story, can we cooperate flexibly in very large numbers.[229]

There are all kinds of stories that shape society today – stories about what it is to be human, about how women and men should behave, and how society should be organized. These stories vary from culture to culture, but there are also some fundamental stories that shape all of Western civilization and, in the 21st century, global civilization.

Just as with any social system, the for-profit economy in which we live is underpinned by a certain set of assumptions, beliefs, and worldviews. Business is based on shared stories about the economy - a common economic myth.[230] What is the overarching story around which we organize our businesses and the economy? What narrative keeps our current economic system going?

Even people who have never taken a single economics course know this story, because most of us have internalized the narrative of the for-profit system. These are some of the fundamental beliefs of modern Western civilization and so we learn them as children, through our schooling, from our parents, and even from fairytales, comic books, cartoons, and video games. We're exposed to this story daily through the news, TV shows, films, magazines, and advertisements.

In short, the story goes something like this:

> *Evolution is a process of different organisms competing for scarce resources. Thus, evolution is driven by competition and survival of the fittest. That's why human nature is mostly selfish, greedy and competitive and people are primarily motivated by their ability to profit; to make a personal gain. In light of the fierce competition of evolution, it is only rational to always be looking out for your own self-interest and trying to maximize your individual gain.*
>
> *Not only are we are all separate individuals who compete with each other, but we are also separate from our environment. And because we have free will, we can use nature as we like. We earned that right, by evolving to be clever enough to take ourselves out of the food*

---

228 (Ref: Harari - primary sources?)
229 (Ref: Harari, *A Brief History of Humankind* online course; Storytelling Species article (primary sources: Gazzaniga; Gottschall?))
230 Ref: Harari, A Brief History of Humankind course

*chain and develop modern civilization. Due to that same cleverness, we will always be able to innovate our way out of any environmental problem we might create.*

*Inequality is just the way of nature. Not everyone can win the game. There are winners and losers; that's how survival of the fittest works. Besides, inequality is actually good for the world because it spurs us into action. Without the threat of poverty, most people would be too lazy to do anything and they would just try to leach off the few naturally hard-working, productive people in society. This is because we all seek to maximize pleasure and rewards and minimize effort and pain. Without inequality to motivate us to work, the economy would fail, civilization would descend into chaos and most of us would suffer greatly.*

*Progress also requires inequality, because inequality motivates people to move up the ladder of competition that leads from one social stratum to the next. The economic winners are those who accumulate the most because they are the most clever, competitive, rational, hard-working people. The more money, assets and power you accumulate, the more successful and prosperous you are. Wealthy people set a standard for the rest of us to aspire to and provide an example for us to follow.*

*Wealthy people are very important to society for many reasons. In trying to turn a profit and become richer, they set up new companies and invest in new technologies, thereby creating jobs and fuelling innovation. Their profit-seeking behavior benefits society. The economic growth they create ensures that wealth trickles down to other parts of society. The pie is always growing, so everyone can always get a bigger piece, if they work hard enough. Thus, the Invisible Hand of the market can resolve almost all social, economic and political problems. As Gordon Gekko famously said in the movie Wall Street, "Greed is good."*

*If you work hard, you too can rise to the top. You deserve what you get, whether you're rich or poor, because we live in a meritocracy. If you are poor, that's because you're lazy and aren't trying hard enough. If you've had bad luck in life, you need to pull yourself up by your bootstraps, because capitalism offers everyone the opportunity to succeed if you only try. And if you're rich, then you've done well and should continue to accumulate more in order to provide a role model for the rest of society.*

Chances are you've been steeped in this story for most of your life. You might agree with all of it, parts of it, or none of it, but every idea in the for-profit story is undeniably familiar. The vast majority of us in Western society have been exposed to it since we were young children, so it can be shocking to look at it as a story, rather than as the indisputable facts of life. It's like a fish suddenly understanding that it's surrounded by water.

The for-profit story encompasses the fundamental theories on which our global economy is based: rational, self-interested firms maximize profits; rational, self-interested consumers maximize their 'utility'; the decisions of these actors drive supply and demand; the market moves; and resources are allocated fairly almost as if by an invisible hand.

When this story was first being developed by the early modern economists and philosophers, it was a story of freedom and prosperity. In Anglo-Saxon society in the 18$^{th}$ and 19$^{th}$ centuries, the for-profit story was part of a radical movement to go beyond the social barriers that had long kept people stuck within their class of birth.

The individualism, competition and meritocracy of the for-profit story were a source of inspiration and freedom for many people. It enabled lots of everyday people in the Western world to understand and express their individuality for perhaps the first time in history. Through the emergence of new business models and banking standards, non-aristocrats were able to take out loans and start up their own businesses. Over time, a true middle class emerged and millions were raised out of poverty through modern business, making it much easier for them to meet their needs. And through this rapid emergence of small and medium-sized firms, innovation occurred at a faster and faster rate. Many of these businesses did a great job of allocating resources, goods and services. The for-profit market delivered on many of its advocates' promises (depending on what gender, race and nationality you were).

However, we always need to remember the importance of context. Firstly, it must be acknowledged that the wealth accumulated by early capitalists and the rise of a middle class in the first industrial countries was only made possible by slave labor, worker exploitation, colonialism and the genocide of indigenous populations and appropriation of their lands. When resources and labor are virtually free, it's quite easy to produce affordable goods and services - and to get rich doing so.

Aside from the moral failures that enabled the emergence of capitalism, there is also the wider historical context. In a world in which there was no middle class, only an aristocracy and peasants, the capitalist system worked wonders. In a context in which average people had very little choice or freedom in the way they could live their lives, the for-profit story empowered many of them to work hard, start up their own businesses and contribute to their communities by providing goods and services that were never before available. Before the Industrial Revolution kicked into full gear, resources were plentiful, production was low, and consumption was minimal. The market was wide-open and there was enough room for everybody to participate in growing the economy at an unprecedented rate. In a context in which the sky was the limit to the expansion of the market, the for-profit system worked miracles for many.

However, this 250 year-old economic system is no longer suitable for today's radically different world. The for-profit system and the for-profit story are outdated and are now causing far more problems than they're resolving, driving what is now increasingly referred to as 'un-economic growth'. In essence, **the for-profit system can't solve today's major problems, because it *is* the problem.**

Whereas in the 19th and early 20th century, for-profit businesses were limited in size and scope by the communications and transportation technologies of their time, in the current globalized context, for-profit business naturally leads to a concentration of wealth and power that is proving too difficult to keep in check. Now we have levels of inequality that the early 20th century capitalists could never have foreseen. (One report revealed in early 2016 that the wealthiest 62 individuals on the planet own as much wealth as the poorest 3.6 billion.[231]

In the old days, the for-profit story of self-interest was balanced by communities holding their local business owners and managers accountable for their actions. Business owners were both revered and held responsible for their behavior. These days, supply chains have become so globalized, business management so anonymous, and production so distant from where

231 (Ref: Oxfam report)

products are consumed, that most companies have hardly any accountability to local communities at all. The greed and speculation at the heart of the for-profit story are going unchecked, and are leading to the very demise of the capitalist system.

You only have to take a quick look around to see that our global, capitalist economy is not delivering on its promises for most people. In rich countries, people are working longer hours than in previous decades[232] and many in the 'developing' world must work more hours than those in the 'developed' countries, without even managing to meet their basic needs. Real wages for the average worker in most places have stagnated or declined since the 1990s.[233] Public and private debt all over the world continues to rise at worrying rates.[234] Global unemployment has risen in much of the world over the last decade due to economic crises, which resulted in part from the stock market's boom-and-bust cycles and the bursting of speculative bubbles. The International Labour Organisation expects unemployment trends to continue to worsen, leading to 200 million jobless people in the world, for the first time on record, by the end of 2017.[235] And, perhaps worst of all, there's no real safety net for far too many people.[236] The middle class in Western economies has shrunk to such a degree that people can no longer afford to consume at levels that keep the economy growing as fast as economists would like[xx]. As a result of declining consumption in the West, many companies are seeking to expand and tap into the consumers of the so-called emerging economies. But who will provide the cheap labor to raise low-income countries out of poverty?

Not only are we bumping into what are likely the social and economic limits of capitalism, but we began exceeding ecological limits to economic growth several decades ago and now face enormous environmental crises that are only exacerbated by the growth-based economy.

It is becoming very clear from every angle that the for-profit system is in a troubling decline. Capitalism has reached its 'use by' date and a growing number of people can sense the fundamental flaws in the for-profit story. There are certainly holes in the plot, but what are those holes? And how do they lead to the dysfunction we're seeing in today's economy?

## The Profit Motive and Homo economicus

The for-profit story and our economic system are completely interdependent. The for-profit story supports the capitalist economy and vice versa. This means that we cannot have truly systemic change without also reshaping the underlying stories that support and validate the system.

The most obvious way in which the for-profit story and economic behavior mutually reinforce each other is through the profit motive; the notion that economic behavior is the result of individuals seeking to accumulate as much financial wealth as possible. This manifests in capitalists investing in whatever will give them the highest return on investment. It can be seen in the job market, with job-seekers looking for jobs with the highest salaries. It also plays a role for some employers, who will try to pay employees as little as possible without losing them altogether. And it can be observed in shoppers always on the lookout for the best

232 (Ref: Extreme Working Hours paper)
233 (Ref: *International Labour Organisation Global Wage Report 2014/15*)
234 (Ref: McKinsey Global Institute report).
235 (Ref: International Labour Organisation *World Employment Social Outlook*)
236 (Refs: CNN (primary source: World Bank?).

bargains. This leads us to **the protagonist of the for-profit world: Economic Man, also known as *Homo economicus*.**

Homo economicus represents the idea that individuals are naturally inclined to make economic decisions based on maximizing their own self-interest. Although the term was not used until the 19th century, it has its roots in the earliest modern economic theories.[237]

In 1776, Adam Smith, the 'father of capitalism' wrote, "It is not from the benevolence of the butcher, the brewer, or the baker that we expect our dinner, but from their regard to their own interest."[238]

In an 1836 essay, early economist John Stuart Mill described man as "a being who inevitably does that by which he may obtain the greatest amount of necessaries, conveniences, and luxuries, with the smallest quantity of labour and physical self-denial with which they can be obtained."[239]

This notion that people base their decisions and behavior on personal gain is pervasive in society. It implies that if you are smart, then you will look at how you can gain as much money as possible (as in the case of investors, shareholders and employees) or how you can lose the least money possible (as in the case of employers and consumers). Paying more than the lowest possible price for something is considered foolish behavior. Yet, if you are a seller, not getting paid the highest possible price for a good is considered equally foolish and even weak. That's the profit motive at work. That's the engine of our economy.

But is this really the best way to run an economy and to motivate economic activity (and behavior more generally)? Is it reasonable to hold as a common belief that the smartest people in society are those who take the most for themselves? That the most intelligent are the greediest? What does that say about generous people?

On closer inspection, it becomes very clear that when private gain is the primary source of motivation in the economy, people tend to make very selfish decisions that may provide a financial gain for them, but at a cost to the wider community. Take the example of an oil drilling company that, in order to save money, doesn't update its safety measures, risking the lives of its employees as well as the health of the natural environment in which the company is drilling. Or think of a car manufacturer that lies about its cars' emissions in order to boost sales and maximize profit.

The profit motive also sets the stage for very unhealthy norms in corporate culture. In publicly-listed companies (i.e. – companies traded on the stock exchange), CEO compensation often depends on share price rather than any other indicators of performance. This detaches performance from real world value creation. In other words, a financial company can be trading toxic financial instruments that harm its own customers, not to mention undermine the entire economy, but if it raises share prices in the next quarter[xxi], the CEO will be handsomely rewarded.

---

[237] (Ref: Investopedia (primary source?))
[238] (Ref: Smith, Adam. "On the Division of Labour," The Wealth of Nations, Books I-III. New York: Penguin Classics, 1986, page 119)
[239] (Ref: Mill, 1836)

This was demonstrated very clearly in the U.S. Congressional hearing of Lehman Brothers' CEO, Richard Fuld, who received over $480 million, mostly in bonuses, from 2000-2008, while the company was going bankrupt.[240] Lehman Brothers was the first big financial company to declare bankruptcy in September 2008, triggering the global economic crisis. These Congressional hearings did an important job in questioning the ethics and fairness of Wall Street CEOs taking so much money home, just because share prices were going up. What the hearings failed to note is that this sort of greed and short-termism is built into the very fiber of the profit motive.

Although it's easy to point fingers and declare these actions unethical, and label these people a few 'bad apples', this sort of behavior is actually just the rational thing for business managers to do, according to the for-profit story. It naturally flows from the for-profit worldview. **If we want systemic change, we need to take a deeper look at the rules of the game, and not get caught up on blaming the individual players.** Perhaps people like Fuld *are* greedy sociopaths, but it's not a coincidence that they're the winners of our economic game. The profit motive was guiding them 100% of the way. They are the epitome of 'rational economic actors' being motivated by the accumulation of private profit and self-interest. Indeed, they are playing by the rules of capitalism very well.

It is a paradox of society to hold the profit motive as a central tenet of business, and then call people who succeed through maximizing personal gain 'greedy', when it would be exceptional for them *not* to be.

Think of Monopoly, the ultimate capitalist board game, where the goal is to win by accumulating the most money and property on the board. What do players do? They follow the game rules and to try to amass as much property as they can. Even the most generous of us can act like insatiable sociopaths when we play Monopoly. It's hard not to when domination and accumulation are the goals of the game.

It's self-contradictory to encourage the Golden Rule - do unto others as you would have them do unto you - but only within the larger framework of the profit motive. This is a recipe for inner conflict, because you can't expect a society built on individual gain to function for the common good. This hypocrisy is particularly evident when business people that play by the rules of for-profit business are praised and envied for their financial success, but at the same time scorned for being greedy and selfish. How can we rationalize the idea that we all should want more material wealth, but we shouldn't harm others by our greed and selfishness?

There is a larger cultural crisis lurking beneath the economic crisis. It's a crisis of the values we hold and the stories we tell.

We've been taught to only see our side of the equation when we sell or buy something, and that we should focus on bettering our own position while seeking to exploit the other's position. Rarely ever are price negotiations and financial transactions centered on creating a reasonable deal for both parties. Much more often, it's about whether *we* made money. The more expensive whatever we are negotiating is, the greater the stakes, which means that we become even more focused on maximizing our own gain or minimizing our losses. This often promotes a certain level of dishonesty relating to major financial transactions. Think of a house for sale: the sellers try to make the house seem as valuable as they can (exaggeration is common) and potential buyers try to point out the flaws in it. The inexperienced or poorly researched buyer will certainly emerge less well off, as will anybody who is thinking about a fair deal for all.

---

240 (Ref: Richard Fuld hearing video)

When the stakes are high, there's also a much higher risk for deals to fall apart due to this element of exaggeration, dishonesty and the resulting mistrust. And when deals do go through, there are often sore feelings or doubts afterwards about not having gotten enough money or having paid too much. Whatever the results, this way of thinking about economic transactions does little to build real trust in society and can do a lot to tear relationships apart. Most of us have seen how even families can be broken apart due to financial mistrust; brothers not speaking because of a business deal that went wrong or cousins going to court over the value of a previously shared family property. These stories are commonplace in the modern capitalist economy.

That's how the system works. You're not supposed to consider the other people involved in an economic transaction. You're not supposed to worry about the previous owner who lost their home to foreclosure, or to feel happy for the customer who might have gotten a really good deal. You're supposed to maximize your self-interest, and it's up to the others to maximize theirs. This way of thinking compromises our ability to be empathic and compassionate with one another.

That is why, in most of the 'developed' world, we often advise each to never do business with a friend or loved one. Keep your personal and business lives separate. The profit motive seems to encourage a kind of schizophrenia. Be a kind, loving person at home. Be a ruthless, selfish person at work. Be fun around your friends; be serious around your colleagues. After all, you're only at work to make money – it's serious. Yet, empathy and compassion are the glue that keeps families, communities, and organizations together; they're pre-requisites for trust and deep social connection.[241]

There is a growing body of research on financial wealth that shows with overwhelming consistency that the rich tend to be more selfish, less empathic, less generous and less compassionate than people who have less money.[242] Again, this is neither surprising, nor a coincidence. Either they have become this way as a result of succeeding in the for-profit world, or the only people who can succeed in the for-profit world are those who are less empathic and more selfish. It's likely a combination of both. In any case, the data allows us to make a strong argument for moving beyond the profit motive and the privatization of profit.

## Incredible Inequality

Aside from the selfishness and greed promoted by the for-profit story, there is another very good reason we need to move on from the for-profit system: it is self-destructing. This self-destruction is evident in the three global crises of the 21st century: inequality, declining

[241] (Refs: Frans De Waal; Emma Seppala)
[242] (Refs: Piff & Keltner, 2012; Piff & Keltner, 2010; New York article; Daniel Goleman, 2005, Sydney Morning Herald article)

wellbeing and ecological collapse. And these crises of capitalism are very much rooted in the mechanism of the wealth extraction siphon.

## The Wealth Extraction Siphon

What started out as a game of 'who can amass the most wealth' has led to a situation in which the system is starting to collapse in on itself. This is an inevitable consequence of the for-profit system. **Capitalism has always contained the seeds of its own demise – it just had to run its course before this became evident[xxii].**

The wealth extraction siphon works by taking wealth out of the real economy and putting it into the elite economy. The wealthiest people in American society gain 47% of their income from the real economy via wages and salaries (16%), business income (17%), and interest and dividend payments (14.1%).[243] Trickle-down theory proposes that the wealth will trickle down to lower-income households from this elite group, as they spend money back into the real economy, fueling the wealth circulation pump. But do they put that back into the real economy? Not exactly. Economist David Harvey points out that many modern-day capitalists are choosing to reinvest much more of their capital in financial assets than in production and labor, because that's where they get the greatest financial returns on investment.[244]

Data from 2015 shows that the super-rich in the U.S. put 57% of their wealth directly into the elite economy of stocks, securities, derivatives, bonds, hedge funds, and mutual funds to personally profit even more.[245] And it's questionable how much of the remaining 43% actually makes its way back into the real economy. Eighteen percent of super-rich income is invested in real estate,[246] much of which is spent on luxury property, which just moves money from the hands of one super-rich owner to another, keeping the wealth in the elite economy. It can be argued that even when the super-rich buy property in order to rent it out in the real economy, they are actually priming the wealth extraction siphon even more, because they will be receiving rent from tenants who work in the real economy and they will put most of the payments they receive into the elite economy. So their investments in real estate are more likely to strengthen the siphon than to strengthen the pump. The other 27% of their wealth is kept in cash and bank deposits, some of which is spent into the real economy.[247] Many of their purchases, like $20,000 handbags, multi-million dollar yachts and vacations on private islands, keep wealth in the elite economy. In any case, **the super-rich take more from the real economy than they put back in.** This is the wealth extraction siphon at work. It is a self-perpetuating mechanism, because once you've bought equity in a successful company, the payments just keep rolling in from that investment, with very little or no work involved, allowing you to buy more equity in other companies. In other words, the for-profit world drives unhealthy levels of inequality because its accumulating tendencies are greater than its circulating tendencies.

---

[243] (Ref: CNBC report, Tax Policy Center report)
[244] (Ref: David Harvey talk)
[245] (Ref: Capgemini World Wealth Report 2015, Business Insider report on Capgemini report)
[246] (Ref: Capgemini World Wealth Report 2015, Business Insider report on Capgemini report)
[247] (Ref: Capgemini World Wealth Report 2015, Business Insider report on Capgemini report)

A very important aspect of how the siphon functions is that people in different income brackets spend money into the real economy at different rates.[248] A person living on a low income might spend 100 percent of their money back into the real economy, and a person living on a moderate income might spend 85 percent of it back into the real economy, saving 15 percent for retirement or a rainy day. But what small percentage of their wealth do billionaires spend back into the real economy in a given year?

Nick Hanauer, one of the richest people in the U.S., gave us an idea when he said in his TED talk, "There can never be enough super-rich people to power a great economy. Somebody like me, who makes hundreds or thousands of times as much as the median American… I don't buy hundreds or thousands of times as much stuff. My family owns three cars, not three thousand. I buy a few pairs of pants and shirts a year, like most American men. Occasionally we go out to eat with friends. I can't buy enough of anything to make up for the fact that millions of unemployed and underemployed Americans can't buy any new cars, any clothes, or enjoy any meals out."[249]

One report published by Pricewaterhouse Coopers in 2015, titled *Billionaires: Master architects of great wealth and lasting legacies*, determined that 83% of billionaires who decide to sell their businesses and 'cash out' put that money into financial and portfolio investing (i.e. the elite economy), rather than entrepreneurship.[250] This leads to the fact that 94% of all long-term capital gains in the U.S. go to the richest 20% of the population, and 47% go to the richest 0.1.[251]

Some super-rich people get the majority or all of their income from the elite economy, never really interacting with the real economy at all. The portfolio and financial investing of the wealthy only serves to expand the 'rentier class', people who don't work because they can live off of income gained from their assets.[252] In the U.S., the rentier class's wealth and power has increased significantly in recent decades. Between 1973 and 1985, the rentier class received 16% of domestic corporate profits, but by the 2000s this number had increased to 41%.[253]

Some of these millionaires and billionaires are growing their wealth exponentially through financial markets. What else can be expected when private wealth is seen as the primary driver of economic activity, and most big businesses hold profit maximization as their top priority?

A statistic that people throughout the world have come to know very well since the Occupy Wall Street movement began in 2011, is that 1% of the U.S. population has 40% of the nation's wealth.[254] This is where the term 'The 99%' came from – it refers to the 99% of Americans who have only 60% of their country's wealth. In fact, it is now estimated that the

[248] (Ref: Ha Joon Cha and the failure of trickle-down?)
[249] (Ref: Nick Hanauer, TED Talk)
[250] (Ref: PWC, *Billionaires*)
[251] % (Ref: Center on Budget and Policy Priorities, Tax Policy Center report)
[252] (Ref: David Harvey's RSA talk, minute 10; Extreme Wealth is Not Merited)
[253] (Ref: "The Quiet Coup: The Take-over of the U.S. by the Finance Sector", by Samuel Johnson, p?)
[254] (Ref: Stiglitz article)

richest one-tenth of 1% of Americans own almost as much wealth as the bottom 90% combined.[255]

The Walton family, for instance, owns 1.6 billion shares in Walmart, which delivers more than $3 billion to them annually. Similarly, the famous investor, Warren Buffet, receives more than $2 billion in dividends every year, and many others like Larry Summers and Bill Gates make hundreds of millions of dollars every year just from dividend payouts. And that doesn't even include capital gains they make from trading assets or the interest payments they receive on other financial assets. These numbers refer strictly to the money paid out of company profits at the end of the year.[256] It's easy to see why capital gains and dividends have been the largest contributors to rises in inequality in the U.S. (Hungerford, 2013). This can only happen in a for-profit system.

It's not just a U.S. phenomenon, though. The entire bottom half of the U.K. only owns 2-3% of the wealth[257] and, likewise, the bottom half of the global population together possess less than 2% of the world's wealth.[258] Inequality is at a historic high now,[259] with more than 80% of the world's population living in countries where income differentials are.[260] Even in 'emerging' economies, like China and Brazil, where we were supposed to see the formation of a large middle class, we are seeing enormous inequality persist. The World Bank and the Chinese government both concede that the wealth gap in China is among the world's largest.[261]

A quick look at history will show just how dramatically inequality has risen in the past 50 years.

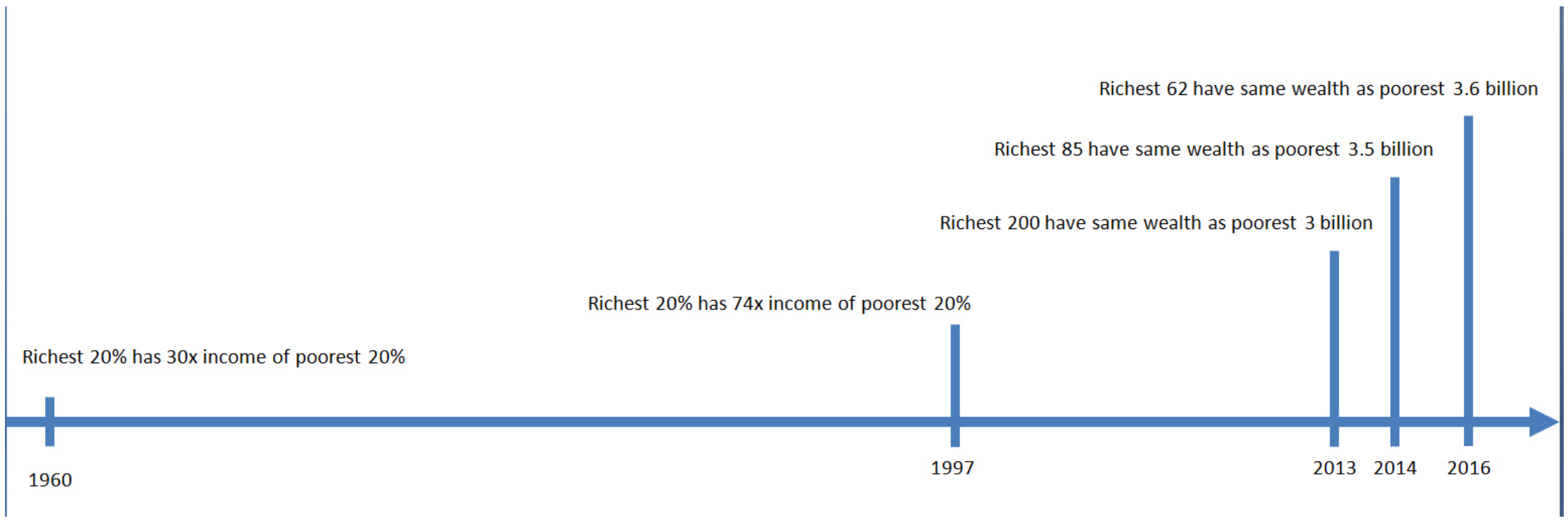

## Timeline of Global Inequality

In 1960, the richest 20% of the world's population had 30 times the income of the poorest 20%.[262] In 1997, nearly forty years later, they had 74 times as much (Ibid). In 2013, the international charity Oxfam reported that the richest 200 people in the world owned as much wealth as the poorest 3 billion[263]; a shocking statistic. However, in 2014, they updated this

255 (Ref: Saez & Zucman, Wealth Inequality in the US since 1913)
256 (Ref: Major Executives Getting Rich Off Dividends)
257 (Ref: Toynbee article)
258 (Ref: ZeroHedge article (primary source?))
259 (Ref: Pew Research Center?)
260 widening (Ref: UN Report)
261 (Ref: Social Enterprise Emerges in China)
262 (Ref: The Economist; primary source: United Nations' *Human Development Report?)*
263 (Refs: Al Jazeera and Oxfam)

number, showing that 85 people owned as much wealth as the poorest 3.5 billion.[264] That's a big change in just one year! The follow-up Oxfam report published in 2016 estimates that now the 62 richest billionaires have as much wealth as the poorest 3.6 billion people in the world.[265] This latest report also found that the richest 1% of the world's population own more wealth than the other 99% of the people on the planet combined! (Ref: Ibid) Incredibly, there are single suburbs, like Mosman, Australia, whose combined taxable income is more than some countries' GDPs.[266]

These super powerful individuals are not only using their wealth to generate even more for themselves (as any good capitalist should), but a great number of them are also evading taxes. They do this by keeping their wealth in 'tax havens', places that are happy to help the super-rich avoid paying taxes, by offering exceptionally low or no taxes on wealth and also by protecting banks from having to report their clients' wealth. Well-known tax havens include Luxembourg, Bermuda, Switzerland, the Cayman Islands, and more recently some U.S. states, such as Delaware, Nevada and South Dakota.[267] It is estimated that a global total of between $21 and $32 trillion of super-rich individuals' wealth is sheltered in tax havens.[268] For comparison, the total U.S. GDP in 2014 was about $17 trillion.[269]

This tendency for inequality to grow is exacerbated by the for-profit story - that the rich are good capitalists who will use their wealth to benefit society, an idea that leads to the wealthy receiving all kinds of special privileges and advantages, making it even easier for them to accumulate more wealth. In his talk, one-percenter Nick Hanauer also pointed out that the incredible difference between the 15% tax rate that capitalists pay on carried interest, dividends, and capital gains, and the 35% top tax rate on work that regular Americans pay is difficult to justify without a bit of deification. (Ref: Nick Hanauer talk)

From the for-profit worldview, it is justifiable that the wealth of the world's 7 richest people is more than the combined GDP of the 41 most heavily indebted low-income countries in the world, consisting of 567 million people[xxiii].[270] Many successful capitalists will say that this needs to be fixed somehow, admitting that it is unsustainable (indeed, Nick Hanauer wrote a memo to his 'fellow zillionaires' called: 'The Pitchforks Are Coming… For Us Plutocrats'[271] but the for-profit worldview continues to justify itself. This is just a temporary glitch; the market will get things back to normal eventually. Yet, **without addressing the wealth extraction siphon, we will only be able to make small adjustments within an economic system that inherently leads to more inequality.**

The statistics above are shocking. They show us that our world's resources, including the basic resources needed to live, are not being distributed fairly. Nearly 800 million people are undernourished and 2.5 billion don't have access to sanitation,[272] while a few dozen billionaires have more wealth than the rest of the world combined. Even if you're in the middle class in Europe or the U.S., odds are that your real wage (your wage in relation to the cost of living) has stagnated or declined while the super-rich have been multiplying their wealth.[273] It is clear this system is failing to allocate goods and services as efficiently as we need it to. This level of inequality is simply unacceptable, inhumane, and a dangerous threat to social cohesion. And **these numbers would actually be much worse if it weren't for great efforts to counteract the siphon** with progressive taxation and philanthropy**.** The extreme inequality that has developed despite such efforts shows that wealth redistribution strategies will never be able to outpace the exponentially widening income gap inherent in a for-profit economy.

---

264 (Fuentes-Nieva & Galasso, 2014)
265 (Ref: An Economy for the 1%).
266 (Ref: Postcode income bigger than GDP)
267 (Ref: Bloomberg tax haven article)
268 (Ref: Tax Justice Network Report)
269 (Ref: World Bank data)
270 (Ref: Global Heritage Fund)
271 (Ref: Hanauer article),
272 (Refs: World Food Program, World Water Day (UN))

In order to gain a better understanding of current economic trends, economist Thomas Piketty painstakingly analyzed massive amounts of historical tax return data. In his book, *Capital in the 21st Century*, he describes that capitalism has an inherent tendency to increase inequality because returns on capital exceed the rate of economic growth and so any economic growth benefits only those at the top.[274] Piketty also notes from his data that we have entered an indefinite era in which returns are (and will continue to be) higher than the rate of economic growth. This means massive inequality is written into the very fabric of capitalism in the 21st century.

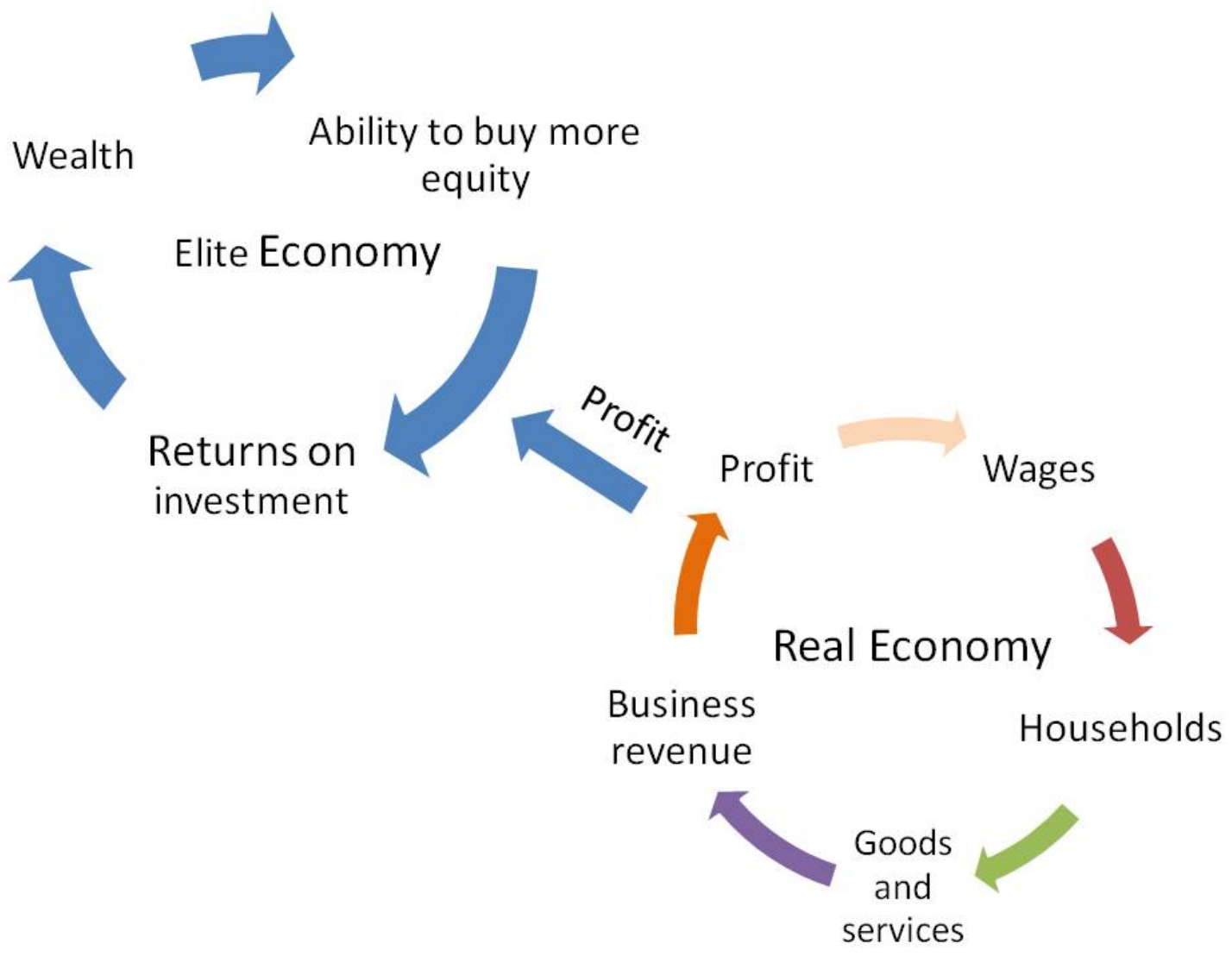

**Wealth Extraction Siphon in More Detail**

## The For-Profit World Always Wants More

Some don't see Picketty's conclusion as a problem, because they aren't convinced of capitalism's requirement for growth. Nobel Prize winning economist Robert Solow notably stated, "There is no reason at all why capitalism could not survive with slow or even no

273 (Ref: Congressional Budget Office, *International Labour Organisation Global Wage Report 2014/15*)
274 (Piketty, 2014)

growth... There is nothing intrinsic in the system that says it cannot exist happily in a stationary state."[275]

Unfortunately, this is simply not true. In order for capitalists to amass more and more wealth, economic activity must continuously expand. The economy must grow in order to simply generate more of a surplus each year (as a result of firms maximizing profit). But **the for-profit system also requires economic growth in order to compensate for the surplus taken out of the real economy.** The only way that a business and an economy can continue to compensate for the privatization and extraction of wealth is by expanding.

Some might argue that the for-profit system does not require growth because businesses can compensate for their profits being privately distributed by reducing costs. In essence, they lower their expenditures, rather than expand, to maintain or increase their profit margins while staying the same size. Aside from the ethical issues that are often part and parcel of strategies to reduce costs (like moving factories abroad in order to exploit cheap labor and take advantage of lax environmental and labor laws), costs can only come down so far; it's not a viable strategy in the long-term. The for-profit system requires growth. And this brings us to another crisis of capitalism: growth at all costs.

## Money Creation

Another symptom of this economic system's self-destructive tendencies that is closely tied to inequality and concentration of wealth and power is debt. Growing inequality is driving more people to take out more debt, which puts poorer people in an even more difficult financial position and it puts the wealthy people who benefit from debt in an even more advantageous financial position. In essence, the modern banking system itself is built on debt, which, in the hands of for-profit banks, acts as a siphon that sucks wealth from the poorest to the richest people in society. Usury in the for-profit economy is leading this system to its own destruction.

In the for-profit economy, fractional reserve banking means that banks can create money out of nothing and loan it out to people who don't have much money to create private profit for people who already do have a lot of money; the shareholders, owners and investors of for-profit banks. And, of course, the whole system promotes consumers to take on more debt for the sake of consumption, in the name of economic growth and profit maximization. In fact, debt is so ubiquitous, that it's estimated that 35-40% of the costs of any product are actually interest on debt.[276]

Thanks to the exponential growth of compounding interest, world debt is at US $199 trillion, having risen by $57 trillion since 2007.[277]

Fractional reserve banking refers to the common practice of banks only being required to keep a fraction of the loans they make in their reserves. So, as a hypothetical, if the fractional reserve rate in Australia is 5%, an Australian bank that loans out $100, only has to actually have 5 of those dollars in its reserves. It effectively creates the other 95 dollars when it makes the loan. Then when the $100 is paid back by the borrower, the slate is wiped clean and the 95 dollars ($100 minus the reserve amount of $5) that were created no longer exist. However, the interest that must be paid on the loan was also created. So, if the interest on the loan was 3%, compounding annually, and it takes the borrower two years to pay back the loan, then the total interest would be about $7. That extra $7 did not exist in the system before the loan was made and, thus, it was created as debt. Advocates for ending the fractional reserve system argue that more debt-based money must be loaned into the system in order for that extra $7 to be paid back to the bank because that extra money doesn't exist in the system[278]. So, interest on debt requires more debt to be created, which requires more debt to be created in order to pay off that new debt, in an infinite spiral of growing debt and interest. This means that there will always be more debt in the system than there is money to pay back the debt, which puts pressure on banks and borrowers to continuously keep growing the money supply through creating more debt.

---

275 (Ref: Stoll, 2008).
276 (Ref: Margrit Kennedy)
277 (Ref: RT article about McKinsey Global Institute study)

So profit-maximization is the goal that drives the system of money creation through debt.

The compound interest is piling up and must get paid down by borrowers. Banks that pay out dividends and give returns to owners and shareholders need that interest to be paid down in order to deliver it to the people at the top.

This is largely motivated by the desire to make the bank owners and shareholders even richer, the sign of a successful for-profit company. However, the bank owners and shareholders gaining from the interest paid on loans contributes to worsening inequality in society, as the wealthy gain from the interest payments of people in the working class. The more inequality there is in society, the more people in the low-income bracket go into debt, sometimes in order to consume to compensate for their low financial standing, and sometimes just to scrape by. The more people go into debt, the more the wealthy stand to gain.

Money creation by for-profit banks in the for-profit system creates a situation of scarcity, because much of the interest being repaid by borrowers is distributed to bank owners and shareholders and is, thus, siphoned up into the elite economy. This means there's less liquidity in the system (basically less money in the system), so there is a necessity to keep loaning more and more money into existence in order to keep the economy going and growing. In fact, this debt-based money in the for-profit context is another form of pressure on the economy to grow, because all of the interest extracted must be compensated for somehow – it must be loaned into existence.

# Crisis of Wellbeing

For-profit companies employ many different strategies to try to meet their endless need for expansion. Over the years, companies have become more and more manipulative in their production and advertising methods, enticing consumers to constantly buy more products. This focus on consumption has not only taken over the economy, it's taken over our lives, lowering our levels of wellbeing. And it seems there are no limits to where companies will try to extract profit.

---

278 If this all sounds confusing, don't worry. It is. There are lots of resources on the Internet that explain fractional reserve banking, if you're interested. Perhaps it's worth mentioning that we used to be very skeptical about fractional reserve banking ourselves.

## Race to the Bottom

In 2011, Australian journalist Stuart Washington wrote an article for the *Sydney Morning Herald* about an $8 pedestal fan. Why would a journalist at a top-knotch national newspaper write an article about a fan?

In the article, cleverly titled 'Absurdly low prices fan the flames of a flawed economy', he shares the story of when he visited a friend's home and his friend pointed to a fan on the floor and asked him to guess how much it cost. Stuart guessed $80, maybe $90. His friendly proudly revealed that he had paid $8 for the item, because it was discounted from $12 at the shop.

Stuart felt an instant sense of unease. How could a fully-functioning pedestal fan, that was doing a great job of cooling the room, cost $8? He felt that something wasn't right. How could $8 possibly cover the cost of the metals, plastic, labor, transportation, packaging, marketing and sales that went into getting that fan into the store from which his friend bought it?[279]

Mr. Washington's article goes on to blame bad pricing for short-term behavior, but if you zoom out, it's a bit more complex than that. What drives the bad pricing?

When *Homo economicus* is guided by profit-maximization in the for-profit market, the ruthless competition inherent in the capitalist system brings about a race to the bottom.

Producing a lot of something costs less per unit than it does to produce a little of that same something. For instance, it might cost a lot to manufacture just ten pedestal fans in your basement. But if you're manufacturing 100,000 fans per year, you can invest in a factory and it will cost a lot less to manufacture each fan, because you'll have assembly lines where workers or robots put the pieces together very efficiently in a continuous flow. An assembly line could put together dozens of fans in the same time it would take you to put one fan together at your house. If you produce 500,000 fans at your factory next year, each fan will cost even less to make. This is called *economies of scale*. Companies that make cheaper products generally outcompete their peers who sell the same products at higher prices, so economies of scale have enabled large companies to outcompete smaller producers in many industries.

Mass production isn't the only way that companies have reduced costs, though. Survival of the fittest in the past few decades has meant that, in many sectors, only the producers who moved to countries where labor is cheap, safety standards are lax, and environmental damage doesn't cost anything can prosper or even survive. Manufacturers who don't move abroad have higher costs and can't compete.

In the last decade or so, producers have also been lowering costs by cutting corners and using cheaper materials.[280] Prices have fallen, but so has quality.[281] Paul Midler, author of *Poorly*

---

279 (Ref: The $8 Pedestal Fan)
280 (Ref: Global Supply Chain Quality Management).
281 (Ref: Poorly Made in China by Paul Midler?, Global Supply Chain Quality Management?).

*Made in China* calls this phenomenon 'quality fade'.[282] The pedestal fan Stuart's friend bought for $8 is probably going to fall apart or stop working a lot quicker than a fan made ten years ago, which would fall apart quicker than a fan made 20 or 30 years ago. But that's not a concern for most manufacturers. Poor quality is a positive thing for them, because it means that people will have to buy a new fan sooner rather than later, which means that sales and profits will rise. The competition to produce the highest-quality products has given way to the competition to produce the cheapest products. Why? This comes back to the *Homo economicus* consumer, always looking to be clever and save a buck in the short-term. (Of course, in light of current levels of inequality, this also has to do with the fact that most consumers simply have less money to spend.)

We can see the trend to produce cheaper, lower quality products in almost every manufacturing sector, from electronics and toys to clothes and shoes. It's even reached the point where food products, medical equipment, and construction materials - where we'd hope quality would always be the highest priority - increasingly come from unregulated factories overseas[283] and are cause for serious health and safety concerns.

This race to the bottom has co-evolved with a phenomenon called planned obsolescence. This is when companies intentionally design products to break or become obsolete before their time, in order to keep consumption up and boost profits.

The 2010 documentary about planned obsolescence, *The Light Bulb Conspiracy*, features many examples of this phenomenon. It starts with a printer that stops working. The printer's owner, Marcos, takes it to a few different repair shops, but is told by each one that it wouldn't be worth fixing, and that it would be cheaper to buy a new one. The end of the film reveals that the printer was programmed to stop printing after 1000 prints and Marcos was able to get it started again by simply resetting the counter to 0. This story probably resonates with most of us very well. Living in the 21st century, the line, "You might as well get a new (insert name of object here). Repairing it will cost more than buying a new one," has almost become a mantra.

Communications technology is one sector that is basically designed for planned obsolescence. Companies could give us the most current version of the mobile phone, tablet, or computer they've developed, but we won't see the latest version for several years. Instead, they offer us a slightly different version every year, in order to make last year's version seem obsolete and spur consumers to buy the latest adaptation. This doesn't only work on consumers who are concerned about keeping up with the latest trends and looking up-to-date. Technology companies create less-than-optimal software for their products on purpose, so that next year's edition can resolve the issues they intentionally created in this year's edition and the consumer can look forward to having a better experience each year… as long as they can afford to buy a new phone or tablet every year.[284]

---

282 (Ref: Poorly Made in China by Paul Midler)
283 Ref: Ibid?, Global Supply Chain Quality Management?
284 (Ref: Sustainable Electronics article?; Bodies of Planned Obsolescence?; Environmental Impacts of Planned Obscolencence?)

As economic commentator Umair Haque puts it, "Most businesses are still serving up the economic equivalent of fast food: negative-impact goods and services that fail to make people, communities, and society tangibly better off."[285]

Immoral as it may seem, this is common practice and considered a perfectly legitimate strategy to maximize profits in the for-profit world. Many large companies generate a surplus by exploiting workers, destroying ecosystems, and selling low-quality products that break-down soon after they're purchased, mostly in order to generate profits for owners and investors and to increase their share price. Large numbers of people are being taken advantage of in order to line a few individuals' pockets.

In capitalism, markets are supposed to efficiently assign prices to goods and services through negotiations between buyers and sellers based on the information they have. However, in the for-profit world, it's often in the sellers' self-interest to manipulate or withhold information related to their products. This keeps costs low, but means consumers have imperfect information when it comes to their choices. Consequently, prices do not reflect the actual cost of goods and services, especially in terms of negative effects on ecosystems and human health.

Many clothing companies in high-income countries produce their clothes in factories in low-income countries with very unsafe, unfair and unhealthy working conditions. In some of these factories (also known as sweatshops), workers are essentially modern-day slaves. The companies do this so that they can sell clothes at cheap prices and make large profits. However, few consumers have access to information about how clothes are made and by whom, so their choices can't take that into account when comparing their purchasing options, even though it's an important human rights issue. The same goes for the environmental impact of products; it is nearly impossible for the average customer to find out what the environmental impact of a certain product is. In fact, just the opposite is often true: green marketing tactics ('greenwashing') are used to make customers think that products were produced in a more ethical, environmentally-friendly way than they really were.

Incomplete or concealed information about how things are produced means that consumers can rarely make fully informed choices, as capitalism assumes.

## Dominator Model

In places where the for-profit story has taken the strongest hold, we are witnessing a very steep decline in wellbeing.

In the early 2000s, U.K. public health researchers Richard Wilkinson and Kate Pickett collected and analyzed long-term data on a wide variety of social problems including homicide, drug use, crime, mental illness, suicide, infant mortality, obesity and life expectancy, in 20 'developed' countries. They compared the data to levels of socio-economic equality in those countries and found that there is a very strong correlation between inequality and social problems. The countries with more inequality, like the U.S. and the U.K., do far worse on all of the wellbeing indicators than countries with less inequality, like Japan and Sweden. Wilkinson and Pickett concluded that more equal countries have higher levels of wellbeing.[286] And this, of course, makes sense, but is inequality really the root-cause of all of this, or does it go even deeper?

[285] (Ref: Haque p. 20)

In the 1970s, cultural historian Riane Eisler described in her book, *The Chalice and the Blade*, the difference between what she identified as 'dominator cultures' and 'partnership cultures'. The latter are cultures that value cooperation and equality. They honor and uphold those who give and contribute to the wider community. In contrast, the dominator cultures uphold those who are most effective at conquering and controlling others. Accordingly, dominator cultures tend to be unequal and violent. Somebody has to be at the top and somebody has to be at the bottom.[287]

Eisler argues that for at least 30,000 years of human existence, partnership cultures were the norm, but at around 5000 BCE, dominator cultures arose quite rapidly and conquered the partnership communities. And we have been living in a dominator world ever since.

The worldview of 'only the strong survive' justifies and encourages the domination of enemies and the 'weak'. In the dominator mindset, the way we know when someone or something is weak is when we have been able to conquer it, just like the European conquerors who went all over the world, subjugating indigenous people in Australia, Africa, the Middle East, Asia and the Americas. Did the conquerors feel bad for the people they slaughtered or enslaved? According to the dominator worldview, the weak will inevitably be conquered because it is simply nature's way. This way of organizing society naturally leads to the concentration of power in the hands of the top dominators.

Eisler illustrated that the same 'dominator model' of politics and the economy that encourages individuals, companies and countries to dominate each other also encourages the domination of women and minorities as well as the domination of nature. Indeed, the quest of the Scientific Revolution, which started in Europe in the 1400s, was to conquer nature. Many famous figures of the time, like Descartes, talked about taking nature apart in order to understand and manipulate it. Scientist and philosopher Francis Bacon, in an extreme example, even talked about torturing nature to make her reveal her secrets,[288] and it's no coincidence that he used the word 'her'. Eisler explains that anything seen as caring, giving or nurturing is seen as weak by the dominator mindset and so it has to be controlled. As women are the givers and nurturers of society, because they give birth, breastfeed and take care of small children, they have been identified as weak in dominator cultures and are thus oppressed.[289] However, the oppression is not limited to women. Men who are perceived as caring or nurturing or otherwise 'weak' are also oppressed.

Part of the drive to dominate is based on the quest to be the all-powerful leader, but perhaps a more significant part of the dominator drive is based in fear of being dominated, which has translated into a fear of the 'other'; a fear of the unknown and of not having control. This manifests in sexism, racism, xenophobia, genocide, religious oppression, abuse, and bullying. One of the core principles of the for-profit world has been if you can dominate some 'other', then you should, or they might dominate you. This narrative has motivated much of the oppression we've witnessed in the for-profit era. It alienates us from nature, each other, and ourselves.

---

286 (Wilkinson & Pickett, 2009)
287 (Ref: The Chalice and the Blade)
288 (Ref: Critiques and Contentions article)
289 (Refs: The Chalice and the Blade)

The for-profit story is a manifestation of the dominator narrative, and capitalism is a dominator economic model. It encourages businesses and people to conquer each other and, as such, it naturally results in high levels of inequality, no matter how much we try to counteract its dominator tendencies with taxes, regulation and appeals for philanthropy. The for-profit culture values the act of taking over the act of giving. This is why it has led to an economy of fear and scarcity. In many ways, it's a self-fulfilling prophecy.

Capitalism also puts high value on the parts of the economy that dominate, conquer and expand. It doesn't put much value on the parts of the economy that nurture and conserve. The 'caring economy' and 'women's work' are simply not valued in the for-profit world, not only in a cultural sense, but also in a financial sense.[290] Daycare workers, teachers and nurses are often undervalued for the services they provide in fulfilling basic social needs. People who work or invest in sectors that are based on extracting as much value as quickly as possible and then moving on to the next project (i.e., sectors like drilling, mining, construction, manufacturing, advertising, marketing, gambling, arms trading, finance and banking) are those who make the most money in the world.

Understanding that our economy functions through domination helps us answer Stuart Washington's question about how a pedestal fan can cost $8. Raw materials needed for a fan, like metal, can be as cheap or expensive as the owner of the land where the steel is mined wants it to be. After all, the land isn't charging them any rent – they own it, they've conquered it. Mining and drilling companies can charge whatever it takes to outcompete the other mining and drilling companies and keep large orders coming in. Likewise, the enforcement of a minimum wage is questionable in China, where the fan was probably assembled, and workers' rights are sub-standard and often overlooked, making labor as cheap as factory owners want it to be. The managers dominate the workers and the owners dominate the managers. So, what really allows the fan to be produced and sold for $8? In short, **the workers, the ecosystems and future generations are paying for unacknowledged costs in ways that can't be measured in dollar amounts.** All because they found themselves at the losing end of a game of violent domination.

Perhaps it is the capitalist impetus to dominate others in order to extract private profit that is responsible for inequality as well as the other social problems Wilkinson and Pickett studied. The for-profit story has served to make us feel ever more separate and isolated from each other. Not only is it a widely ingrained story that encourages greedy, selfish behavior, but it also pits us against one another in a dog-eat-dog economy, where the value of a person is determined by how much they take rather than how much they contribute. The competition, greed, selfishness, individualism, consumerism and workaholism that are inherently part of the for-profit system have, to a great extent, eroded our relationships, our communities and our self-esteem. They've left many of us feeling unworthy of love and support, and have encouraged us to perpetuate the for-profit myth in our social relations. All of this has an effect on our mental and physical health.

## Consumerism

[290] (Ref: Riane Eisler; Ariel Salleh)

One way in which the for-profit system has eroded our relationships with each other, as well as with nature, is that it seeks to commodify and marketize everything it possibly can. The growth imperative of the for-profit world fuels the idea that everything should be a commodity. After all, markets do get saturated and if there were a limit on the number and size of markets, growth would end as soon as everyone had all the houses, TVs, and cars they can handle, and capitalists would no longer be able to maximize their profits. However, as all good economic and business students know, all you have to do to expand the market is create a new niche. There are two ways of doing this: pursuade or force people to start paying for something they weren't previously paying for; or manufacture a new 'need'.

In the case of the first strategy, Charles Eisenstein provides an excellent exposé of the many ways that the market has taken over more and more of our lives. He points out that we used to cook most meals at home, but more and more often we pay for our food to be prepared for us. (In fact, Americans now spend more money at restaurants than at grocery stores for the first time ever.)[291] We used to tell each other stories as a form of entertainment (think of families having story-time or each village having its own storytellers). This has been replaced to a large extent by stories that we pay to read or watch in best-selling books and blockbuster movies. Childcare used to be done by parents, family members and neighbors, free of charge. Now we routinely drop our children off at daycares, where we pay strangers to care for them. This trend of paying for things that we wouldn't have imagined paying for just a generation or two ago might be fine in some areas of life, but it becomes absurd with the advent of 'professional cuddlers' and 'friends for hire' (both of which truly exist).[292]

Looking back through history, we can see that this is the latest stage of an ongoing 'enclosure of the commons'. For most of history, the fields of England had been regarded as communal property and were left open for all to use, but in the 1400s the Parliament began to close the land off to the public with fences so that landlords could use them for private benefit.[293] The landlords wanted to be able to farm more efficiently rather than having to worry about the community's crop rotation schedule, and they wanted higher rent from people working the land (Ref: Ibid?). The ongoing enclosure of the commons, in all its forms, is necessary to fuel the growth required by the for-profit economy.

The commodification of a range of goods and services is both a cause and effect of the for-profit world. We now have to pay for things that used to occur as transactions based on trust and friendship. Often these commodities – such as child care and fast food – are bought because people are time-pressed from working longer hours and have longer commutes to and from work.

The more time we spend at work, or getting to and from work, the less time we have to satisfy a variety of needs and desires, like exercising, preparing healthy food, caring for friends and family members, participating in leisure activities, and even simply resting.

The second strategy for market expansion is to create a 'need' where there isn't one, or even to create an entire market where there isn't one. How do you do this?

---

291 (Ref: Bloomberg article)
292 (Ref: professional cuddlers and rent a friend)
293 Ref: Common Rights to Land in England, English Peasantry and Common Fields

In the movie *The Wolf of Wallstreet*, the protagonist, Jordan Belfort, is giving a seminar to teach people how to become better at sales. He hands a pen to a member of the audience and says, “Sell me this pen.” The man starts to describe all of the wonderful features of the pen. Jordan takes the pen away, interrupting the man’s sales pitch. He hands it to the next man in the row of chairs, “Sell me this pen.” The man starts to describe what you can do with the pen, like writing down your life’s memories. He takes the pen away, interrupting the second man’s sales pitch. He hands it to yet another man, looking a bit frustrated. Earlier in the movie, he had asked one of his salesman friends to sell the pen to him and the friend responded, “You want me to sell you this pen? Do me a favor. Write your name down on that piece of paper.” Jordan responds, “I don’t have a pen.” The friend sets the pen down and said, “Exactly. Supply and demand, my friend.” He created a sense of urgency and need that the protagonist had not previously had. He manufactured a need.

*The Century of the Self* documentary series, by BBC filmmaker Adam Curtis,shows how mass consumerism was a carefully planned, manufactured trend that emerged in the U.S. in the 1920s. The documentary describes the overproduction worries after the end of World War I. Mass production had been hugely successful during the war and continued to roll out products at an incredible pace, but there was the fear of a looming overproduction crisis. Until then, companies had advertised their goods based on their usefulness, like the men who described the virtues of the pen and how it could be used.

Curtis explains, “Goods like shoes, stockings, even cars were promoted in functional terms, for their durability. The aim of the advertisements were simply to show people the products’ practical virtues, nothing more… There was no American consumer. There was the American worker. And there was the American owner. And they manufactured, and they saved and they ate what they had to and the people shopped for what they needed. And while the very rich may have bought things they didn’t need, most people did not.”[294]

This sounds exactly like the vision that many people have today of what sustainable consumption is: most people bought only what they needed. Why did it change?

Corporations realized they had to change the way most Americans think about products. A leading Wall Street banker at Lehman Brothers, Paul Mazur outlined what exactly was needed in an article he wrote for the *Harvard Business Review* in 1927. “We must shift America from a needs, to a desires culture. People must be trained to desire, to want new things even before the old had been entirely consumed. We must shape a new mentality in America. Man’s desires must overshadow his needs.”[295]

This idea took hold strongly among economists and the business community. In 1955, economist Victor Lebow wrote along similar lines in the *Journal of Retailing*, “Our enormously productive economy…. demands that we make consumption our way of life, that we convert the buying and use of goods into rituals; that we seek our spiritual satisfaction, our

294 ( Refs: Century of the Self)

295 ( Refs: Century of the Self; Harvard Business Review 1927 (See: Inventing Times Square, Note/Reference 50)).

ego satisfaction, in consumption… We need things consumed, burned up, replaced and discarded at an ever-accelerating rate."[296]

The results of this line of thought can be seen all around us in the ubiquitous advertisements that appear on walls, TV screens, bus stops, highways, as well as in buses, trains and airports. Unless you live in one of the few places that have banned advertising in public spaces, you can't avoid advertisements however hard you try. In 2011 alone, the advertising industry spent $464 billion marketing the consumer lifestyle worldwide.[297]

We live in a world in which everything is a brand – even cities, cultures and people - and so we have become numb to ads, hardly even noticing them. Yet our subconscious recognizes certain brands and logos over others when we are shopping for food, clothes, and other everyday items.

Marketing in the 20th century worked so well that it successfully created a culture of shopping. Shopping has become a pastime, a therapeutic hobby for many of us. If we're anxious, we go shopping. If we're bored, we go shopping. If we want to hang out with friends, we go shopping. When we travel, how do we experience a different place and culture? We go shopping. In fact, it's become very rare in many parts of the world for people to socialize in ways that don't involve paying for goods or services.

We shop in order to identify ourselves, we shop as a pastime and we shop to make ourselves feel better. All this shopping requires money. How do we get enough money to shop as a leisure activity? We must work even more.

The *Story of Stuff's* Annie Leonard dissected this work-watch-shop treadmill.

> "…do you know what the two main activities are that we do with the scant leisure time we have? Watch TV and shop… So we are in this ridiculous situation where we go to work, maybe two jobs even, and we come home and we're exhausted so we plop down on our new couch and watch TV and the commercials tell us 'YOU SUCK' so we gotta go to the mall to buy something to feel better, then we gotta go to work more to pay for the stuff we just bought so we come home and we're more tired so you sit down and watch more T.V. and it tells you to go to the mall again and we're on this crazy work-watch-spend treadmill."[298]

High levels of consumption not only require a lot of money (or debt), and nature's resources, but also a lot of time. No matter how wealthy we are, each of us has a budget of 24 hours in a day. We can spend most of that time playing, creating, connecting, resting, and enjoying life or we can spend it accumulating and managing money and material goods.

In much of the 'developed' world, there is a tacit game of one-upmanship about how many hours we work – in many cases, this becomes a status symbol, a way we show how valuable and important we are. This is creating a work culture in which it's looked down on to leave the workplace at a reasonable hour, to take vacation time, and even to get a healthy amount of sleep[xxiv].

---

296 (Ref: Victor Lebow, *Price Competition in 1955*)

297 (Ref: Erik Assadourian's talk)

298 (Ref: Story of Stuff)

Research conducted by The Australia Institute found that each year Australians work almost 110 billion Australian dollars' worth of *unpaid overtime*.[299] It's gotten so bad in the land down under that there's even a reality TV show that encourages people to take their annual paid leave time from work.[300]

To add injury to insult, the wealth extraction siphon is so efficient that when employees of for-profit firms work more paid or unpaid hours, they're putting themselves in a worse position, because the more they work, the more profits their employers make, the more money is siphoned out of the real economy, and the lower their real wages will be in the long-term.

Perhaps most sadly of all, much work in the modern economy involves doing small, meaningless tasks (often paperwork) in order to maximize profits for a company that then gives those out as dividends to distant owners, who don't really need more money and didn't work to earn it. Turning a profit is the primary reason the companies that employ many of us even exist. In addition to working long hours, this feeling of meaninglessness in one's work can be exhausting and dispiriting. It prompted Professor of Anthropology at the London School of Economics, David Graeber, to pen an article called 'On the Phenomenon of Bullshit Jobs', in which he explores the links between growing inequality and meaningless work. A Gallup poll in 2012 found that only 1 out of 8 people, globally, are engaged and feel motivated at their jobs.[301]

In this system, education is seen just as a means to getting a job (albeit a job one feels disengaged from). In his well-known 2006 TED Talk *'Do Schools Kill Creativity?'*, Sir Ken Robinson pointed out that our education system is based on an industrial way of thinking.[302] Our schools were created in the image of mass industrial output. It's the carryover of the for-profit ideology; everything supports the single-bottom-line. Modern schools are structured like assembly lines. Each year in school is like another phase in the assembly line, where knowledge is put into our minds like new parts are put into a car. The entire goal of these industrial schools is for each student to come out into the 'real world' and start a career. Some might go on to be capitalists, some might try to work their way up the corporate ladder, and some might stay on the bottom rung. But the idea is that if you are successful in the assembly line schools, you will be successful in creating a career. And the whole point of having a career is to accumulate as much power and wealth as possible. There's very little sense of deeper meaning built into our economy.

Most ecosystems have a lot more play and leisure than our economic system allows for. As social creatures who learn through play and who have a basic need for creativity, we are deprived of important opportunities to learn and express ourselves. Rest is an integral part of the equation too. That's why, when people work 50, 60, or even 80 hours per week (which has become quite common), they often experience 'burn-out' or a break-down, resulting in mental and physical illness[xxv].[303]

---

299 (Ref: The Australia Institute report)
300 (See: No Leave, No Life)
301 (Ref: Gallup)
302 (Ref: Ken Robinson talk)
303 (Ref: Leiter & Maslach, Handbook of Health Psychology, 2001)

Yet the original promise of progress and technological advances was that we would all have more leisure time, as our basic needs could be met with less work and increased productivity via technology. However, in a for-profit system which necessitates incessant growth and in which the surplus is extracted to the elite economy, this can't happen for most of us. Instead, technological advances are used for, and actually geared toward, the encouragement of ever more consumption and work, such as the technology that goes into planned obsolescence. Another example is the amount of technological innovation that's geared towards marketing and advertising, like High Definition billboards.

Although it might appear that there is more abundance than ever, many people feel a sense of scarcity because there's such intense competition for jobs, money and status. Not everyone can make it to the top, and extreme inequality means there's a struggle to keep from falling to the bottom - because the bottom means not having enough to make ends meet. The constant threat of scarcity is built into the modern global economy.

Most people have some sense of cognitive dissonance, a sense that something isn't quite right, in terms of the work-watch-shop treadmill, and many feel little or no sense of purpose in work.[304] Many people feel trapped and unable to live differently. The fear of poverty, shame and social exclusion is more than enough to keep them from hopping off the treadmill.

It is no wonder that mental and physical health are declining.[305] Since the 1950s and 1960s, happiness hasn't increased in either the U.S. or Australia, despite large increases in the average financial wealth of citizens in both countries.[306]

What's the answer to this in the for-profit world? It's another opportunity to turn a profit, by treating these symptoms of a much deeper, systemic problem. We throw anti-depressants, anti-anxiety medications and a whole host of other psychotropic pharmaceuticals at it. Ironically, some of the most popular anti-depression medications, like Prosac and Paxil, list 'suicidal thoughts' as a side-effect.

Psychotropic drugs might play an important role in helping individuals deal with depression, but what might happen if more of our efforts and ingenuity went towards resolving the major socio-economic drivers of feelings of isolation and emptiness?

The for-profit system also encourages 'quick fix' thinking and compulsive behavior.[307] When we don't have time for ourselves, we don't address feelings of anxiety and stress in healthy ways. Instead we turn to things that make us feel better instantly, but only for the short-term. This paves the way for a culture that encourages compulsive behavior in seeking instant gratification, like retail therapy, binge eating, drug use and Internet porn; reinforcing the vicious cycle of consumption and feelings of alienation. And this vicious cycle increases levels of violence, homicide and suicide in society. People feel disenfranchised and become apathetic from being on the work-watch-shop treadmill for too long.[308]

---

[304] (Ref: Gallup 2013)
[305] (Ref: Bruce Levine article, How Stress Influences Disease, Incidence and Prevalence of Chronic Disease)
[306] (Ref: Easterlin)
[307] (Ref: 'Why capitalism makes us sick' talk by Gabor Maté, and Andy Fisher)
[308] (Ref: Newsweek article linking suicide epidemic to societal problems)

These increased levels of stress, anxiety and isolation also manifest in physical health crises.[309] People eat unhealthy foods, because they don't have enough time or energy to cook. Nor are most people given adequate information about healthy nutrition because most healthy foods (like fresh fruits and vegetables) aren't big brand names, so they aren't advertised nearly as much as unhealthy, processed, convenience foods. Nor do many people have enough time to be active and get exercise. Sedentary jobs have replaced work that we used to do to meet our basic needs at home, like care-giving, food production, and making things for the household – all of which involved physical movement. Many of us attempt to compensate for this by going to the gym and paying to satisfy our need for physical activity, when we should just be able to burn off calories by being naturally active in our free time.

While people are focused on maximizing consumption and generating the income to sustain it, there is a corresponding retreat from involvement in civic and community life, and a weakening of social cohesion. Time poverty not only threatens our health and that of our connections with people we care about, it undermines the wider community fabric.

## Profit Maximization

Companies are continuously manufacturing 'needs' and expanding the markets into every area of our lives. This business behavior is socially and ecologically destructive in many ways, but the profit motive can be especially morally dubious when it comes to certain sectors, like healthcare.

Imagine your last trip to the dentist. When you look around the room, you might notice a lot of different brand names. A brand name on the monitor where the dentist shows you what your back teeth look like, a brand name on the mouse she uses to control the camera and screen settings, a brand name on the powerful lamp she uses to light up mouths while she works, a brand name on the reclining chair you're sitting in, a brand name on the filling material she's about to apply to your teeth, and a brand name on the mouthwash she gives you for rinsing. Think about how much all of the equipment and supplies in the room must cost. Now think about what amount of those costs went into private pockets as profit dividends to owners and investors of all those companies. If those businesses were all oriented toward the mission of keeping people healthy and didn't let profit margins interfere with that mission, how much more affordable would healthcare be?

We would argue that it is unacceptable that healthcare companies (including those that provide health insurance, medication, and medical equipment and supplies) are even allowed to be for-profit. In the U.S., the costs of healthcare and dentistry are such that many people simply can't afford to visit professionals for their health problems - even in one of the most 'developed' countries on Earth.

In a for profit world, a healthy population is bad for the bottom line of the for-profit health industry for the very same reasons that durable products are less desirable than planned obsolescence – because the profit from perpetuating a market is needed . When people are less sick less often, healthcare CEOs are under pressure to find ways to generate more profits.

---

309 (Ref: How Stress Influences Disease, Incidence and Prevalence of Chronic Disease, 'Why capitalism makes us sick' talk by Gabor Maté)

For-profit healthcare companies have a clear incentive to keep people sick and that's probably why they focus on medication rather than prevention.

In many countries around the world, universal healthcare systems are heavily subsidized by the government and offered to people at a much cheaper cost than in the U.S. However, even nations with universal healthcare have a version of a for-profit healthcare system, because the public hospitals and clinics buy the bulk of their medicine and equipment from for-profit companies. This means taxpayers' money in those countries is still going into the private pockets of wealthy business owners.

Most people who have worked in the medical field are familiar with 'drug reps'. Pharmaceutical companies send representatives to clinics, hospitals, emergency rooms and even to intensive care units. They give doctors pens, notepads, food, free medicine and even offer to send them to far-away conferences, expenses paid. This is done in order to gain the doctors' favor and persuade them to prescribe certain drugs more often, which leads to doctors sometimes (perhaps more often than we'd like to think or they'd like to admit) prescribing unnecessary medication.

Why does this happen? It is a strategic business tactic that helps pharmaceutical companies increase their profit margins. And it works! Pharmaceutical companies are some of the most profitable businesses listed on the U.S. stock-exchange.[310]

It's no wonder that the healthcare sector has consistently grown its profit margins: healthcare is a basic human need. Products that provide for basic needs will always be in high demand. Whether or not private business owners should be able to enlarge their bank accounts by exploiting basic needs is a different question.

In 2015, the Wall Street Journal published a story on how there's a new kind of deal-making in the pharmaceutical sector, in which companies buy up drugs that they see as 'undervalued' and then dramatically raise the price. The price for one blood pressure drug rose over 500% to $806 per vial overnight, due to an ownership change. This is mostly happening with drugs that are deemed to be very valuable because they lack alternatives, so companies can charge as much as they want without impunity; at least in the U.S. where the Food and Drug Administration doesn't take pricing into account in its regulations. As a result, the prices of medicine have risen 120% there over the last several years.[311] Clearly, *Homo economicus* is at the helm of the pharmaceutical industry.

Even people's personal lives are being turned into a way of maximizing profit. Some companies, like Facebook and Google, do this by selling their clients' personal information to other companies. There is also a growing market of companies that profit from bullying and humiliation. Gossip magazines and websites sell the humiliation of celebrities, as well as regular, everyday people. This trend has expanded greatly with the emergence of the Internet and social media.

---

310 (Ref: Market Watch)
311 (Ref: WSJ story)

In a recent talk about this phenomenon, Monica Lewinsky, who was publicly humiliated for years after she was part of a U.S. Presidential scandal, said it very well, "A marketplace has emerged where public humiliation is a commodity and shame is an industry. How is the money made? Clicks. The more shame, the more clicks. The more clicks, the more advertising dollars… all the while someone is making money off of the back of someone else's suffering."

Perhaps one of the most concerning examples of the for-profitization of a sector is that of prisons. Private, for-profit prisons, commonplace in the U.S., are morally hazardous for many reasons. They have a financial incentive (and, not to mention, a fiduciary duty to their owners and investors) to keep their facilities as full as possible. This means it's in their interest to have a lot of crime and violence in society, and to encourage the criminalization of more and more acts. As a result, the U.S. has the highest incarceration rate in the world[312] and the vast majority of prison inmates are serving time for drug offenses (a large portion of which are marijuana-related).[313] Many for-profit prisons have even tried to set up deals with local justice systems to create some sort of quota of arrests and prison sentences to keep the jails full.[314]

As profit-oriented businesses, these prisons are constantly looking for ways to deliver higher profit margins. Like any other business, this means increasing sales, decreasing costs and expanding to stay competitive. In addition to taking tax-payer money in the form of government contracts, for-profit prisons often generate revenue by producing products like automobile license plates, clothing and furniture inside the prison and selling them on the outside market. Because incarcerated populations are considered not to have many rights, they can be paid as little as 12 cents per hour.[315] The prison owners are making money from slave labor. And, of course, this slave labor force is mostly made up of the most vulnerable people in society,[316] as the vast majority of prisoners and their families come from low-income neighborhoods with inadequate school systems.[317] **Vulnerable people's lives are quite literally being destroyed in the name of profit.**

Following this for-profitization trend, a growing number of addiction treatment facilities are for-profit, a $35 billion industry in the U.S.[318] who are now profiting from another vulnerable group, those struggling with substance abuse.

The for-profit market is already cashing in on the Syrian refugee crisis. G4S, the world's largest security firm notorious all over the world for running prisons, is expanding into the very lucrative business of providing shelter for asylum seekers.[319] And G4S isn't the only for-profit to move into this 'hot new market'. A for-profit Swiss company called ORS Services allegedly profited to the tune of $99 million in 2014, from taking care of migrants in

312 (Ref: Tsai & Scomegna)
313 (Ref: Government data; FBI data)
314 (Ref: April Short article)
315 (Ref: TIME article)
316 (Ref: TIME article)
317 (Ref: Punitivity, p. 67)
318 (Refs: Forbes article))
319 (Ref: Guardian article, Open Democracy article)

Switzerland, Germany and Austria.[320] Unlike similar programs run by an NFP contractor, those profits would have ended up in private bank accounts.

While these examples may sound like shocking outliers, or fringe conspiracy theories, they are more central to the current economy than many of us realize. In fact, the wealth extraction siphon also plays a central role in the debt crises of the world. Take the ongoing Eurozone debt crisis. In the mid-1990s, the Eurozone leaders put pressure on for-profit banks to be more competitive with banks in the U.S. and the U.K. As a result, for-profit German and French banks offered low interest loans to governments in countries where the money was needed, mostly in southern Europe. They loaned out billions of Euros to these governments.[321]

Two of Germany's biggest banks, Commerzbank and Deutsche Bank, loaned $201 billion to Greece, Ireland, Italy, Portugal and Spain, according to calculations by Business Insider. And two large French banks, BNP Paribas and Crédit Agricole, gave loans totaling $477 billion to those same countries.[322]

When the Irish, Spanish, Portuguese and Greek governments received bailout money from the European Union and the IMF in the 2010's, those for-profit banks got a large portion of the money.[323] And those countries' citizens have been enduring harsh austerity measures that have pushed many people into poverty and despair, in order for the bailout loans to be repaid to for-profit banks.

In effect, the Eurozone bailouts have contributed to enriching the wealthy owners and shareholders of for-profit banks and corporations at the great expense of everyday European citizens. Public funds are going to for-profit, private coffers. This can only make sense in light of the for-profit story and the corresponding ethic of 'never enough', which legitimize and encourage this kind of business behavior. Through that lens, these are great business strategies that will result in benefits for the wider community via the Invisible Hand. **If only the Invisible Hand were a bit more visible**.

# Ecological Crisis

We are now in a precarious position in many ways. Our capitalist economy is not only undermining social relations and individual wellbeing; it is ravaging the planet's ecosystems. The life-sustaining systems of Earth are collapsing at an alarming rate. There are real ecological thresholds that we have passed and if we continue to disregard these thresholds in order to achieve an ever-growing economy, modern civilization will experience a very painful crash.

## Ecological Limits to Growth

In the late 1960s, a group of scientists convened in Rome to strategize about how to make the increasingly global economic system more long-sighted and sustainable and the Club of Rome was born. Shortly afterwards, the Club commissioned a group of systems scientists at

320 (Ref: Bloomberg article)
321 (Refs: find primary refs from CorpWatch article)
322 (Ref: Eurozone Profiteers article)
323 (Ref: Guardian article)

the Massachusetts Institute of Technology to analyze the long-term impacts of human systems on the planet's ecosystems.

In 1972, the Club of Rome published the group's findings in the *Limits to Growth* report. They had created a computer model of the global economy's interactions with the natural environment, using real data, in order to explore the impacts of a growing human population on the planet's biosphere.[324] Their investigation concluded that if we continued on the path of pursuing economic growth, and if we didn't do something to slow population growth, we would soon hit ecological limits. Ecosystems would go into rapid decline and human systems would begin to collapse as a result. This process of collapse would entail food shortages, fresh water shortages, disease and political instability (Ref: Ibid).

At the time of its publishing, the *Limits to Growth* report was met with a mixed reception. A nascent environmental movement embraced its findings, while many economists and politicians dismissed them as ludicrous, stating that human ingenuity will always find ways around natural limits. However, since the study was published, scientists have been keeping a close eye on the various indicators that were used in the model and The Club of Rome's 'standard track' is right on the mark for the years 1970 - 2000.[325] This scenario predicts a global collapse of ecosystems by 2050.

The logic is simple. We live on a finite planet and we only have so many resources. If our economy outgrows our planet, if we outstrip the foundations for life, then we're in real trouble. In fact, we have already passed the ecological limits to growth. According to the Global Footprint Network, it was around 1980 that we started using more resources than the Earth can regenerate each year. **We now use 1.5 planets' worth of resources every year, but we only have one planet**.[326] If everyone in the world consumed like the average American, we would be consuming 5 planets' worth of resources each year (Ref: Ibid). If our ecosystems can't even keep up with current rates of consumption, how can we expect the Earth to go on supporting even more consumption year after year, especially as 'emerging economies' aspire to consumer lifestyles? If the economy grows at the generally accepted rate of 2% per year, it will be 40 times bigger at the end of the century, but Earth will be the same size it's always been.[327]

We are entering a period that natural resource expert Richard Heinberg refers to as 'Peak Everything.'[328] Peak Everything is an extension of the idea of peak oil, which was developed by a geologist named Marion King Hubbert in the 1950s. Hubbert calculated that at some point in the early 21st century, we would reach the point at which half of the world's natural petroleum reserves had been used; it's called *peak* oil because it takes the shape of a mountain peak on the charts.[329] He looked at the examples of Ohio and Illinois to explain the principle of peak oil. Those states had both experienced a boom of steadily rising oil production until they peaked and then experienced a slow decline in production. At some point, the oil reserves were so small and so out-of-reach that it was no longer worth the effort and cost of

324 (Ref: Limits to Growth)
325 (Refs: Turner, 2012; CSIRO, 2008; GAiA - Ecological Perspectives for Science and Society, 21(2), 116-124)
326 (Ref: Global Footprint Network)
327 (Jackson, 2009)
328 (Ref: Heinberg)
329 (Ref: Hubbert, 1956)

drilling in those states.[330] Based on his observations of this general trend and the amount of known American petroleum reserves at the time, Hubbert correctly predicted that U.S. oil production would peak in the early 1970s.[331] As a result of declining production in the U.S., a greater percentage of oil comes from unconventional sources each year, because the easy-to-access oil is gone.[332] Given Hubbert's correct predictions in the past, we have great reason to believe that his prediction about the world's peak in oil is accurate as well.

This might not sound too alarming at first glance, but oil is absolutely integral to keeping our modern society going. Over 400 gallons of oil are used to feed each American every year.[333] This is due to our globalized industrial food system in which agriculture has become highly mechanized and food is frequently imported from far away. All that transportation, not to mention the fuel required by the machines and fertilizers used to produce and process the food, requires a lot of oil.

Some analysts believe that we have already reached the peak and that, in the coming decades, remaining oil will become so difficult to access that it will be prohibitively expensive.[334] World discoveries of new oil fields peaked in the 1960s and the discoveries since then have been smaller and further offshore, so it's easy to imagine a corresponding peak in production in the near future.[335] This is particularly scary, as based on current rates of consumption, it's estimated that in the next 20 years, we'll consume more fossil fuels than in all of previous history.[336]

Peak Everything refers to a similar pattern with most of the basic resources we rely on to meet our needs, including fresh water, fertile soil, metals, and even fish, just to name a few.[337]

We're also facing the potential of peak fresh water.[338] Underground aquifers of fresh water that took thousands of years to accumulate are being drained at phenomenal rates, fresh water is being polluted, and climate patterns are changing, all of which leaves a growing number of people (and other species) with less fresh water, a basic building block of life. Eighty percent of the world's current population lives in areas where water security is threatened.[339]

Topsoil, needed to grow our food and keep ecosystems healthy, is also said to be peaking - we're using it faster than nature can replenish it. Much of the planet's topsoil has become degraded and eroded by intense farming practices and some of it has essentially died, losing the micro-organisms that make it fertile, due to drought and desertification of the land.[340] Ninety percent of the world's arable land is already exploited[341] and the United Nations' Food and Agriculture Organization estimates that 25 % of agricultural land is highly degraded, while another 8 % is moderately degraded.[342] At the same time our ecological footprint is expanding, our biological capacity and ability to produce is declining.

---

330 (Ibid)
331 (Ref: Heinberg, End of Growth
332 (Ref: Breaking Energy article)
333 (Ref: Ross Jackson)
334 (Ref: Heinberg, End of Growth; Miller & Sorrell)
335 (Ref: Miller & Sorrell)
336 (Ref: Dick Smith's book)
337 (Ref: Richard Heinberg, Peak Everything)
338 (Ref: Pacific Institute)
339 (Ref: Nature09440)
340 (Refs: Guardian article, Earth Island Journal)
341 (REF: Dick Smith's book)

Additionally, we have been mining more copper, nickel, and gold than we've been finding and many researchers believe we might have already passed the halfway point in exploiting the Earth's reserves of these important metals or that we will reach Peak Metals.[343]

Just as we're having to drill deeper and deeper to find oil, so are we having to look deeper and deeper to find fish.[344] This is because we're fishing faster than the fish can reproduce. Based on current trends, world fisheries are expected to collapse by 2050.[345] Many communities have already experienced fishery collapse, such as the collapse in North Atlantic cod stocks in 1992.[346] That fishery was closed indefinitely in 2003 (Ref: Ibid). Ongoing fishery collapses could be catastrophic for the 1/5 of the world's population that relies on fish as their primary source of protein and the 660 million people who depend directly or indirectly on fish for their income.[347] Fishery collapses could also have dire direct and indirect consequences for ecosystems all over the globe.[348]

Our ever-growing economy is making global resource scarcity a very real possibility, yet we continue to extract 60 billion tons of resources annually – and growing.[349] That's 50% more than 30 years ago.[350] **Humanity uses nature's services 50% faster than the Earth can renew them,[351] but the extractive nature of the for-profit market requires constant expansion, at all costs.**

The ecological impacts of the for-profit economy are not only about resource extraction and the inputs that we need for our activities, but also the pollution and waste we create and how it affects all of the other beings we share this planet with. Recycling is an after-thought, a 'damage control' strategy in our linear economy. This means that resources go into one end of the system, then they are processed, consumed, and the waste is thrown out of the other end of the system. What happens to the waste at the end of this linear production-consumption process is critical, because rather than being reused in our economy or processed into something that ecosystems can use, most of it is dumped directly into the biosphere (i.e. air, water and land).

Climate change might be the first thing that comes to mind when thinking of our economy's effects on nature. All ten of the hottest years on record have been since 1998.[352] 2015 is now the hottest on record and, based on trends, 2016 will likely pass that.[353] Many of us have been experiencing this in increasingly strange weather patterns that manifest in both unusually hot and unusually cold weather as well as stronger storms. In 2012, a town called Needles in California experienced the hottest rainstorm ever - 115 degrees Fahrenheit, 46 degrees Celsius.[354] Rainfall and snowfall records in the Eastern U.S. over the last years have been off the charts, not to mention the numerous polar vortexes that are coming to be the norm in North American winters.[355] Australia has experienced increasing outbreaks of bushfires in record-hot summers.[356] The UK and Northern Europe have seen unprecedented heat waves and floods, as well as severe winter storms with hurricane-force winds.[357] Tropical storms, typhoons and hurricanes are also increasing in number and strength, affecting communities all over the world; especially island nations like the Philippines (Ref: Klein?).

---

342 (Ref: Reuters article)
343 soon (Ref: The Coming Copper Peak; SBS article; On the Materials Basis of Modern Society, Sverdrup & Ragnarsdottir)
344 (Ref: Millenium Assessment, p.8)
345 (Ref: UN report)
346 (Ref: Millenium Assessment, p. 12)
347 (Ref: Ibid; World Ocean Review)
348 (Ref: Impacts of Biodiversity Loss on Ocean Ecosystem Services)
349 (Ref: Living Planet Report, 2012)
350 (Ref: Ibid)
351 (Ref: Ibid)
352 (Ref: UNEP; NASA)
353 (Ref: NASA website)

By narrowly focusing on climate change, many activists, organizations, governments and businesses have kept our attention on greenhouse gas emissions, or even just carbon dioxide emissions. This has allowed *Homo economicus* to find ways of monetizing carbon dioxide and in many cases financially profiting from climate change via trading schemes and carbon mitigation programs, while completely missing the point that **climate change is a symptom of the growth imperative of the for-profit system**. We must move beyond single issue, linear thinking and look at climate change in the context of much bigger, systemic issues in order to address the root causes.

As our consumption grows, so does the amount of waste and pollution we produce. We have produced more plastic in the last 10 years than we did in the entire last century, which means we have also created more plastic waste than in the entire last century.[358] Plastics never fully biodegrade and are wreaking havoc on the planet. There are masses of plastic waste collecting in the oceans where currents meet. The biggest is known as the Great Pacific Garbage Patch. As fish and birds eat these pieces of plastic and they are, in turn, eaten by their predators, the plastic spreads throughout the food chain. Although many plastic-related deaths have been observed among marine and bird species, it is hard to say what the long-term consequences of plastic in the oceans will be.[359]

Pollution isn't the only thing harming the planetary community of life. Forests all over the world are disappearing. What were once vibrant forests in much of northern China are now deserts, due to unsustainable logging.[360] As a result, places like Beijing increasingly experience sandstorms that would never have occurred fifty years ago.[361] Lush forests like the Amazon are also being cut down to make room for beef cattle ranches and to grow crops to feed the cattle, which is largely fueled by rising meat consumption and a profit-hungry meat industry.[362] In addition to people in high-income countries eating unsustainable amounts of meat, consumption of meat is increasing very fast in 'developing' countries as it is identified

354 (Ref: Hottest rain article)
355 (Ref: The Atlantic article; Phys.org article)
356 (Ref: Climate Council)
357 (Ref: Guardian article; Independent article)
358 (Ref: timsilverwood.com)
359 (Ref: Eriksen et al., 2014)
360 (Ref: Guardian article about deforestation in China)
361 (Ref: South China Morning Post; China.org article)
362 (Ref: Guardian article; WorldWatch Institute)

with wealth, prosperity and the lifestyle of the 'developed' world.[363] When forests are cut down, ecosystems are destroyed and habitats for entire species can be lost.

All of this has contributed to the dying off of species at an incredible rate. It is estimated that we are now losing 150- 200 species per day.[364] Most environmental scientists agree that human activity is causing the sixth mass extinction event in the history of the planet.[365] The last mass extinction was 65 million years ago, when the dinosaurs died out. Aside from this being a very sad situation, it is also frightening because we don't know what the effects of this will be on human society, but scientists are certain that it puts us in a dangerous position, as meeting our basic needs for food and water requires functional, healthy ecosystems.[366]

Despite resource shortages, ecological collapse and decreasing levels of wellbeing, we just keep trying to grow our economy to satisfy the growth imperative of the for-profit system. The inherent contradiction of capitalism is that the economy needs to grow, but **overconsumption of resources threatens the human species' existence in the long-run and isn't even making us happy in the short-run.**

Furthermore, despite some shifts in mainstream business, the profit motive still encourages business managers to cut environmental corners in order to reduce costs and deliver high profit margins to shareholders and investors, as well and large bonuses to themselves.

## Population Growth

The growth of the human population only exacerbates ecological destruction, as the more resources that are used by humans, the fewer resources there are for other species. The more the human consumption expands, the more habitats it destroys in the process. Indeed, human population growth is an important contributing factor to the Sixth Mass Extinction and is causing us to use up the Earth's resources even faster.

In 1800, before the Industrial Revolution, the world's human population was about 1 billion people.[367] By 1960, it had reached 3 billion. In the last 50 years, it has more than doubled to 7.3 billion people in 2016.[368]

---

363 (Ref: WorldWatch Institute book)
364 (Ref: Guardian article)
365 (Ref: Stanford News; The Sixth Extinction)
366 (Ref: Cardinale et al., 2012; Diaz et al., 2006)
367 (Ref: U.S. Census historic estimates)
368 (Ref: U.S. Census)

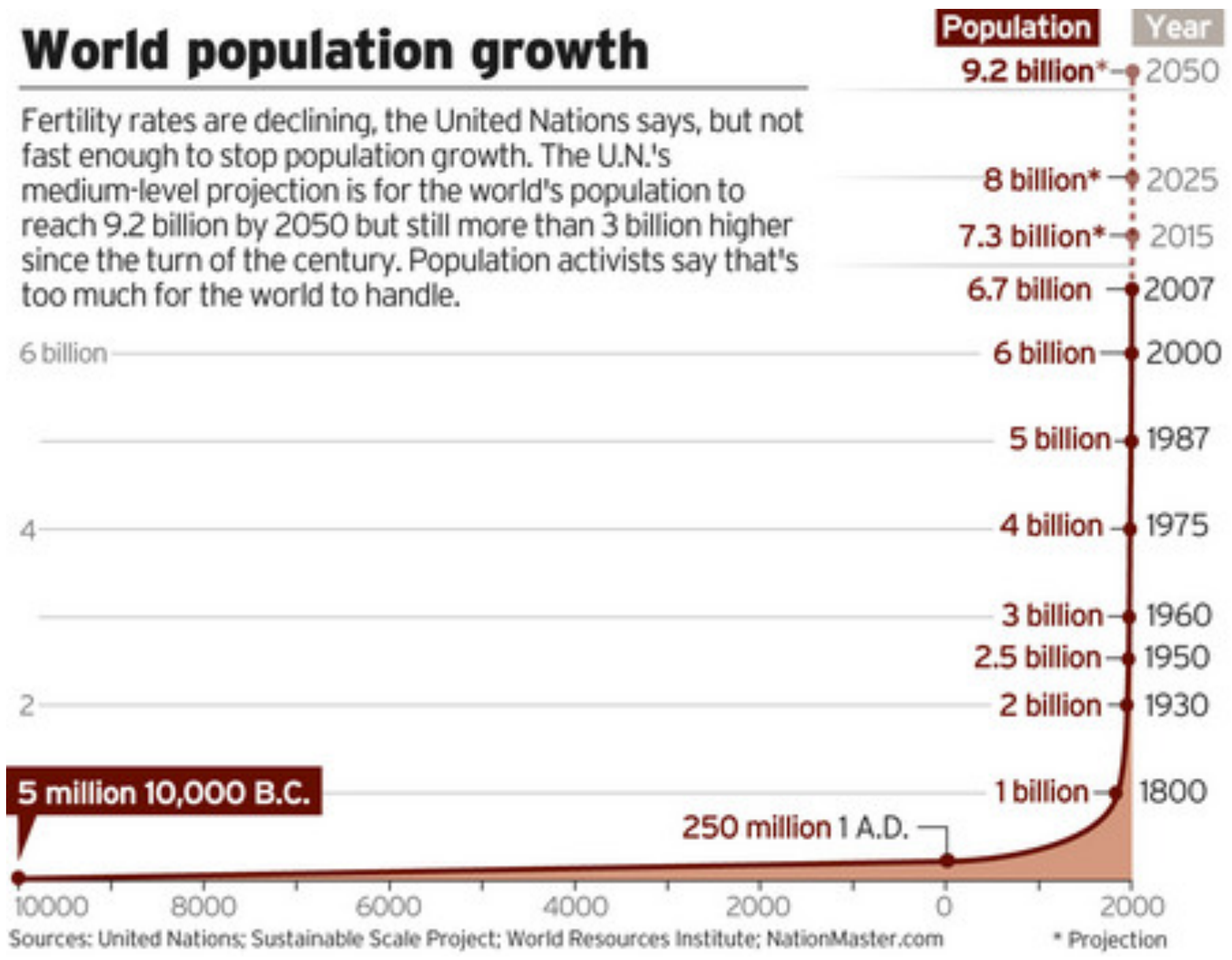

## Human Population Growth

The global population is growing by just over 1%, each year.[369] This sounds harmless enough, but it's also exponential growth – most of us know how this works with compounding interest on loans or investments, but it works the same way with population. At the beginning of 2016, we started with 7.3 billion people on the planet, which means that the population will grow by roughly 73 million people this year.[370] But in 2025, we will have an estimated global population of 8.1 billion,[371] which means that, if the rate of growth is still 1%, the population will increase by 81 million people. That's ten million more people being added per year than in 2016, but it's still 1% of the population. One percent doesn't sound like much, and that's why it can be deceiving. In fact, Professor of Physics at the University of Colorado once said that humanity's greatest shortcoming is our inability to understand the exponential function. Every year, we have more people to start with and so we grow by a greater number. A major study in 2014 found that there's a 70% chance that our population numbers will continue to rise to 11 billion in 2100.[372]

What is driving this unsustainable population growth? It is a very complex issue that involves a wide variety of factors, including inequality, poverty, education levels, employment, access to healthcare and social norms. It's true that the highest rates of population growth are in the 'developing' world and so it is often framed as an issue that only 'developing' countries need to address. However, there are clear connections to the for-profit system. Extreme global inequality means that the people at the bottom have a lot less choice in how they live their lives. They have less access to education, employment opportunities and healthcare. This is especially true for women in low-income regions.

369 (Ref: U.S. Census)
370 (Ref: U.S. Census)
371 (Ref: USA Today)
372 (Ref: Gerland et al (2014) 'World population stabilization unlikely this century')

The marginalization of women due to the dominator mentality plays an important role here. Women in many places do not even get to think about how many children they want (or if they want children) because they are expected to do what their husbands and fathers want them to do. This control of women's reproduction might seem limited to the 'developing' world, but it's not. Political debates rage on in the U.S. about women's reproductive healthcare rights. When 40 percent of all pregnancies in the world are unwanted or unintended, the importance of women's control over their own reproductivity cannot be understated, especially not in the context of overpopulation.[373]

It must also be mentioned that an average child in the U.S. will cause 13 times as much environmental damage throughout their life as an average child in Brazil.[374] So, any serious discussion of overpopulation and resources must also take into consideration economic and political domination and the resulting inequitable distribution of wealth and opportunities.

In the 21st century, population growth is also often equated with economic power via economic growth. More people mean more markets and more consumption, which is good for the bottom line (in the short-term at least). Population growth is one of the most reliable ways a country, city or region can grow its economy and show high GDP numbers. We can see this in political rhetoric today. High birth rates are encouraged by many governments as a driver of political and economic power. It's bad to be small when the name of the game is domination.

The global population growth *rate* is declining due to increasing access to family planning, including methods of contraception, as well as more access to education and diverse employment opportunities (factors which do a lot to lower birth rates). However, high inequality is a looming threat to progress made on this front. Trying to tackle overpopulation in the for-profit world is like trying to swim against a riptide.

## Techno-Fixes

A great number of people believe that capitalism's innovative potential can and will come to the rescue of our ecological crisis and allow us to keep growing the economy. The 'substitutability' assumption in the field of economics leads many to believe that we will be able to substitute scarce resources through innovation. However, a study led by Yale Professor of Chemical Engineering, Thomas Graedel, found that there is no known suitable substitution for 62 metals integral to current economic activity, many of which might become scarce in the relatively near future.[375] The substitutability assumption also begs the question: how do you substitute things like fresh water or complex marine ecosystems?

Others believe that efficiency is the key. Innovators the world over are working hard to find ways to 'decouple' economic growth from ecological destruction – that is, to grow the economy while using fewer natural resources and causing less damage to ecosystems. Major advancements have been made, and resource efficiency is part of what we need to do. However, advances so far have only been in terms of *relative* decoupling, not absolute decoupling. This means that we have found ways to lower our ecological impact *per unit of production*, like lowering the amount of energy a car requires. But when the goal is to sell

373 (Ref: Guttmacher Institute)
374 (Ref: Scientific American article)
375 (Ref: Graedel article)

more and more cars, and our global population of 'consumers' is expanding, the overall - or absolute - amount of resources we're using still increases.

The former leader of the U.K.'s Sustainable Development Commission and author of *'Prosperity Without Growth'*, Tim Jackson, makes it clear that the levels of innovation that would enable an absolute decoupling of economic growth from ecological destruction are totally unrealistic.[376] In the case of carbon dioxide emissions alone, Jackson calculated that we would need a decrease of carbon dioxide emissions by 11% per year to reach our reduction goals by 2050, which requires a rate of innovation 16 times faster than all of the innovation we've had since 1990 combined.[377] As economist Sir Nicholas Stern points out, "It is difficult to secure emission cuts faster than about 1% per year except in instances of recession."[378] And that's focusing on carbon emissions, without taking into account other forms of pollution and usage of nonrenewable resources.

This argument has been proven true, as global GDP grew by 44% between 2004 and 2014, and the consumption of fossil fuels increased by 19%, resulting in a 22% increase in global carbon dioxide emissions, all while governments and businesses were trying harder than ever to decrease emissions through efficiency and the increased use of renewable energy sources.[379]

Despite concerted and sustained efforts to become more eco-efficient, we only continue to increase our negative impact on the environment, because our profit-oriented economic system tends to use any efficiency gains to expand the overall scale of production – this phenomenon is known as 'Jevon's Paradox.'[380]

Some people are still betting that the transition to an information economy or a service-based economy can decouple economic growth from ecological impacts. It is often claimed that the emerging service economy in the world's largest cities is resulting in de-carbonization. On the surface of things, this may seem true, but the service economy usually includes the financial sector, which is largely the elite economy, and the growth of the elite economy requires the growth of the real economy of goods and services, in order to compensate for the wealth siphoned away. Furthermore, the claim that electronic services are naturally more eco-friendly than paper-and-pen services is not as cut-and-clear as it may seem. The world's information-communications technology systems, for instance, use 10% of global electricity generation every year, so going online does not necessarily mean going 'green.'[381] In fact, the global IT industry contributes about the same amount of carbon dioxide emissions as the global aviation industry.[382]

Similarly, we hear numbers about the decreasing carbon footprint of entire nations that continue to grow their GDP. However, these numbers are not 'trade corrected' – they don't account for the carbon emissions of the goods a country imports and consumes. **Countries that claim they are lowering their ecological footprints while maintaining economic**

---

376 (Jackson, 2011)
377 (Ibid
378 (REF: Stern)
379 (Ref: Turning Point report, p. 10)
380 (Ref: John Bellamy Foster, Brett Clark and Richard York - Capitalism and the curse of energy efficiency: the Return of Jevons Paradox, Monthly Review, 2010/11/01)
381 (Ref: Mills report)
382 (Ref: Gartner study)

**growth are commonly 'outsourcing' their environmental impacts**, as more production and waste streams are moved abroad.[383] In most high-income countries, the majority of goods are produced abroad.[384] In outsourcing the dirtiest aspects of production, the 'developed' countries have outsourced the carbon emissions on which their economies depend. Instead, the carbon emissions of countries like Vietnam, China and Bangladesh, where much of the production occurs, have risen significantly.[385] Although we can change the way we count each country's emissions, global emissions continue to increase. The total negative impact of human activity on the planet's natural systems continues to rise, and that affects us all. Greenhouse gas emissions increased twice as fast in the first decade of this century than they did in the three previous decades.[386]

Furthermore, information and certain services are rapidly becoming 'zero marginal cost' items.[387] This means that, once the initial product has been created, the price of producing an additional unit of that product is close to zero. The music and publication industries are the most obvious examples of this. With dramatically increased accessibility to advanced technology and software, it might cost a few thousand dollars to produce a new music album or a new book, but the cost of producing additional units of the album or book in electronic format is almost nothing – so the marginal cost is effectively zero. The price to the consumer of accessing the product, often via the Internet, can be very low or even free, with producers making profits largely from advertising revenues, if any profits are made at all. In the internet and digital era, information and service economies can't support the kind of economic growth that the for-profit system requires.

Often, efforts to become more eco-efficient fly in the face of a for-profit system that actually encourages massive inefficiency in the name of growth and profits. There are millions of examples of just how insanely inefficient global trade has become. Despite being an agriculturally fertile country, 78% of Australia's apple juice concentrate comes from China and 87% of its frozen orange juice comes from Brazil (Ref: Dick Smith). Apples grown in the U.K. are often shipped halfway around the world to be waxed, then returned to the U.K. to be sold.[388] Seventy-five percent of the apples in New York City come from the west coast of the U.S., even though the state of New York produces far more apples than the city's residents consume.[389]

But it's not only apples and oranges. It's everything. England imports more than 100,000 tons of milk each year, then exports roughly the same amount.[390] Even in Mongolia, a country with 10 times as many milk-producing animals as people, shops carry more European dairy products than Mongolian ones.[391]

Why all of this redundant trade? Because in the for-profit economy, the pricing mechanism includes costs and margins for sellers and buyers, but fails to include social and ecological costs. Companies increase profits by doing things as cheaply as possible, so if it's financially cheaper to ship apples around the world to be waxed than to pay a bit more for the apples to be waxed locally (due to cheaper labor), that's the 'rational' thing to do; it doesn't matter to that company if their actions create inefficiency in the wider system. Instead, the price for this ecological inefficiency is passed on to the environment, underpaid workers, local economies that are deprived of employment and business, and future generations that will have to try to repair the damage.

---

383 (Refs: Carbon emissions outsourced developing countries; The Myth of Decoupling)
384 (Ref: UNCTAD)
385 (Ref: Trends in Global CO2 Emissions)
386 (Ref: IPCC Report)
387 (Ref: Rifkin, Zero Marginal Cost Society)
388 (Ref: Economics of Happiness)
389 (Ref: Bill McKibben)
390 (Ref: Norberg-Hodge)
391 (Ibid)

**Within the growth paradigm, gains in efficiency will never be enough**. Growth and the drive to maximize profits will always cancel out whatever efficiency we have gained. But the more important question is: Why should we try so hard to maintain this growth-based economic system when it's not making us any happier, it's creating unacceptable levels of inequality and it's eroding the ecological stability of the planet?

# Economic Instability

Ecological instability is intricately connected to the economic instability of the for-profit system. The story of palm oil illustrates these connections very well. Ancient forests in some of the world's most precious biodiversity hotspots in Indonesia are being cut down in order to cultivate palm oil, a very versatile oil that is increasingly added to a wide range of products including lipsticks, cookies, shampoos, instant noodles, and detergents.[392] Palm oil is even used as a biofuel.[393] In clearing the forests to plant trees for palm oil, a great many species' habitats are destroyed. This includes some of the most exotic, and even endangered, species in the world, like orangutans, rhinos, elephants and tigers.[394]

But palm oil is not just an environmental problem. Indigenous people are often forced off the land where their families have flourished for many generations.[395] Palm oil producers are also notorious for violating the rights of their workers, exploiting the local people who work on their plantations.[396] Because so much profit is to be had from palm oil, it has been relatively easy for companies to bribe and lobby government officials at the local and national level in order to let them produce palm oil in the cheapest ways possible, paying no attention to environmental impacts and the rights of workers and indigenous peoples. In the end, the business owners and investors make a lot of money, workers are paid very little and indigenous populations are left landless, contributing to even more inequality.

The story of palm oil is emblematic of the for-profit economy: **business owners' mandate to maximize profit leads to species loss, exploitation, inequality, monocultures, oligopolies, corruption and political capture.** This is completely unsustainable in ecological, economic and social terms. In fact, the concentration of wealth and power is one of the most unsustainable aspects of this for-profit economic system; it is *economically* unsustainable.

## Concentration of Power

392 (Ref: WWF)
393 (Ref: Ibid)
394 (Ref: WWF article on illegal palm oil)
395 (Ref: Palm Oil and Indigenous People in Southeast Asia)
396 (Ref: Guardian article)

In 2011, a group of complexity scientists in Switzerland published an article that explored the question 'Who Controls the World?'[397] This article analyzed data on 43,000 transnational corporations. Interestingly, they discovered that there is a core in this network of corporations - 36% of them make up 95% of the total operating revenue of the whole group. That alone is an incredible level of concentration, but even more astounding, their research showed that the top 737 shareholders of these 43,000 large transnational corporations have the potential to control 80% of total corporate value. That's 0.123% of the 600,000 shareholders in those 43,000 companies, controlling 80% of the value. And it's even more concentrated as you go deeper into the core of this cluster of companies. The 146 top players have the potential to control 40% of corporate value. That's 0.024% of all shareholders in the network that have the potential to control 40% of the total value of those 43,000 corporations![398] This highly concentrated network translates to extreme amounts of wealth and power in the hands of very few, via shareholder voting rights. Part of the way this power of shareholders is magnified is by corporations either partially or wholly owning other corporations. This means **the people with the largest stake in the largest companies are unbelievably powerful.**

In order to grow and remain competitive, big for-profit corporations often acquire successful smaller companies. Some well-known examples of this are Unilever buying up the wholesome Ben and Jerry's ice cream company, Coca Cola buying the humble Odwalla juice company, General Mills buying the organic Cascadian Farms cereal company, and Groupe Danone (the parent company of Dannon Yogurt) buying most of Stonyfield Farm's yogurt company. This way, huge corporations are able to market their products as different, smaller brands, even though it's all under the control of one large company.

The market is becoming further concentrated all the time. Take 2015's mega-merger between Kraft and Heinz to create one of the world's biggest 'food empires.'[399] Along the same lines, many people might be surprised to learn that Perrier bottled water, Gerber baby food and Friskies cat food are all owned by Nestle. Not only are such enormous companies able to market their products everywhere because they have such large marketing budgets, which helps them maintain high profit margins, but they can afford to lobby politicians to make policies that will allow them to profit even more. In the U.S., they can also financially support political campaigns that will be good for their bottom line, thanks to the Supreme Court ruling on Citizens United.

All of these examples are big, publicly traded corporations. But what about Mom and Pop shops; family-owned, local, small- and medium- sized businesses? Surely it's fine to have a for-profit world composed of those benign players?

There are a few reasons why the scenario of having only benign for-profit players is becoming less and less possible. One of them is the inevitability of a for-profit system to evolve into what we have now. Companies must dominate other players and they must grow. In order to do this, successful businesses purchase smaller companies and merge with other large, powerful businesses.

---

397 (Ref: Vitali, Glattfelder & Battiston)
398 (Ref: Ibid)
399 (Ref: Washington Post article)

This leads to enormous concentration of wealth and power, which if left unchecked, just keeps increasing at an almost exponential rate. As the biggest companies get bigger, they can afford to buy even more small companies, which makes them even bigger and more able to buy up other companies. Their larger size gives them an even greater advantage over small, local companies, as they can offer lower prices and advertise more extensively. The result is that more people buy from the large, transnational corporations, which gives them the ability to expand even more.

This is precisely the trend that has led to the phenomenon we've seen over the last few decades of small, local businesses dying out at an increasing rate and large mega-corporations taking over the market. Many other analysts have pointed this out before, but it is essential to understand that this is the nature of the *for-profit* economy. Expansion and concentration are inevitable in this system. That's why the revenue of the largest 1000 publicly traded corporations represents 80% of global industrial output.[400]

It is also important to note that Mom and Pop shops[xxvi] are not immune from the lures of the for-profit system. Many small companies operate according to the profit motive as well. Because many small businesses adhere to the *Homo economicus* rationality, they often sell out to larger companies in order to maximize their gain. Large companies give them an offer they can't refuse. For the few that do hold out and stick to purpose-driven values, they usually get outcompeted by the big players anyway, because they don't have the same budget for advertising, they can't offer the same level of consistency and convenience that the big companies offer and they can't offer the same low prices, because of the 'economies of scale' principle. For-profit economies naturally result in larger corporations, homogeneity of products and services, and highly concentrated wealth and power - and these are forces that locally-owned, purpose-driven shops can hardly reckon with in the for-profit context.

What many small businesses do offer is a more personalized experience in both product and service, alongside the fact that your money is much more likely to stay in the local economy, benefitting the local community.[401] Yet, these benefits do not align with the values of for-profit consumers. It's just not seen as rational to pay more for products (even if it means purchasing from your neighbor), when you're supposed to focus on low prices, as mandated by *Homo economicus*. And, of course, the inequality created by the for-profit system also means that a lot of people simply can't afford to pay more for local, ethical products.

It's not that these locally owned shops are part of the problem because they're for-profit, but rather they're (perhaps unknowingly or unintentionally) aligned with a story that doesn't support the values in society that would motivate consumers to choose them over a bigger competitor. **'For-profit' represents an ethic and underlying worldview, not just a business model.**

This tendency for concentration of ownership is particularly dangerous when it comes to the flow of information in society. In 2009, alternative news outlet Mother Jones published an infographic showing a timeline of the mergers and acquisitions in the U.S. media[xxvii]. In 1983,

---

[400] (Ref: Gabel M. & Bruner H., Global Inc.: An Atlas of the Multinational Corporations, New York, 2003)

[401] (Ref: Amy Cortese article) 'Local community' in this book refers to the level of a village, town, or city; or a part of a city. In some cases, it could be a cluster of smaller areas within a city, or a cluster of villages, towns or cities that share the same bioregion.

26 different companies controlled most of the media in the U.S. By 2006, after many mergers and acquisitions, just 8 companies controlled most of the U.S.'s media, each with a market value of over $40 billion.[402]

These media sources are all in business to maximize private profits for a relatively small number of individuals by creating TV shows, movies, news channels, newspapers, magazines and advertising for billions of people. This concentration of power in a few hands is very dangerous given that they are guided by the profit motive. Considering these companies are such an influential source of information for the general public, they have an enormous amount of power and control over human consciousness. Thus, they are able to manipulate what people think, how they feel and how they act. Remember the power of stories to create our reality? These companies are, in effect, reality-makers.

This concentration of information outlets reveals another internal contradiction of capitalism. One of the main principles of the capitalist market is that people will make rational choices based on accurate information, which is a key element of the Invisible Hand making self-interest work out for society's benefit. Yet, the expansion and concentration of the market leads the main players to skew information for their own benefit;[403] and this concentration of the market is an inevitable mechanism of the for-profit system. The accurate information that the for-profit system requires doesn't exist in the for-profit system.[xxviii]

In 2014, Oxfam published a report entitled, 'Working for the Few: Political Capture and Economic Inequality.'[404] The paper described how inequality has led to the huge levels of undue influence that wealthy individuals and companies have on political processes, otherwise known as political capture.

One vivid illustration of political capture is the revolving door phenomenon. This refers to when a person moves between professional positions in the public sector and the business or lobbying[xxix] sector, in a manner that resembles a revolving door. This can allow for-profit companies to have undue influence on policies and legislation through their 'revolving door' connections.

One study found that 56% of the revenue generated by private lobbying firms in the U.S. between 1998 and 2008 can be attributed to individuals with some sort of federal government experience – meaning that they have good connections and personal relationships with policy-makers in the federal government.[405] A study that ranked the top 50 lobbyists in Washington D.C. found that 34 had federal government experience.[406] Yet another report found that the revolving door is much more common and widespread in the U.S. than most experts realize because there are loopholes in legislation, allowing lobbyists to effectively hide some of their employment history.[407] Transparency International's work on the topic shows that revolving doors are rampant all over the world.[408]

---

402 (Ref: 'And then there were eight', also check out: Bagdikian, Ben The Media Monopoly, 7th edition, (Boston: Beacon Press), 2004)
403 (Ref: Information Asymmetry and Power in a Surveillance Society?)
404 (Ref: Working for the Few)
405 (Ref: Revolving Door Lobbyists)
406 (Eisler, K. *Hired Guns: The City's Top 50 Lobbyists*, 2007)
407 (Ref: Revolving Door Lobbyists and Interest Representation)
408 (Ref: Transparency International report)

The concentration of wealth and power in the hands of the elite has led to forms of political capture other than revolving doors. The case of Citizens United versus the Federal Election Commision in the U.S. provides a perfect example. Citizens United is a conservative lobbying group that wanted to make sure corporations and lobbying organizations can put as much money as they want into political campaigns, and they won the case in 2010.[409] Now in the United States that is just what's happening: corporations, including lobbying organizations, are putting massive amounts of non-taxable money into political campaigns, including 2016 presidential campaigns.

What's driving all of this? The bottom line: maximizing private profit. It is an inevitable outcome of the for-profit story and the profit motive acting as the glue that holds our society together; a rather weak glue it turns out. It might seem morally corrupt, but these people are doing the 'rational' thing in seeking to maximize their personal gain.

Other examples of how extreme concentrated power is negatively impacting democracy include the current TransPacific Partnership (TPP) and Transatlantic Trade and Investment Partnership (TTIP), trade agreements that have been negotiated behind closed doors, away from public scrutiny. Large transnational corporations are pulling the strings in these agreements that seek to, among many other things, allow corporations to take governments to private arbitration courts when they feel a government policy or decision is impacting their profit margins.[410] This clause in the trade agreements is called the Investor-State Dispute Settlement (ISDS) provision and it's nothing new - ISDSs have already caused a lot of damage, especially for Canada, which has been sued by corporations under the ISDS provision more than any other country.[411]

Around 63% of the claims against Canada in these cases have involved challenges to environmental protection legislation and resource management programs that the corporations say interfered with their investor's profits.[412] As a result of these corporate tribunals, Canada had to loosen its national environmental regulations.[413]

In this way, trade agreements like the Trans Pacific Partnership (TPP) and Transatlantic Trade and Investment Partnership (TTIP) give for-profit corporations much more power in the market and in politics, in many cases allowing them to side-step regulation. Politicians use the for-profit story to explain to their citizens that it's worth their while to incentivize for-profit business by means of setting up such trade agreements, export processing zones and tax havens.

This concentration of power is a self- reinforcing phenomenon. Capitalism is a system based on the contribution of capitalists – those who have the capital to invest. **By deciding where to invest their money, capitalists have the power to decide which sort of projects and businesses deserve to exist and which don't.** And their investment decisions generally come down to which companies and entrepreneurs are seen as most likely to boost their return on investment.

---

409 (Ref: Story of Citizens United)
410 (Ref: BEUC report)
411 (Ref: CCPA report)
412 (Refs: Huffington Post article; CCPA report)
413 (Refs: Ibid)

When all the investment power is in the hands of the few, the economy has a tendency to result in a very homogenous market, and this creates enormous systemic risk. Think of all of the chain shops and big box stores that have begun to dominate the global landscape and that almost everything you can buy in the average supermarket, almost anywhere in the world, is produced by just a handful of enormous companies.

As any good biologist knows, the lack of diversity in a complex system means a lack of resilience. A shock that affects just one or two of those huge companies can have major ramifications for the entire system; as the world experienced in 2008. The current economy is rigid, centralized and hierarchical, when it needs to be flexible, distributed and networked.

## Financialization

The speculation inherent in for-profit business can wreak havoc on the economy, too. When profit can be privately distributed, it creates a sense of excitement in the people who may receive that profit. If a company achieves high profits, then the owners or investors can receive more money. If the company doesn't generate any profits, they receive nothing. **This attachment of personal gain to the profits of a business, in the greater context of the for-profit story, lays the perfect foundation for a casino economy.**

The stock market is one of the most common ways profit is privatized. People buy shares in a company listed on the stock market. These publicly listed companies often deliver a certain percentage of their profits to their shareholders at the end of the year in the form of a dividend. Although most of us have come to think of the stock market as a means of investing in a company, it's not really investment. As we mentioned earlier, only when new shares are sold for the first time, as in Initial Public Offerings or new share issuance, is money actually put into the bank accounts of companies. Most transactions on the stock market involve the trading of existing shares between traders on the secondary markets, not buying up new shares. This means most stock market activity doesn't directly touch companies at all - it occurs in, and is restricted to, the elite economy. The trading of stocks is about the possibility of a gain for the investor, whether via dividends or capital gains (from selling the share for more than you bought it for). If a company raises its profit margins, it is seen as more valuable and can be traded for more money; this is known as shareholder value, which plays a major role in how companies are managed and is the primary and overriding concern of most companies' management and board. It doesn't matter much *how* shareholder value is increased; whether through producing a high-quality product, or by using sweatshops with child laborers in Vietnam and throwing toxins into their local drinking water supply cuts costs. We know this because we see it over and over in how many companies are operating right now. The only ethical guidance embedded in the stock market is that of helping shareholders make a personal gain from businesses[xxx].

This was bluntly stated by pharmaceutical CEO, Martin Shkreli, who came under fire for raising the price of a life-saving medicine by 5000%, making it much more expensive for patients. When asked by a journalist if he would choose to act differently if given the chance to go back in time, he replied that he would have raised the price even more because, as the CEO of a company traded on the stock market, his number one duty and obligation is to maximize shareholder value. He summed up the for-profit system beautifully when he said, "No one wants to say it. No one's proud of it. But this is a capitalist society, a capitalist system and capitalist rules and my investors expect me to maximize profits, not to minimize them or go half or 70%, but to go to 100% of the profit curve that we're all taught in MBA class."[414]

Bonuses tied to share prices create a very strong incentive for managers to focus on the short-term expectations of the market, rather than creating long-term value in the real economy. This has paid off for them - while real wages for most American workers have increased little if at all since the early 1970s, wages for the top one percent of employees have risen 165% and wages for the top 0.1 percent have risen 362%![415] In fact, the amount paid out in Wall Street bonuses in 2014 ($28.5 billion) was more than double the income of all 1 million Americans working full-time on minimum wage that same year ($14 billion).[416]

Then there's the arcane world of even riskier business. It seems that traders in the elite economy just can't get enough financial gambling. More and more complicated financial instruments have been concocted and at a faster and faster pace since the first derivatives were created in the 1980s.[417] A derivative is part of a full share, which means two shareholders can split the amount of risk that one shareholder would normally take. Until the U.S. Commodity Futures Modernization Act (CFMA) of 2000, derivatives were mostly used to 'hedge' a company's financial risk (Ref: Ibid). This means that a derivative was just a way for a company to share the risk of either losing or gaining money in the future. For instance, an American company that does business in Europe might make a contract with a bank to lock in a certain exchange rate for Euros to U.S. dollars for a year. This means that the bank will gain or lose money if the exchange rate goes wild, rather than the company. The bank has effectively taken this risk of losing money due to a fluctuating exchange rate away from the company, but in the case of favorable exchange rate changes, the bank stands to gain money. This contract can then be traded or sold to another person or organization who would like to take on that risk. Until 2000, derivatives in the U.S. were regulated by government agents and lawmakers and were required to provide a form of safety, or 'hedging', to companies. No purely speculative derivatives were allowed. After the CFMA was passed in 2000, though, derivatives were deregulated and financial speculation became rampant. Why was the CFMA passed? Perhaps part of the answer lies in political capture.

Each step in the evolution of financial instruments has taken this financial trading further away from real economic activity, allowing traders to gamble in the elite economy to their hearts' delight with higher risks - because the value of derivatives fluctuates wildly, they can win big or lose big. Because some financial instruments are so distant from real economic activity, it encourages traders to do irresponsible things that they probably wouldn't do if they had to deal with the actual people whose lives they're betting on - like the mortgage holders who lost their homes in 2008.

All of this has created a financial economy that is vastly larger than the real economy of goods and services that most of people think of when they hear the word 'economy'. Analysts estimate that the combined face value of all derivatives is more than a quadrillion (1,000 trillion) dollars; greater than 14 times the entire world's annual GDP.[418] The total value of all the stocks on the New York Stock Exchange, for comparison, is roughly $15 trillion.[419] This growth of the financial sector, financial speculation and the number and complexity of financial instruments is often referred to as the financialization of the economy.

---

414 (Ref: Martin Shkreli interview)
415 (REF: Krugman)
416 (Ref: Institute for Policy Studies, *Wall Street Bonuses and Minimum Wage*)
417 (Refs: The Man Who Gave Us Derivatives and Heinberg's The End of Growth)

This financialized sphere of the economy is now so detached from reality that in 2012, the New Scientist reported that a multi-billion dollar submarine fibre optic cable link extending 15,600km would be built between London and Tokyo via the Arctic Ocean, which is now conveniently accessible due to the retreat of sea ice.[420] This cable would cut the 'friction' between the two cities from 230 milliseconds to 168 milliseconds, and the article noted that 'reduced transmission time will be a boon for high-frequency traders who will gain crucial milliseconds on each automated trade'. One can imagine that this will facilitate more of the consumption that creates the greenhouse emissions that melts the sea ice.

The financial takeover of the economy is dangerous because investments in financial assets are most often very short-sighted, so the flows of capital tend to fluctuate erratically, causing asset bubbles to inflate and burst. And when such a big bubble bursts, it can cause a lot of damage to the real economy, because companies and their owners that might have lost a lot of money in the crash will try to make up for their losses by cutting costs, which often entails lay-offs and wage cuts. If they can't pay for their losses, these companies will have to shut down, which also puts people out of work.

The banks and financial institutions play a huge role. The total combined assets of the top six banks in the U.S. were equal in value to about 65% of the U.S.'s GDP in 2011[xxxi].[421] If these banks are selling more financial instruments than they can pay off in the case of a bubble bursting, they might face bankruptcy and, with so much of the economy's money dependent on them, that poses a serious systemic threat to the whole economy.

When banks pull back, liquidity in the system dries up and the economy grinds to a halt, which can be really painful for the people in the real economy who don't have extra money laying around, but rather require steady wages from the companies that are now downsizing to pay off their loans that are being called in. The effect of this on ordinary people includes loss of livelihoods, inability to afford medical care, and loss of homes when they are foreclosed on. This is where the global economy is at now - an era in which we can expect to see bursting asset bubbles, credit crunches, and liquidity crises with increasing regularity, as trust in the system continues to erode, and as the social and environmental upheaval the system is generating feeds back in, causing further instability.

## The Market and the State

The most common suggestions for addressing all of this dysfunction in the economy focus on the role of the market versus the role of the state.[422]

---

418 (Ref: Time article)
419 (Ref: Ibid)
420 (Ref: New Scientist article)
421 (Ref: Politifact calculations)
422 (Refs: Ostrom, 2009; Underhill, 2009)

From this perspective, the market and state are seen as being complementary forces that simultaneously constrain each other. The market is thought to be the optimal force for producing and exchanging goods and services, while the state forces self-interested players in the market to contribute resources necessary for maintaining the basic conditions for public health and wellbeing.[423] Thus, the state constrains the market's self-interest, while the market limits the state's centralized control. This binary way of looking at economics has restricted most economic thought to the market-state spectrum, with both sides assuming that the market must be for-profit.

At one extreme of the spectrum is free market capitalism, with the government playing a minimal role, while at the other extreme is state socialism, with the government controlling most economic activity.[424] As with any spectrum, there are also a variety of mixes of state and market along the spectrum. The majority of critical economic decisions have traditionally revolved around what the optimal mix of market freedom and state control is, in terms of balancing productivity and efficiency with meeting the needs of the wider population.[425]

## State-Driven Responses

The state-driven side of the economic spectrum advocates for a greater role of the government in regulating the market, and increasing taxation in order to address issues of inequality and environmental damage caused by a market that's a bit too free.

For instance, after thoroughly analyzing the dynamics of wealth inequality and economic growth, Thomas Picketty's solutions focus on raising tax rates for the wealthiest slices of society. Whilst tax systems vary widely with regards to whom, how much, and what they tax, one of the main aims of taxation around the world is to fund governments' provision of essential services and safety nets for the most vulnerable people in society, and balance out wealth inequality. This is why taxation seems like a natural solution to inequality.

Although taxation is an essential part of keeping any modern society functioning, even the most progressive taxation, functioning in the most philanthropic society, can't counteract the ever-increasing speed of the wealth extraction siphon. In order to do so, tax rates would have to be so high that there would no longer be an incentive to start or invest in a business and make money. Starting a business wouldn't be worth the risk, given the limitations placed on a potential return on investment, so the market would stall. High tax rates are fundamentally at odds with the profit motive that guides capitalists and for-profit corporations.

Seeking to resolve inequality by means of taxation also belies an acceptance that the economy should inherently create such negative outcomes, as if they're a natural part of the economy and the best we can do is mitigate these unavoidable effects.

Not only does taxation not get to the root of the problem, seeming to accept the idea that an economy has to be socially and ecologically destructive, but it also just can't happen in the current context. In fact, tax policy is moving in the exact opposite direction; the biggest corporations and wealthiest people are enjoying reduced taxes, thanks to tax havens and the ability to lobby their ideas into policy.

---

423 (Ref: Ostrom, 2009)
424 (Ref: Arnold, 1996)
425 (Ref: Ostrom, 2009)

Furthermore, higher tax rates don't necessarily lead to governments receiving more tax income. Data from the Internal Revenue Service of the U.S. government shows that when the top individual tax rate was 91% of income in the 1960s, total individual tax revenue accounted for about 8% of GDP. Likewise, when the top individual tax rate was 70% of income, the total individual tax revenue was about 9% of GDP. Now that the top individual tax rate is almost 40% of income, the total individual tax revenue is about 8% of GDP.[426]

It is also worth noting that, for a long time now, a great number of Fortune 500 companies and other global giants have managed to minimize, or entirely avoid, the taxes one often assumes they pay. Only 10% of Federal 2014 tax revenue in the U.S. came from corporations.[427] In the U.K., corporate taxes accounted for only 7.5% of total tax revenue in 2014.[428] And in Sweden, widely acknowledged as one of the most socialist countries in the world, corporate tax as a percentage of total tax revenue was actually less than in the U.S., at 6.1% that same year.[429]

Yet another approach often cited to help resolve our crises is more regulation[xxxii]. More regulation of the financial sector, more environmental regulations, more employment regulations, more laws, more rules and more policies are seen as a way to reign in a destructive free market.[430]

The main problem is that regulation is usually designed and operated for the benefit of the industries involved, as the for-profit story tells us that profit margins are an indication of the health of the economy. So it is very difficult, if not impossible, to pass regulations that would have a detrimental impact on corporations' bottom line.[431] As with taxation, regulation is a top-down approach that requires political will to enact new laws, rules and policies. And in the current context of political capture, the odds are stacked against regulations that favor the common wealth over the privatization of wealth.

**Corporate taxes and regulation can be seen as ideologically at odds with the maximization of private profit.** Business owners and entrepreneurs receive conflicting messages that they should seek to maximize their financial gains via for-profit business, but that they should also give a large portion of their financial gains back to society or that they should forgo potential profits in order to operate in ways that benefit the broader community.

The most powerful business people in the world are not going to allow themselves to be subjected to higher taxes and regulation for the sake of resolving the very social and environmental crises that they're creating in order to be in that position of wealth and power in the first place. The for-profiteers would be shooting themselves in the foot to encourage more taxation of themselves. So, through revolving door tactics, lobbying and financially supporting candidates that favor low tax rates for the rich, society's wealthiest few will make sure that their taxes won't go up.

The 'more regulation' approach can easily lead to a bureaucratic nightmare and concentrate power in the hands of the bureaucrats and government officials.

---

426 (Ref: Heritage report)
427 (Ref: Center on Budget and Policy Priorities)
428 (Ref: National Statistics)
429 (Ref: OECD revenue statistics calculator)
430 (Ref: Graeber, The Utopia of Rules)
431 (Ref: NY Times Alperovitz article)

People on the more extreme end of the state-driven side of the spectrum believe that many of these problems can be resolved by the state having ownership of enterprises in the industries it seeks to regulate. The idea is that if the market is not adequately meeting society's needs, then the government is the only other actor that can provide essential goods and services, such as healthcare, education, infrastructure, and energy. Yet, while there are examples of some successes, especially at the municipal level and in countries that have focused on developing state-owned enterprises[xxxiii]; in most countries the state has generally proven too large, centralized and bureaucratic to allocate resources more efficiently than the private sector,[432] and corruption within large bureaucracies remains an ever-present challenge[xxxiv]. Furthermore, **even in countries where taxation is high and there are a large number of government-owned enterprises, inequality continues to rise**.[433]

As we illustrated at the beginning of the first chapter with the thought experiment about the shape of the economy, most of us have some sort of intuition that a healthy, sustainable economy has the circulation of wealth built into it, as part of how it naturally operates. Taxes, regulation and technology aren't enough to balance out the negative effects of the for-profit system because they are afterthoughts; they're meant to mitigate the symptoms of a disease that has grown from the very way we've organized our economy. This is why despite widespread calls to increase taxation and regulation, the crises of inequality and ecosystem collapse continue to worsen. A systemic problem requires a systemic solution and state-driven responses are just band-aids that the for-profit system bleeds through.

## Market-Driven Responses

Proponents of the market-driven side of the spectrum argue that the state concentrates power and is prone to inefficiency and bureaucracy, so state-driven responses create more problems than they solve. Instead, this side of the spectrum calls for market-driven responses to the ecological and inequality crises of the 21st century. Such responses entail voluntary efforts on behalf of business owners and managers to take more than profit into account in their operations. Ideas such as 'corporate social responsibility', 'conscious capitalism', 'shared value capitalism' and 'triple bottom line' business put forth visions of a for-profit market that balances the focus on profit-maximization with concern for social and ecological wellbeing.[434]

One of the first attempts to make for-profit business better aligned with social and environmental goals is known as corporate social responsibility (CSR). It was a very promising concept when it first gained popularity in the 1990s. It is the idea that corporations should take more responsibility for caring for society and the environment, rather than just focusing on profit.

There aren't any strict rules in the realm of corporate social responsibility, but there are best practices and CSR experts can be hired to put together a program for your company. Planting trees, setting up playgrounds in disadvantaged neighborhoods, cleaning up a local park or beach and sponsoring charitable events are common CSR activities. The idea is that such projects not only help the community, but they also give your company a reputation-boost.

432 (Ref: Boycko et al., 1996),
433 (Ref: OECD, 2011)
434 (Ref: Mackey & Sisodia, 2014)

Today, a great number of companies around the world create CSR reports, quantifying the positive environmental and social impacts of their CSR programs. And they use these numbers to improve their public image. "We saved the environment 1000 tons of carbon dioxide emissions this year by planting 1000 trees", for example. Unfortunately, they don't measure how much their business activities *cost* the environment and society. As more companies use CSR as a ploy to enhance their marketing strategies, more observers become disappointed with the lack of deeper change in corporate behavior.

'Triple bottom line' is another popular idea and goes a step further than CSR. It is a business philosophy that seeks to incorporate social and environmental concerns into the mission of the business, thus the triple bottom line is often described in the catch phrase 'people, planet, profit'[xxxv].

Seeking to solidify the triple bottom line approach and make it tangible for businesses, various initiatives and frameworks have sprung forth, including B Corp certification, shared value, and conscious capitalism. These are voluntary initiatives that companies can engage with in an effort to incorporate the triple bottom line approach into their business operations.

Governments have even created legal entities to encourage triple bottom line thinking. These include benefit corporations, flexible purpose corporations and low-profit limited liability companies (LC3's) in various U.S. states, C3 companies in Canada, for-profit social enterprises and social cooperatives in Europe, and Community Interest Companies limited by shares in the U.K., all of which are for-profit legal models that require companies to have a social mission in addition to their profit-maximization mandate.

At first glance, the triple bottom line might seem like a really good idea. After all, why shouldn't the aims of businesses and their shareholders align with the health of the whole living community? But in light of the fact that businesses are embedded in society and society is embedded in ecosystems, **it is misguided to treat profit as if it is on an equal footing with people and the planet.** Without people, there's no profit. Without the planet, there are no people. There is an obvious hierarchy of priorities that is not acknowledged by the triple bottom line approach. And, again, profit for what or for whom? Profit is still seen as a goal in the triple bottom line philosophy.

For-profit businesses are set up to financially benefit private owners and so have an inherent incentive to prioritize their owners' financial concerns over the concerns of other stakeholders (Ref: O'Toole & Vogel, 2011). This is the for-profit legal model. Environmental health and wellbeing are an afterthought in for-profit business, not woven into the fabric of the legal structure. These businesses take environmental and social concerns into account only if they are also able to generate high profits for their owners, because **it is antithetical for them to sacrifice shareholder value for environmental or social concerns**.[435] Attempts at triple bottom line business unfortunately still tend to result in single bottom line thinking when it comes to difficult decisions in which profit, people and planet do not all naturally align.

At best, these models can only marginally rectify the negative social and environmental impacts that these businesses have. At worst, they can be misleading and destructive.

[435] (Ref: Crane et al., 2014).

As scholar John Erehnfeld put it, “What firms are doing is reducing *un*sustainability, but this is not the same as creating sustainability.”[436]

Disillusioned with these corporate attempts to resolve social and environmental issues, some observers have come to see cooperatives as the best and only alternative to business-as-usual, because they are democratically-owned and run by their members. Cooperatives do have a critical role to play in the evolution of a sustainable economy, yet **we might be missing something very fundamental by grouping all cooperatives together by assuming they all treat profit in the same way.**

It’s important to understand that cooperatives also fall along the for-profit to NFP spectrum. Not all co-ops are created equal. Some producer cooperatives are formed solely in order to maximize profit for their members. For instance, there are many agricultural cooperatives whose members are large for-profit farms. In these cases, the cooperative’s goal is to negotiate the best prices for all of its members’ products, in order to maximize their profits, which can then be distributed to private owners. Clearly, such cooperatives remain very much for-profit and their only point of distinction from their corporate counterparts is that they adopted the 8 cooperative principles, meaning their operations are likely more democratic than a conglomeration of for-profit corporations where the biggest generally wields the most power.

Although there have been some attempts to describe the potential for a worker-led economy dominated by worker-owned co-ops, this is just another flavor of capitalism, as worker-owned co-ops still operate according to the profit motive and private ownership of businesses. The distribution of profit to workers shows that this business model is still rooted in the for-profit story, albeit tempered by the democratic aspects of the cooperative model. So while it has many strands and variations, overall the worker-owned co-op might not be as radically different from business-as-usual as many people assume.

The not-so-orthodox Marxist, Ellen Meiksins Wood, has pointed out that even workers who own the means of production will have to respond to the market’s imperatives. They will have to compete and accumulate and to let uncompetitive enterprises and their workers go under.[437] They fall into the for-profit economy category and will face the same challenges.

Many of these new business models and initiatives play an important role in accelerating the movement of the market in the not-for-profit direction, but they’re not enough to address our crises on a systemic level.

## The Nonprofit Enabler

The focus on the market versus the state distracts us from seeing how the for-profit economy systematically creates poverty and inequality, regardless of how free or constrained the market is. Instead of getting to the source of the disease, band-aids are applied and the symptoms medicated. Another band-aid used as a remedy for inequality and poverty is the realm of charities, NGOs (non-governmental organizations), international aid, and development loans.

---

[436] (Ref: Erehnfield article)
[437] (Ref: Meiksins Wood)

A great deal of development loans and aid are geared towards supporting the for-profit economy, and charities and NGOs serve to relieve for-profit guilt and maintain the guise that the Invisible Hand really does work, as long as we help it along.

We've all heard the grim statistics about global poverty - 22,000 children die each day due to poverty,[438] 800 million people are undernourished,[439] and 2.5 billion don't have access to adequate sanitation.[440] There are thousands of good-hearted organizations and people all over the world trying to 'relieve poverty'. They work to get food, water, medicine and other basic goods to people living in poverty. There are programs that help people in poverty get micro-loans to start their own business. There are plenty of good intentions and tons of resources being poured into the problem of global poverty.

However, most of these efforts are grounded firmly in the very system from which inequality stems in the first place. When aid and loans are given in the for-profit context, they are formed in accordance with the for-profit story. They often end up being a form of evangelizing, to bring the 'less developed' into the modern, consumerist, for-profit world. In many cases, aid can be 'tied', with the benefits flowing back to the donor nation.

Despite all of the negative impacts of the for-profit economic system, many of the leaders in 'developing' countries have been convinced of the urgent need to participate in the global capitalist market. Their rush to become 'developed' is costing them their traditional cultures, family ties, community connection, and the stability of their ecosystems, all in the name of exports and economic growth. People in rural areas who previously lived off the land are encouraged to leave their communities in a quest to become workers and consumers in the modern economy. The for-profit world convinces people that they are poor and backwards and they are promised a future of prosperity, material wealth and dignity. In more violent cases, people are forced off of the land, when it is taken away from them by big businesses, as has happened recently in Ethiopia, Cambodia, Indonesia, and Kenya.[441]

The aid given to the 'developing' world will never be enough in light of the massive extent to which low-income countries are indebted, often as a result of colonialism and having a late start in the for-profit game, but also through lending from institutions, the interest payments on which dwarf the amount of aid received. In 2010, total external debt owed by 'developing' nations was $4 trillion,[442] while the total official development aid they received was $128 billion.[443] This means that for every $1 in aid a 'developing' country receives, more than $31 is owed on debt repayments. Clearly this debt maintains dependency. Many of these debts are 'odious' – that is, the funds were siphoned off by corrupt leaders, and never benefited the people who remain saddled with paying off the debt.

This is illustrative of a much wider trend. Traditional nonprofits, charities, and NGOs in the aid sector (that are largely if not solely reliant on donations and grants) are part of a larger mechanism in the for-profit economy; what we call the *Nonprofit Enabler effect*. Nonprofits in the for-profit system (as distinct from NFP enterprises) are seen as vehicles for delivering social goods, funded by private money, rather than taxes. Because it's private money, the biggest donors and sponsors have a lot of control over the mission of the nonprofits they support and how that mission is carried out. In fact, most nonprofits can only exist because the for-profit system sees them as valuable. Nonprofit initiatives that don't complement the values and priorities of the for-profit system have a much harder time finding sources of funding in the for-profit world than initiatives that can make a clear link between their strategy and increased profits or economic growth.

[438] (Ref: UNICEF)
[439] (Ref: World Food Program)
[440] (Ref: World Water Day (UN))
[441] (Ref: In Ethiopia; In Cambodia; In Myanmar; In Indonesia; Kenya)
[442] (Ref: World Bank report)
[443] (Ref: OECD report)

**The relationship between for-profits and nonprofits is much like the relationship between an addict and an enabler.** The addict wants ever more of their chosen drug and the enabler makes the addict feel okay about it and entitled to it, while cleaning up the messes that the addict makes. This is exactly what charity-dependent nonprofits do in the for-profit economy. For-profit businesses, focused on maximizing profit for a relatively few privileged individuals in a short amount of time, generate social and environmental crises, and charities receive funding, largely from for-profit businesses and business owners, to clean up the social and environmental crises that the for-profits themselves make.

This creates an unhealthy co-dependent relationship between for-profits and nonprofits. Nonprofits can't exist without the will and generosity of for-profits (and often have to guilt-trip the for-profits to get the resources they need). And for-profits feel okay about contributing to inequality and ecological distress because they are able to maintain the idea and image that they're doing more good than harm, by giving money to nonprofits to address those issues.

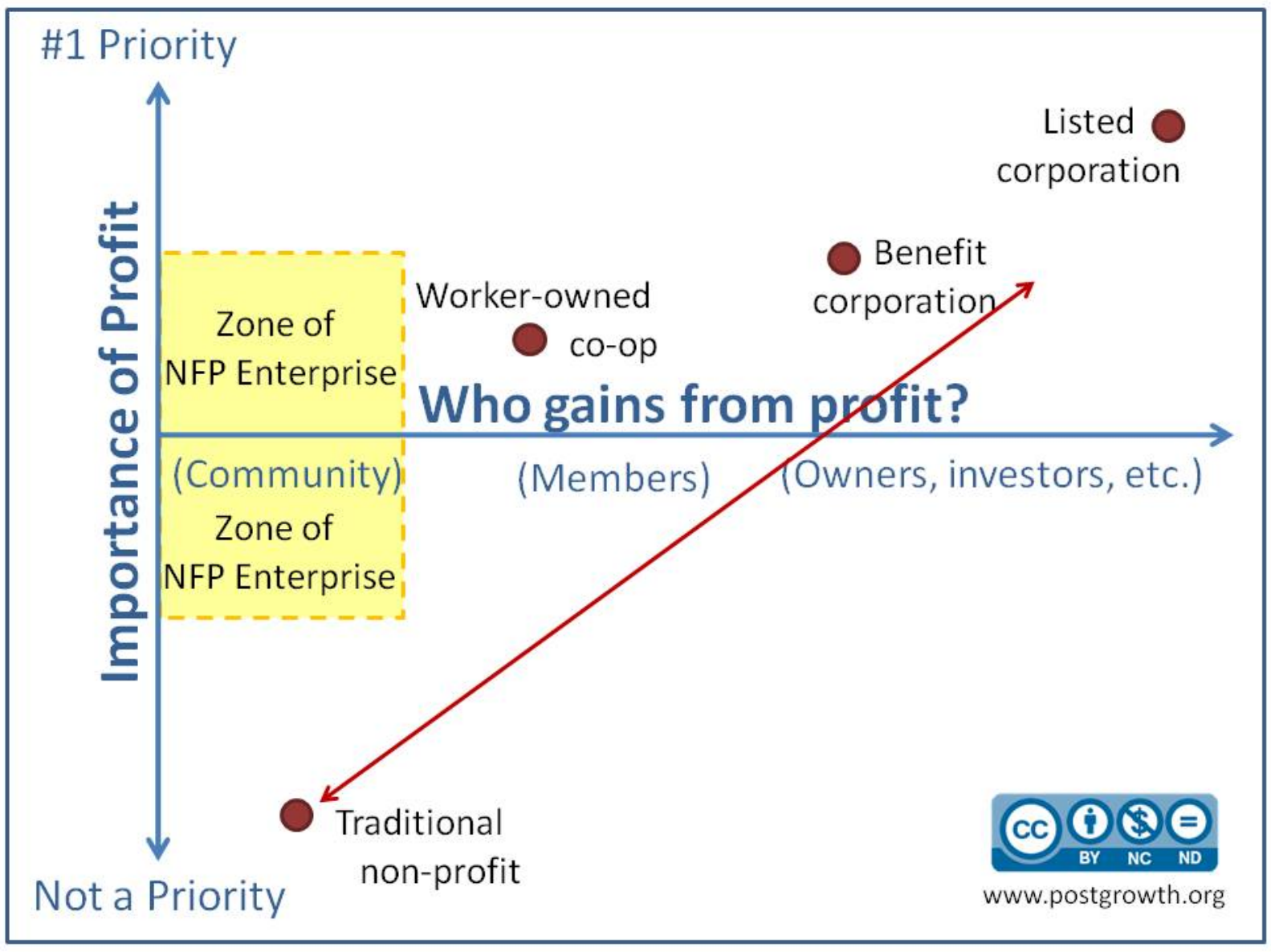

## Nonprofit Enabler Effect

Who better to comment on this phenomenon of the Nonprofit Enabler (also known as the 'charitable-industrial complex'), than Peter Buffett? His father, Warren, is the world's fourth richest person. As Peter, who has worked in the world of philanthropy for decades, and has never accepted a dollar of inheritance from his father, says, "The more lives and communities are destroyed by the system that creates vast amounts of wealth for the few, the more heroic it sounds to 'give back.' It's what I would call 'conscience laundering' — feeling better about accumulating more than any one person could possibly need to live on by sprinkling a little around as an act of charity."[444]

One of the most obvious examples of the Nonprofit Enabler phenomenon is the Gates Foundation. In 2016, Bill Gates is once again the richest person in the world, a title he's held for 17 of the last 21 years, according to Forbes Magazine.[445] His net worth is estimated at $87.4 billion and he guesses that he makes about 6% of his total wealth in income every year, so his wealth is growing *exponentially*[xxxvi]. He and his wife are the directors of the Gates Foundation, whose website says, "From poverty to health, to education, our areas of focus offer the opportunity to dramatically improve the quality of life for billions of people."[446]

A lot of the critiques of the Gates Foundation come in the form of, "Well, Mr. Gates, why don't you just put 86 of your 87 billion dollars into poverty relief and development aid?" This seems like a great solution, given that $30 billion per year could prevent 860 million people from being malnourished.[447] The for-profit story gives a clear answer to this question: a good capitalist would *never* do that – it would interfere with the mystical mechanism of the Invisible Hand. His role in our economy, as a capitalist, is to reinvest his money to make more money and to accumulate ever more wealth and it is up to his discretion as to how he will use it. He doesn't *have* to give a penny of it away. It's only out of the generosity of his heart that he gives as much as he does. His wealth will have positive effects through the businesses he invests in, regardless.

We are not judging Bill Gates' character here. We are instead looking at the systemic roots of the Nonprofit Enabler phenomenon: the for-profit story and how it shapes and guides the for-profit system.

Many people intuitionally sense that the Gates Foundation is something of a contradiction. They can feel that it's a problem that Gates is the richest person in the world and at the same time, is trying to alleviate global poverty. But this contradiction can only be articulated by stepping out of the capitalist paradigm. If you buy into the for-profit story and you live by it, you can't really argue that Bill Gates is *too* rich. What does 'too rich' mean when the goal of the game is to accumulate private wealth? After all, hasn't he created tens of thousands of jobs since he co-founded Microsoft? Anywhere you try to draw the line of 'too rich' would be arbitrary and would contradict the rules that govern the capitalist economy. If you think that growth can be sustained, then there's no problem with Gates having a higher annual income than the 40 poorest nations in the world combined,[448] because we can just grow the economic pie and they can have more *and* he can have more. No problem. In fact, from this worldview, he's an inspiration.

The same can be said of all of the super-rich people who have signed up to the 'Giving Pledge', in which billionaires like Mark Zuckerberg pledge to give most of their wealth to philanthropy. According to the for-profit worldview, these people are to be looked up to as the world's top industrialists and it is only through their generosity that that the rest of the world should have any access to the incredible wealth they've amassed.

---

444 (Ref: NY Times)
445 (Ref: Time)
446 (Ref: Gates Foundation)
447 (Ref: FAO)
448 (Ref: Forbes)

But in order to truly address poverty and inequality, we have to bring into question the acceptability of for-profit business and economic growth in the first place. When we do this, we can see the addict-enabler relationship between for-profit businesses and charities for what it is.

Another important aspect of this co-dependent relationship between nonprofits and for-profits is that it has led to a sense of desperation among many nonprofits to receive funding from whomever, however. This can easily lead to an organization drifting away from its mission. Charities often don't care which companies sponsor them; they just want money. This can result in bizarre situations like the World Wildlife Fund being sponsored by Coca Cola, a company whose plastic bottles are destroying ecosystems all over the world. A tragically ironic corporate sponsorship is that of Wells Fargo - a colossal American bank that illegally processed home foreclosures via a technique called robo-signing during the economic crisis, illegally kicking people out of their homes - supporting Habitat for Humanity, an American charity that builds houses for people who are in need.[449]

Of course, these partnerships are great for improving the reputation of the for-profit sponsors, but they can also represent a dangerous form of mission drift for the nonprofits involved. Strategic but morally hazardous sponsorship is just a natural consequence of an economy that revolves around profit maximization.

There are also plenty of nonprofit organizations that play an even more central role in the for-profit system by seeking to maximize the profits of their for-profit members. Examples of this include chambers of commerce, mutual funds, associations that lobby on behalf of for-profits, various professional associations, various sports associations, and global organizations like the World Economic Forum.

One such association that increasingly suffers from the Nonprofit Enabler syndrome is the National Rifle Association (NRA) in the U.S. Some have noted that since the NRA has gotten more of its funding from for-profit weapon manufacturers' donations than from membership dues and program fees, it has become more oriented around the demands of large gun producers than the needs of hunters and individual gun owners (whom it was set up to serve).[450]

The Nonprofit Enabler dynamic sees nonprofits playing the 'submissive caretaker' role in an economy that's largely geared towards domination. This is why NFP enterprise offers such an exciting path forward. It not only breaks with the for-profit story, but it also breaks the unhealthy enabling relationship between for-profit business and charities. Of course, not all nonprofits are enablers, and not all NFP enterprises avoid enabling behavior, but the Nonprofit Enabler effect plays such an important role in maintaining the for-profit system that it must be addressed in any discussion about going beyond capitalism.

# Time to Move On

The paradigm of economic growth and for-profit business, which must expand at all costs, has become too expensive. Capitalism is failing to deliver higher levels of wellbeing for most people, and in fact we see just the opposite happening. The for-profit system is decimating the Earth's ecosystems too quickly for them to recover, sending us into the Sixth Mass Extinction, which represents a threat to the very survival of humanity. And the inequality and concentration of power created by the wealth extraction siphon have reached unbearable levels, even in the 'rich' world.

---

449 (Ref: Wall Street Journal)
450 (Ref: Business Insider)

**It is painfully clear that we can't grow on like this.** All of the inherent contradictions of the for-profit system are coming to a head and we are reaching the end of the capitalist paradigm. It's time to take the next step in our economic and social evolution.

# 4. Stirrings of a Not-for-Profit Story

## *The seeds for a not-for-profit economy are being sown*

In 1935, Dr. Scott Williamson and his wife, Dr. Innes Pearse, set out to find out what the causes of good health are. In order to investigate this issue, they opened the Pioneer Health Centre in a part of London called Peckham. Over the course of the next 15 years, the center provided the setting for one of the most interesting social experiments of all time; what is now commonly referred to as the Peckham Experiment.[451]

Over 1,000 local families used the health center, which had sporting facilities as well as a cafeteria that served organic food from a local farm.[452] In the registration process, participants agreed to let researchers evaluate and monitor their activities and health over time. Aside from the researchers noting the activities of the participants, the center provided minimal rules and supervision.

Researchers noticed that the first 6- 18 months were quite chaotic, as participants learned to be more proactive and self-organize as a community. However, over the following 12 years, it was noted that there was a high level of cooperation and creative collaboration, there was no bullying, there was very little interest in competitive games, and there was a general increase in levels of health and wellbeing. Social ecologist Stuart Hill concludes that these successes were due to the supportive environment that the Pioneer Health Centre provided, as well as the freedom that participants had to be spontaneous and receive non-judgemental feedback.

The Peckham Experiment offers a couple of very important lessons when it comes to laying the foundation for a more sustainable economy. First, that wellbeing is a dynamic process that is constantly being shaped by myriad factors within and around us. Secondly, under the right conditions, human nature can be extremely cooperative and altruistic.

Renewed interest in the Peckham Experiment indicates a growing willingness to think about humans in more complex terms. All of this research is changing the story of human nature and opening up a whole new potential for understanding human societies and the role of the economy.

## New Stories for a New World

Stories create the world in which we live. They are shared beliefs about the way the world works, why everything is the way it is, what humanity's role is, and what our shared goals should be. They are the mental maps by which we navigate the world around us. These shared narratives, worldviews and belief systems shape our values and influence our feelings and our actions. As a result, our collective stories have a profound effect on the conditions in which we all live. Stories are also the way we pass on our shared beliefs, values and social norms, through generations, and from one society to another. In essence, our shared stories are the blueprints for how we organize our communities, states, governments, businesses and

451 (Ref: Stuart Hill, The Peckham Experiment into Health Ecology: An old study with modern implications)
452 (Ref: Ibid)

organizations. Not only do they guide us in how to structure our communities and organizations, but they tell us *why* we should organize them in one way versus another. This is why any transition to a healthier economy must be rooted in a healthier collective story.[453]

The for-profit story and way of seeing the world has become very destructive. The view of human nature as mostly greedy, individualistic, competitive and acquisitive has led us to build an economic system based on fear and scarcity. The profit motive and *Homo economicus* have wreaked havoc all over the globe.

Luckily, very different stories have been emerging and re-emerging all over the world as more people realize that the for-profit story doesn't accurately explain their feelings, behavior and experiences and, worse yet, that it is harming collective wellbeing. These stories form an overarching narrative that paves the way for a profoundly different economy. The realization that the for-profit economic system is built on an outdated, incomplete narrative about what it is to be a human on this planet is creating space for a new story to flourish.

## The Story of Interconnectedness

The overarching new story is rooted in a deeper understanding of ourselves and our world; an understanding that everything is interconnected as part of a complex, dynamic planetary system. We call it the Story of Interconnectedness.

Modern philosophers, scientists, researchers, religious leaders and artists have been exploring and trying to communicate a deeper understanding of the complexity of the world for millenia. Many believe that this more complex[xxxvii] worldview is the next big paradigm, which is starting to overtake the mechanistic worldview which has dominated for the last three centuries.[454] This new worldview begs us to go beyond seeing everything in black-and-white, binary terms. It's about moving from a mindset that sees the world as a machine, to seeing the world as an integrated, complex whole which is much more than the sum of its parts; a system that is self-organized; a system in which everything is interrelated and 'feedback loops' abound; a system that is constantly evolving in unpredictable ways because new patterns and trends are always emerging and interacting.

Our complex planetary system is constantly evolving due to dynamic flows and relationships, and because of this constant change it cannot simply be managed or controlled. In fact, part of the reason this story is gaining popularity is because our efforts as a 'dominator' civilization trying to control and manage the complex systems of our world have failed. From our planet's ecosystems to our economic system, it is becoming increasingly obvious that the status quo approaches are failing both people and the planet.

Renowned physicist Fritjof Capra has called this new worldview Ecological Consciousness. Environmental scientist Donella Meadows called it Systems Thinking. Philosopher Charles Eisenstein calls it the Story of Reunion. Philosopher Ken Wilber calls it the Integral Perspective. However it is named, it is a story that is eager to dive into the depths and richness of complexity and the interconnectedness of all things on Earth.

---

[453] (Ref: Yuval Harari "Sapiens"; Eisenstein "Ascent of Humanity"; Eisler "Chalice and the Blade")
[454] (Ref: Fritjof Capra "Web of Life"; Sahtouris "Earthdance"; James Lovelock Gaia Theory)

One of the assumptions on which the for-profit story is built is that evolution is based on the survival of the fittest. We're taught that the basic mechanism of biological evolution is ruthless competition and whoever survives gets to pass on their genes. This understanding of biological evolution emerged side-by-side with economic thinking in the 1700 and 1800s, and that explains to a large extent why our economic system is rooted in self-interest.

However, in recent decades, biologists, sociologists, philosophers, and economists alike have increasingly questioned this story. And much research has proved that in most natural systems, **competition happens within a greater framework of cooperation**. Cooperation is particularly important for a hyper-social species like humans.[455]

We can see this in today's global open-source movement. Open-source is a fundamentally different way of creating and designing technology, relying on the knowledge and skills of a large community instead of just a handful of specialists. The open-source community functions as a sort of 'gift economy', in which innovators contribute ideas, time and energy to the community freely and are rewarded by receiving feedback and help from community members.[456] If someone's contributions are deemed worthy, they will have a greater say in the decision-making process of designing a product or service, but work is largely done for the benefit of the community, not out of competitive, self-interest.[457] Open-source is an ethos of 'share and share alike'. It's based on the notion that we can maximize the potential benefit of our ideas by letting others use and build on them.

In light of new findings, the for-profit story of a rational, self-interested, profit-maximizing *Homo economicus* is dissolving into a much more complex picture of human nature. The work of Professor of Economics and Psychology at the University of British Columbia, Joseph Henrich, and his colleagues has dispelled the myth of a static, rigid human nature. They have been looking into the WEIRD phenomenon, where most economic research and thus the bulk of economic and business policy is based on the assumption that everyone in the world behaves like WEIRD people (Western, Educated, Industrialized, Rich and Democratic). Henrich and his colleagues argue that experiments about economic behavior and human nature have traditionally been conducted by researchers at rich, Western universities, where the subjects of the experiments are often the students at those universities. Henrich and his team took the same experiments (mostly games that engage people in economic-like choices) to populations all over the world, from communities in Siberia to small tribes in the Amazon. They found great diversity in the way people responded. Their robust cross-cultural research proves that there is no set human nature, but rather that human nature encompasses an extremely wide range of characteristics.[458]

In addition to Henrich's exploration of economic behavior beyond WEIRD people, behavioral research tells us that our institutions and social environment play an extremely important role when it comes to which aspects of human nature we express in our actions. The famous Milgram and Stanford prison experiments, for instance, show that people's willingness to inflict harm and shame on others increases in settings of social hierarchy and stress.[459]

---

455 (Ref: Martin Nowak, Sam Bowles "Cooperative Species", 'The Penguin and the Leviathan'; Hodgson's Morality and the Economic Man)

456 (Ref: Maclurcan and Radywyl)
457 (Ref: Ibid)
458 (Ref: Henrich, 'The weirdest people in the world')
459 (Ref: Philip Zimbardo, *Psychology of Evil TED Talk*)

Extensive research in the field of behavioral economics tells us that humans are not the cold, calculating, rational actors that the for-profit story describes.[460] And **a growing body of research shows that people are motivated more by purpose than profit**. Dan Pink's analysis of research on motivation and work indicates that once we have enough money to not have to worry about meeting our basic needs, the majority of us are motivated by purpose, mastery and contributing to something larger than ourselves at work; not financial gain.[461]

Economic analyst Jeremy Rifkin has gone so far as to suggest that *Homo empathicus*, the empathic human, is taking an ever larger role on the economic stage and *Homo economicus*, the Economic Man, is starting to bow out.[462]

This new perspective of evolution and human nature allows us to acknowledge our interconnectedness. In addition to recognizing that human nature is much more complex than we had previously understood, we are simultaneously coming to understand that our planet is extremely complex as well and we are interconnected with all things on Earth. We are deeply embedded in our planet's ecosystems, so **what happens to our ecosystems also happens to us.**

This understanding clearly indicates that we have a responsibility to ensure that human societies operate in ways that let ecosystems flourish; we have a duty to be ecological stewards. The story of interconnectedness reminds us that Earth is our one and only home. It is beautiful and full of wonder. We are not just objective observers of nature. We do not only participate in nature, we are *part* of nature. We must take care of the planet so that it is suitable for all of us, including other species and the future generations that will inherit it. We are not owners of this magnificent Earth, we are the care-takers.

This story represents a planetary consciousness – an awareness that we all share this planet and so we must think of the planetary consequences of our actions. It is often said that this planetary consciousness began when the first astronauts went out to space and saw just how small and how isolated Earth really is. The astronauts brought back their photos, and people all over the world were shocked when it also occurred to them just how important it is for us to keep our unique little planet healthy. It was the first time that modern civilization comprehended in such a direct way the boundaries of our planet and that life on Earth is actually quite fragile, floating through an infinite amount of space that cannot support life. This is when the term 'Spaceship Earth' was coined. It was also around this time that former NASA scientist, John Lovelock, developed the Gaia Hypothesis, a theory that describes Earth as a self-regulating organism.

It's easy to see how this contrasts with the for-profit story, which is based on the notion that humans are so clever that we deserve to do with other species what we want and we can easily innovate our way out of any problem. The new story responds by saying that human beings are a very young species and we need to pay greater attention to the evolutionary scale of things, or risk extinction by hubris. Scientifically measurable phenomena like climate change and the sixth mass extinction are reinforcing this story, proving to more and more people that we're all in this together.

---

[460] (Ref: Ariely "Predictably Irrational"; Kahneman "Thinking Fast and Slow"; Richard Thaler "Misbehaving")
[461] (Ref: Pink, *Drive*)
[462] (Ref: Jeremy Rifkin, Empathic Civilization)

The story of interconnectedness also manifests in very practical ways, like the rise of new policies. Mother Earth laws and ecocide laws, for instance, are aimed at making destruction or harm to ecosystems punishable by law, as a very serious criminal offense. Not only does this new legislation protect nature, but it is making more people aware of just how negative an impact we're having on the Earth, as well as the natural world's inherent rights to health and wellbeing. These frameworks are some of the first efforts to give nature a legal voice in the human economy and modern policy.

The new story humbles us, as just one species of millions, and this humility allows us to learn from the intelligence of nature. One example of learning from nature is an approach to research and design, called 'biomimicry'. A whole new generation of designers, architects and engineers are modeling everything from household products to apartment buildings after organisms and ecosystems found in nature. It is a way of learning from the complexity and inherent wisdom of nature. A great example of biomimicry is a turbine designed in the shape of a shark's tails and fins in order to capture the power of tides for electricity generation.[463]

This way of thinking has also lead to new approaches to economic activity, such as the 'circular economy' – designing the economy in a way that uses the waste of one process as the fuel for another process, like ecosystems do. A great example of the circular economy concept in action is the Baltic Biogas Bus project, which seeks to help cities in the Baltic region set up processing centers to convert municipal organic waste into biogas that they can then use as fuel for public transportation systems.[464] This helps cities deal with their organic waste in a more sustainable way, cut down on fossil fuel use, and provide healthier, more affordable public transportation to their citizens.

These ideas are not as new as they may seem. They are grounded in timeless wisdom from ancient and indigenous cultures and are being re-born through lessons learned from modern, globalized civilization. One traditional example is Ubuntu, a philosophy of some southern African cultures, which refers to 'the belief in a universal bond of sharing that connects all of humanity.'[465] Nobel Peace Prize recipient Archbishop Desmond Tutu is dedicated to spreading Ubuntu around the world. He says, "You cannot be human on your own. You are human through relationship. You *become* human."[466] In fact, he says we are so interconnected that "whenever you de-humanize another, whether you like it or not, you yourself are dehumanized."[467]

Another excellent example of this story is the concept of 'buen vivir' in Latin America. Buen vivir, directly translated to English, means 'good living', but the concept runs much deeper than that. Buen vivir is the Spanish translation of Sumak Kawsay, a phrase in the language of the native Quechua people of the Andes, which refers to 'living in harmony with ourselves, our communities and our natural environment'.

These beliefs foster compassion and empathy. We are interconnected, so we must take care of each other and in order to do that, we need to understand how others feel. These are not just nice thoughts. Recent research into our brains in the field of neuro-science indicates that we are 'soft-wired' for empathy.[468] We have 'mirror neurons' in our brains that make us feel the pain, joy, enthusiasm or disappointment of those around us.[469] This is an automatic, involuntary response that researchers believe has evolved in humans to allow for our very cooperation-based evolution.[470]

---

463 (Ref: BioPowerSystems website)
464 (Ref: Baltic Biogas Bus project website)
465 (Ref: Wikipedia article)
466 (Ref: Desmond Tut, Templeton Prize 2013 video)
467 (Ref: Desmond Tutu talk 2007 Semester at Sea video)

Everything on Earth is interdependent, so what happens to one of us happens to us all, in a very real sense. Understanding our deep interconnectedness also means acknowledging that we're *all* affected by social issues like inequality. Richard Wilkinson and Kate Pickett's landmark book on the effects of inequality on public health, *The Spirit Level*, describes how levels of inequality are a major determining factor in whether a society thrives or not. Their data shows that even the richest people in highly unequal societies have lower levels of physical and mental health and encounter higher rates of social problems, like crime and violence, than the richest people in more equal societies[xxxviii].[471] For instance, infant mortality rates among the rich are higher in unequal societies like the U.S., compared to more equal societies like Sweden.[472] This illustrates that if my community is not doing well, I'm not either, and vice versa.

This idea is one that nearly all wisdom traditions embrace, and that more of us are coming to learn through our own experiences with modern consumerism. A growing number of people are living lifestyles that focus more on human connection and less on material accumulation[xxxix] - even some of the least likely people in the world.

In 2007, Tom Shadyac, the director of many popular American comedy films, including *Ace Ventura* and *Bruce Almighty*, started making a documentary about something more existential, called *I Am.*

The film takes viewers on two parallel journeys sparked by one life-changing event: Tom nearly died in a mountain biking accident. Facing his own death made him think about what really matters in life. The film tells the story of Tom's personal journey from a lifestyle of Hollywood glamour to one of material sufficiency. Tom had achieved all the markers of success in modern society: fame, fortune, mansions, limousines, the nicest furnishings that money can buy, and he was sure that he would feel happiness beyond belief. But sitting in his mansion alone, he realized that rather than feeling overjoyed at all of his success and achievement, he felt a sense of emptiness.

Realizing that it's not what you take but what you *contribute* that matters, Tom seeks out a better understanding of the human journey, at large. He started by asking two essential questions, "What's wrong with the world?" and "What can we do about it?"

In Tom's quest for answers, he has conversations with some of the leading thinkers of our time, including Desmond Tutu, David Suzuki, Lynne McTaggart, Noam Chomsky, Daniel Quinn and Elisabet Sahtouris. These conversations, though they cover a wide array of topics, all come back to one essential point: **humanity's main underlying problem is that we believe we are separate, when in fact, we are all tightly connected to each other and to**

---

468 (Ref: Rifkin, *Empathic Civilization*)
469 (Ref: Giacomo Rizzolatti, *Mirror Neurons*)
470 (Ref: Ibid)
471 (Ref: Wilkinson and Pickett, *The Spirit Level)*
472 (Ref: Ibid)

**the living systems that we inhabit**. This is why Tom's material success didn't make him happy: it only served to further separate him from others. The moral of Tom's story is that true wellbeing comes from the opposite of individual accumulation; true wellbeing comes from strengthening the community and sharing.

This deeper understanding of human nature, motivation, and wellbeing, leads to a re-defining of prosperity and progress. Another documentary, *Ancient Futures*, tells this story by exploring how the Himalayan village of Ladakh has changed as outside influences have brought the culture of individualism, consumerism and materialism to it. The film shows how the community bonds that once kept everyone happy and healthy disintegrated when people started to believe they were poor and must compete in the modern economy to become richer. As a result, everyone in the community has become less happy, despite many individuals having more material wealth. Social ailments like homelessness, mistrust and crime emerged for the first time in this remote community.

The film's director, Helena Norberg-Hodge, a Swedish researcher who experienced this transition in Ladakh first-hand over a period of twenty years, points out that this case study has a lot to tell us all about what development really is. Is development about having more material possessions, but at the cost of more homelessness and crime, and less trust in society? Is it about losing our sense of connection in order to produce more consumer goods and grow the GDP?

As a result of the collective realization that the answer to these questions is 'no', different ways of measuring progress are being developed. On the national and global level, tools have emerged to help us measure wellbeing and true prosperity, rather than just the size of each country's economic output or income. Often described as Gross National Happiness metrics (as opposed to Gross National Product), these tools include the Human Development Index, Genuine Progress Indicator, Happy Planet Index, and the Social Progress Index[xl]. In 2011, the United Nations published the World Happiness Report, in which countries are ranked according to their wellbeing, across many different indicators. That same year, Canada developed the Canadian Index of Wellbeing. And in 2014, the United Kingdom launched its own national wellbeing statistics system.

The story of interconnectedness tells us that progress is about having greater wellbeing and meeting our needs in healthy, balanced ways. **Progress means increasing our quality of life, not the quantity of goods.** This story redefines what it means to prosper and be well. It seems a lot of ancient wisdom is being revived by humanity's twentieth century journey of learning the hard way that money can't buy happiness.

Highly-acclaimed author Malcolm Gadwell provides a thorough exploration of the ubiquity of what is known as the 'inverted U curve' in his book *David and Goliath*. He describes how, in most social phenomena, there is the tendency for more of a good thing to make us feel or do better, but only up to a point. After a certain point, more of that same thing actually makes us feel or do worse. So it's like an upside-down U.

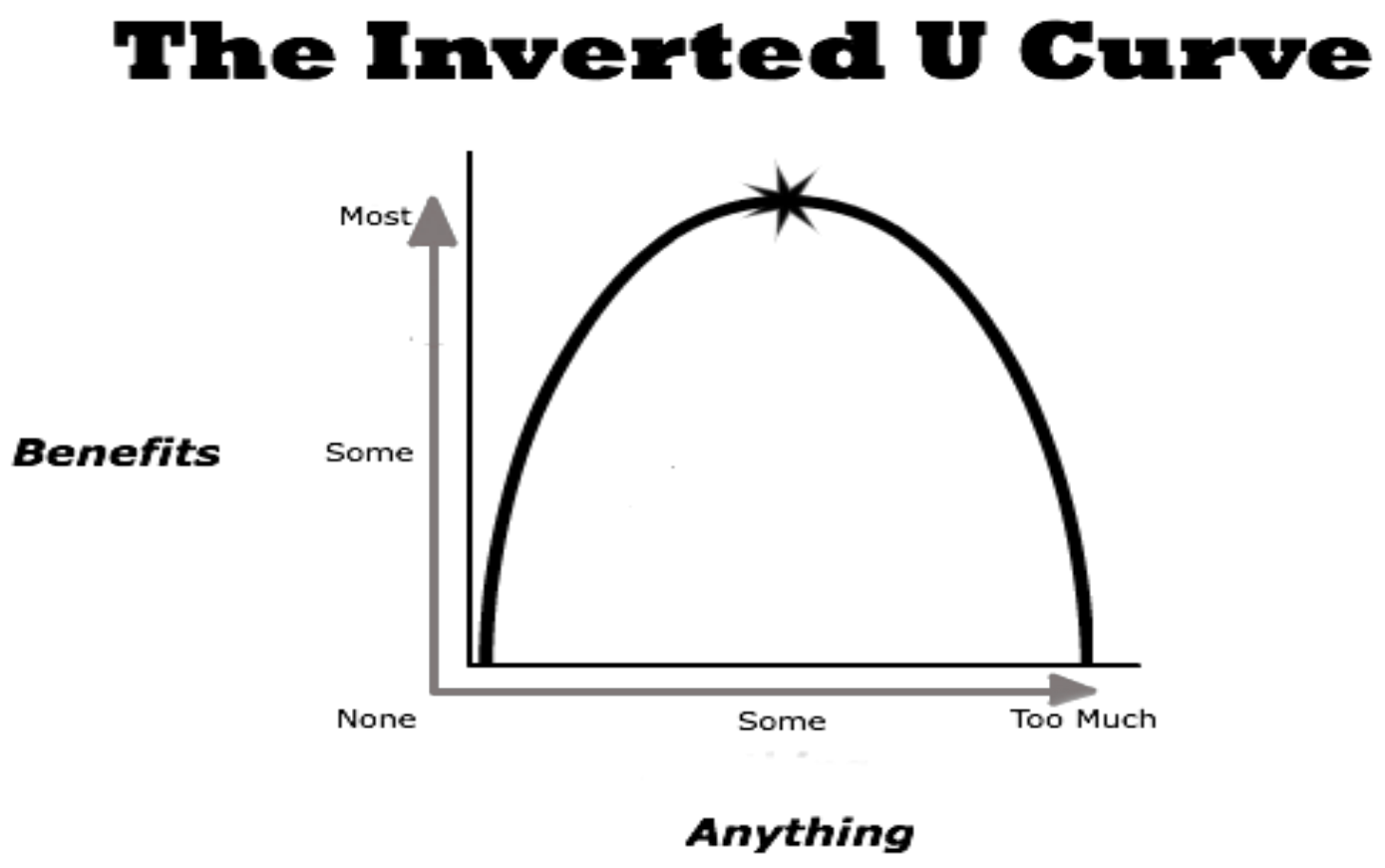

**The Inverted U Curve**

The curve goes up as we begin getting a bit of something we perceive as good, then, at a certain point, the line flattens out, once we've had *enough* of that good thing. After that, the line goes down when we start getting too much of a good thing. This is very easy to understand in terms of food. If we're starving, a bit of food can do a lot to improve our health, and more food will continue to do so until we have taken in sufficient nutrition. If we've eaten enough food, then more food actually starts to harm our health. This principle of enough applies to a whole host of things.

For Gladwell, an interesting example is the effect of money on parenting. He cites research which found having too little money makes parenting difficult, because parents must spend so much time and effort just trying to put food on the table that it's difficult for them to find the time and energy to satisfy their children's other needs. Having more money can make parenting easier for a person with limited financial means. But after a certain income level, parenting actually starts to become more difficult again (Gladwell claims that the ideal income level is about $75,000 per year in the U.S.). This is because in higher income brackets, people often have so many business endeavors and social appointments that they don't have the time and energy to also satisfy their children's emotional needs. The art of living well is to find the happy medium.[473]

This idea is aptly expressed in the old proverb, 'Everything in moderation'. You *can* have enough of a good thing. In Swedish, the idea is succinctly captured in one word – 'lagom' – which roughly translates to 'just the right amount'. In Chinese culture, 'xiaokang' is an ancient Confucian term that refers to a situation in which people live comfortably with modest means.

The principle of enough has also been verified by longitudinal studies which have found that happiness is not correlated with higher income levels after a certain point. Known as the Easterlin Paradox, the phenomenon was discovered by economist Richard Easterlin in 1974.[474] Easterlin found that levels of self-reported happiness rise along with a country's income up to a point, after which there is no correlation between happiness levels and national income (Ref: Ibid). The Easterlin Paradox only seems paradoxical in the context of the for-profit story, which promises that more material wealth will always make you happier. But Easterlin's findings make perfect sense in light of complexity and the story of interconnectedness. **Happiness is more complex than just having more stuff.**

[473] (Ref: Gladwell, "David and Goliath")
[474] (Ref: Easterlin Paradox Revisited)

## Universal Needs

Easterlin's research is very closely related with important work that Chilean economist Manfred Max-Neef did in the area of human needs in the 1990s. He conducted a ground-breaking, cross-cultural investigation into what people really need in order to thrive and be well. He sought to find out whether or not there are universal needs that all humans on the planet have, regardless of background, economic standing or culture. This contrasts greatly with the well-known hierarchy of needs that psychologist Abraham Maslow developed in the 1940s, which insinuates that high-income people have different needs than low-income people and that there is a linear progression up through different levels of needs.[475] Through his findings, Max-Neef identified nine universal human needs: subsistence, protection, affection, understanding, participation, leisure, creation, identity and freedom.

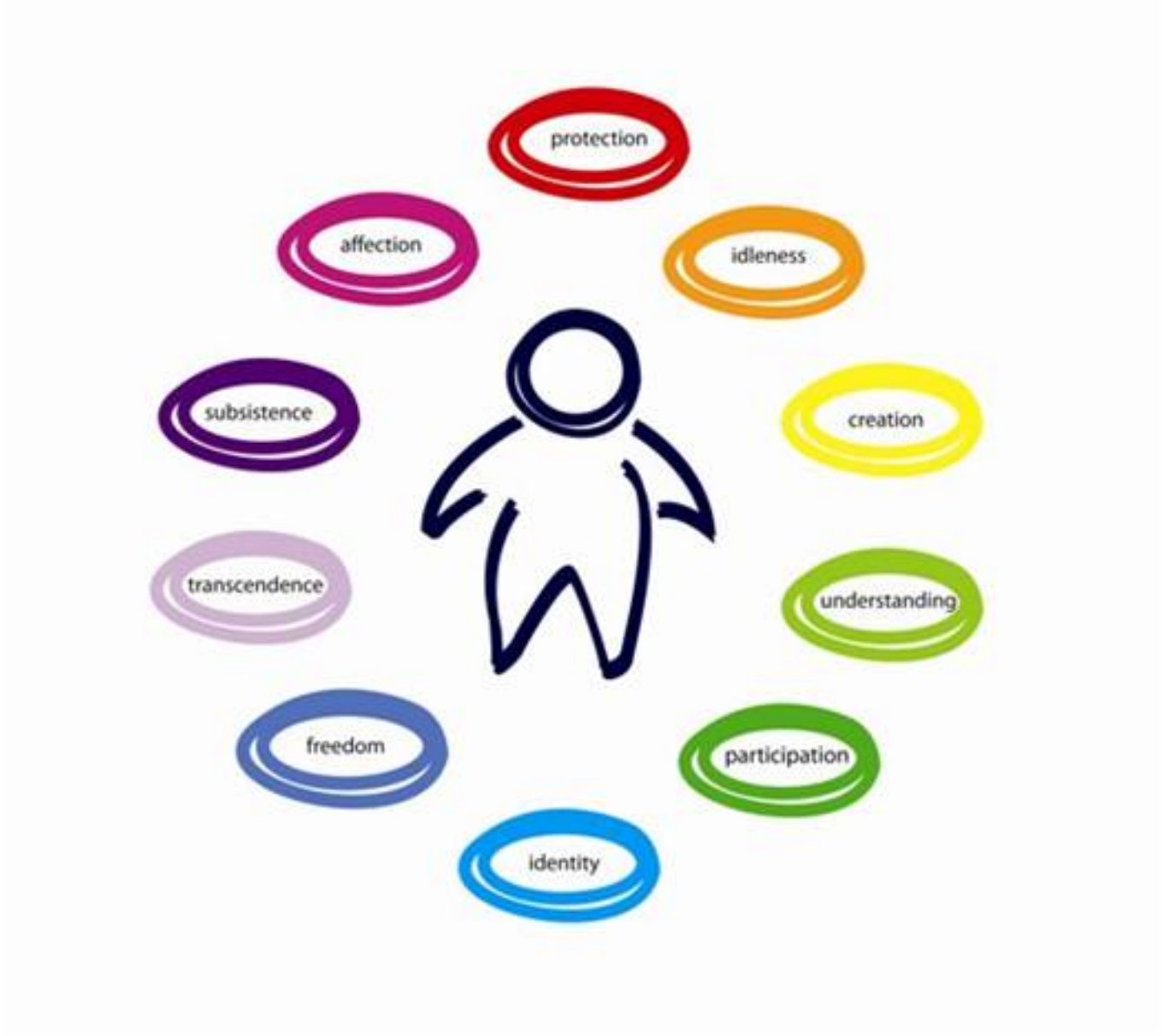

**Universal Human Needs**

Max-Neef went on to create a very useful way of thinking about these needs and how we satisfy them. He envisioned our needs as a system, in which all needs are interdependent and interactive. For example, food and exercise are not really needs in themselves, but are better thought of as satisfiers of the needs for subsistence, participation, and leisure – they can satisfy many needs at once.

Max-Neef categorized the ways in which we satisfy our needs into five different types: singular, synergistic, pseudo, inhibiting and violators. *Singular satisfiers* satisfy one need at a time. For example, eating a meal alone meets the need for subsistence. *Synergistic satisfiers* satisfy more than one need at a time. For example, a community cook-out might satisfy the needs for subsistence, participation and leisure. *Pseudo-satisfiers* give us the false sensation that they are satisfying our needs, when they really aren't. For example, drinking a caffeinated

475 (Ref: Maslow, Hierarchy of Needs)

beverage might satisfy your thirst temporarily, but caffeine can actually contribute to dehydration. *Inhibiting satisfiers*, which actually inhibit meeting other needs, by over-satisfying one need. For example, watching television can satisfy your need for leisure, but it might actually keep you from meeting your need for creativity, understanding, identity, and participation. And *violators*, under the pretext of satisfying a need, actually make it harder to satisfy that need and often also violate our ability to meet other needs. For example, censorship, which is meant to meet our need for protection, actually violates our ability to meet the need for understanding, participation, leisure, creation, identity, and freedom, which in turn makes us feel unprotected. The optimal situation is obviously to meet most, if not all of our needs through synergistic satisfiers, and to avoid pseudo-satisfiers, inhibitors and violators.[476]

Due to the tendency for market expansion in the name of maximizing profits, advertising in our current consumer-driven, growth-based economy, leads us to believe that most or all of our needs can and should be met through the market. This has led many of us to use pseudo-satisfiers, inhibiting satisfiers and violators on a regular basis (like buying stuff we don't need and can't afford via 'retail therapy') and unsurprisingly we are left feeling unsatisfied, which leads us to buy more stuff in a desparate effort to fulfill our needs. And so the work-watch-shop treadmill goes on and on, in a vicious cycle.

The story of interconnectedness tells us that not all needs can or should be met by buying goods and services. In fact, most of our needs can be better met *outside* of the market, by things that can't be paid for, like strong personal relationships, a sense of community, connection to nature, free time, self-reflection, and open spaces to explore.

Initiatives like Transition Towns and the Slow Movement are helping people meet their needs through creating stronger communities and local networks of collaboration. Places such as community gardens and tool libraries enable us to connect with others in our community and share resources, meeting our needs outside of the market.

The new story replaces the elusive Invisible Hand with a much more visible mechanism of social benefit. Rather than the vague idea that society will benefit as a result of everyone seeking to maximize their own immediate self-interest, society benefits from everybody trying to fulfill their needs in the healthiest, most synergistic ways possible. When people have a clearer understanding of wellbeing, universal needs and our deep interconnectedness, society benefits.

# Corresponding Shifts in the Economy

Environmentalist Paul Hawken believes that we have entered an era of 'blessed unrest.'[477] The negative conditions of inequality and the destruction of our planet's biosphere have given way to a collective sense of frustration, dissatisfaction, anxiety, and an unrelenting urge for a deeper sense of purpose in work and life. A growing number of people feel a deep desire to align their actions with their visions of how the world could be; a more beautiful world. This feeling of unrest creates an openness to new ways of thinking and doing things. When people are no longer comfortable, change doesn't seem to be as difficult. This is why it's a *blessed*

476 (Ref: Max-Neef's needs)
477 (Ref: The Blessed Unrest)

unrest; **it is from this unease that great creativity, innovation and collaboration are springing forth**, creating a whole new economy in the process.

## Changing Demand

An increasing number of people are orienting their lives more towards meaning and purpose, away from money and materialism. Dan Pink has been tracking this trend for decades. He points out that the best selling non-fiction book in the U.S. for the first decade of the 21st century was *The Purpose-Driven Life*, a book about how to live a life of greater purpose and meaning. He also points out that this parallels the trend of a growing number of people meditating and doing yoga in the U.S. in recent decades.

Pink says, "We have been liberated by (financial) prosperity, but not fulfilled by it, and so you have this wide-spread democratization of the search for meaning, this wide-spread democratization of self-realization, self-actualization… It's unprecedented and it's not going away."[478]

People all around the globe are making changes in their personal lifestyles in order to align with the values and story of interconnectedness[479] and one of the ways this impacts the economy is through more conscious consumption. Everyday people are increasingly thinking about the impact of their shopping patterns on ecosystems and people around the world.[480]

The data clearly shows that we want to buy products from companies that have a low ecological footprint and don't use sweatshops or slave labor. For instance:

- 77% of U.K. consumers refuse to buy products or services from a distrusted company , while 91% choose to buy from trusted companies.[481]
- 2/3 of UK adults believe it is unacceptable for shareholders to profit from running health services, children's homes, police services, and care homes for the elderly and disabled people.[482]
- 81% of consumers worldwide believe that businesses have a responsibility to address social and environmental issues,[483] and 72% say business is failing at it.[484]
- 2/3 of global consumers prefer to purchase products and services from businesses that 'give back to society' and 55% are willing to pay more for products and services from such companies (interestingly, this number grows to 63% or more in Latin America, the Middle East and the Asia-Pacific region).[485]
- When consumers were asked to name the most important trait a company could have, the answer "kindness and empathy" rose by 391% between 2008 and 2012.[486]
- Between 2005 and 2011, the brand attributes 'mysterious', 'confident', 'sensuous', 'trendy', and 'glamourous' became less important to consumers, while the attributes of 'kind', 'high quality', 'friendly', 'socially responsible', and 'leader' became more important.[487]

---

478 (Ref: Dan Pink talk)
479 (Ref: Buying into Social Change)
480 (Ref: OECD, Promoting Sustainable Consumption)
481 (Ref: Edelman's Trust Barometer 2009)
482 (Ref: The Shado State report)
483 (Ref: Cone/Echo Global CR Opportunities study 2011)
484 (Ref: Accenture report)
485 (Ref: Nielson, Global Socially Conscious Consumer; Nielson, Doing Well by Doing Good)
486 (Ref: The Rise of Mindful Consumption)
487 (Ref: Chip Walker talk)

The Millennial Generation[xli] is part of the driving force behind this transition. The Brookings Institution found that the Millennial Generation is composed of people "who collectively favor companies that embrace the values of good citizenship". In the U.S., Millennials vest overwhelming levels of trust and loyalty in companies that support solutions to social issues and are 89% more likely to buy from those companies.[488] Another study found that 75% of Millennials around the world are willing to pay more for sustainability-oriented products and services.[489]

It is glaringly obvious that 21st century consumers are craving a whole new baseline for the businesses that provide their goods and services.

## Shifting Workforce

Our needs, desires and expectations are also shifting when it comes to work. Decisions of what kind of work to do and for whom are moving in the direction of having more purpose, more self-direction, and more mastery.[490]

This is especially evident in the Millennial Generation who are, more than previous generations, drawn to jobs that allow them to feel that they are contributing to something bigger than themselves, something positive for the larger community.[491] This makes sense because a large proportion of these young people have grown up painfully aware of the environmental, social and economic crises that the world faces, and often don't see any advantage in following the same sort of life path as their parents.

A study by the Brookings Institution called 'How Millennials Could Upend Wall Street and Corporate America' found that **sixty four percent of millennials would rather make $40,000 a year at a job they love than $100,000 a year at a job they think is boring**.[492] Even very for-profit-minded magazines like *Forbes* and *Fortune* are catching onto this, with article titles like 'Millennials Work for Purpose, Not Paycheck' and '3 things Millennials want in a career (hint: it's not more money).'[493]

And this is not just limited to American Millennials. A 2009 survey of 100,000 people in 34 countries all over the world found that 88% of respondents are more likely to want to work for a company that is considered ethically responsible.[494] The survey concluded that social responsibility is key to attracting top talent.

In addition to the desire for more purposeful work, there is a growing interest in more democratic and decentralized workplaces. In general, workers flourish when they have more self-direction and they can participate in decision-making at the job.[495] As such, it seems that networks are to the twenty-first century, what hierarchy was to previous centuries. The Internet and digital technology like computers, tablets and smartphones have enabled people to work together in all kinds of new ways. This trend manifests in interlinked blogs, wikis, social media, creative commons licensing, online sharing networks, peer-to-peer lending,

488 (Refs: Brookings article about Brookings report)
489 (Ref: Nielsen Report 2015)
490 (Ref: Dan Pink, Drive; Fritjof Bergmann)
491 (Ref: The Deloitte Millennial Survey, Brookings Institution)
492 (Ref: Brookings article about Brookings report)
493 (Refs: see hyperlinked text).
494 (Ref: Kelly Services Survey 2009)
495 (Ref: Communicating for Results: A Guide for Business and the Professions; Workers Participation in the Decision-Making Process)

citizens' journalism, freely available videos and presentations, 'good news' magazines, and big data. Complementing the networks enabled by digital technology are physical spaces that allow people to do valuable work together, such as co-working spaces and makerspaces, where people who usually work alone can share a physical space as well as equipment. This enables them to share resources, cut costs and do their work in an environment that can help them develop their ideas and strategies through interactions with others. As self-organized, self-managed networks become more common, the whole economy starts to take a different shape.

## Transformation of Business

The business world has also been co-evolving with the changes in our shared story. For example, many mainstream companies have seen an advantage in moving into the 'ethical', 'social' and 'green' business spheres in response to more conscientious consumers and, in some cases, as a result of a generation of genuinely concerned business leaders.

Today's global market includes a wide range of business responses. Some companies distribute a portion of profits to charities. Ben & Jerry's ice cream company donates 7.5% of pretax profits to its charitable foundation[496] and Clif Bar, a company that makes energy-dense snacks, is proud to tell its customers about its commitment to environmental and social causes and that it gives 1% of its revenue to.[497]

There are also examples of companies that are legally registered as for-profit and have private owners, but they voluntarily give 100% of their profits to a sister charity or foundation. Examples include the Belu bottled water company in the U.K., Quartiermeister beer brewery in Germany, and Dick Smith Foods in Australia.[498]

This is part of a much larger trend in business leadership and investment. One global survey found 77% of business leaders agree that "the greatest innovations of the 21st century will be those that have helped to address human needs more than those that have created the most profit."[499] And investors increasingly see a connection between community wellbeing and corporate wellbeing, which is why social and sustainable investing has been on the rise since the mid-1990s.[500]

The early 2000s saw the emergence of new triple bottom line business models designed to incorporate social responsibility and eco-friendliness into companies' missions. These include many social enterprises, social cooperatives, and community-oriented companies. Social cooperatives in Europe have a social mission and are limited in the percentage of profits that can be privately distributed (between 20% and 35%).[501] Similarly, community contribution companies in Canada, also known as C3 companies, have a 40% cap on distribution,[502] and community interest companies limited by shares, in the United Kingdom, can privately distribute up to 30% of their profits.[503]

---

496 (Ref: Ben & Jerry's website)
497 charity (Ref: Values-Centred Entrepreneurs, p.120)
498 (Refs: Belu website; Quartiermeister; Dick Smith Foods)
499 (Ref: Tony Wagner, Creating Innovators)
500 (Ref: Do investors care about sustainability)
501 (Refs: Social Cooperatives in Greece; Social Co-operatives in Italy))
502 (Ref: Centre for Social Enterprise website)

Indeed, the business landscape is changing rapidly. Research in the early 2000s found that 1 in 3 of all businesses in development in the United Kingdom aimed to be social enterprises, 1 in 3 entrepreneurs had primarily social motives, and 1 in 3 MBA students[xlii] did not see profit as a top priority.[504] In 2011, almost 60% of all new higher education and training opportunities globally, concentrated on 'green careers.'[505]

This phenomenon is truly international. A new focus on social entrepreneurship is taking hold in Asia, Africa and Latin America.[506] Business schools in India are increasingly incorporating social goals into the curriculum[507] and younger generations in China are entering the world of social enterprise at a growing rate because they believe that there should be innovative approaches to solving social problems and are eager to act on that belief.[508] Echoing Green, a global social enterprise incubator, reports that about one-third of the applications that it receives come from Africa; after the U.S. and India, the three top countries for applications are Kenya, Uganda and Nigeria.[509] And social cooperatives in Latin America are helping more communities meet their needs for employment, social services and environmental protection.[510]

Much of this is clearly driven by the younger generations' desire for more purposeful work and their urge to use business as a means of addressing global issues, however more than 12 million Americans aged 44-70 also want to start a business or nonprofit that has a positive social impact.[511]

This shift includes the rise of certification schemes, as evidenced by the rapid emergence of schemes such as the Fairtrade certification for fairly traded products; the Marine Stewardship Council certification for sustainable seafood; the Forest Stewardship Council certification for sustainable wood and paper products; the plethora of national organic certifications for organic agriculture; and the myriad sustainable business certifications all over the world. The B Corporation is one of the most well-known certification schemes for socially-oriented for-profit businesses. There are currently over 1600 certified B Corps in 47 different countries[512] and this number has grown very quickly since 2011, when there were just 500.[513]

More recently, new ways of measuring business success that relate even more directly to a company's social impact and ecological footprint have emerged, such as The Balance Sheet for the Common Good. This means of evaluating business goes well beyond a singular focus on profit and that's why it's been growing in popularity since it was first developed in 2011.[514] Nearly 2000 companies are now measuring their progress with this balance sheet, which assesses businesses across 17 indicators, including human dignity, cooperation and solidarity, ecological sustainability, social justice, and transparency.[515]

---

503 (Ref: CIC Regulator)
504 (Refs: Rebecca Harding's research for Delta Economics *Social Entrepreneurship in the UK 2008*; Global Entrepreneurship Monitor United Kingdom 2004, also by Dr Harding; Dr Rory Ridley-Duff)
505 (Ref: AASHE report 2011)
506 (Ref: The Emergence and Development of Social Enterprise Sectors)
507 (Ref: Social enterprise rises in India)
508 (Ref: The Social Enterprise Emerges in China)
509 (Ref: Forbes article)
510 (Ref: The Emergence and Development of Social Enterprise Sectors)
511 (Ref: Encore Entrepreneurs report)
512 (Ref: B Corps homepage)
513 (Ref: B Corps Our History page)
514 (Ref: 'History' page Balance Sheet for the Common Good)

Aligned with the ecological stewardship embedded in the story of interconnectedness, more businesses are organizing around zero waste principles and using cradle-to-cradle design and closed loop production, which helps them reuse as many resources as possible and avoid creating waste.[516] At eco-industrial parks, such as Kalundborg in Denmark, manufacturing plants, such as a pharmaceutical producer and a plasterboard manufacturer, use each others' waste as inputs to make new products.[517] A growing number of companies, including big names like The Dow Chemical Company and Ford, are also using life cycle analysis to understand and improve the environmental impact of their products.[518]

Companies like Seventh Generation in the U.S., which sells a range of household products from disposable diapers to dishsoap, use the Iroquois Native American concept of planning for seven generations ahead to design their products.[519] A great number of businesses, including the major home improvement supply company B&Q in the U.K., use the concept of bioregionalism – which proposes that we meet most of our needs from local, renewable and waste resources - to guide their social and ecological efforts.[520] Similarly, the Global Footprint Network has helped businesses, investors, credit-rating agencies and risk analysts use the ecological footprint as a tool to measure environmental impact and risks.[521]

In response to the increasing desire for purpose, participation and more equality, we are also witnessing the rapid emergence of cooperatives around the world. More than 1 billion people worldwide are members of cooperatives and it is estimated that the cooperative sector employs over 250 million people.[522]

While the four countries with the most co-op members are the U.S., India, Japan and Iran, countries with smaller populations also have impressive numbers.[523] For instance, about 70% of Quebec's population is a co-op member. In New Zealand co-ops account for 95% of the dairy market, and in Denmark co-ops account for more than 35% of the consumer retail market. Co-ops generate 3% of Uruguay's GDP and 63% of Kenya's population derives its livelihood from cooperatives. [524] All over Africa, cooperatives are on the rise in finance, education, housing, and consumer retail, as well as cottage industries.[525] Ethiopia has 3,400 housing co-ops and Egypt has over 4,300 consumer co-ops.[526] Credit unions in the U.S. are currently growing at two times the rate of population growth.[527]

There is also the resurgence of worker co-ops, which is perhaps the clearest way we see the business world changing in response to the growing desire for democracy and flexibility at

515 (Ref: Balance Sheet for the Common Good)
516 (Ref: Closed Loop, Cradle to Cradle, Circular Economy, and the New Naturephilia)
517 (Ref: IISD)
518 (Ref: Life Cycle Management)
519 (Ref: Seventh Generation About page)
520 (Ref: Bioregional.com)
521 Ref: GFP Footprint for Finance
522 (Ref: ICA 'Co-operative facts and figures)
523 (Ref: ICA 'Facts and Figures')
524 (Ref: ICA 'Co-operative facts and figures)
525 (Ref: Coop Africa Working Paper No 1; Cooperatives and the Sustainable Development Goals)
526 (Ref: Cooperatives and the Sustainable Development Goals)
527 there (Ref: CUNA "Facts, Fallacies and Recent Trends")

the workplace. Mondragon in Spain has become known world-wide because it is such a successful worker co-op, inspiring many others to start their own worker-owned cooperative businesses.

There's also a widespread a resurgence of support for local businesses. This is especially evident when it comes to food. The number of Community Supported Agriculture (CSA) farms has also grown rapidly, especially in the U.S. CSAs are arrangements in which consumers buy fresh produce directly from local farmers, which has many different kinds of benefits for both the consumers and the farmers. The first two CSAs there appeared in 1986 and by 1990 the number had grown to about 90. It is estimated that today there are about 6,000, which means there's been exponential growth of nearly 25 percent annually since 1990.[528] Farmers' markets In the U.S. have also experienced a revival. Their numbers have increased five-fold since 1994.[529]

All of these types of initiatives are collaborating and co-creating more often all the time, spreading their influence even further, for example the World Council of Credit Unions, the International Cooperative Alliance, and Fairmarket, an online marketplace created by worker co-operatives in the UK and Germany.[530]

## The Spiral of Business Evolution

It would be a great underestimation to view these trends as just some small, temporary bubble of 'feel-good' business practices. Evidence shows that there's an increasing advantage for local, ethical, green and democratically-run businesses in the market.[531] As more people seek to lead eco-friendly, ethical, purpose-filled lifestyles, the companies that flourish will be those that are mindful of how they carry out their business.[532] In essence, businesses that can show that they treat their employees well, obtain their materials from ethical sources, and use their profits for social good will have stronger advantages.

In the 1990s, as the ecological crises and global inequality started becoming clearer to people, corporate social responsibility (CSR) and socially responsible investing emerged to deal with concerns that traditional ways of doing business were not addressing.[533] These trends gained a lot of traction and many companies with CSR strategies outperformed their peers who were not participating in the CSR movement.[534]

Today, social enterprises in the U.K. are outperforming regular businesses in terms of growth and impact.[535] During the 2009 economic crisis in the UK, social ventures increased by 50%, proving more successful than small business start-ups.[536]

In Canada, co-ops are creating jobs at nearly five times the rate of the overall economy and are generating 11% more income for their employees.[537] Similarly, in India's dairy industry,

528 (Ref: Community Supported Agriculture, M. Ernst)
529 (Ref: Bridging the Gap)
530 (Ref: Co-operative news, ICA website, World Council of Credit Unions)
531 (Ref: Nielsen Global Survey; The Business Case for Purpose; Ethical Consumer Market Reports; Financial Times article; What do we really know about worker cooperatives)
532 (Ref: Good Must Grow survey)
533 (Ref: Embedded Sustainability; Laszlo Evolution of Business ; Hawkins, Lovins & Lovins)
534 (Ref: Laszlo; Natrass & Altomare, *The Natural Step for Business: Wealth, ecology, and the evolutionary corporation*; Hawkins, Lovins & Lovins)
535 (Ref: State of Social Enterprise Survey)
536 (Ref: Best, 2010 – ask Donnie)

research shows that cooperative members enjoy higher and more secure incomes than non-members.[538] And in Ethiopia, agricultural producer cooperatives have better incomes, more savings and reduced business costs, compared to non-cooperative producers.[539] In fact, it's been found that worker-owned co-ops are simply more productive than conventional companies.[540]

In the global financial sector,, social banks, including microfinance, development banks, credit unions, consumer coops, mutual banks and B Corp banks, achieved a growth in net income of 15.7% between 2007 and 2012, while conventional banks suffered a net income decrease.[541]

Sustained economic success isn't as simple as capturing market share, though. Attracting talented employees who add value to the company plays a key role. Studies have found that workplaces that provide a high level of engagement for employees have an important advantage in harnessing the talents and energy of their employees.[542] This gives most co-ops an automatic advantage. John Lewis Cooperative in the U.K. is an excellent example of a competitive worker co-op. It has an inter-equity model for salaries that ensures that the CEO will never make more than 70 times as much as the lowest paid employee. Each of the 76,500 permanent staff is a partner in the business. The business enjoys annual revenues of over £8.2 billion and its mission is to keep its employees happy.[543] This concern for employees is why worker co-ops are gaining popularity, and a competitive advantage, very quickly.[544]

The increasing advantages of ethical, social, local, green and democratic businesses might indicate that we are in a time of transition between economic paradigms. Traditional for-profit business is no longer adequate and is causing more problems than it is resolving. Different business models are emerging as part of an effort to better meet the challenges and needs of society in the 21st century. The rapid rise of participation in ethical business, green companies, social enterprise, and co-ops is a sign that consumers, workers, business leaders and policy-makers no longer see profiteering as the business world's only objective.

Where is this evolution of business leading? We hypothesize that businesses that prioritize purpose over profit will increasingly outperform those that don't. Each step in this evolution is a step away from the for-profit ethic in business and a step towards the *not*-for-profit ethic. Each step is a step away from the profit-motive and a step towards the purpose-motive; a step away from maximizing private self-interest and a step towards maximizing collective wellbeing. For-profit companies embodying more of an NFP ethic are already outcompeting traditional companies[545] (Forbes magazine is even advising for-profits to operate more like NFPs[546]). So it is not a very big stretch of the imagination to postulate that NFP enterprise, as a business model, might very well outperform other business models in this time of metamorphosis and usher in a not-for-profit economic era.

**We predict a shift in the primary business model of the economy from the for-profit firm to the NFP firm in the 21st century.** We see this as the next natural step in the

---

537 (Ref: Cooperative Difference)
538 (Ref: Cooperatives and the Sustainable Development Goals)
539 (Ref: Ibid)
540 (Ref: The Nation article)
541 Ref: Social Banks and Their Profitability
542 Ref: Gallup 2013, State of the Global Workplace, p. 4
543 (Ref: BBC article)
544 (Ref: Marjorie Kelly, *Owning Our Future*; Gar Alperovitz; It Takes an Ecosystem)
545 (Ref: Areal & Carvalho, Bloomberg Business article, The World's Most Ethical Companies)
546 (Ref: Forbes article)

evolution of the economy that is occurring through market dynamics, based on social and cultural changes in response to changing economic and environmental conditions. We call this the Spiral of Business Evolution[xliii].

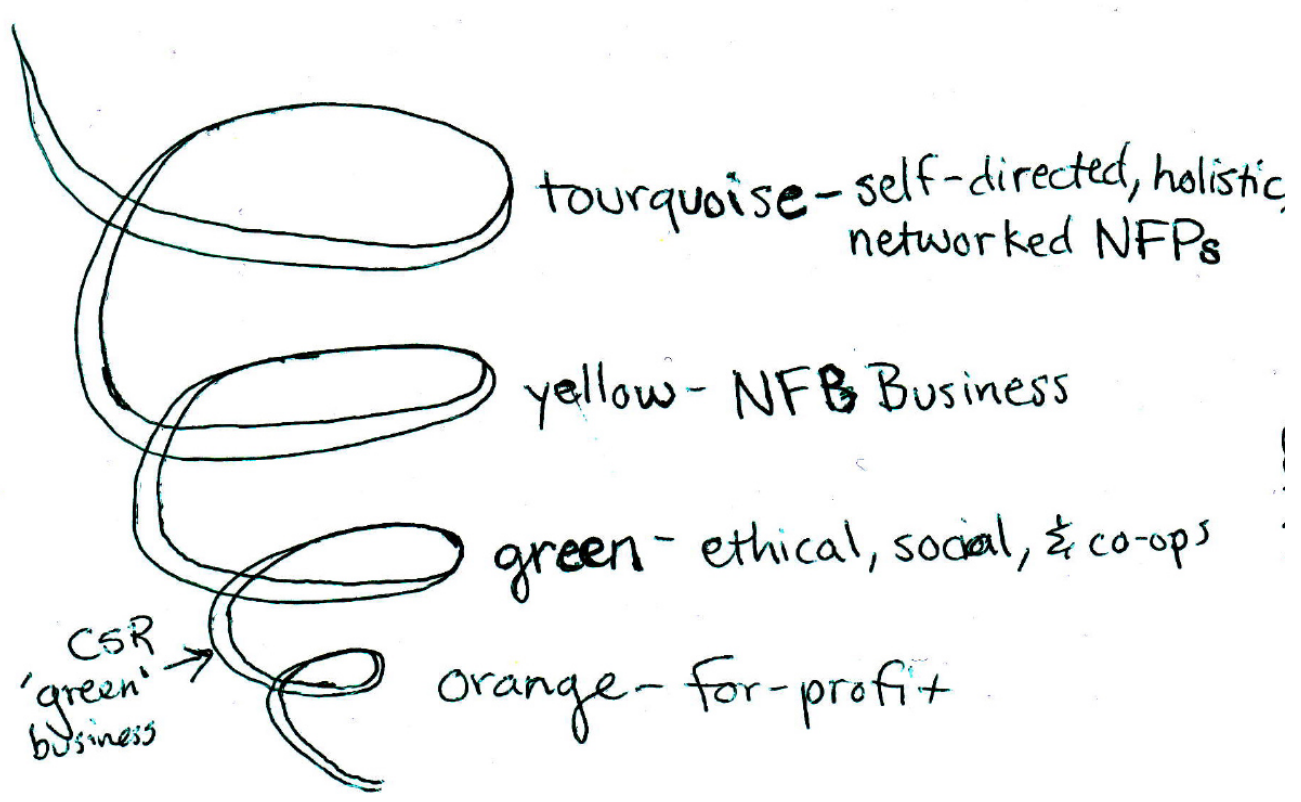

### Spiral of Business Evolution

This idea suggests that different business models have been dominant at different points in history because they are deemed the most appropriate for meeting society's needs. But as society's needs and challenges change, they might become obsolete or even destructive. When this happens, they lose support and more relevant business models, which adapt the best aspects of the previous business models in order to more effectively meet the new needs, gain support.

Business model preference rested with the for-profit firm up until the 2000s largely because capital-raising mattered so much in the pre-Internet era. Companies that held proprietary knowledge and were organized in hierarchical, centralized ways had the upper hand. It was an era in which externalizing social and environmental costs was either unknown or just accepted as a necessary evil; and where worker rights were commonly traded for the promise of a higher salary or financial returns in the future. But we live in a *very* different world now. Negative social and environmental impacts of business are becoming less and less tolerated. And the digital era has lowered the costs of starting up and running a business, allowing for new ways of offering goods and services, marketing them, and raising capital. In other words, the for-profit firm no longer has the upper hand.

Up to now, most new business models, like social business, social enterprise and the community interest company, have placed little or no importance on distinguishing whether companies are for-profit or NFP. An increasing awareness of this important distinction is the next step; the next turn in the spiral. In addition to being socially responsible, environmentally-friendly, democratically-run and locally-based, another major advantage exists for a business to have in the twenty-first century market: being not-for-profit.

It's easy to take for granted that the global economy will always revolve around for-profit business, as that's how it has been throughout our lifetime, and because NFP business is so nascent. But it's also important to remember that it has only been that way for about 200 years, which is not a long time in the larger scheme of things. The economy has never stopped evolving and now the crises of the for-profit, capitalist economy are birthing a new post-capitalist era. We are currently seeing signs that the NFP firm is starting to outperform the for-profit firm and many NFP enterprises already regularly out-compete for-profit counterparts.[547]

---

[547] (Ref: McLeod Grant and Crutchfield, 2007, pp. 32-41)

The nonprofit sector in the U.S. has generated more employment and higher wages in recent years, while the for-profit sector has been shedding jobs and decreasing wages.[548] During the recession, from 2007- 2009, NFP employment grew in 45 of 46 U.S. states, while for-profit employment declined in 45.[549] From 2000 to 2010, nonprofit revenue in the U.S. grew at a rate of 41%[xliv], which more than doubled the growth of GDP during that same period.[550] At roughly the same time, the growth of nonprofits (including NFP businesses) greatly outpaced the growth of for-profit businesses, at 25% compared to 0.5%.[551] Globally, from the late 1990s to the mid-2000s, the nonprofit sector's contribution to GDP grew at an average rate of nearly 6% per year.[552]

In the 16 countries that have registered national accounts of their nonprofit sectors with the UN[xlv], which are located in all regions of the world, the following was found:

- the nonprofit sector employs 7.4% of the workforce on average;
- in 6 of the countries, nonprofits account for 10% or more of the workforce;
- nonprofits account for 4.5% of GDP on average;
- the nonprofit sector is growing as a proportion of the global economy[xlvi]; and
- fees and charges, not philanthropy are the major source of income for nonprofits.[553]

In Kyrgyzstan, 85% of the nonprofit sector's income comes from fees for service and that number is 67% in New Zealand and 59% in Japan.[554] Importantly, employee compensation makes up a much higher percentage of nonprofits' GDP contribution compared to compensation in the for-profit sector.[555] This is part of how the nonprofit sector keeps the wealth circulation pump going: rather than a large percentage of an organization's revenue going to owners and investors who put most of it into the elite economy, the money goes to employees who are much more likely to spend it back into the real economy.

The NFP advantage is even clearer when one looks at data about specific segments of the economy, such as community-owned businesses that are outperforming their peers. Community cooperative expert in the UK, Peter Couchman says "In a climate that has seen commercial village shops close at a rate of around 400 per year, and the stalling growth for the major retailers, community-owned shops are reporting not only an increase in sales, but are continuing to open in a challenging climate… Community-owned shops succeed where commercial ventures have failed because they engage with the whole community. When the owners are the customers, the business can directly respond to consumers' needs in a way that larger retailers just aren't able to."[556] In fact, a study of community-owned shops in the United Kingdom in 2011 found that they have a 95% survival rate, more than double the average business survival rate of 46.8% in the UK.[557]

Credit unions in the U.S. provide another powerful example. A survey of the top ten banks (by total deposit size) and the top ten credit unions (by total membership) found that the credit unions outperform their for-profit counterparts in many different ways. The credit unions

548 (Ref: Johns Hopkins CCSS, Holding the Fort)
549 (Ref: Ibid)
550 (Ref: Rifkin, The Rise of Anti-Capitalism)
551 (Ref: New York Times article )
552 Ref: The State of Global Civil Society, p. 11
553 (Ref: The State of Global Civil Society)
554 (Ref: The State of Global Civil Society, p. 8)
555 (Ref: The State of Global Civil Society, p. 8)
556 (Ref: Community-owned shops see massive sales growth)
557 (Ref: Community-owned shops see massive sales growth)

scored significantly better than the for-profit banks in terms of offering: free checking accounts, discount student accounts, free or discounted senior checking accounts, free business checking accounts, saving rates above the inflation rate, free ATMs, and more ATM machines.[558] Studies show that credit unions are taking a bigger market share of mortgage and car loans, specifically.[559] And while trust levels in U.S. banks at the end of 2013 were at 35%, credit unions received trust ratings of 58%.[560] Perhaps that is why the Bank Transfer Day initiative has been able to persuade about 400,000 people in the U.S. to move their money from big banks to credit unions.[561]

Of course, NFPs are outperforming their for-profit counterparts sooner and faster in some sectors than others. Because NFPs have greater advantages in social sectors, like education and healthcare, NFPs are expanding their market share in these sectors first.[562] For instance, they are taking up an ever-larger share of the market in health care, education, social assistance, arts and entertainment.[563] In the U.S., NFP hospitals are even more profitable than for-profit equivalents.[564] Not-for-profit leisure centers in the U.K. are significantly outperforming for-profit health and fitness chains.[565] And mutual and cooperative insurers have been significantly outperforming the rest of the market on the global stage since the financial crisis began.[566]

However NFP enterprises are also emerging in more diverse industries, where they haven't traditionally had any market share, like manufacturing, software and natural resource management. This is because **the NFP way of doing business actually has significant advantages in the market of the 21st century in terms of business ethics, finances, employment, and innovation**[xlvii].

# Not-for-Profit Business Advantages

Not-for-profit companies all over the world are increasingly recognizing their advantageous position as NFPs. Take HealthPartners, the largest consumer-governed NFP healthcare provider in the U.S. In 2011, its revenue was just under $3.9 billion. At the White House Community Leaders Briefing, they touted the advantages of being NFP, saying, "Because of its non-profit status and cooperative roots, HealthPartners has a unique governance structure that allows it to re-invest in care and services that will improve members' health. The structure also helps to generate better ideas, and allows HealthPartners to be more agile than other organizations. Rather than answering to shareholders, HealthPartner's board is governed by members – meaning, for example, that during the recent economic downturn the organization could remain focused on high quality affordable health care and an exceptional experience, rather than profit margins. In fact, its goal is to maintain a two percent profit

---

558 (Ref: Nerdwallet article; U.S. News article)
559 (Ref: Market Watch article)
560 (Ref: CUNA slides)
561 (Ref: New York Times article)
562 Ref: John's Hopkins, Urban Institute, Impact of Economic Downturns on NFPs?
563 (Ref: Ibid)
564 (Ref: Sesana, L. (2014) 'Why Nonprofits are the Most Profitable Hospitals in the US', *Arbiter News*, 8 April. [http://www.arbiternews.com/2014/04/08/why-nonprofits-are-the-most-profitable-hospitals-in-the-us/])
565 (Ref: SE100 (2014) 'Sector focus: Leisure, sports, arts and culture', *The RBS SE100 Quarterly Data Report*, Quarter 2. UK. [https://se100.net/data/report-july-2014])
566 (Ref: ICMIF (2014) 'Global mutual and cooperative market infographic 2014', International Cooperative and Mutual Insurance Federation. [http://www.icmif.org/global-mutual-and-cooperative-market-infographic-2014])

margin to cover reinvestment in facilities, programs and services. Most publicly traded organizations, in contrast, aim for profit margins of 15-20 percent."[567]

## Ethical Business

Not-for-profit businesses are in an enviable position in the more ethical, eco-friendly, socially-responsible market place of the 21st century. By law, they are mission-based and invest 100% of their profits back into their mission. Because of this, they are more likely to care about whether or not their supply chains are ethical, about their environmental impact, and about their employees' wellbeing. They are more likely to take into consideration the concerns of the local communities in which they operate and to give their employees the chance to express themselves. And they are not at all likely to sacrifice any of these concerns in the name of profit-maximization, like for-profit businesses are. For-profit companies that are trying to incorporate ethical business practices can easily drift from a social mission due to the pressure to maximize profits or generate financial value for owners and investors. **It's not that for-profit companies can't do work that has a deeper purpose; rather it's that the profit motive often distracts them from that deeper purpose.** NFPs don't have that built-in distraction. That's why, in many parts of the world, people tend to trust NFPs over for-profit organizational forms and the public generally has a high level of confidence in NFPs[xlviii].[568]

CBHS is an insurance fund in Australia that clearly sees its NFP status as an advantage in the market. On their website, they say, "In not-for-profit organisations, resources are geared toward goals other than the bottom line. In CBHS, the central goal is to service our members and improve members' satisfaction and offer greater benefits and thus value in the context of health insurance."[569]

Similarly, Nationwide Insurance Company used its lack of private owners to appeal to customers in a 2013 commercial that featured the line, "Just another way we put members first, because we don't have shareholders." Marketers at that Fortune 500 not-for-profit company clearly believe that 'no shareholders' is a strong selling point[xlix].

Food Connect in Brisbane also recognizes this. Their website proudly proclaims:

"… All our profits go into doing more good. We differ from conventional, profit-driven businesses, in the following ways:

- we work to bring economic, social and environmental benefits to society
- we reinvest all of our trading surplus into fulfilling our mission

Our customers come to us because they want to support a local, connected, embedded, engaged and wealth distributive (not wealth concentrating) food system."[570]

This ethical advantage clearly translates into a financial advantage as well. As Rogue Credit Union, in Oregon says, "In the context of a national economy still reeling from several years of downturns and setbacks, our financial performance was overwhelmingly positive, due to the loyalty and commitment of our members."[571]

---

567 (Ref: NCBA White House Community Leaders Briefing)
568 (Ref: Handy et al.; University of San Diego, citing Rosenthal and O'Neill)
569 (Ref: CBHS website, 'Not-for-Profit: How We Help You for Less)
570 (Ref: Food Connect website)
571 (Ref: Rogue Credit Union Annual Report 2014)

In Japan, the Seiketsu Club Cooperative (a consumer co-op) also sees value in spending more in order to provide local organic produce, ethical products and recycling services.[572] They have also invested in measuring their environmental impact and conducting life-cycle assessments of their products and services, in order to learn how they can be more environmentally-friendly.[573]

American consumer food cooperatives are outperforming for-profit grocery stores when it comes to a wide array of environmental and social indicators. They work with more local farmers and producers, carry three times as many locally-sourced products, donate more than three times as much income to charities, sell twenty-four times as many organic groceries, spend six percent more of their revenue on local wages and benefits, have significantly higher rates of recycling, and are more energy-efficient than their for-profit peers.[574]

The German electronics company, Nager IT, provides another excellent example. They are in the business of trying to create ethical electronics; the electronics equivalent of fairtrade coffee. In order to do this, they disclose their entire supply-chain on their website,[575] where you can see which parts of the supply-chain have passed their ethical standards and those that are still ethically problematic or unknown. As an NFP company that exists to create ethical products, it is natural for them to maintain this level of transparency, and their customers expect no less. With no mandate to maximize profit, they can focus on their mission. They even encourage people who look at their website not to buy the products if they're already happy with what they've got, as they don't want to promote unsustainable consumption.[576]

Similarly, Eco-Home Centre, a home improvement supply store in Cardiff, says, "As a not-for-profit organisation there is less pressure on sales and more on providing great advice and getting you the correct product."[577]

Not only is NFP business not fundamentally at odds with environmental concerns, but the NFP ethic also allows people to move on big issues like climate change with less skepticism of companies trying to make a quick buck. One of the reasons there are so many climate change deniers is because so many 'green ventures' and investors really have profited in a big way from carbon trading and carbon mitigation schemes while greenhouse gas emissions continue to rise. Some of them had a positive impact, some did not. In any case, whenever private profit is involved, it's not long before suspicion of profiteering sets in. And for good reason: the profit motive is at the helm. Even for-profits with the best of intentions will face this scenario. Financially independent NFPs are much more able to avoid that suspicion, due to their explicit social mission, the fact that no one can privately profit from them, their high levels of accountability and transparency, and the fact that they're earning their own revenue – they're not just an extension of for-profit sponsors.

Nebraska is the only state in the U.S. where all electricity is delivered by community-owned NFP institutions and so the residents of Nebraska have a direct say in their electricity infrastructure. In 2003, they voted for the development of 200 MW of wind energy by 2010 (96% of vote participants supported the project).[578] It's hard to imagine a for-profit renewable energy project garnering that much support from the public.

---

572 (Ref: Japan: Land of Cooperatives)
573 (Ref: Lewis and Conaty's 'Resilience Imperative' (ask Donnie if you can't find it))
574 (Ref: National Coop Grocers (2012) *Healthy Foods, Healthy Communities: Measuring the Social and Economic Impact of Food Co-ops.* Iowa City, IA: NCG)
575 (Ref: Nager's supply-chain)
576 (Ref: Nager's website)
577 (Ref: http://www.monmouthshiregreenweb.co.uk/rde-ltd-eco-home-centre)

A great number of NFP businesses in the retail sector are even based on repairing and refurbishing products, rather than selling new ones, in order to reuse materials and avoid waste. Pedal Revolution in San Francisco, for instance, is a full-service bicycle shop that reuses, repairs and refurbishes bikes, in addition to selling new ones.[579] The London Re-Use Network refurbishes and sells used furniture.[580] Goodwill Industries International employs people with disadvantages in the job market, to repair, refurbish and sell second-hand goods that are still in good working condition.[581] And there are lots of other NFP second-hand stores all over the world, keeping perfectly good products in circulation, saving people money and lowering demand for new products which require raw materials and energy-intensive manufacturing processes. Not to mention that a growing number of NFP enterprises, like Ocean Cleanup, have started up with the explicit mission of protecting and restoring the natural environment, precisely because the planet's ecosystems are deteriorating so fast.

Of course, not all NFPs are perfectly ethical companies that help protect the environment. There's a big spectrum in the realm of 'not-for-profit'. However, an NFP is much more likely to act in alignment with higher ethical principles than an equivalent for-profit business, for the simple reason that they exist for a social purpose and nobody can financially gain from their activities.[582]

Not-for-profit businesses have less pressure to prioritize quantity over quality, so the latter often prevails. A study of nursing homes in Minnesota revealed that NFP nursing homes do as well or better on quality measures than for-profit homes. According to resident surveys, not-for-profits had higher levels of food enjoyment, sense of safety, sense of adaptation to the living environment, general satisfaction and satisfaction with relationships than their for-profit counterparts. They also had lower levels of regulatory deficiency citations, and administration of antipsychotic drugs that were not prescribed.[583] There is a constant impetus with for-profits to try to maximize the quantity of patients and money coming in, even if that means sacrificing the quality of care. And, while for-profit businesses spend billions dollars on advertising to show how much they care about their customers, connecting with customers is often just part of fulfilling an NFP's mission.

In addition to NFPs holding higher ethical standards for themselves, the public expects more ethical behavior from the NFP business community. In fact, we usually *only* trust NFPs and governmental agencies (which are also NFP) when it comes to the commons and public goods, because we don't trust the profit-motive to ensure basic public welfare. This is demonstrated every time there is uproar over the privatization of water. In the past decade, 180 cities and communities in 35 countries, including Buenos Aires, Johannesburg, Paris, Accra, Berlin, La Paz, Maputo and Kuala Lumpur, have all 're-municipalized' their water systems after water privatization resulted in higher prices, environmental hazards and a lack of investment in infrastructure.[584]

578 (Ref: Nebraska's Community-Owned Electricity System)
579 (Ref: Pedal Revolution website and emails)
580 (Ref: London Re-Use Network)
581 (Ref: Goodwill Industries International)
582 (Ref: Agarwal & Malloy; Horwitz)
583 (Ref: Ben-Ner and Ren)
584 (Ref: Guardian article)

Not-for-profit entities are also often favored over for-profits by local government in the contracting process. This greater trust means that they can usually get more contracts, and contracts of longer duration than for-profit competitors.[585] For example, some research shows that in the UK, for-profit social enterprises have found it more difficult to gain public sector contracts and certain grant funding than their NFP peers.[586] Part of the reason governments favor NFPs over for-profits is because NFPs can often offer better prices,[587] but it's also due to an ethical alignment. It makes sense that citizens would be happier knowing that tax payer money won't go to lining the pockets of for-profit owners and investors.

For-profit companies in social sectors are increasingly coming under fire, such as for-profit universities in the U.S that are benefitting from federal student funding. On average, for-profit colleges in the U.S. receive 72% of their funding from federal programs, which means that public money is likely enriching the private owners of these universities.[588] An example of the trust issues involved with for-profits in education and public funds is the recent revelation involving the U.S. military. After the Department of Defense found that the for-profit University of Phoenix was using misleading advertising tactics, including the misuse of military symbols, it declared that it would no longer allow service members to use public defense funding to attend classes at the for-profit institution. This case shed light on the fact that the university, up to October 2015, had been receiving a larger portion of GI Bill money than any other college in the country, to the tune of $1.2 billion since 2009.[589]

As of 2016, thousands of students have appealed to the U.S. government to forgive their federal student loans because they feel they have been defrauded by for-profit universities.[590] Not-for-profit universities in the U.S. do not get the same kind of bad press for using government funds because it's clear that there are no private owners to benefit from the money, and they have boards to hold them accountable for staying aligned with their mission.

Many Americans are also angry about the Affordable Healthcare Act (also known as 'Obamacare') benefitting for-profit health insurance companies. It's easy to connect the dots: the private owners and investors of these for-profit businesses are profiting from government funds. Furthermore, it's been shown that NFP health insurance companies offer lower prices than for-profit counterparts in the U.S.,[591] which in part is because they don't have to compensate for privately distributed profits. Obamacare might have more supporters if the program only allowed NFP insurance and healthcare providers to receive public funds.

Another clear illustration of the higher ethical expectations to which society holds NFPs is the public outcry about nonprofit CEOs making large salaries. *Bloomberg Business* published an article in 2015 about nonprofit CEOs making more than $1 million per year.[592] The 2014 scandal with the NFL (National Football League in the U.S.) illustrates this even more clearly. People across the country were outraged when it was discovered that the NFL's Commissioner, Roger Goodell, received over $30 million dollars as his salary, pension payment and work incentive payment in 2013, given the organization is a nonprofit trade

585 (Ref: Witesman and Fernandez, Bryce, Salamon)
586 (Ref: Pushing Boundaries study)
587 (Ref: Jang)
588 Ref: Atlantic article
589 (Ref: Business Insider article, The Center for Investigative Reporting)
590 (Ref: Wall Street Journal article)
591 (Ref: Health Pocket article, Obamacare Healthcare Facts)
592 (Ref: Bloomberg Business article)

association and, as such, receives tax exemptions. And this is surely an obscene amount of money for anybody to make in one year, especially at a nonprofit, not least of all because we live in times of such extreme inequality.

However, the critical piece of the story that was missing from most news coverage was that the NFL is not an NFP enterprise. The NFL is a professional association, supported and financed by the 32 football teams that make up the league, 31 of whom are for-profit companies[l]. It's the same with lobbying associations, which are usually nonprofits. Most lobbying associations are created solely to support the interests of for-profit companies. They are not enterprises that seek to be financially self-sufficient in order to achieve a social mission. Such lobbying and professional associations are more accurately seen as instruments of the for-profit world; appendages of the for-profit companies and industries they serve. They are the embodiment of the Nonprofit Enabler phenomenon. It would be impossible for such an obscene salary to be justified in an NFP enterprise.

The NFL case clearly highlights the fact that NFPs are held to much higher standards in terms of having an ethic of enough, paying fair salaries, and being more transparent than their for-profit peers. That this even became a scandal, shocking millions of Americans, is precisely because the NFL is a nonprofit organization. If the NFL was a for-profit company, no one would have batted an eye. Commissioners and CEOs of big for-profit companies routinely make over $30 million per year. That's not shocking; rather it is something we've all gotten used to in the for-profit world, but it's not at all acceptable in the realm of NFPs.

Many NFP managers see this increased scrutiny as a boon rather than a constraint. As Matt Flannery, co-founder of Kiva, an NFP online lending platform says, "People hold nonprofits to a high standard. They scrutinize how you spend every dollar. I'm glad because it makes us stronger."[593]

There is even more confidence in democratically-managed NFPs, like credit unions, which are seen as more transparent than banks because they hold annual meetings that are open to all members. They also issue annual reports and post financial statements every month.[594]

This kind of mission-driven, transparent business is just what conscious consumers are looking for. **We want an economy with more integrity and NFP business is better designed to deliver it.**

## Purpose-Driven Workforce

NFP enterprises are in a better position than for-profit counterparts to fulfill our changing needs and desires not only as customers, but also as participants in the workforce. In the past, big for-profit companies were often seen as able to outperform smaller ones in part because they could hire the best talent, but with changes in what people are seeking from their work, NFPs are becoming competitive in terms of talent acquisition, due to a workforce increasingly driven by the purpose motive.

A great example of this is the story of Doug Rauch, former president of the highly successful American grocery store chain, Trader Joe's. After leaving Trader Joe's, Rauch decided to

593 (Ref: SSIR article)
594 (Refs: Bankrate article)

start up the Daily Table, an NFP grocery store in a low-income neighborhood in Massachusetts. One of the most talented business managers in the U.S. went into NFP business, not because he could make more money that way, but because he had the desire to make a positive contribution to society.[595]

And this is not an isolated incident. One study found that workers who voluntarily moved from the for-profit sector to the NFP sector in the U.S. reported higher job satisfaction, even when receiving lower pay.[596] Another study in 2015 found that 42 million workers in America see their work primarily as a source of personal fulfillment and as a way to help others.[597] It's no surprise that most of these purpose-driven people work in the NFP sector[li].[598] Nor is it surprising that the study also found these purpose-driven workers have higher levels of wellbeing and are generally more reliable and effective at work than their peers.[599] This means there's a double advantage for NFPs: they more easily attract a growing number of purpose-driven workers, and the workers they attract are more likely to add value to the company.

In the absence of private ownership, NFP businesses are also more able to operate with horizontal, networked and participatory management structures and to work in more collaborative ways. Aside from the exception of for-profit cooperatives, the for-profit firm typically necessitates hierarchy. The profit beneficiaries (shareholders, investors, and owners) are on top; below them is the the CEO or president, who's charged with maximizing profits and shareholder value; below that are the managers and employees that must carry out the CEO's strategy. This is inherent in almost all for-profit business structures, not just publicly traded companies, because they exist for private profit. A profit-driven business where just a few people benefit from the profit means it logically follows that those few people have the most power and influence when it comes to decision-making. Even for-profit companies that try to take on more participatory management structures will face hierarchy when it comes to financial decisions, because there is a select group of profit-motivated individuals that have more power in such decisions.

Zappos, an online retailer owned by Amazon, made headlines when it started using the Holacracy management approach, which allows workers to essentially manage the business. But Brian Robertson, the inventor of Holacracy, says that Holacracy provides a set of core principles for shaping your business processes over time but it doesn't account for financial control or budgeting processes.[600] When it comes to difficult decisions, it's likely that Zappos' owners at Amazon will have the final say.

**Not-for-profit businesses, on the other hand, can much more easily be structured in a flatter, more participatory way**, as there are no private owners and no one is expecting a portion of the profits. They especially have advantages when it comes to financial decision-making.

It's no surprise that NFP cooperatives have led the way on this front and continue to be some of the most participation-oriented businesses in the NFP sphere. However, other kinds of NFP businesses are increasingly fostering a participatory work culture where employees are able to take part in decision-making processes.[601] The Sustainable Economies Law Center created a freely available presentation on Worker Self-Directed Nonprofits and how they use sociocracy circles as a tool for workers to manage the center. There are no bosses and, of course, no owners to worry about when making financial decisions. All of the employees truly co-manage the organization.[602] Enspiral is another NFP business that is worker-managed. In 2014, Enspiral created the online tool, Loomio, in order to facilitate discussions and decision-making. Now groups all over the world are using Loomio to make decisions more democratically.[603] Even the NFP enterprise with the largest number of employees in the world, BRAC, has a relatively flat structure and seeks to become even flatter in the future.[604]

---

595 (Ref: Time article)
596 (Ref: Becchetti, Castriota & Pedetri)
597 (Ref: (Hurst, A. & Tavis A. (2015) *2015 Workforce Purpose Index: Work orientations of the U.S. workforce and associated predictive indicators of performance and wellbeing*, Imperative: New York)
598 (Ref: Ibid)
599 (Ref: Ibid)
600 (Ref: Business Insider article)

More democratic decision-making means that women are able to reach higher positions in the company. Nearly half of the members of the Spanish Confederation of Worker Cooperatives are women, 39% of whom have directorial positions in their co-ops, compared to 6% in non-worker managed enterprises.[605] And in Ireland, women's representation on NFP boards is 52% compared to 7% in the for-profit boards.[606] Likewise, 31% of NFP boards in the UK are women, while this number is only 14% in the for-profit corporate sector.[607] This makes sense in the historical context of the nonprofit sector, which has been a traditional stronghold for women's employment, so NFP enterprise is emerging from a different understanding of the power of women in business. That is part of the reason women continue to disproportionately favor work in the NFP sector over the for-profit sector.[608]

The purpose-based nature of NFPs as well as the increasing prevalence of participatory decision-making in NFPs means that there is a higher likelihood that NFP businesses will have more wage equality.[609] This is because there's internal pressure in NFPs to keep salaries within a healthy range and a lower chance of employees allowing big salary gaps in companies that operate more democratically. Greater disclosure of executive incomes in NFPs compared to for-profit companies also contributes to greater wage equality. A study of NFP, for-profit, and government hospitals in the U.S. confirms this, finding that NFPs pay significantly higher wages to most paid positions.[610] In other words, NFPs' median salary is more, but because they pay less at the executive level, their overall wage costs are lower. This creates more pay equality and keeps the NFP business lean, because it's not losing money on inflated management salaries and it doesn't need to pay exorbitantly high salaries because NFP managers are more purpose-driven.

More equal pay in turn helps foster collaboration and prevents the internal competition and resentment to which highly unequal pay can lead. More purpose at the workplace along with better working conditions, and more stable, equal wages naturally leads to lower levels of workplace stress and burnout. It also means lower levels of unwanted staff turnover. Purpose-oriented employees tend to stay at the job for a longer period of time.[611] One study, for instance, shows that average executive turnover at American nursing homes is 7-11% lower at NFP facilities compared to for-profit facilities.[612]

---

601 (Ref: The Participatory Revolution in Nonprofit Management)
602 (Ref: What is a Worker Self-Directed Nonprofit)
603 (Ref: Enspiral talk)
604 (Ref: Managing for Change, p. 125)
605 (Ref: ILO and ICA, 2014)
606 (Ref: Not-for-profits lead the way in women's representation on boards)
607 (Ref: Ibid)
608 (REF : Salamon 2013)
609 (Ref: Leete article)
610 (Ref: James article)
611 (Ref: Hurst & Tavis, 2015)
612 (Ref: Malani and Choi)

However, it's not only the purpose motive, participation and pay equality that attract people to work for not-for-profit enterprises. There's also the opposite force at work; growing resentment and avoidance of for-profit work. Perhaps the most obvious sign of this is the growing resentment towards (and rejection of) exploitative, unpaid internships with for-profit companies.[613] A large number of young people in the U.S., Australia and Europe have gone through one or more unpaid internships, doing unrewarding work for free at a company that distributes profits to owners and shareholders at the end of the year, and perhaps retains intellectual property rights to their ideas, and most of them did not gain a paid position after their internship was over (Ref: Ibid). This is a rising trend. And the people who have experienced this trend first-hand come away feeling used, resentful and angry. And rightfully so.

An increasing number of people are growing tired not only of unpaid internships at for-profit firms, but also of the ever-longer work hours and worsening workplace conditions required by management that is seeking to squeeze profits from ever-declining margins (Ref: Juliet Schor?, Jeremy Rifkin?). As profits decline, companies cut costs by reducing wages, benefits and other 'amenities' that make work more enjoyable for employees. Remember that for-profit wages declined by 1% from 2001- 2010, while NFP wages increased by 29% during that same period (Ref: Urban Institute). The average weekly wage for a worker in the U.S. nonprofit sector was $85 higher than in the for-profit sector in 2012.[614] In addition to having a greater sense of purpose in their work, most NFP employees also tend to make more money than their for-profit counterparts. In an era of increasingly purpose-oriented workers, NFPs are in an ideal position to attract the best workers.

## Financial Advantages

The advantages that NFPs inherently have as mission-focused businesses that attract more socially and environmentally conscious consumers as well as a more purpose-oriented workforce, also lead to financial benefits. Being purpose-based gives NFPs a significant headstart in terms of tax benefits, volunteer labor, not having to worry about dividends, receiving discounts, avoiding financially risky behavior and operating on a leaner basis more generally.

The most obvious financial advantage that NFPs have over for-profits is tax exemptions and benefits. In the U.S. healthcare sector, for instance, tax breaks to NFP hospitals amount to billions of dollars every year.[615] Many policy-makers and economists have argued that tax-exempt NFPs should be legally restricted in the amount and type of business activities they can engage in, because their tax exemptions give them an unfair advantage over for-profit companies in the market.[616]

However, as we mentioned in chapter two, tax exemptions for NFPs are entirely justified and an advantage they deserve for being purpose-oriented, providing goods and services that benefit the community, having no private owners and forgoing the ability to privately distribute profits in order to stay mission-focused.

Another financial advantage is that NFP managers and executives accept being paid less than their for-profit peers, because they get a sense of intrinsic fulfillment from their work, which

613 (Ref: Rallying Cry Against Unpaid Internships, The Growing Culture of Unpaid Internships, Sydney Morning Herald article)
614 (Ref: Morath article)
615 (Ref: Provision of Community Benefits by Tax-Exempt U.S. Hospitals)
616 Ref: Yong, L. and C.B. Weinberg (2004) 'Are Nonprofits Unfair Competitors for Businesses? An Analytical Approach', *Journal of Public Policy and Marketing*, 23(1): 65-79) they value more than making a lot of money.[617] This doesn't mean they are paid a less than adequate salary – just that there is no need to pay them excessively.

In 2013, the median salary for NFP CEOs in the U.S. was just under $120,000,[618] compared to the average CEO salary among the S&P 500 companies of $13.8 million.[619] In 2014, JP Morgan's CEO was compensated to the tune of $27.7 million.[620] In contrast, the Bank of North Dakota spent about half of that ($13.8 million) on *all* employee salaries and benefits, including the president of the bank's salary, that same year.[621] In essence, NFP managers feel rewarded in non-material ways for the work they're doing, so there's no need for their companies to try to motivate them with monetary rewards. Despite the corporate gyms, ringside tickets, teambuilding retreats and sponsored trips to volunteer in the 'developing' world that many for-profits increasingly offer, they just can't compete with NFPs when it comes to the underlying purpose of work their employees do. Being able to pay managers less because they feel a deeper sense of purpose in their work gives NFPs a major financial advantage over for-profits that must spend money on financial incentives to motivate their managers.

Another major financial advantage that NFP enterprises have in the market is that they are able to engage volunteers. Volunteers at NFPs can work full-time,[622] whereas for-profit companies can usually only have part-time volunteers in most countries (if they are even allowed to have volunteers at all).

In the U.S., for instance, for-profit companies can't use interns as substitutes for paid positions, so interns are more often used for simple tasks like getting coffee, scanning documents and making photocopies.[623] In NFP entities, volunteers and interns can be given work with meaning; even formal positions can be covered by volunteers.[624] This can help make NFPs more resilient when finances are low.

Even in cases where a for-profit can benefit from volunteer work, at a charitable event for example, it feels more comfortable to volunteer with an NFP than with a for-profit enterprise, because nobody is privately profiting from your volunteer work, there is a clear social mission, and there is a long history of NFP volunteerism.

Globally, 971 million people engage in volunteer work in a typical year, contributing $1.35 trillion to the global economy.[625] This means that the companies and organizations that engage those 971 million volunteers (the vast majority of which are nonprofits) had lower costs and more outputs than they would have had without all of that volunteer work. This is a major financial advantage for NFPs in the market place and they are increasingly recognizing this.

The YHAis an NFP company that provides recreation and accommodation services in the United Kingdom and claims that volunteer hours are a key financial advantage for them.[626]

617 (Ref: Leete, L., Wage equity and employee motivation in nonprofit and for-profit organizations)
618 (Ref: The Nonprofit Times)
619 (Ref: Glass Door report)
620 (Ref: Charlotte Observer article)
621 (Ref: Bank of North Dakota annual report 2014)
622 (Ref: Nonprofit Organizations in the Sharing Economy)
623 (Ref: US Department of Labor)
624 (Ref: Nonprofit Organizations in the Sharing Economy)
625 (Ref: Salamon, 2011)
626 (Ref: YHA Annual Report)

In 2014, the owners of Scarecrow Video store in Seattle gave up their potential for private gain and turned the company into a not-for-profit. The company was struggling financially, but was able to keep its video library alive and to continue serving the community as a not-for-profit due to the ability to engage volunteers to help run the business. The previous owners said that if they had not made such a move, the business would have likely failed and the fate of the collection of over 120,000 videos would have been in limbo.[627]

Because most NFP board members can't be paid, they are effectively volunteers, and NFPs (especially in their startup phase) are often able to draw on valuable expertise and wisdom from their volunteer board members in a very hands-on way. It's a win-win situation, because NFP volunteers derive a sense of meaning and joy from the contribution they're making to the social missions of their chosen NFPs - otherwise, they wouldn't be volunteering their time and energy.

On top of this, because our world is more tightly connected than ever in both physical and virtual terms, volunteers are more readily able to contribute their time and energy to NFPs across geographical boundaries. Mozilla, one of the most well-known global NFP businesses, has over 10,000 volunteers spanning 87 languages who help with translation, user support, localizing the website, community events and more.[628] This is a huge resource that only purely purpose-driven companies can truly tap into.

This is not to say that NFP enterprises are or should be dependent on volunteers. Rather, we are highlighting this as one of many advantages NFPs enjoy over for-profits. And, like tax exemptions, volunteer work is something NFPs deserve to benefit from, as they exist purely to meet the needs of the community and that is why people want to volunteer for them.

People feel good about contributing to NFPs in these ways and an added incentive is that financial contributions are often tax deductible, meaning that people don't have to pay tax on income, goods, or services they've donated to a NFP.[629] This is a great advantage for NFPs, as their for-profit counterparts are usually on the other end of this, feeling social pressure to donate to charities.

Not-for-profit businesses can also often get discounts for products and services precisely because they are NFP, largely because companies want to support the important social work that NFPs do. For example, Coastshare is an NFP company in the U.K. that only provides services to other NFPs at cost and tax free[lii], giving their NFP customers immediate savings of 20%.[630]

EGive, an Australian company, provides software to small and medium- sized NFPs to assist them with their fundraising, managing memberships, e-commerce, email, conference management, communications, and campaign tools. In the spirit of enabling them to do good work with as few financial barriers as possible, they offer these services for free or at a very low cost. Being an NFP enterprise itself, E-Give puts all of its profit into further developing and supporting its services as well as donating to Environmental Projects Australia, an environmental charity.[631]

---

627 (Ref: Nonprofit Quarterly)
628 (Ref: Mozilla website)
629 (Ref: Charity.org)
630 (Ref: Coastshare sign up)

Not-for-profit discounts are also widely available when purchasing software licenses, website hosting, and even postage in some places. These discounts are so common and so helpful that there are even services, like www.discounts-for-nonprofits.com, dedicated to helping NFPs find offers on products and services. This means that NFP enterprises can often get the goods and services they need in order to run their businesses at lower prices than their for-profit competitors. And, again, this is for a good reason: their work explicitly helps the wider community.

In addition to benefitting from tax exemptions, nonprofit discounts, volunteers and managers who accept lower salaries, the NFP ethic is aligned with the rapidly emerging open-source movement (networks of people who freely share their ideas and work, such as software programs). This allows them to tap into many free products and services via sharing networks.

For instance, CiviCRM is an open-source contact management system specially designed for nonprofits.[632] It was developed because its creators believe that nonprofits shouldn't have to use for-profit, proprietary software that doesn't give them the independence or flexibility to modify it to suit their unique needs. There's even an open-source software package especially for credit unions, called CU-Centric.[633]

Participating in the open-source and sharing economies also enables NFP enterprises to draw on existing best practices and cutting-edge ideas. Although many for-profit companies have also financially benefitted from using open-source software, the open-source community is recognising that the value created by their communities should be accessible on commercial terms for for-profit entities who are profiting from those contributing to the commons.[634] Such an approach could create financial flows to enable and sustain contributors to the commons who wish to move entirely out of the for-profit economy.

As a result of holding community benefit at the heart of business, the NFP structure is also more conducive to collaborating and sharing with other NFPs. This includes NFPs offering each other special discounts, but it also means that they're more likely to combine efforts and share resources, such as physical materials, knowledge, money, and human resources, in order to achieve a common mission. Akhuwat in Pakistan uses places of worship to distribute microloans, cutting down on costs because there's less need for office space.[635] The places of worship are happy to lend their space because they feel Akhuwat is aligned with their own mission of helping people in need. And more NFP collaboration means more NFP innovation. Indeed, research shows that niche groups experimenting with cooperation-based economics are outperforming competition-based enterprises.[636]

This sharing and collaboration also strengthens business viability in times of economic downturn, creating a lot of resilience in the NFP enterprise sector. In fact, one study done by

631 (Ref: EGive website)
632 (Ref: Free Software Foundation article)
633 (Ref: CU-Centric website)
634 (Ref: Michel Bauwens article)
635 (Ref: Akhuwat website)
636 (Ref: Shane Hughes' talk)

the Australian government found that the economic downturn has spurred more collaboration between NFPs, because they must find a way to do more with less.[637]

Although solvency is important for all companies, NFP businesses have a different financial baseline compared to most for-profit companies, because NFPs are never required to make a profit and they rarely feel any pressure to. Not-for-profit companies will never pay dividends or retained earnings to private individuals. **The NFP non-distribution principle means that they don't lose money to anything that is not directly adding value to the company.**

Shareholder financing has been important for businesses in order to raise investment capital, but dividends only serve as a drag on budgets in a future where shareholder financing is no longer necessary. After the initial injection of capital to start a company is paid back, dividends extract value from the company, as well as putting pressure on managers to focus solely on profit. Shareholders are also a source of more bureaucracy for companies to deal with, as they must allocate resources to keep track of dividends and shares.

It is estimated that cooperatives are about 20% more cost efficient than regular companies, because they don't have third party shareholder obligations.[638] As some co-ops are for-profit, paying out dividends to members, this statistic is a conservative estimate. This means that, all things being equal, a co-op is more likely to survive an economic downturn than a regular company. It's easy to extend this logic to NFP businesses, given they have absolutely no obligation to distribute profits. For example, studies have found that NFP nursing homes and hospitals in the U.S. are more affordable and have lower administrative costs than their for-profit peers.[639] Accordingly, getting an NFP to a 'successful level' of business viability is easier because they don't need to make as much profit in order to satisfy foundational goals; they just need to pay wages and other operational costs.

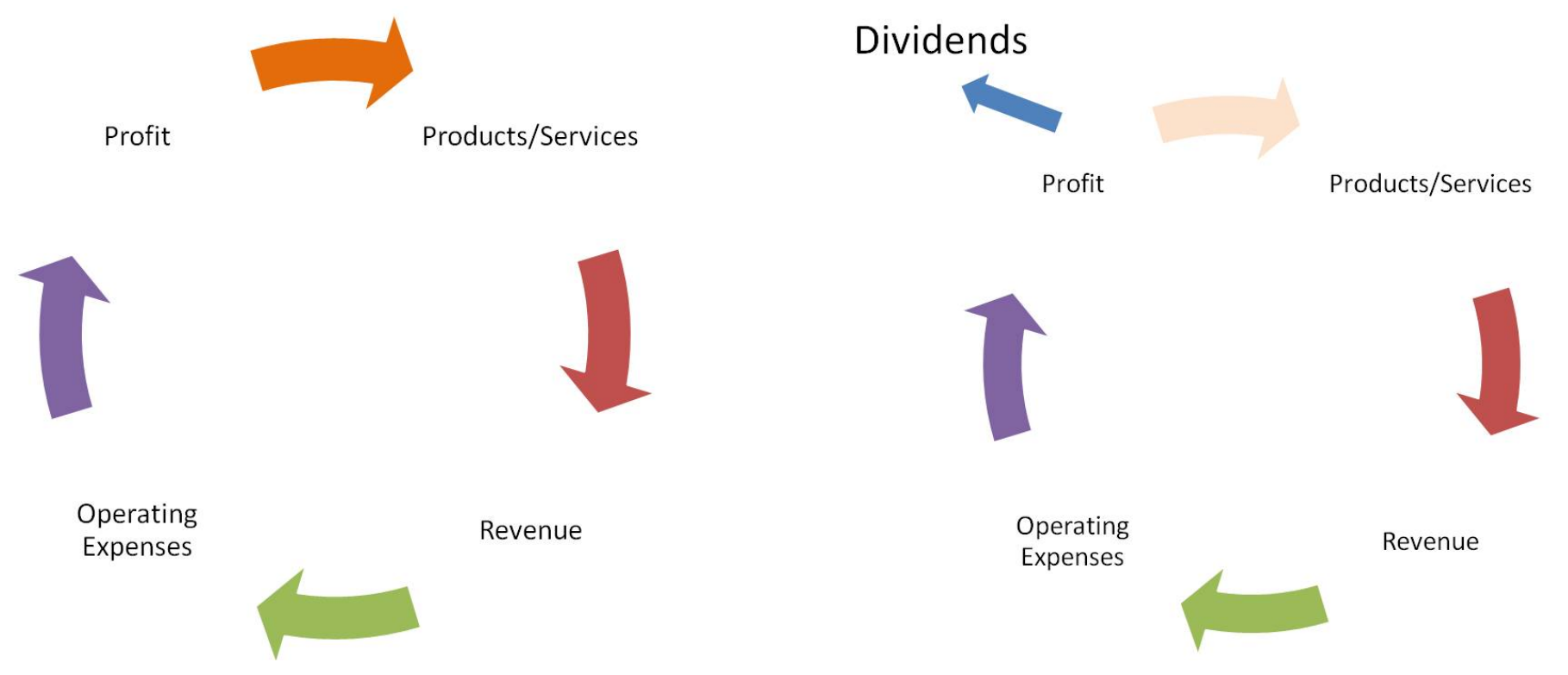

## For-profits must compensate for profit distribution

**As NFPs have a lower cost structure due to lower requirements of return on capital, they have great potential to outcompete for-profit equivalents**.[640] Even more altruistic, outcome-oriented for-profits hold competitive advantages over their profit-maximizing peers, because their preference for output allows them to price their products and services below average cost.[641]

---

637 (Ref: Australian gov study)
638 (Ref: Michael Cook and USDA, 2012 International Year of Coops final conference)
639 (Ref: Nonprofit Organizations and Healthcare, p. 83)
640 (Ref: The Nonprofit Sector and Industry Performance)
641 (Ref: The Nonprofit Sector and Industry Performance)

This extra financial burden isn't limited to companies with shareholders. Obligations to deliver profits to owners can even cause financial drag and conflict in family-owned businesses and other business partnerships. Not-for-profit enterprises don't encounter such issues.

While many traditional nonprofits are heavy and bureaucratic, NFP enterprises are often super-lean. They generate their own revenue and one hundred percent of their financial surplus is reinvested into their mission. **It is truly circular - nothing is lost.** They take care of their operating expenses and every extra dollar is put back into doing what they do.

This is a big advantage compared to for-profits that must constantly expand to compensate for dividends extracted from their profits. Some for-profits even create a lot of extra risk by paying out dividends on a frequent basis, generating expected demands on their cashflow. If the company has a high-revenue month, big payouts are made. Then if the market suffers a downturn for three months, the company may find it is struggling to manage expectations about the shareholder returns.

Even though we see it as a financial advantage that NFPs can't take speculative investment, this has been one of the most prominent arguments *against* using nonprofit legal forms for business. The argument is that NFPs can't attract sufficient investment, because profit-hungry investors have little to gain from them. However in the market of the 21st century, the non-distributional aspect of NFPs means that they're less likely to take on risky debt obligations and that they are more able to grow slowly and mindfully. **They have to build themselves more on proven, real value**, which means that in challenging economic times, they will be less likely to fail. Plus, a community that's benefitting from their work will tend to support such a company, whereas a venture capitalist is likely to cut their losses and run when things go downhill.

Prioritizing social outcomes over profit margins, NFP businesses are less likely to 'leverage'. They take on less external debt or credit to finance their operations and are much less likely to take big financial risks in seeking financial returns, as for-profits routinely do.[642] This means that they are also less likely to go bankrupt or insolvent due to debt. Furthermore, because NFP board directors are generally not paid for that role, their decision-making tends to be more objective and less guided by personal interest, paving the way to more strategic decisions and less exposure to speculative financial risk in the stock market, securitized trading and asset markets. An illustration of this is that U.S. credit unions have less risk on their balance sheets than for-profit banks and are also more careful lenders, approving fewer risky loans.[643] Perhaps that is why in 2009, while big for-profit banks drastically slowed their lending due to the financial crisis, American credit unions actually increased lending, especially to small businesses, keeping the wealth circulation pump going.[644] Critically, taxpayer funds have never been used to bail out a credit union in the U.S.[645] Likewise, the Bank of North Dakota never got involved in financial derivatives and weathered the economic storm of the early 2010s much better than its for-profit counterparts.[646]

There's also a mindset of 'doing more with less' in the NFP sphere. Because these companies are motivated by a deeper sense of purpose, not profit, they have a strong desire to find ways

642 (Ref: Charitable Insolvency and Corporate Governance in Bankruptcy Reorganization)
643 (Ref: Credit Union deposits article and CUNA powerpoint slide 14)
644 (Ref: Institute for Local Self-Reliance)
645 (Ref: Public Service Credit Union)
646 (Ref: Ellen Brown article)

to fulfill their mission no matter what. Many NFPs have had to make do with limited resources and as a result have developed flexible, innovative ways of thinking about financial challenges. Because NFPs traditionally have had leaner budgets, they are more adapted to finding and creating value at lower cost. As Chase Rynd, President of the National Building Museum says, "What serves a lot of nonprofits and museums well is that it's pretty rare for a nonprofit to ever be totally flush. We are used to having tight budgets and still producing great results. I have never, in my entire career in museums, had the staff size that I am supposed to have. There is never any fat sitting around, so when we come to a challenging time, we're already disciplined."[647]

## More Innovative

The NFP structure also gives businesses advantages over for-profit peers in terms of innovation. In an effort to capture more short-term profits, many for-profit boards have increasingly made decisions to cut funding for basic research and bringing innovation to scale because they are not seen as core compentencies and are not central to profit-maximization.[648]

Compared to for-profit companies, NFP employees and managers have greater flexibility and are less fearful when it comes to innovation because they don't have a vested interest in the financial outcome, at least not beyond their ongoing employment. This means they're generally more willing to take risks, experiment and innovate, in terms of organizational structure, technology, and strategy (in fact, many of them have had to, because they are not awash with cash). They're not as afraid of failure as for-profit companies, where owners, shareholders and investors are putting pressure on managers to make sure things work out for their benefit.

The for-profit approach is to only invest when the prospect of financial return exists. This means a significant lack of needed innovation where there is no foreseeable financial gain in the for-profit economy (it's why certain diseases, such as Leishmanaisis which affects about 1 million people, have been neglected for so long). This lack of innovation is very visible when it comes to social and environmental problems, and is a large part of the reason why there are still about 800 million chronically malnourished people in the world,[649] yet each year millions or even billions of dollars are poured into innovating better drugs for things like male sexual enhancement or curing baldness. What we really need is investment in innovation for the sake of addressing serious health, social and environmental issues. And this is what NFPs are able to do.

The world leader in home appliances Bosch is 92% owned by its NFP foundation, making it mostly NFP. Bosch seeks to honor the spirit of its founder by demonstrating social and environmental responsibility.[650] About 40% of Bosch's sales are products that specifically aim to minimize environmental impact and resource consumption.[651] Half of its research and development (R&D) budget is invested into these products, with Bosch investing nearly double the industry average into its R&D, more generally. It attributes this ability to be so innovative to the fact that it doesn't have the obligation to distribute surplus that its for-profit competitors have.[652]

---

647 (Ref: Architect Magazine)
648 (Ref: Forbes article)
649 (Ref: FAO)
650 (Ref: Bosch Philosophy)
651 (Ref: Bosch Philosophy)
652 (Ref: Forbes article, Bosch website)

The Mayo Clinic, a world-renowned hospital in the U.S., has been heralded as one of the greatest innovators in healthcare.[653] It is no coincidence that it's an NFP business. It can afford to invest in things like its Center for Innovation, where employees are encouraged to engage in unofficial activity (i.e., play around and experiment with ideas). There are no owners or investors asking how they will gain from such investments, and that gives The Mayo Clinic and other NFPs a degree of freedom that most for-profits in the health industry can only dream of. And **that degree of freedom often translates into greater innovation; some that's profitable and some that's not, but all of it addresses major issues.**

Not-for-profit businesses can also better maximize what has been called the 'cognitive surplus'; the innate human urge to create and produce. Generally, in a for-profit company, your boss is only interested in your ideas if it can be shown that they either generate revenue or reduce expenditures. But in an NFP, you can suggest any form of internal or external innovation and expect a much more appreciative reception, often irrespective of costing. The fact that there's not a profit-maximization mandate creates a lot more space for out-of-the-box thinking and expression of ideas. This includes more space for employees to question if the goods and services that the business is producing are truly beneficial to society, whereas there is much more of a "if you don't like it, then leave" attitude in the for-profit sphere. Imagine employees at a for-profit corporation's staff meeting openly questioning whether the company's products are doing more harm than good to society.

In terms of companies whose mission is to innovate, it's much easier to stay focused on the innovation itself when you don't have shareholders, owners and investors hovering over you, asking about the bottom line.

In the early 2000s, a man named Joe Justice started Wikispeed, a car manufacturing company in Seattle, with the mission of creating the most fuel-efficient, user-friendly car in the world. By 2011, Wikispeed had actually manufactured a street-legal car that goes more than 100 miles on a gallon of gas (42.5 km per liter).[654] Predictably, his invention attracted a lot of attention, especially from potential investors, but when he started meeting with the interested investors, he felt very uncomfortable with their terms.

As a result, Joe decided to establish Wikispeed as an NFP. Joe says, "Becoming a not-for-profit was a self-defense mechanism so we could focus on what we were trying to do. Shifting to NFP improved our competitiveness by honing our focus. Half of my time was spent listening to venture capitalists pitch. All of them had abusive terms. If the venture capitalist got cold feet they could shut down the project and keep all the IP (intellectual property). Non-negotiable. I wouldn't accept a personal loan on those terms. Team Wikispeed has no debt and is cashflow positive."[655]

So, aside from being aligned with the NFP ethic, a big part of the reason Joe took Wikispeed NFP was in order to have fewer for-profit distractions and to avoid the for-profitization of open-source ideas.

**Without shareholder influence, quarterly reports and obligations to maximize profits, companies like Wikispeed have the ability to be more innovative.** Mozilla has said that being NFP allows them to innovate in ways that truly align with their mission. When you download Mozilla's Internet browser, Firefox, a message tells you, "Thanks for downloading Firefox! As a non-profit, we're free to innovate on your behalf without any pressure to

653 (Ref: Harvard Business Review articles 1 and 2)
654 (Ref: Ref: Joe's TEDx Talk)
655 (Ref: Donnie's emails with Joe)

compromise. You're going to love the difference." And in an article, Mozilla's CEO Chris Beard wrote, "Being nonprofit lets us make different choices. Choices that keep the Web open, everywhere and independent."[656]

Another example of the NFP freedom to innovate is The Fred Hollows Foundation disrupting the eye-glasses industry in the 1990s by innovating an intra-ocular lens that cost 3.5% of the average industry price. This innovation led to high-quality, low-cost lenses being more widely available throughout the world.[657]

Research shows that businesses with clear, motivating missions tend to be more innovative.[658] And this makes sense. How many of the world's greatest innovators were in it for the money? Probably very few, if any. Leonardo Da Vinci, Isaac Newton, Thomas Edison, Albert Einstein, James Watson, Francis Crick, Marie Curie, Nikola Tesla, Jonas Salk, Buckminster Fuller – the best innovators in human history were not driven by the profit motive.

Although intellectual property and proprietary knowledge are often seen as drivers of innovation, providing the incentive for investment based on the potential for a return through licensing fees, the rise of the purpose-motive and open-access to knowledge is rapidly becoming an even better driver of innovation. Kickstarter, arguably the world's most successful crowdfunding platform, also sees the value of keeping private profit out of innovation (despite being a for-profit corporation themselves). When explaining why Kickstarter would not be moving to an equity-based investment model for the way people can support projects via its platform, CEO Yancey Strickler said, "We believe the real disruption comes from people supporting things because they like them, rather than finding things that produce a good return on investment."[659]

It can also be argued that because NFP entities have a more motivated workforce, this translates into enhanced performance in terms of innovation. Contrary to the popular notion that performance and innovation stem from good management, studies in 2011 found that high-performance organizations were distinguished from low-performance organizations by high scores in fairness in the work place, job satisfaction and wellbeing.[660] Based on all of the advantages described above, it's arguable that NFP businesses are better able to meet all of the criteria for high performance.

## More Accessible

As NFPs can more readily benefit from lower costs in terms of goods, services and wages, and because they have no one expecting a cut of the profits, they can pass these savings on to their customers.

This is another part of the reason NFPs get tax deductions – in order to make their social services as accessible as possible. These discounts and lower costs allow them to offer lower prices, which in turn enables them to compete in the market with for-profit equivalents.

Customers of public utilities in the U.S. pay 14% less than customers of private utilities.[661] This makes sense, considering that CEOs at for-profit utility companies are paid 25 times as much as CEOs at public utilities, and because public utilities are NFP and purpose-driven, they pass those savings on to their consumers.[662] Credit unions in the U.S. also consistently offer lower fees, lower rates on loans and higher yields on savings than their for-profit competitors and can afford to do so because they are NFP.[663] The Bank of North Dakota likewise has proven to be safer for depositors and has enabled public infrastructure costs to be cut in half.[664]

---

656 (Ref: Chris Beard blog post)
657 (Ref: Fred Hollows Foundation website)
658 (Ref: McDonald, 2007)
659 (Ref: Kickstarter not moving to an equity model)

660 (Ref: EY Annual Review)
661 (Ref: Gar Alperovitz article; Public Power report)

In the U.S., for-profit colleges charge far higher tuition than comparable NFP and public colleges, but spend much less on instruction for each student.[665] This is because they usually spend more on marketing than on teaching, in an effort to maximize short-term profit for owners.[666]

Many NFPs already use this advantage to advertise. Ebico, the UK's first not-for-profit energy provider, states "Ebico Ltd aim to offer a competitive, fairer deal for domestic electricity and gas to British households."[667] Groupe SOS in France states that because it is an NFP, the company is able to provide an efficient healthcare program based on the needs of individuals and communities, accessible to all, regardless of income.[668]

Public Performance Partners, an American firm that aims to help public services function more effectively and efficiently, say on their website, "Public Performance Partners is a not-for-profit enterprise because it allows us to provide first rate services to public entities at greatly reduced rates."[669]

Even in cases where NFP prices are higher than those of for-profit competitors, NFPs are generally able to attract more loyal customers (including government agencies) who are willing to pay a higher price, given they know any financial gains go towards better wages, more reliable products, and strengthening communities, rather than shareholders' pockets.

Another important strategy that NFPs use to ensure accessibility is sliding-scale fees through cross-subsidization. Sliding-scale refers to a pricing strategy wherein the customers who can afford to pay more, do pay more, enabling the company to offer lower prices to people with lower incomes, who can't pay quite enough to cover the basic cost of a product or service. If an apartment costs $400 per month to maintain at a housing co-operative, a person who has more money might pay $600 per month which enables another person to rent a similar apartment for $200 per month. This approach is already quite common in NFP enterprises.[670]

Essentially, because they are mission-driven, NFP enterprises find ways of making sure that their prices are affordable for the people who need their products and services the most.

## Seeds of a New Era

What the story of interconnectedness, the changing business preferences and the increasing advantages in the marketplace point to is the possible emergence of an entirely new economic

662 (Ref: Gar Alperovitz article)
663 (Ref: My Credit Union.gov; CUNA slides 22-26)
664 (Ref: Ellen Brown article)
665 (Ref: Degrees of Deception; Bloomberg article; Economist article)
666 (Ref: Ibid)
667 (Ref: Ebico website)
668 (Ref: Groupe SOS website)
669 (Ref: Public Performance Partners website)
670 (Ref: Dennis Young, *Financing Nonprofits: Putting Theory into Practice*)

era. The demand for transparency, accountability and trust-worthiness is moving well beyond the social sectors. As for-profit scandals and stock market bubbles continue to fill the headlines, and as inequality worsens, species disappear and the climate changes, the demand for more ethical modes of doing business will continue to rise. This gives NFP businesses a clear advantage over for-profit competitors. Not-for-profit forms of business will be more competitive in the 21st century because that's where demand is heading, and that's where social and environmental limits are pushing the economy.

As the world makes a transition away from the profit motive and towards the purpose motive, we can expect even more social innovation to occur in terms of NFP financing, capital-raising, ownership structures, and types of businesses. This innovation will only serve to further solidify the NFP advantages in the market.

Of all the amazing things being done in terms of building a healthier economy in which we can all thrive, one critical piece of the puzzle is almost always overlooked or underestimated. People working to build up new economic alternatives often point to the fact that many great things are happening – our shared stories are shifting, technology is improving, communities are coming back together, environmental awareness is growing - but we still lack a blueprint or a framework for how the economy can be organized in a fundamentally different way. **What has been missing is not agreement on core values, but rather agreement on a core operating model.**

# 5. Workings of a Not-for-Profit Economy

**A market economy can be innovative, viable and not-for-profit**

In 1946, American director Frank Capra made a classic film about the importance of caring for one's community. Initially dismissed as simplistic and idealistic, *It's a Wonderful Life* tells the fictional story of George Bailey, whose family runs the local 'Building and Loan', a not-for-profit financial institution which uses member money to provide affordable housing loans in the small town of Bedford Falls. But the bank is under constant threat from wealthy business mogul Herbert Potter, who seeks control of the housing market through his local, privately-owned bank. In an unfortunate turn of events, Bailey's uncle misplaces a large sum of the Building and Loan's deposits, leaving the company on the verge of bankruptcy.

With a despairing Bailey about to take his own life, an angel visits him, showing what Bedford Falls would be like, had he and the Building and Loan not existed. Now called Pottersville, the town is suffering from high unemployment and homelessness, while its economy is based on shady activities, like… Gone is the thriving community Bailey and his bank helped create. Shocked by this vision, Bailey chooses to live, although he remains unsure how to pay the bank's debts. The film ends with the community's generosity ensuring the Building and Loan stays afloat.

Pottersville speaks to our modern reality. Driven by the ethos of maximizing private gain, our for-profit system is failing, with faith in trickle-down economics misplaced. While capitalism, in all its forms, has certainly delivered stunning infrastructure, and numerous scientific and social advances, the underlying principle of accumulation has reached its expiry date. Further economic growth cannot deliver widespread social and ecological wellbeing.

*It's a Wonderful Life* highlights a counter-narrative that has existed within the capitalist story all along. Bedford Falls was able to share the wealth generated by communally-backed banking activities because there were no private owners extracting the company's profits. In Bailey's town, prosperity meant nothing unless it was shared. The Building and Loan highlights the restorative power of not-for-profit banking within a market economy. It begs the question: **what if not-for-profit businesses were *at the heart* of the economy?**

## Introducing the NFP World

**The NFP World involves an innovative market economy, centered upon not-for-profit forms of business and the not-for-profit ethic.** In the NFP World, wealth and power is decentralized, yet the system operates with interconnected efficiency. Individual choice and freedom are privileged, yet the greater good prevails. Social services and safety nets have expanded, but government and taxation have contracted. Entrepreneurialism and competition help drive gains in innovation (along with innate human creativity and problem-solving), but businesses are less ruthless and aggressive. Institutions are highly accountable to the wider community, and there is a deeper sense that not everything that matters can be counted. Leisure and wellbeing have increased, but per capita consumption has reduced.

Idealistic? Yes. Removed from our present reality? Indeed. But we'll soon explore the four mechanisms that underpin the magic of the NFP World model and how the transition to this world is not only more simple than we realize, but could already be underway.

Paradoxically, free market *and* state-led economic thinking heavily influence the NFP World model. Yet our approach is not merely an iteration on the mixed economy, or a new 'third way'[671]. Rather, the NFP World transcends the centrality of both private and public ownership. And while a diverse range of familiar structures remain in the NFP World, and strong variations between approaches are expected across geographies and cultures[672], the NFP economy has a fundamentally new base: not-for-profit business and a thriving community economy, which refers to the goods and services gifted and traded between community members without the use of money (including family life and the unpaid caring economy).

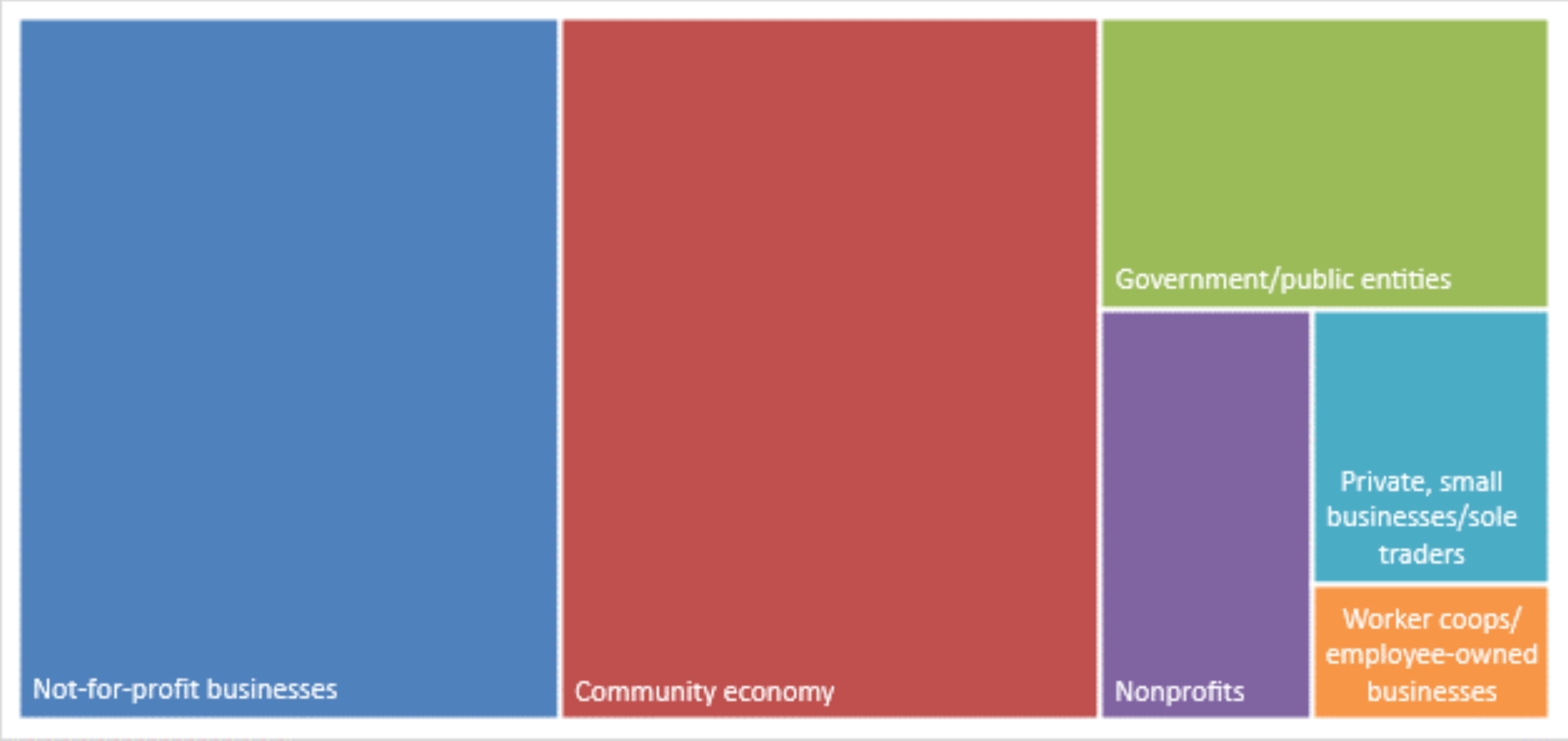

### Contributors to the Not-for-Profit Economy

Alongside the community economy, NFP businesses provide the biggest contribution[673] to the NFP economy. Including start-ups, established organizations, subsidiaries, joint ventures and a range of cooperatives[674], NFP businesses cover all sectors, from retail to manufacturing, pharmaceuticals and tourism. They are largely locally-based and most are small to medium-sized. Many are 100 percent financially self-sufficient, sometimes choosing to avoid

671 E.g., 'third way' or worker-owned 'cooperative economy', which can still be profit-motivated. Certain existing proposals come close to addressing both of these concerns. See, for example, the solidarity economy, Distributism, Third way – e.g. Poland - worker coops (focus on unions); Commonwealth pluralism (Alperovitz). These are often presented as alternative economic models, but in all of these models, profit-maximization and private ownership of business, often through worker co-ops, are still a possibility.
672 The NFP World acknowledges that human nature is complex, that economies can and should take different shapes according to different contexts. For example, some countries may be more state-focused, choosing to publicly fund more services via taxation. Religious states or monarchies might also deal with this in different ways.
673 Measured in terms of outcomes enhancing social and environmental wellbeing.
674 Purchasing co-ops, for example, largely act on behalf of NFP clients, rather than for-profit producers in the NFP World.

charitable status and, therein, federal tax exemption. For those NFP businesses that are 'charities', some draw on philanthropic support, grants and government assistance to supplement their income-producing activities[675].

Complementing the market, the community economy continues to make a significant social contribution[676]. Sharing is a natural cornerstone of family and community life, from peer-to-peer production, to collaborative consumption, bartering, and mutual support networks. Along with domestic labor, unpaid caring work continues in an expanded form, with its contribution to wellbeing more appropriately acknowledged. In this sense, there is an increase of non-market activity in the NFP World. While sole traders[677] and NFP services remain available to assist with tasks such as childcare, cooking, cleaning, and tutoring, families have less need (or desire) for paid support, given adults have more time available to participate in caring activities themselves, because employees work fewer hours at formal, paid jobs.

This is strengthened by labor saving innovations as well as more supportive community and extended family arrangements. The grey economy of 'under the table' payments forms an insignificant portion of the community economy, given how freely wealth circulates in the NFP World.

The public sector maintains important roles in the NFP World. The state continues to run services such as government, the judicial system, the police and fire brigades. Healthcare and education are also generally publicly funded (although NFPs may run these services). Government has also developed its entrepreneurial capacities, participating more fully in the NFP market. However, government is smaller and more decentralized, with many functions having been transferred to NFP businesses, municipalities, or de-marketized, reducing the state's influence over daily lives.

Traditional nonprofit activities continue, but their contribution is relatively small. These include donor- or grant-dependent activities that are without a business model or subsidization via other income-generating activities in areas deemed of public importance that people are willing to fund fully, such as research centers. Nonprofits may also exist for activities that don't require funding at all, such as organized rallies, or unincorporated associations that pursue shared interests with minimal cost.

As it remains legally possible to operate as a for-profit company, there is also a relatively small, for-profit component to the economy. Privately-owned small businesses, including sole proprietors, artisans, freelancers, and small family-owned companies are generally viable forms of business, integrated into the NFP World[678]. However, it proves financially and socially advantageous for incorporated entities of any significant size to be NFP, because profit margins are so low and society demands the transparency and mission-driven aspects of NFP business. Worker cooperatives (and to a lesser extent employee-owned businesses) are a broad exception, having remained somewhat competitive with NFP businesses. There is still for-profit trading in precious metals, bonds, and housing, but it is greatly reduced and less socially acceptable, given heightened awareness of interconnectedness and the detrimental effects that speculative trading can have on the entire economy.

---

675 If this sounds like too much dependence on outside financial help, recall that many for-profit companies are supported by subsidies of various kinds in our present economy, such as corporate tax breaks.
676 Although many of its contributions can't and shouldn't be measured in economic terms.
677 Also known as self-employed and freelancers.
678 Via what Sarah Horowitz from the Freelancer's Union calls 'the new mutualism' <REF>.

One area not shown as a contributor to the economy in the diagram above remains nonetheless: the exploitative, black market. This includes the trade of drugs, alcohol, weapons, counterfeit items and pharmaceuticals, migrants, organs, illegal gambling, sex trafficking, loansharking and money laundering. However, it is waning. Alongside the rise of associated rehabilitative programs, most addictive substances are now decriminalized and law enforcement increasingly relies on highly effective crowdsourced data. The tighter family ties and community connection allowed for by the NFP World economy, as well as a greater sense of purpose at work and more leisure time mean that feelings of isolation, alienation and emptiness in society have decreased, which leads to lower rates of addiction and crime. It is also easier and much less taboo to seek help for mental health issues, which can go a long way to stopping addictive and criminal behaviors before they start.

Relationships exist between the various contributors to the NFP economy. Governments enter joint ventures with NFP businesses, bringing new meaning to the term 'public-private partnerships'. Small-scale producers are members of NFP producer cooperatives. Local communities assist in managing state-owned natural resources, and community members support NFP or nonprofit projects through volunteering.

But the NFP World is much more than a shift to not-for-profit forms of business. While the separate parts that constitute the NFP economy are familiar, together they form a *fundamentally* different operating system, focused on advancing human and ecological wellbeing. This guiding ethic is reinforced by new narratives, values, and ways of organizing economic activity. It involves a whole new baseline for business and a new standard for the economy. The NFP ethic enables business practices to align with values like generosity, justice, compassion, caring for nature, and human rights.

We believe an NFP World is possible by 2050. Our overarching assumption is that an effective, post-capitalist economy must ensure:

1. systemic efficiency (i.e., minimal or no waste);
2. wealth circulation and decentralized access to resources; and
3. human wellbeing within ecological limits.

The NFP World possesses these three characteristics in the forms of the lean society model, the wealth circulation pump, and the cycle of wellbeing. Combined, these characteristics form the foundations for how finance, the corporation, the market, the state, and global relations function in the NFP World.

Let's explore the first of these four characteristics, the lean society model.

# The lean society

In manufacturing, lean refers to *making obvious what adds value to the whole system by reducing everything else*. In the NFP World, only parts of the system that produce value (defined as social and ecological health) remain.

The market should be effective and efficient. Not efficient in terms of for-profit outcomes - basically just the high-speed churning out of consumer goods – but rather efficient in terms of doing more with less (known as eco-efficiency or closed-loop production). In this way, the NFP market also gears work towards activities that create real value.

The market is more integrated, collaborative and efficient (e.g., closed-loop). De-monetization removes the inefficiencies associated with market transactions, also known as transaction costs (e.g., time, travel, middlemen, interest, and taxation).

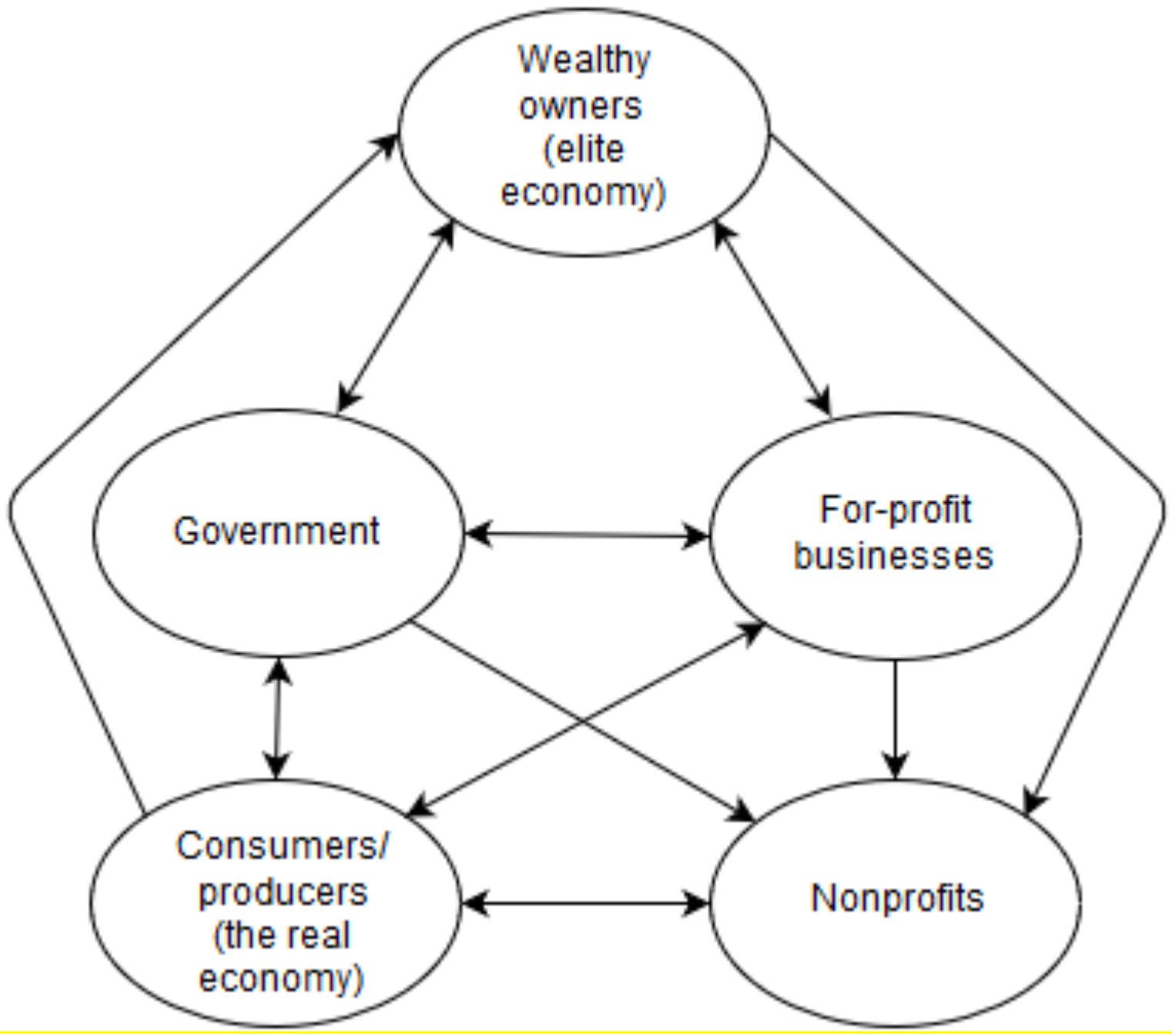

**The wasteful, for-profit economy**

Let's take a look at the for-profit economy to better understand this point. The for-profit economy's systemic waste comes from needing the existence of nonprofit and public institutions, as well as high levels of taxation, merely to counteract the negative externalities (especially inequality) created by enterprises driven by profit maximization. In this system, competition for grants and philanthropy by nonprofits are a major waste of resources. In fact, the wealth used to apply for grants might sometimes be more than the final amount allocated.

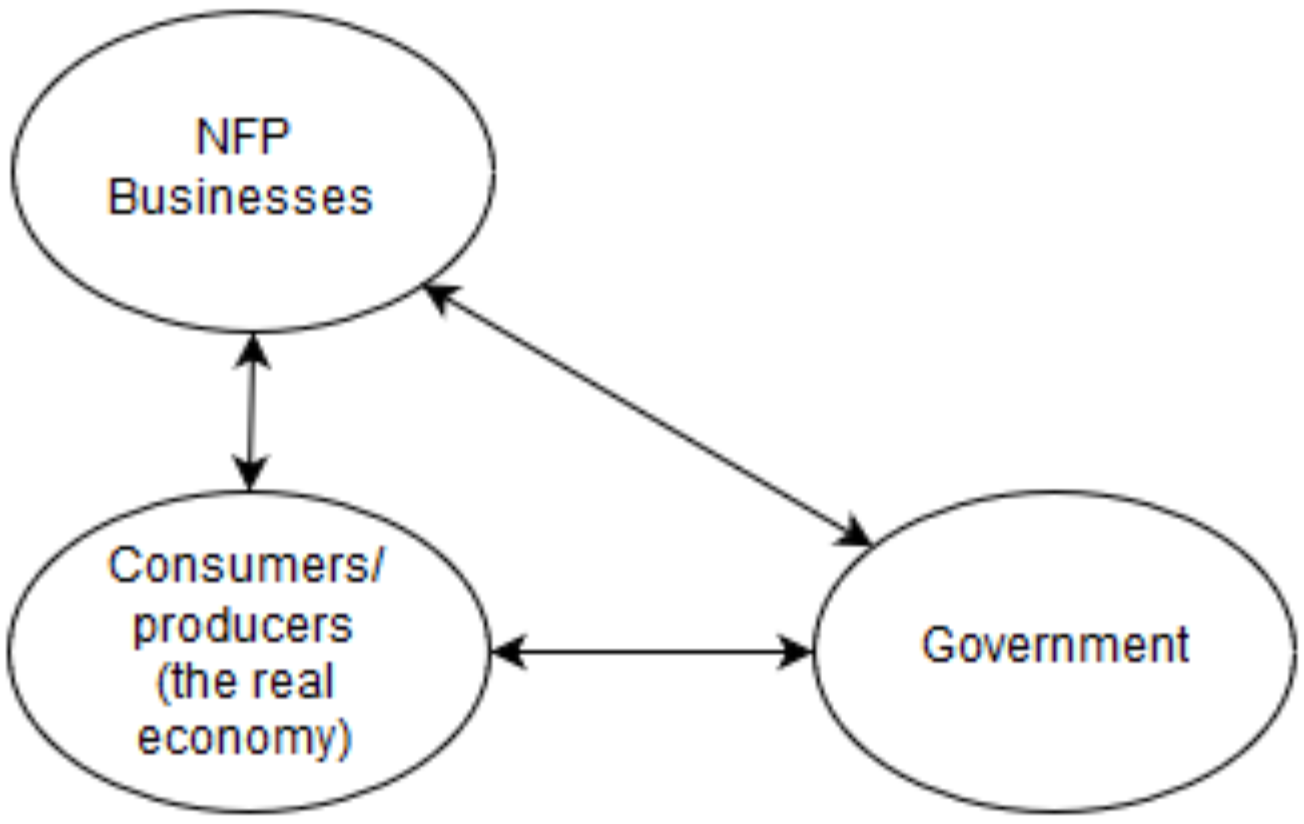

**The Lean, NFP Economy**

The NFP economy's systemic efficiency comes from having no 'for-profit middle man' between capital flows and meeting social needs [679].

By eliminating the siphon, we actually eliminate a huge number of inefficiencies that existed in the system, including market inefficiencies. By reducing the waste in the system, by reducing the redundant, cross-cancelling activities (and making social and ecological value creation more central), there are massive efficiency gains. This results in saved resources on many fronts. For example, the battle against environmental policies by for-profit companies in the current economy wastes a lot of energy that, put to social outcomes, could be really powerful. Imagine if all of the money and time that companies currently spend on lobbying against taxes and regulations (in order to cut their costs and generate more profit) was instead spent on taking care of forests, cleaning up beaches, or even helping people find work that gives them a sense of purpose.

The NFP economy's saved energy comes from merging entrepreneurialism and social outcomes through business structures that circulate wealth throughout the real economy, thereby avoiding the social disruption that comes from wealth extraction via profit-seeking. The NFP World model is also simpler in this way, because there are fewer conflicts of interest.

The NFP World is a lean economic model that seeks to find the most efficient ways to meet needs, so superfluous aspects of the economy that either don't aid this mission or actually hinder it (like speculation through financial markets) do not survive long, as they're potentially harmful to the whole system and they're simply not needed.

# The wealth circulation pump

Modern economic theory is built on scarcity. In capitalism, scarcity-based thinking justifies the wealth extraction siphon and the narrative of 'never enough'. But it's rare for us to consider what happens to an economy built on abundance and the notion of 'enough'. In doing so, we discover that, when scarcity is removed, health is restored to the economy's critical organs: finance, businesses, the market, the state and international relations.

How is this possible? **In the NFP World, the circulation of wealth**, derived from purpose-driven economic activity, **is self-perpetuating**. In other words, distribution is built into this market system. The restricted ability for private gain, via the predominant NFP form of business, means that everyday economic activities fuel the *wealth circulation pump*.

[679] The unrelated business, subsidiary model means there is some extra energy is needed to ensure the fulfilment of social needs – money needs to be transferred over from a separately governed company, but it's still a lot more efficient than the FP-to-NP via the Government flow of capital and the Nonprofit Enabler effect.

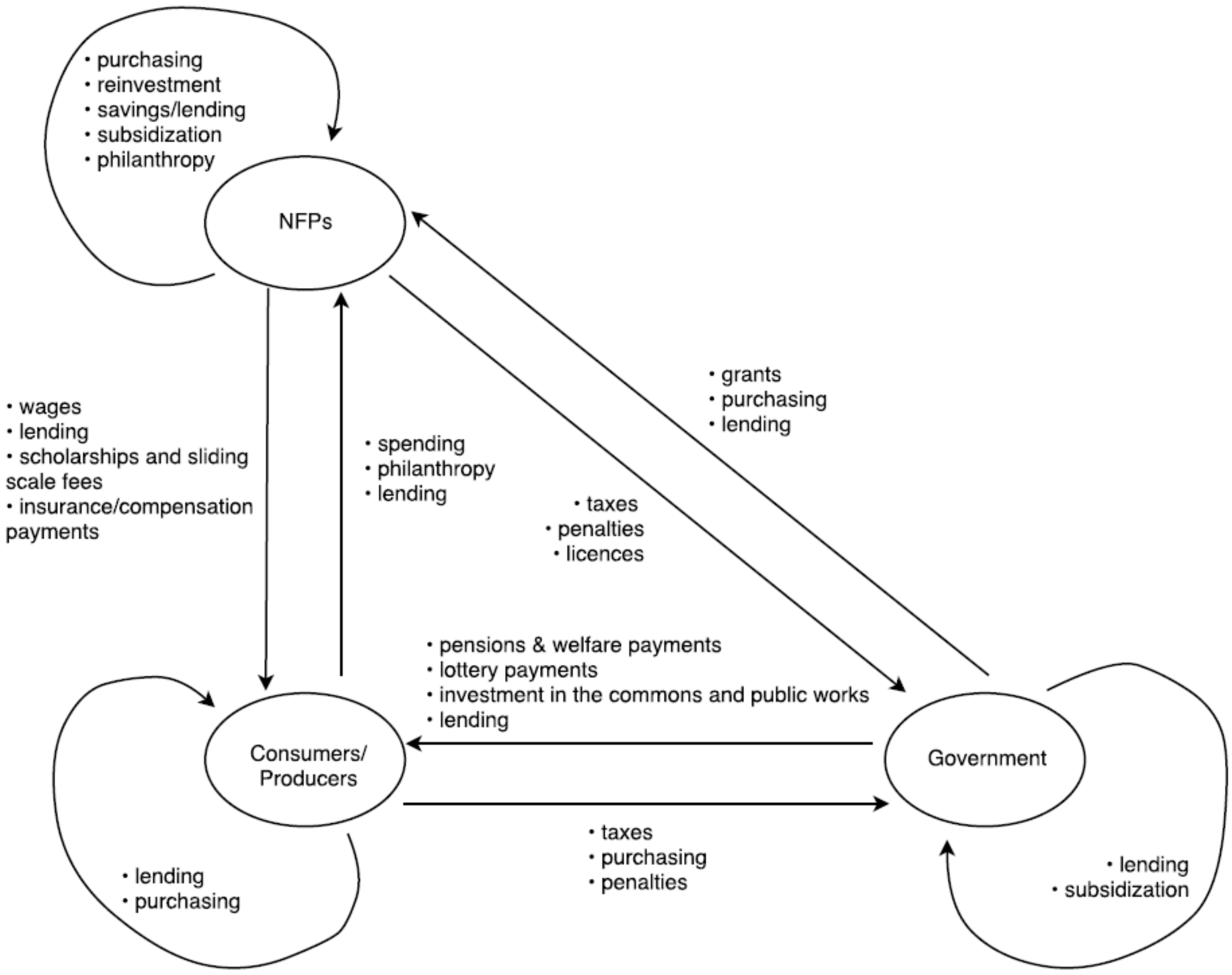

**The Wealth Circulation Pump of the NFP Economy**

When you purchase goods and services in the NFP economy (including from a for-profit subsidiary of a NFP), your money goes in two directions: internal operating expenses and profits.

Internal operating expenses, include payments to external providers who are also NFP, thereby fueling the wealth circulation pump, as well as wages. Wages also fuel the wealth circulation pump, because: 1) the extractive impact of owners is absent; 2) wages more accurately reflect the contribution of workers; and 3) companies operate in more participatory ways that maintain limits on wage disparities.

Wages then either go towards spending, saving, or philanthropy. Spending fuels the wealth circulation pump, even through activities such as purchasing insurance, luxury goods and services, or participating in public lotteries.[680] As for savings, people invest their money in a NFP bank, the bond market, or assets. Money that is saved in an NFP bank enables the bank to provide both commercial loans and personal loans. Commercial loans strengthen the NFP infrastructure for the wealth circulation pump and personal loans enable the fulfillment of individual and family needs. Money that is invested in the bond market provides capital to governments or NFP companies. Upon maturation, these savings are available for further spending or investment, making another round through the wealth circulation pump.

---

680 http://www.icnl.org/research/journal/vol12iss4/art_5.htm and http://fivethirtyeight.com/datalab/what-percentage-of-state-lottery-money-goes-to-the-state/

And money that is invested in assets – such as houses, vehicles[681], and other property - allows individuals and families to create a sense of physical and financial security.

When people spend into philanthropy, then the money is directed towards social purposes through nonprofit and NFP organizations in ways that once again prime the wealth circulation pump.

In addition to wages, other operating costs that an NFP enterprise has to cover might include: taxes, which of course go to support things that society deems important such as infrastructure or pension funds, and other forms of social security for employees. They might also need to invest in more machines, furniture or other physical capital, which they would buy from other NFPs. All of these operating costs keep the wealth circulation pump operating. One operating expense that might not contribute to the wealth circulation pump is the rent that some NFPs have to pay for their space. A private person could own and rent the space out. In the current economy, private real estate ownership is a major source of wealth for some, so this is an important concern. However, in the NFP World, most commercial space is owned by community land trusts, non-equity housing and building cooperatives, and by NFP businesses themselves, which means the rent paid would also contribute to the wealth circulation pump.

The second thing that might happen with the money you spend at an NFP business is that it contributes to the business's profits. Financial surplus can be used by an NFP business to:

a. further its mission, including the creation of publicly accessible capital and infrastructure, as well as supporting socially-beneficial[682] goods and services;
b. keep as savings, with a NFP financial institution;
c. invest, directly or indirectly in other NFPs; or
d. give as philanthropy to other nonprofit or NFP entities.

As you can see, whenever an NFP business spends money on operational costs or uses its profits for any of the above activities, it helps generate real value in the economy. Through the majority of possible interactions, wealth circulates, rather than accumulating in an elite economy. Value creation stems from the expansion of individual and community assets necessary for living, greater wellbeing, developments in tangible and intangible social infrastructure, and enhancement and expansion of the commons. Without private business ownership, wealth can circulate in service of greater wellbeing for everyone.

The power and efficiency of this wealth circulation mechanism can be seen in parts of the existing, for-profit system that are increasingly run by NFPs. A recent assessment of the year-long impact of NFP credit unions in the U.S. State of South Dakota, for example, found that $15,775,244 was returned in direct financial benefits to the state's 251,140 credit union members – the equivalent of $63 for each member or $120 per member household. This means there was $15 million more circulating in the economy, from $21 million in profits. This amounts to a lot more than would have been paid to the state via business tax.[683] And it

681 Of course, widespread access to safe and efficient public transportation is an important part of any sustainable society, but some people might still need to invest in vehicles.
682 Again, the government decides what constitutes 'social benefit' in its legal descriptions of NFP business structures. These definitions can and should include ecologically-beneficial activities.
683 <REF>.

means that the $15 million was not extracted to the hands of a few wealthy owners. It contributed to more equality rather than more inequality.

Thus, **a distributional mechanism is built into the very DNA of the not-for-profit economy, leaving very little need for redistribution.** It is not an afterthought or a strategy to compensate and mitigate the undesirable side effects of a for-profit market. The wealth circulation pump doesn't require trickle-down economics, government stimulus, or the belief in an Invisible Hand that mysteriously turns self-interested accumulation into social good. Any form of consumption drives wealth around the system. But this is different to what Keynes advocated because the pump is based on existing needs, not manufactured demand. This means there's less need for neo-Keynesian intervention (prices/incomes) and monetarist policy (interest rate adjustments). The wealth circulation pump in the NFP World changes everything by introducing abundance and real liquidity into the system.

It is worth noting that not all wealth circulates freely in an NFP World; some aspects of the economy continue to have a siphoning effect. Some forms of capital gains still exist. These include earned income via small business and sole trader profits, royalties, patents, license fees and trading in areas like commodities. Unearned income stems from interest on loaned capital, rents on land, and asset appreciation (such as property or collector items). When inheritance involves very large amounts, or is passed down to those who are already wealthy, it can have an accumulating effect within the economy. Fraud and exchange through the black market present other avenues for wealth accumulation.

| | |
|---|---|
| **Circulating** | Philanthropy; taxation; NFP earned income; government earned income; scholarships and sliding scale fees; investments in NFPs; investments in the commons; public works; welfare payments; insurance payments; compensation payments; penalties; low income finance; licenses; rent paid to NFP entities |
| **Semi-circulating**<br>**(can accumulate as savings, but without intent to accumulate)** | Wages; pensions; asset appreciation |
| **Both circulating and accumulating** | Public lotteries; capital losses |
| **Semi-accumulating**<br>**(some desire to accumulate)** | Capital gains; interest on private loans; inheritance; private rent; sole trader profits; royalties; patents; license fees |
| **Accumulating (clear desire to accumulate)** | Black market exchange; fraud |

**Mechanisms of Wealth Circulation and Stagnation in the NFP World**

Additionally, a number of factors reduce the threat of these stagnating forces in the NFP World. To begin, the combined size of the forces that centralize wealth has been drastically diminished in the transition to the more equal NFP World. As the NFP story took hold and rampant accumulation of wealth went from being exhalted as a sign of success to being socially unacceptable[684]. In fact, the NFP World would not exist without a strong foundation of social norms, values, and shared goals, grounded in the NFP story of social-ecological wellbeing outlined in the previous chapter. This means that the rich now have less capital for investment. There are also significantly fewer avenues for capital accumulation, now that securing business equity is not an investment option. And the financial returns on lending capital as debt are contained via social pressures, the abundance of investment capital, and, in some cases, legal restrictions (for instance, on the amount of interest one can charge on a loan). Similarly, wealth from asset appreciation is less of an issue, thanks to reduced inflationary possibilities. And, given there is no elite economy in which accumulated wealth can circulate, those with greater wealth in the NFP World more readily spend it back into the real economy (and of course there is much less inequality in the NFP World due to the wealth circulation pump, so the wealthier households only have a small factor more wealth than poorer households).

Moreover, with an ethic of enough having replaced the fear-based, scarcity mentality, and in a condition of greater socio-economic equality, individuals don't feel compelled to accumulate for accumulation's sake. Rather, their focus is on converting financial prosperity into greater wellbeing, their own, the wellbeing of others, and the wellbeing of the planet. Public beneficence, not private accumulation, is now the celebrated form of individual and business success. And if that isn't enough to discourage hoarding, capital accumulation via capital gains is financially disincentivized via taxation.

In practice then, **the NFP World maintains healthy levels of equality because its circulating forces are greater than its accumulating forces** (see table above). Thus, the return to a highly unequal world is unlikely, because there's widespread awareness of the siphoning effects that create an elite economy and the damage that kind of economy does to people and planet; and thus there is much attention placed on minimizing the siphon and counter-balancing it by enhancing the circulation of the wealth circulation pump.

The wealth circulation pump acts universally (both within and between countries), equalizing the lack of freedoms and the fulfillment of the conditions for a healthy, happy life. This acceleration of wellbeing is most noticeable on the level of local economies. Countless communities had been negatively affected by the globalized wealth siphon of the for-profit era. These local economies see the greatest improvement in the NFP World because wealth now stays local. National and regional taxation ensures certain communities don't become exorbitantly wealthy. With millions of NFP enterprises and the NFP ethic running the economy, we expect to see a lot less war-profiteering, inhumane working conditions, and exploitation of poorer people. Yet, of course, some international and domestic inequality remains.

---

[684] A much more detailed exploration of a possible transformation pathway from the current for-profit economy to the NFP World is outlined in the next chapter.

# The cycle of wellbeing

There is ample time, plenty of money, low stress, and an easier life thanks to greater systemic productivity in the NFP World. This is how its underlying wealth circulation pump plays out. It starkly contrasts with the experience of the scarcity-driven for-profit world and its underlying wealth extraction siphon: no time, not enough money, too stressed, and lost jobs due to profit-maximizing measures and labor-displacing productivity gains.

In the NFP World, prosperity is defined as a high level of wellbeing, rather than material and financial wealth. It is of course acknowledged that material and financial wealth can be means to achieving high levels of social-ecological wellbeing, but they are seen as a means rather than an end. This means that there is a level of not enough material wealth, which is not good, but there is also a level of material wealth that can be detrimental to social-ecological wellbeing. Actors in the NFP World, whether individuals, households, businesses, cities or nations, seek the level of material and financial wealth that provides for optimal human and ecological health. As such, **the 'work-watch-shop' spiral of despair has been replaced by the 'work-rest-zest' cycle of wellbeing.**

The cycle of wellbeing stems not only from what the NFP World actively creates, but also from what it simply doesn't do. The NFP World doesn't systematically create inequality. It doesn't encourage ecological destruction, war, exploitation, workaholism, and consumerism in the name of profit (as described in chapter 3). It doesn't tear communities and families apart for the sake of productivity and competitiveness. In the NFP World, businesses rarely market their goods and services in ways geared towards making people feel inadequate. We are no longer exposed to thousands of ads per day. Items don't break so easily (and intentionally). Fewer maxed-out credit cards are offered at cash registers. And fewer bad loans are lent. Basically, there is drastically less pressure to buy products and services we don't really need or want. In the absence of the detrimental impacts of the for-profit world, beautiful things are given the space to blossom, driving the virtuous cycle of less work, more leisure and greater wellbeing.

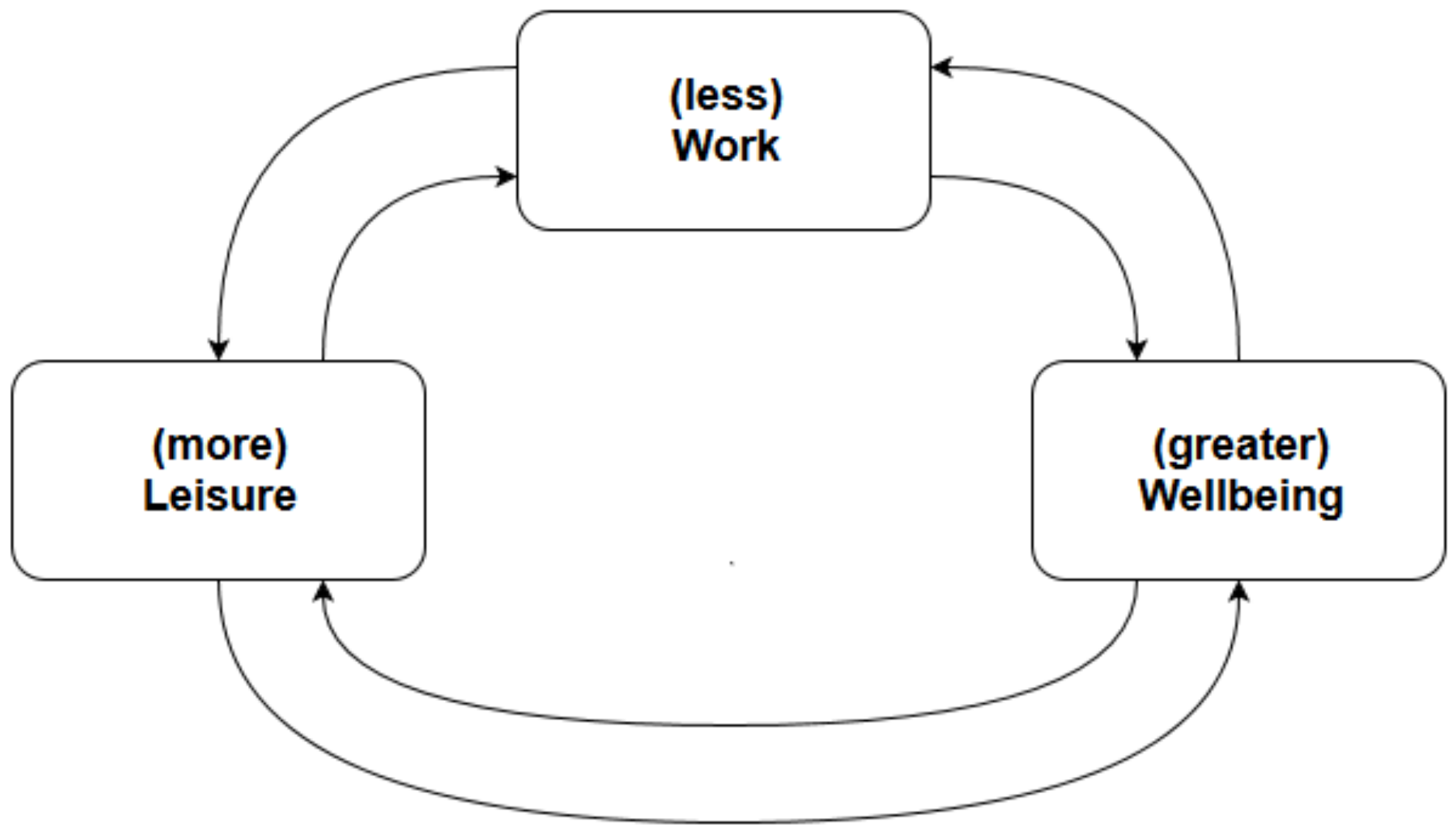

**The Cycle of Wellbeing in the NFP World**

**There is less work that *needs* to be done in the NFP World.** Greater localization and sharing means that needs are increasingly met outside the market, reducing the need for money and formal employment. Businesses no longer manufacture 'needs' to keep their profit margins (and the economy) growing.

Rather, the needs that can't be met outside the market, are efficiently addressed through the socially-oriented market. Thanks to the lean society model, the efficiency with which NFP enterprises respond to social and environmental challenges reduces the negative externalities that are often created by inefficient processes. Increased attention on environmental wellbeing leads directly to greater human wellbeing, improving conditions such as air, water and soil quality, ensuring more livable environments and nutritious foods.

Many NFPs produce goods and services that emphasize beneficial, long-term outcomes, which in turn, reduce the amount of work required by society. In healthcare, for example, there is a greater focus on preventative medicine, rather than the allopathic approach that dominates the for-profit healthcare model. This sensitivity to long-term considerations sees ecological design become the norm, as witnessed by extended product lifecycles through built-in resilience in manufacturing, and a focus on energy and waste reduction. This is a major shift from the short-termism of the for-profit world[685], and leads to significant, long-term cost savings as well as significantly less of a need for work[686]. It also results in a totally different experience for salespeople, increasing their wellbeing and intrinsic motivation when they sell goods that are durable and serve a deeper purpose, versus those they know will soon break and serve profit maximization.

In the NFP World, less work increases wellbeing rather than undesired unemployment. In the for-profit world, less work equals an unemployment problem. No need to create useless jobs in the NFP World to prop up employment in the market, because of the wealth circulation pump and needs being met outside the market, through community and family connections. In the NFP World, capacity of production is purposefully underutilized. While this will undoubtedly shock all kinds of economists, it's less shocking when you realize why: it's to ensure greater leisure and healthy ecosystems. Unemployment from labor surplus is a pronounced problem in systems based on scarcity. In the NFP World, the workweek is much shorter, so paid work is more easily spread out among the population and everyone spends less of their life doing paid work. 'Full employment' is less critical (a term that has often been used as a smokescreen for socially and ecologically harmful policies). While work remains an important mechanism for advancing social and ecological wellbeing and priming the wealth circulation pump in the process, the pump itself ensures people no longer need to work just to survive. Costs of living are lower. There's more average wealth, thanks to more equal pay and no wealth extraction siphon. But there's also greater security, training and support for the unemployed. **In the NFP World, every person's basic wellbeing is looked after, not just because of the moral imperative, but because the economy is built around it.**

There is enough work (although the amount is always falling). In particular, there is enough meaningful work (and adequate employment opportunities) to satisfy anyone seeking to be employed. The legacy of the for-profit world's destruction means that there is a great deal of generative work that needs to be done, socially and environmentally. That said, the full-time workweek of 25 hours [687] continues to fall over time, due to rising labor productivity. Combined with a shorter average working week of 15 to 25 hours[688], the NFP World offers enough work for full-time employment and adequate incomes[689], but not so much that work proves overwhelming. With wealth staying in local communities, jobs are more local. There is less travel time and less migration (due to less of a need to move for work), and thus less pressure that work-related travel puts on services and infrastructure.

---

685 Less than 1% of healthcare budgets in the U.S./Australia, for example.
686 See, for example: and: http://www.who.int/macrohealth/background/en/.

Importantly, people want to work longer into their lives (rather than retire as early as possible) because they derive a sense of purpose from their work as almost all work supports a social mission. Also because the working week is shorter and they've had more leisure time throughout their career (as compared to now), there's a more balanced approach to work throughout one's lifetime (rather than having to front-load one's life with too much work and retire from work all together later in life). In essence, people no longer feel burnt out by paid employment by the time they reach 65 years of age. Furthermore, because they've been living in a society that ensures optimal health, people tend to be in better physical and psychological shape in the later years of life, as compared to what is currently the norm in for-profit societies.

**With less work needing to be done, the elusive goal of greater leisure is finally reached.** The quality and amount of leisure increases because the NFP World creates the space needed for human flourishing. Working less gives us the freedom to be and to do.

Given the chance, and when basic needs have been met, people embrace this freedom, prioritizing quality time with family, friends and oneself over buying more stuff, and working more.[690] As the Peckham experiment shows, when we are stress-free, we have no shortage of crafts, hobbies and sports with which we enjoy engaging. There is time to explore healthy ways of meeting needs, outside the market, thereby reducing the need for work. We also have more time to connect with nature, with the NFP World allowing us to truly appreciate the incredible gift of life and biodiversity on this amazing planet.

With less work, we have more time in our lives for self-reflective processes. We have time to think about what is really important to us, what we value, what we prioritize, dream about, desire and need, and how we can best manage these considerations. Less work gives the space for us to learn and to follow our hopes, aspirations and passions. This raises the chances of finding meaningful work. And as experiments that provide all members of society with a basic income have shown[691], people still choose to work even when they don't have to, given the intrinsic rewards associated with contributing to something greater than yourself and the extrinsic rewards when that value is publicly acknowledged and celebrated. Income still serves as a motivator of work, but it's not at all the only source of motivation. Overall, we see a more satisfied and insightful population, which drives productivity gains both in the workplace and in communities.

There is also more time to be creative. Being an artist isn't restricted to just those who can find some way of making money from their creativity. Rather everyone has the time to undertake creative endeavors, such as writing, drawing, painting, sculpting, woodwork, designing, dancing, playing music, and acting. These are all important forms of expression that contribute to higher levels of mental health.

---

687 If labor is not spread evenly, a maximum number of hours of paid work might need to be instituted.

688 This does not consider domestic unpaid labor, which is likely more.

689 NEF paper on 21 hour work-week

690 (Ref: Schor

691 For example, basic income experiments…

Less work also creates space for us to connect, collaborate, help each other and understand our interdependence more deeply. As there is a lot less fear of being dominated, people experience and explore human connections more richly. We have more time to share our daily struggles with each other, and to receive the support we need to enable our ongoing personal growth. Importantly, we have more time for our loved ones. More time, energy and financial security exists to raise children and provide care for family members, accompanied by greater community support and the assistance of friends and extended family. **In the NFP World, the community ensures that babies are fed, held and loved, and elders are honored, respected and cherished.** And with people having more free time and the purpose-driven culture encouraging them to contribute to the greater good, the NFP World has higher rates of volunteerism, which loops back to relieve pressure on the market.

**The better balance between work, rest and play creates improved physical, emotional and spiritual wellbeing**[692]. But the reverse is also true: greater wellbeing improves the quality of our work, leisure and rest. When the for-profit pressures fade and the NFP era flourishes, most people will feel relatively deeper levels of life-satisfaction, joy, self-confidence, empathy, compassion, security, autonomy, connection, purpose, gratitude, and empowerment - in essence, a zest for life! With improved physical and emotional wellbeing, we relate more positively to others, making work and leisure more enjoyable for us all. We are more able to act in cooperative, empathic, mindful, and creative ways that contribute to the whole.

But greater wellbeing also reduces the amount of work needing to be done in the first place. Combined with financial equality, greater physical, emotional and spiritual wellbeing minimizes homelessness, mental and physical illness, addiction, violence, crime and incarceration[693]. This reduces the burdens on society, such as the costs associated with running rehabilitation and correctional facilities as well as homeless shelters and safe houses, and therein the amount of work that *needs* to be done. At the same time, having greater widespread health and wellbeing increases the available labor force, ensuring even less work per citizen, to achieve desired social outcomes.

Greater wellbeing reinforces contentment, based on an ethos of 'enough', allowing us to feel more deeply fulfilled. As we'll soon explore, this reduces our levels of consumption and, therein, the amount of work needing to be done to match market demand.

Benefits stemming from the cycle of wellbeing are magnified by various factors. The strongest is productivity gains, stemming from logistical improvements, technological advances, social innovations and enhanced workforce skills. Whereas improved efficiencies in the for-profit world often spell labor displacement and unemployment, **in the NFP World, increased productivity multiplies the positive shifts already occurring through the cycle of wellbeing.** When productivity gains drive wealth circulation within a relatively equal society, it makes paid work and volunteering more efficient (creating more time for leisure);

---

692 <E.g., the data on sleep and napping>. Not sure we need refs here – isn't it just common sense?
693 Wilkinson and Pickett's book The Spirit Level shows just how far more equality alone can contribute to better physical and mental health outcomes.

caregiving can become a little easier, and leisure more pleasurable[694]. Society benefits, as a whole, from high levels of ongoing education; open source design; mapping, geocoding and the Internet of Things; certain technological automation[695]; increased built-in resilience; and greater reuse, repair, and recycling via closed loop manufacturing. Associated job losses are counteracted by a safety net that has been strengthened by the additional wealth circulating through the system thanks to the productivity gain. What's more, with stronger, connected communities, people are cognizant of steering work towards their neighbors who might be recently unemployed, knowing this benefits everyone. The rebound effect[696] and efficiency shifting is minimized, because people are more aware of environmental impacts and ecological limits, and there's also less personal and social drive to use disposable income to fund mindless consumption.

As the global population naturally declines, from its plateau of 9 billion or so people, more space is made for gains in personal and ecological wellbeing to accelerate. This is in direct contrast to capitalism, in which the necessity of economic growth demands a constantly expanding population (consumer base). Thanks to modern medicine and the cycle of wellbeing, the population continues to age, however, this doesn't add pressure to the demands on the labor force, thanks to: gains from efficiency, savings being shared, better circulation of financial surplus in the common economy, livelihoods that depend less on money, and people wanting to work well into old age in purpose-driven positions.

Our environment remains under pressure. Even in a less consuming NFP World, 9 billion people demand a large throughput of resources.  People in the NFP World experience the catastrophic impacts stemming from ecological overshoot during the for-profit era. This undermines the cycle of wellbeing to some extent, but the NFP World also gives us a greater chance to ensure the conditions for the countless species with which we share this planet to flourish. The absence of the growth imperative allows efficiencies to actually increase. Overall, the planet has more breathing room to heal, and for other species to flourish. As such, the NFP World also creates the space for ecological stewardship and sustainability.

# The paradox of enough

But doesn't increasing wealth and leisure result in greater consumption and even more damaging environmental impacts? With more money to purchase things, don't humans use more resources? At first, this might happen to some extent, as part of a 'liberation effect' as societies achieve more economic equality (we will explore this in Chapter 6). However, in the longer term, the shift from an economics of scarcity and 'never enough' to one of abundance and 'enough' actually increases our ability to thrive in a world with biophysical constraints. This is the paradox of enough: **in an economy in which wealth concentrates, society consumes more but most people feel like they do not have enough material wealth. In an economy in which wealth circulates, society consumes less but most people feel they have enough material wealth.**

---

694 Think of what open access of movies has done for entertainment.
695 There are many parts of the caring/social economy that you can't automate.
696 Or Jevons Paradox

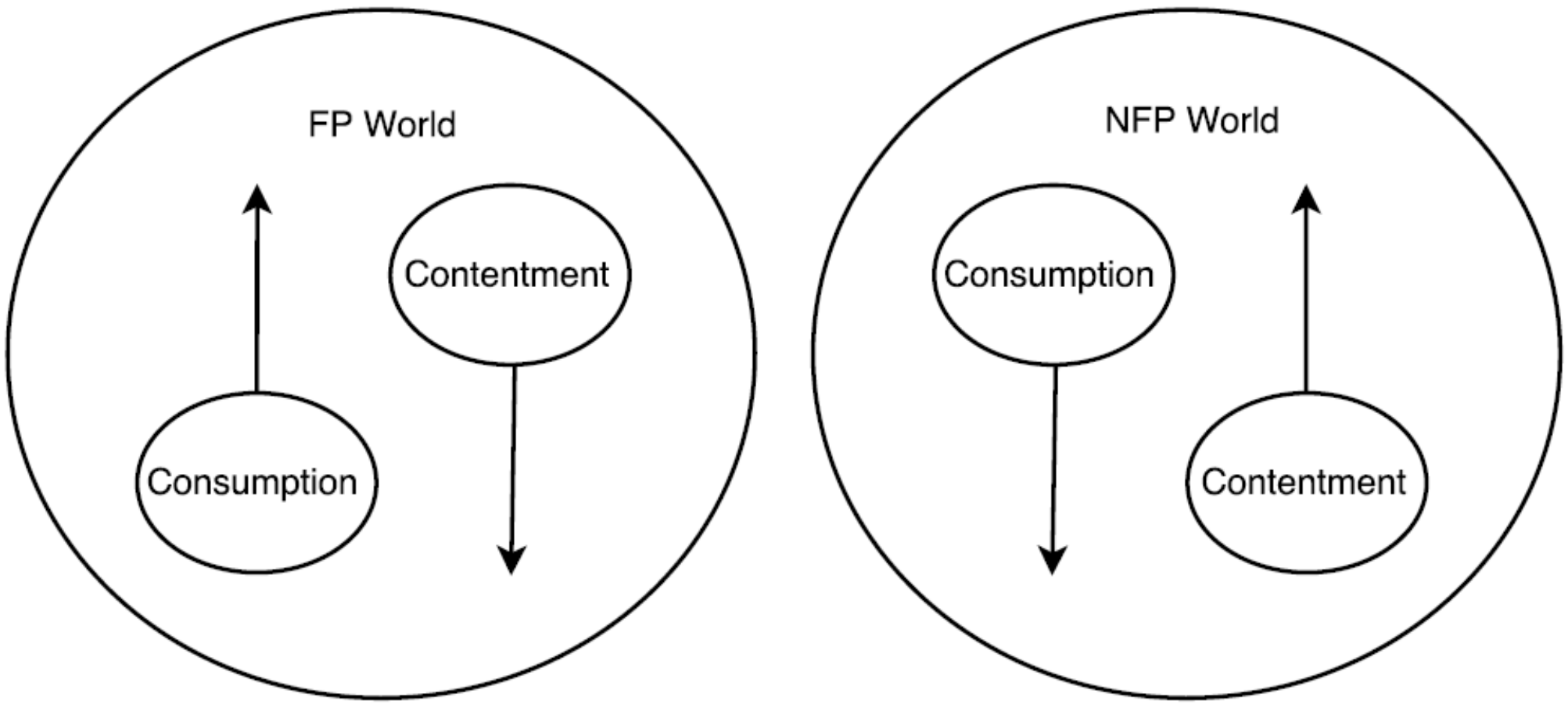

**The Paradox of Enough**

The main reason that per capita consumption in the NFP World enables a non-growing economy[697] is that the conditions that previously drove overconsumption have disappeared. Thanks to the end of the wealth extraction siphon, the economy no longer requires constantly expanding economic output or debt to compensate for the removal of wealth from the real economy. For the first time, our economy operates in harmony with the natural environment, rather than behaving as if we have infinite resources. With needs met in highly efficient ways, there is actually less overall economic activity. As we'll soon discover, the money supply is much smaller than in the for-profit economy, with an end to the superfluous, manufactured activity that is required by the creation of the money supply by for-profit banks.

The wealth circulation pump has reduced social stratification, envy, widespread economic insecurity, and hoarding, which enables the cycle of wellbeing and spreading a new story of prosperity, in which people experience the contentment associated with enough time, money, and rest. Under such conditions, people have the psychological, social and economic freedom to choose to buy less. They understand that endless consumption is not only unfulfilling, but has also been very harmful to social relations and the natural environment. Instead, they seek endless personal and interpersonal development. People appreciate what they have, rather than focusing on what they don't have. There is a sense of having enough material stuff. Businesses reflect this understanding in their practices. Greater transparency about the ecological impacts associated with products and services further spurs conscious consumption. Building on the trends of the early 21st century, people increasingly want to know the origin of the products they are purchasing, with re-localized production having assisted in significantly reducing environmental impacts. Manufactured needs, wasteful production, and mindless consumption are things of the past. Without profit as a goal, doing damage to humans and other species is unjustifiable, so 'economic externalities', like air pollution, are more naturally internalized. Even if a company does not automatically take its full ecological impact into consideration, when it learns of its mistakes from its stakeholders, community and public officials, it is much more willing to change its ways to align with optimal ecological outcomes than the for-profit businesses of the past. This is because harm to ecosystems eventually harms humans, too, so it does not help NFP businesses to achieve their mission. In other words, businesses in the NFP World have a lot more reason to think about

697 Also known as a 'steady-state economy'.

their near-, medium-, and long-term ecological impacts than businesses in the for-profit economy, simply due to their purpose and missions.

The market also no longer seeks to commoditize *everything*, because there is a widespread realization that many needs are best met in non-material ways, including stronger personal relationships that allow for greater rates of re-use and sharing. The willingness of the market to trade economic growth for the fulfillment of needs also facilitates more closed-loop systems design, reducing the ecological impacts of trade by slowing resource throughput, thereby enabling a steady-state, non-growing economy.

When the paradox of enough combines with the lean society model, the wealth circulation pump and the cycle of wellbeing, we see the makings of a new economy. But when it gets to the details of how the NFP economy actually works, what we discovered in this regard was nothing short of breathtaking. When we first began developing the NFP World model, we thought that the emergence of NFP business would make many of our existing structures obsolete. Nothing could be further from the truth. Almost everything that is unhealthy with our present, for-profit economy, becomes healthy under NFP conditions, from fractional reserve banking, to compounding interest, debt, advertising, lobbying and even corporate bailouts!

We begin our deeper exploration with what sits at the heart of any economy: finance.

# The abundance of NFP finance

When all banks are not-for-profit[698], something profound happens with money creation, debt and interest. Recall that, through for-profit banking, money is constantly extracted from the real economy to the elite economy. Thirty-five to forty percent of the money we pay for goods and services goes to bankers, financiers, and bondholders via interest[699]. The scarcity this creates in the real economy demands an ongoing expansion of debt by the average citizen (and government), who can't afford to fully service their loans without taking on further debt[700].

---

698 The range of banks considered NFP include: credit unions, savings and loans associations, building and loans associations, community banks, central banks and public banks. No matter what form they take, though, all NFP banks share the same core traits: they exist solely to meet their customers' or members' needs; they must reinvest 100% of their profits into that mission; and they are without private ownership (in the sense of appropriation rights).
699 <REF: Ellen Brown>.
700 Known in economics as the 'impossible contract'.

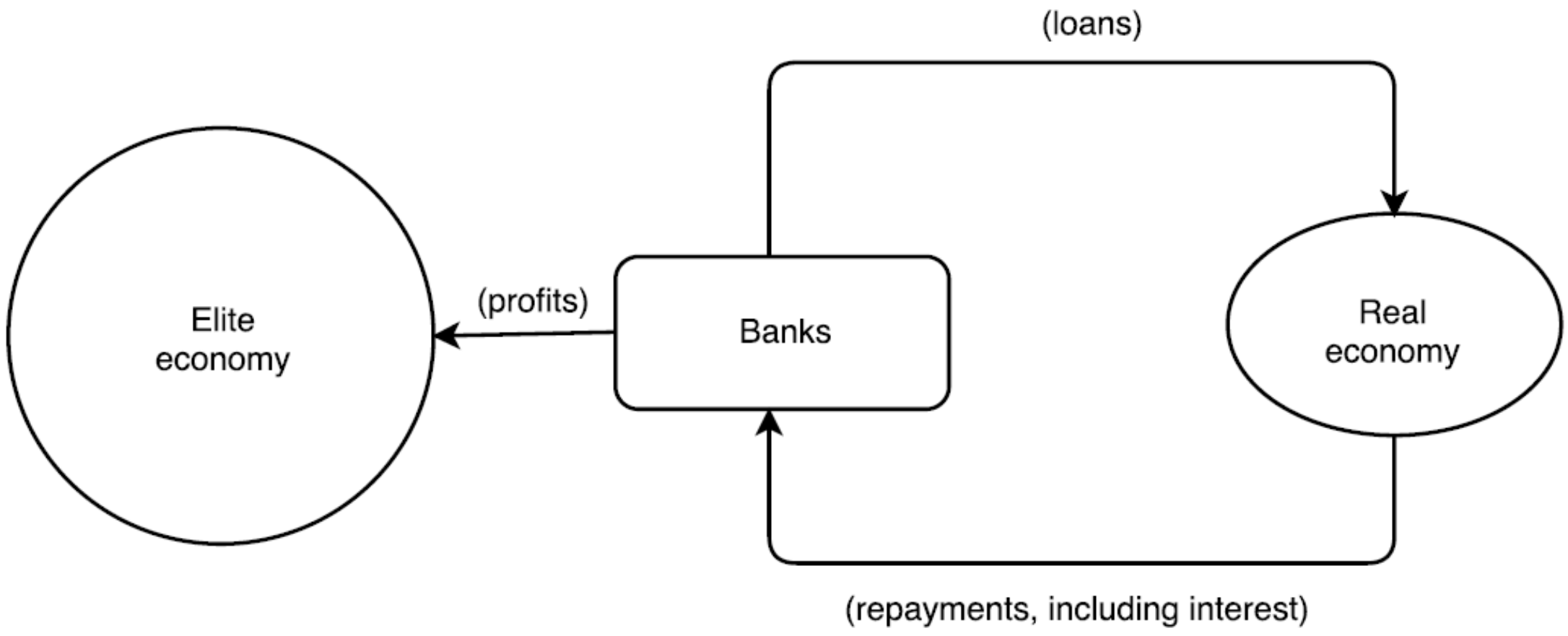

**The Expansionary Tendency of Debt in the FP Economy**

However, when all banks are NFP, then any financial surplus in the banking system (stemming from fees, interest on loans or returns on external investments), returns to the real economy via the wealth circulation pump. Because average levels of household, community and government wealth are much greater in the NFP World, debts are more easily and quickly reconciled, similar to what happens in *It's a Wonderful Life*[701]. And in a world focused on addressing social-ecological needs, with more needs met outside the market, the reduced costs of living make taking on debt less necessary in the first place and, with higher rates of socio-economic equality, debt is also easier to pay off. Combined, these factors create financial abundance, removing the expansionary imperative that debt creates in a scarcity-based, for-profit economy.

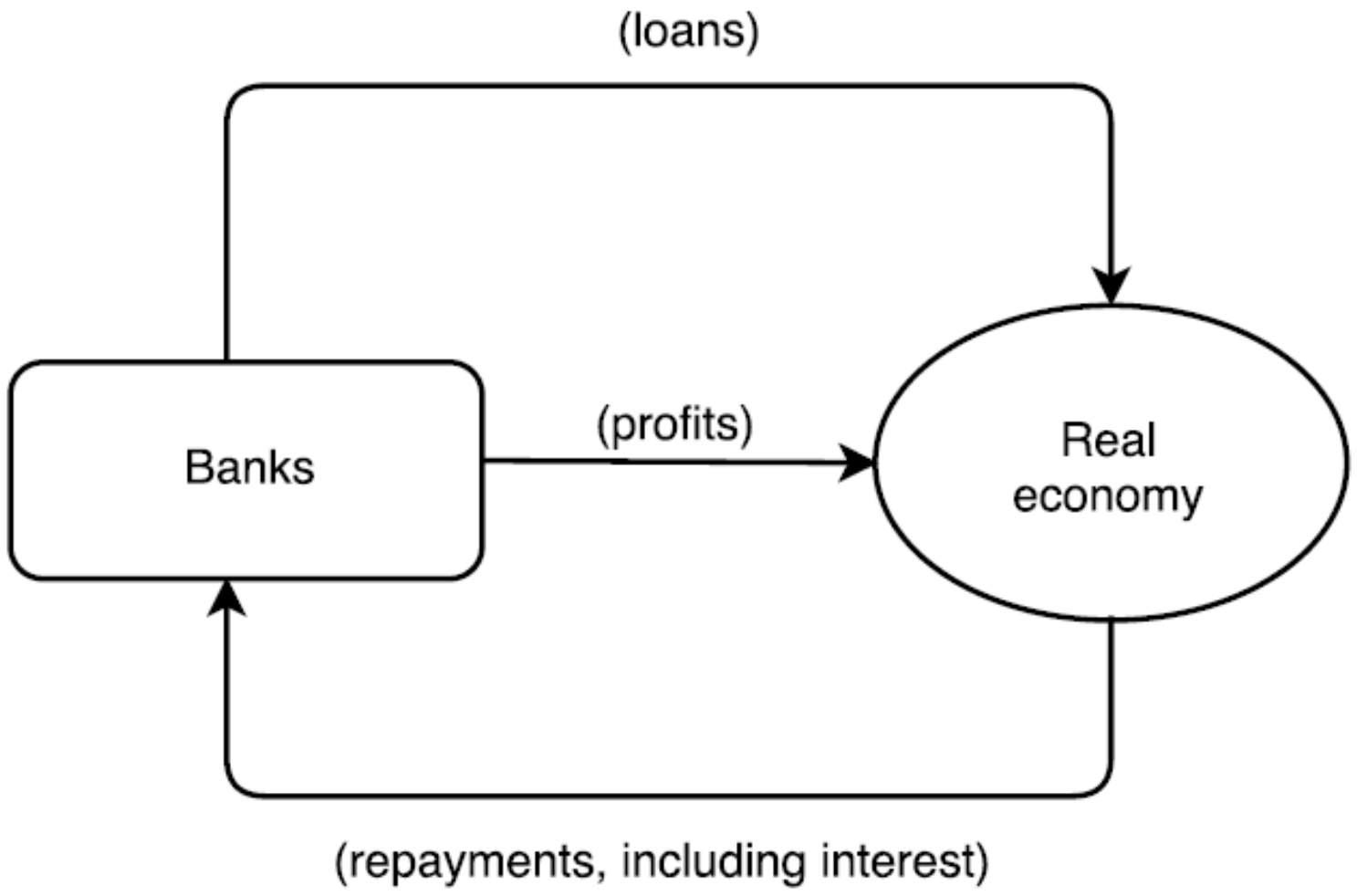

**The Reconciling Tendency of Debt in the NFP Economy**

The method by which money is 'created' doesn't change in a NFP World. Not-for-profit banks loan money into existence, as debt, just as all banks do in the current for-profit economy. But while debt remains an undesirable, albeit more temporary burden for the indebted, private and public debt is positive in the NFP economy. Not only does it promote financial liquidity and the ability for households, communities, businesses and governments to finance the servicing of needs, it also represents social assurance, trust, interdependence and community confidence in the economy and the ability of debtors to repay their loans.

[701] Additionally, there is the advantage of retained earnings and federal support for credit unions to assist with 'runs on the bank'.

Fair interest on loans remains an important component of a healthy, NFP banking system. Compounding interest discourages reckless borrowing and encourages the borrower to repay loans efficiently. Interest also rewards the lender for taking a risk (like a fee for the service of providing the loan), and compensates for any inflation within the economy, although, as we'll soon see, this holds less relevance in the NFP World. Yet, the motivation to gain personally from interest payments disappears when the bank isn't shareholder-owned. Rates are calculated to account for administrative costs (including wages of bank employees), levels of risk, and the development of capital reserves. Nothing more. No profit-maximization and no wild speculation, with the hopes of getting rich at the expense of debtors. Along with the abundance of capital available for investment, as well as market competition to provide the highest quality services to customers, the NFP approach to banking makes unfair interest rates unlikely. But, if needed, rates can be regulated for upper maximums, as in many U.S. states, where it is illegal to charge more than 10 percent annual interest on consumer loans.[702]

Furthermore, because NFP banks only exist to provide high-quality financial services to their customers, they are responsible lenders. With stronger ties to the community and borrowers, they're able to assess risk and work out terms and policies that ensure lower default rates. Not-for-profit banks don't intentionally give bad loans and this can be seen in our present economy in which credit unions offer fairer interest rates and were not nearly as tied to bad mortgages as for-profit banks in the 2008 crash[703]. And their not-for-profit ethic ensures NFP banks don't sneak in new fees or seek to extract as much wealth from customers as possible. No one would benefit from unnecessary fees but the customers, when they get a rebate at the end of the year, so it makes no sense to charge them in the first place.

Combined, these features restore banking to its naturally valuable function: providing households, businesses and governments with a secure place to store savings, and with the loans to start projects and businesses or address personal needs. **An entirely NFP banking sector keeps the circulation of financial resources moving throughout the economy, for the benefit of all.**

When banking is entirely NFP, many of the critiques of the modern money system fade. There is no need to create a moneyless society (e.g., the Resource-Based Economy proposed by the Zeitgeist movement), or shift the money creation process from banks to governments (as proposed by organizations like Positive Money). Or even to see an end to central banks, such as the U.S. Federal Reserve. Nor is there a need to return to the gold standard (Bretton Woods system) or develop a 100% reserve banking system. And there is no need for negative interest rates (like demurrage currencies proposed by Charles Eisenstein and others), non-interest banking (as practiced by JAK Bank), or an end to compounding interest.

---

702 (Ref: Lectic Law Library)
703 Ref: Credit Union deposits article and CUNA PowerPoint slide 14, My Credit Union.gov).

In the NFP World, money is restored to its original purpose as a store of value and a medium for exchange. Fiat money (money with no physical resource underpinning its value) enables widespread trading. The fractional reserve banking system provides valuable liquidity. And compounding interest on loans ensures the efficient servicing of debts. When the interest is used to benefit the wider community, it is not a bad thing.

That said, alternative approaches to finance can happily co-exist within a NFP system. The JAK Member's Bank in Sweden, for example, having successfully conducted interest-free, full-reserve banking since 1965[704] would operate just as fine (and might gain even more members) in an NFP World. And interest-free, Islamic banking (Sharia compliant) can continue just as it is presently practiced in much of the Islamic world, as long as there is no private equity involved.

Indeed, the NFP banking system is complemented by a suite of approaches to monetary transactions. Bitcoin and its ethical counterpart, Faircoin, may be long gone by the time we have an NFP World, but their descendent crypto-currencies enable an important aspect of the economy, with technologies emerging from the block-chain that provide a highly secure and transparent ledger for many forms of economic transactions.

For communities that have difficulty attracting or retaining adequate wealth, community currencies like the Brixton Pound offer a practical way to enhance connections within the community, while encouraging local production and consumption and, thus, building resilience into the local economy. And the Local Exchange Trading System (LETS) continues to offer residents effective ways to exchange goods and services with their neighbors. In fact, most new economy ideas and innovations that exist today are compatible with the NFP World model, because with the shackles of profit-maximization and private business ownership gone, purpose-driven innovation is embraced and allowed to flourish.

Inspired by leadership in Germany, Brazil, Russia, India and China[705], public banking is one such innovation that is crucial to the NFP World. Public banks are financial institutions run by governments at the national, regional, county or municipal levels. While variations exist[706], most public banks have similar functions:

- they offer a repository for government revenue;
- they provide the government, or specific industries with low cost capital to develop infrastructure;
- they have a mandate to lend counter-cyclically, i.e., when there is a block in liquidity; and
- their profits offer a substitute for tax revenue.

Recognizing the value of a diverse, decentralized banking system, public banks work in tandem, not competition, with private-sector NFP banks. As public banking specialist Ellen Brown notes, public banks help collateralize community bank loans and assist with regulatory compliance. The U.S.'s only public bank, the Bank of North Dakota, Brown says "…directly supports community banks and enables them to meet regulatory requirements such as asset-to-loan ratios and deposit-to-loan ratios… [I]t keeps community banks solvent in other ways, lessening the impact of regulatory compliance on banks' bottom lines."[707] An ecosystem of NFP entities managing money creation minimizes both the statist problem of concentration of power and corruption, as well the capitalist problem of for-profit banks expanding the money supply to extract greater wealth for owners through lending.

---

704 (Ref: Wikipedia article and emails with JAK).
705 As of 2016, approximately 40% of all banks in the world are public banks <REF: Ellen Brown>
706 See: http://republicirelandbank.com/?page_id=2

With public banks playing a supportive role, and the wealth circulation pump ensuring financial liquidity, central banks are less necessary. However, their newfound independence[708] makes them a valuable check and balance for the wellbeing of the NFP economy, and they remain the lender of last resort for large, private-sector NFP banks and public banks. But, rather than focusing on how to boost consumption and GDP growth, their mandate for ensuring liquidity is driven by the desire to improve social wellbeing. In this sense, they still assist with inflation rate targets by helping to manage the money supply and setting central interest rates[709], but they do so with a different end goal.

Deflation is certainly less of a concern in the NFP World. Higher wages, adequate employment opportunities, non-marketized ways of meeting needs and the ongoing impacts of the wealth circulation pump (and absence of the elite economy) make deflation highly unlikely in the NFP system[710]. But aren't the conditions we've outlined a recipe for inflation? Here, the overall reduction in the money supply (debt) changes everything. Due to economic abundance, a bank's creation of money, as debt, only increases the money supply temporarily. This is because this debt can be efficiently serviced, often without the need to create more debt in the process[711]. And when the economy doesn't have to incessantly grow, there's less need to expand the money supply via quantitative easing and other such measures[712], reducing the possibility of inflationary pressure. A naturally declining population, less speculation, more socio-economic equality, and the drastically smaller size of the global economy (in terms of both money and production), also means less vulnerability to irregularities or market failures, further offsetting inflationary pressures. Dramatic boom-and-bust cycles end because asset values don't have a natural tendency towards over-valuation.

Additionally, the steady circulation of wealth, along with considerable re-localization of manufacturing and trade, greatly reduces the volatility of prices and market demand. With a reduction in the overall number of hours employees work to earn an adequate paycheck, and NFP companies caring more about employee wellbeing as an internal aspect of their social missions, employment and wages are more stable, contributing to reduced variability in costs. This means inflation is contained, and deflationary and inflationary spirals are extremely rare and less severe in the NFP World.

## Investment and savings

---

707 <REF>.

708 As for-profit banks disappeared, so too did their influence over central banks. In any system, the independence of central banks is crucial, primarily because monetary policy is more reliable when it is able to be distinct from the fiscal policies of an incumbent government.

709 Yet central banks have less influence over public banks, given public banks rarely have liquidity problems and, therefore, don't often borrow from central banks. This makes the central bank less able to guide interest rates within a nation.

710 Especially given that the wealth extraction siphon and the scarcity it created, no longer exists.

711 There are fewer delinquencies, fairer interest rates, and better payback times.

712 Via, for example, a central bank buying securities in the open market, lending new money to NFP banks or buying assets from NFP banks, or a government selling bonds to spur growth.

In the NFP World, social entrepreneurs and governments still pitch proposals for investment. It is still 'make or break' for many projects, with pressure to generate revenue as soon as possible. Yet, less overall investment is required in the NFP World, given the economic growth imperative has disappeared, and capital requirements for most forms of business startup are incredibly low (relative to today's levels).

**Debt has replaced equity as the primary vehicle for financial investment**[713]. Loans and bonds with fixed rates of return continue to drive new NFP business development, expansion and working capital, given individuals can't hold equity in an NFP[714]. Government bonds fund public works and community development, while private lending provides individuals and families with money to meet their needs.

In the NFP World, the ethic of enough is connected with the desire to do good. **Understanding of the 'circle-round' effect has replaced faith in the 'trickle-down' effect.** People now seek *social returns* on their investments much more than profit-maximization. Because NFPs are purpose-driven and most are able to offer tax deductions on donations, people are generally comfortable providing monetary and in-kind support to help a new NFP business start if they're convinced that the organization will serve the community's needs; all without the desire for a financial return on investment.

That's not to say people don't invest to ensure a more financially secure, comfortable future for themselves and their families. They do. It's just that, **in the NFP World, the need for personal wealth creation from investments (and savings) has decreased dramatically**. In retirement, for example, less savings are needed due to the reduced costs of living. Retirement is more affordable in the NFP World. Fewer activities require money because of the increased prevalence of caring, sharing, and collaboration in the community (de-marketization). Communities and families have a greater capacity to care for one another free of charge (with less time spent working for pay). As community ties become stronger in the NFP World, the wider community is more prone to look out for its elderly members. There is also less of a need for money because there are much lower levels materialism in society to drive unnecessary consumption.

Retirement is also cheaper because healthcare, medicine, medical equipment, and other basic necessities are priced for accessibility rather than profit-maximization. There is greater access to public and private services; an increased number of not-for-profit organizations offer services specifically for the elderly, at accessible prices. On top of less of a need for money, people want to work longer into old age in the NFP World, because they work for businesses and orgs that they believe in. They are not feeling burnt out from working all their lives at jobs they don't really enjoy or care about, as is a common occurrence in the for-profit economy.

Another aspect of the NFP World that reduces the need for savings is that the elderly are generally healthier because they've enjoyed their life's work, they've led a more balanced life, they've likely had a better diet and more time to exercise, and they've had less debt and less stress. The healthcare system is focused on prevention and healthy lifestyles rather than medication. Through the NFP system, gains in automation and technological efficiency have translated into real increases in quality of life, particularly support for the elderly (e.g., real time remote monitoring of health, connected to tele-medicine). So while people may live longer, they'll be less of a burden on the healthcare system. There is also a stronger public safety net and more volunteerism, because of the purpose motive, the not-for-profit ethic, the story of interconnectedness, and people having more free time. Volunteers can play a big role in meeting the needs of the disproportionately large elderly population[715].

---

713 Equity-based investment can still happen through personal investment in physical assets and small businesses/sole proprietorships.

714 Because they are not-for-profit institutions, rather than private individuals, governments, nonprofits and NFP enterprises, can have equity in other NFPs. And NFP banks may use equity as collateral for loans; but, if there is a default on the loan, then any equity will be held by an NFP institution (i.e., the bank).

In terms of investment, lending (as compared to equity-based investment), generally offers long-term stability for borrowers and the economy, given it is a source of fixed income and is less prone to speculation. While a spectrum of risk remains[716] (allowing those less concerned about the certainty of their long-term savings to support more risky ventures that may have a higher rate of return), destructive speculation is largely absent; and the spectrum of possible returns is narrower. This is especially due to the abundance of investment capital and because, overall, most ventures are averse to reckless risk-taking, now that profit-maximization is no longer the purpose of business development. In fact, with the average individual having more money, a 0% interest rate on savings at the bank becomes justifiable, given less debt in the system means less overall lending.

Future security through investments, however, has a new ally: the triple dividend. Imagine you invest in an NFP that is developing a new retirement village in your neighborhood; three kinds of returns accompany that investment: a fixed financial return, either from revenue associated with the village's development, or from other streams of revenue the NFP uses to service its loan; a social return, where you feel good about strengthening services for the elderly within your community; and a personal return - in your older age you can move into the village.

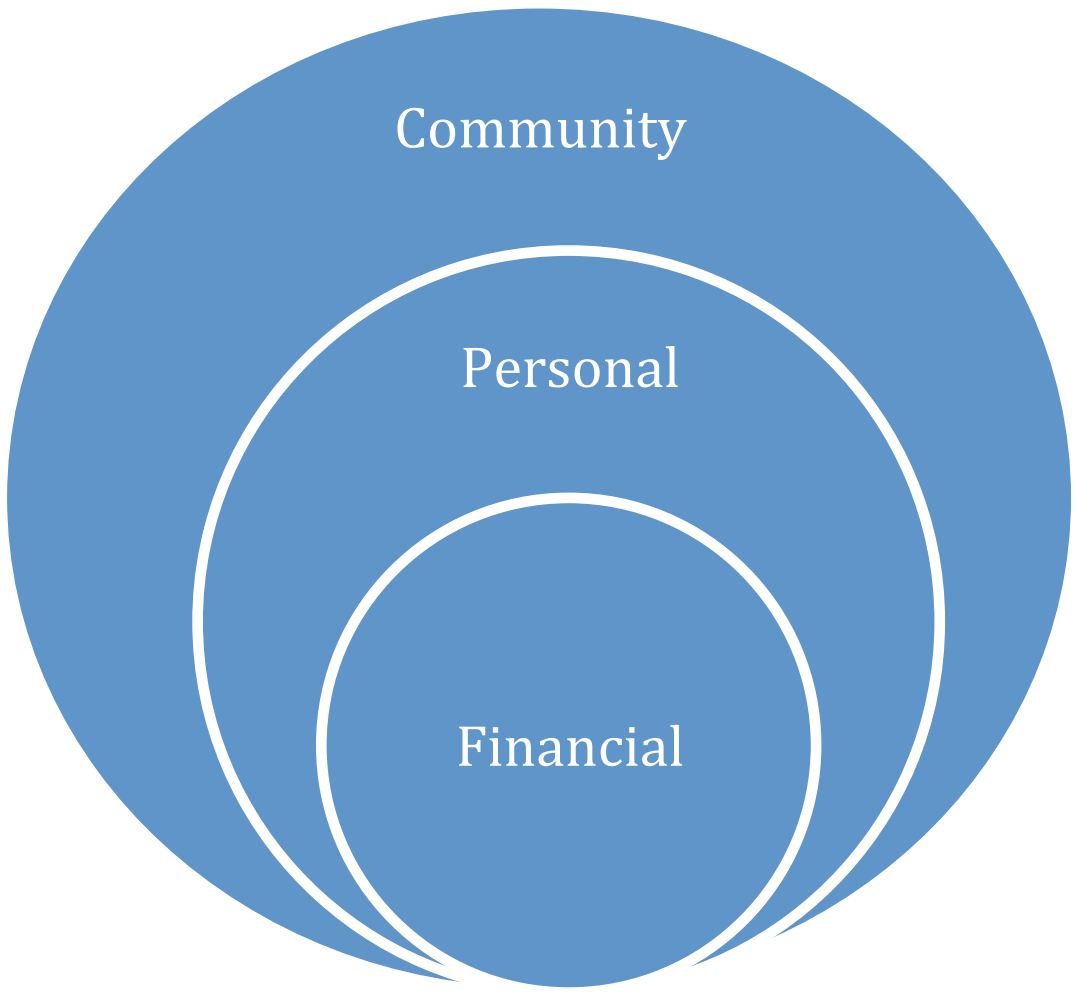

**The Triple Dividend Investment in the NFP World**

---

715 For example, 'Meals on Wheels' (with this company, in Australia's NSW, cross-subsidized by its own insurance agency).

716 There are obviously still levels of risk associated with investment in the NFP World (a loan or bonds may be issued for a heavily indebted company, a company losing money very fast, a company entering a risky market, or an otherwise poorly managed company). But most things are less risky and more insured in the NFP World.

In lending to an NFP, investors are truly invested in the outcome (ensuring high levels of local investment in the NFP World). That is why, in the NFP World, people are more likely to match their financial investments with contributions of time and energy. Indeed, businesses and governments regularly connect with and provide updates to their investors, in part because these people are often wonderful patrons and ambassadors for the projects in which they are invested. However, for NFP projects in which an investor's stake is purely financial, the investor's ability to influence decision-making is legally restricted[717]. This is in stark contrast to our present, for-profit world, in which the owners of capital often have voting rights and a greater ability to influence a company's direction than the workers or the company's board (with shareholder decisions and expectations typically favoring profit-maximization at any cost). Thus, **in the NFP World, investment capital is restored as a tool, rather than a master.**

If your investment in an NFP does not go well, it's a totally different situation to when that happens with a for-profit investment. This is clear when one thinks of the good that might have been achieved in the process as well as the fact that your money went to other NFPs (and didn't get extracted). In the NFP World, "I lose, we win!" replaces what we're so used to "I lose. I LOSE!".

Ironically, while the money supply is smaller in the NFP World, savings are more widespread. And while younger people still need to borrow more than the older generations who, in turn, provide most of the capital for lending, it is much easier (and maybe even common) for average people to transition to being a lender, rather than a borrower, as they grow older in the NFP World.

The wealth circulation pump provides citizens with a greater capacity to save, particularly due to higher 'take home pay' and reduced costs of living. With greater average wealth, governments have greater reserves, and more people can afford small to medium-size investment risks on projects with social outcomes. And, thanks largely to the legacy of the digital revolution, the power of investing is widely dispersed.

Through what mechanisms do investments in communities, public goods and NFP businesses occur? As part of the community economy, lending circles, such as the Tanda in Latin America[718], remain a widespread, informal means of lending for small-scale personal and business loans. But digital lending platforms, like Kiva Zip[719], are now pervasive, making peer-to-peer exchanges fast and simple. These digital platforms, especially those intersecting with cryptocurrencies[720], make the pooling of micro-investments possible, such that you might lend just a few cents to hundreds of organizations, with the simplest of instructions. Other, standard avenues continue to exist, including microfinance and small-scale lending provided by NFP financial institutions, and foundations offering program related investments[721]. At the larger end of the direct lending scale, commercial banks continue to use customer deposits for personal lending and business investment.

[717] And strong 'conflict of interest' policies remain in place.
[718] https://en.wikipedia.org/wiki/Tanda_(informal_loan_club)
[719] https://www.huffingtonpost.com/author/kiva-zip
[720] Because the absence of fees makes small transactions viable.
[721] https://www.irs.gov/Charities-&-Non-Profits/Private-Foundations/Program-Related-Investments

Crowdfunding is commonplace, whether it is the use of online platforms to raise funds for projects, loans to a cooperative from its members, or Community Public Offerings[722] that enable non-accredited investors to support local NFP ventures. Collaborative approaches for seeking investment such as collective impact, whereby NFP businesses raise money as a group by demonstrating how they will collectively address a social challenge in an integrated way, are the norm.

Given the absence of individual ownership in companies, **bond markets have replaced stock markets.** Primary markets (the place in which a company firsts lists for trading), assist NFP corporations and federal, state, county and municipal governments to raise capital[723] for expansion. Such markets simultaneously disperse risk while giving investors a chance to discern which socially- and environmentally-focused businesses they'd like to see flourish. Secondary markets (where bond trading happens), provide citizens and NFP companies with additional fixed-income opportunities. While some physical infrastructure continues to support the regulation of trading, most activity happens through online markets.

Many aspects of the financial world that have been labeled as undesirable or unethical since the great crash of 2008, are actually integral parts of a healthy economy in the NFP World. This includes things like investment banking and derivatives. Lambasted for driving insatiable greed in the for-profit world, **investment bankers have a social function in the NFP World**, assisting NFP companies to raise capital, managing fixed-income investments for clients, and brokering mergers and acquisitions between NFP corporations.

Without stocks as an underlying asset, derivatives (especially futures and hedging) not only become less dangerous, they actually provide a stabilizing force in the NFP economy. Take, for example, the funding mechanism for Community Supported Agriculture (CSA). In a CSA arrangement, members pay the farmer for their produce before the crop is sewn. If the farmer's harvest is bountiful, members reap the rewards via large produce boxes. If the crop is small or destroyed, the customer suffers a small set-back, receiving reduced or no produce (although the market provides many substitute options). But the farmer, who might otherwise have gone bankrupt, keeps the pre-payment and can afford to sew crops the following season. While there is a speculative aspect to these kind of futures, they add great value to the NFP economy, providing a form of insurance for businesses whose processes rely on risky variables and the need to budget for fixed costs[724]. To ensure that derivatives remain primarily steered towards this beneficial function, regulators ensure that any underling assets can be assessed for their real market value.

While debt is the main instrument in NFP financial markets, private accumulation can still happen via equity investment in derivatives, currencies, commodities and land. However, with the NFP World meeting people's social needs and replacing the cultural narrative of 'never enough' with 'enough', the profit-motivated, speculative behavior that we currently witness in financial markets is no longer common or socially acceptable. And the wealth circulation pump ensures private gain is more readily returned to the real economy.

[722] https://hatchoregon.com/im-seeking-funding/raise-capital-public-offerings/

[723] No matter who owns a bond, the NFP company or government still has the original money raised (it hasn't been extracted via an IPO to original founders, for example).

[724] For example, fluctuations in currency or interest rates, although the likelihood of these fluctuations is reduced with the re-localization of trade.

With a new ethic, most currency exchange happens due to need, not speculation, such as hedging currency risk for an NFP business. The residual ability to gain from differences in currency (i.e., arbitrage) has shrunk significantly, due to:

- reduced overall international trade;
- a global reserve currency;
- reduced currency differentiation and fluctuation;
- the rise of global digital currencies and reduced exchange costs;
- greater regulation of the foreign exchange markets, including a high-frequency trading tax; and
- banks and other companies that offer exchange services being NFP.

Commodity trading tells a similar tale, with one key difference: open-source distributed manufacturing, utilizing locally-sourced substrates for production, sees a major reduction in international commodity trading. Local agricultural farms and seasonal diets are more common, due to increased awareness of environmental problems, as well as taxation and transport costs.

Reduced speculation is most visible in the mainstream approach to private property. Thanks to the wealth circulation pump and purpose-oriented ownership frameworks (e.g., community land trusts, numerous NFPs providing assistance with housing affordability, and NFP banks providing more affordable mortgages), home ownership is widespread and property ownership as a purely speculative investment is minimal. Land, including commercial real estate, is increasingly owned by NFPs and the state, ensuring accessibility while reducing speculation in property markets. And because there is dramatically less speculation in the housing market, prices are more stable, better reflecting real value, rather than the inflated (or deflated) expectations of investors.

Given everyone is seen as interconnected and part of a greater whole, private land ownership and belonging to the rentier class is a lot less popular. The majority of people view the private profiting from rent as socially unacceptable, but the rental market still seen as fine, especially when property is owned by NFPs. With less speculative asset investment, renting becomes more service-based, and remains an important step for many to home ownership (or for others, it offers more choice in their lifestyle), and there are social movements that put pressure on people or NFPs who are charging exploitative rent. Housing affordability brings a lot more status and admiration in the NFP World than holding onto the land and living off the rent. Where necessary, there are also rent controls.

## Public finance

**The NFP World requires significantly less taxation than the for-profit economy**, allowing business and personal income taxes to be much lower. It does this thanks to the efficiency gains from the lean model (very little need to redistribute wealth), and the reduced burdens on the system thanks to the cycle of wellbeing.

**Lean model with built-in affordability**

In the for-profit world, taxes are needed to pay for the burdens the market can't, and often won't, bear. The NFP market meets many needs in much healthier ways, allowing for a true partnership role between the state and market. Public finance benefits from important new efficiencies in the NFP World, and the costs to government of addressing for-profit business externalities, such as cost-cutting layoffs or environmental pollution to maximize quarterly earnings, are also greatly reduced. Closely linked, the subsidies to these companies have disappeared while, as we'll soon discover, costly government bailouts of companies 'too big to fail' are much less likely and become a totally different equation in the NFP World.

With not-for-profit businesses at the heart of the economy, there is less pressure on governments to fund (and provide) public goods, especially in the social services field. Networks of efficient, decentralized NFPs have replaced the heaviest, most inefficient government-run programs in the NFP World, and they address all segments of the population, leaving few gaps in the market. As a result, fewer people rely on public services, and the taxes that governments previously levied to fund nonprofits are no longer as necessary.

The privatization[725] and marketization of certain services that were previously considered public goods is possible because more people are able to pay, with mechanisms for accessibility built into NFP businesses themselves to assist those with limited financial means. And governments still serve the important role of keeping NFP enterprises accountable and making sure that they provide essential services in an efficient manner.

But governments are also more enterprising in the NFP World, using self-generated revenue as a means to reduce the level of taxes required to fund public services. This revenue is often generated within a service that may involve sliding scale models for fees[726]; for example, cafeterias within public hospitals, gift shops in public museums, or laundry cleaning by state-run prisons. However, cross-subsidization is common for services where extensive revenue is harder to generate directly, such as policing, fire services, the military, and the management of national parks (although each of these involves greater community volunteering and management). Key revenue sources for governments include public banks; municipal utilities[727] such as water, electricity, gas, Internet, phone, waste and sanitation; cemeteries; gas stations; housing and commercial real estate[728]; cafes, pubs and restaurants; liquor stores[729]; and car sharing programs. These come in addition to more traditional forms such as parking meters and road tolls; business licenses, permits, fines and penalties.

---

725 NFP privatization, of course.

726 In most systems you have to show proof of income because it is assumed that nobody wants to pay more than others for the same product. That's the voice of Homo economicus again. It's actually surprising how many people are happy to pay a bit more when they know it will allow less fortunate people to also have access to something. This has been widely documented in participatory pricing case studies (Ref: Norms, Moods and Free Lunch; Shared Social Responsibility).

727 "In Los Angeles, for example, the Department of Power and Water contributes about $190 million per year to the city's revenues." <REF>.

728 U.K.

729 Pennsylvania and Sweden

A smaller role for government as a service provider means less money is needed to support government bureaucracy. And with many of the processes that government oversees, such as taxation, penalties, licensing, medical payments and transportation, the automation of systems has reduced costs dramatically. Reduced costs translate into less need for governments to take on debt, which means less taxation is needed in order to repay debts.

The cycle of wellbeing makes the burden on the public purse much less costly. The burden on the healthcare system, for example, is greatly reduced when people work less, have more leisure time, and the food, health, and education industries are run by purpose-driven companies. With wealth circulating through the economy, NFP insurance can provide a strong private-sector safety net. Low levels of unemployment also improve public health. There is also much greater security, training and support for the unemployed in the NFP World, through both NFPs and the government. Overall, the taxes required to fund welfare payments is massively reduced. The expanded community economy also removes significant pressure from government in areas such as childcare, where families and communities are more able and willing to look after children, reducing the need for subsidies and tax credits.

While in the present economy, there are many social services that the for-profit market can't or won't bear because they aren't profitable enough and those fall on the state or charities to provide. In the NFP World, these social services are readily taken up by the purpose-driven NFP market or are supported by the financial surplus of NFP businesses via philanthropy to charities.

**Taxes**
Thanks to the wealth circulation pump, most people have more money with which to pay taxes. Consider, for example, that inequality costs the U.K. Government up to £32bn, annually, in welfare payments[730].

Approaches to taxation continue to vary greatly between nations. Adequate liquidity in the system enables governments to more easily discriminate with taxation policies. However, the general shifts in *what* gets taxed are almost universal. Tax exemptions remain for NFP businesses, with tax on for-profit activities maintained to discourage profit-maximizing behavior, and encourage NFPs to have mission-driven subsidiaries[731]. Thus, given most companies are NFP, revenue from business taxation[732] is reduced, although employment taxes on all businesses remain (e.g., social security payments, healthcare levies and worker's compensation in the U.S.).

[730] http://positivenews.org.uk/2014/positive_perspective/15042/lots-less-solve-income-inequality/.
[731] In countries such as the U.S., in which unrelated business income remains taxed, this is the only way the income from such subsidiaries could qualify for tax exemption.
[732] Including, amongst others, business income taxes, payroll taxes, commercial land taxes and goods and services taxes).

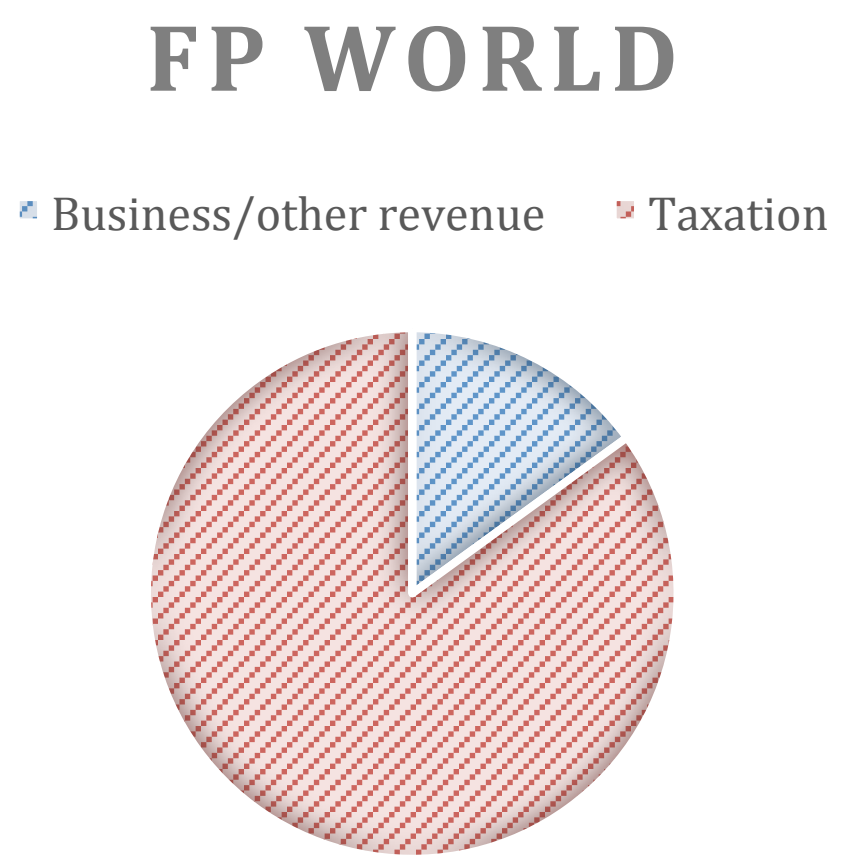

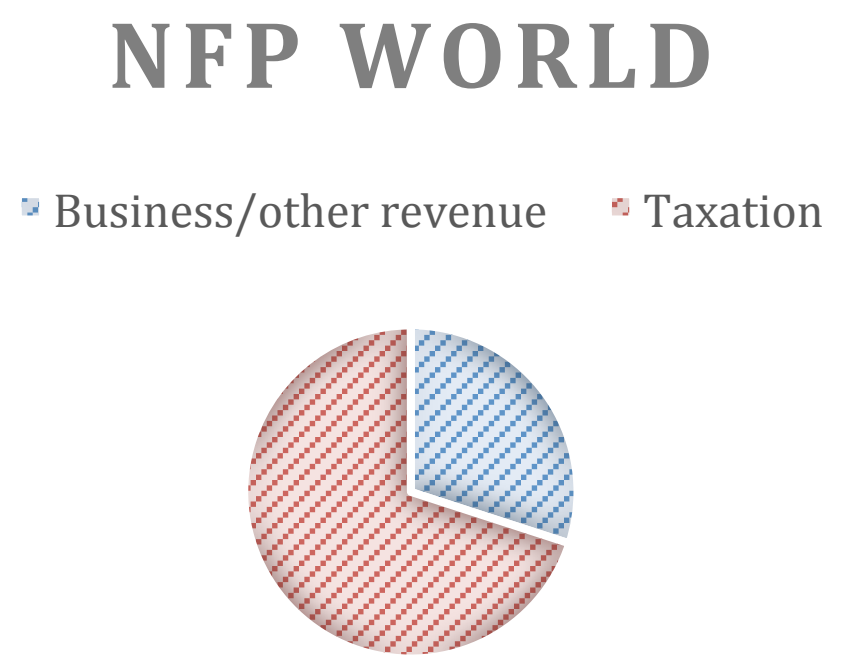

**Government Revenue Requirements in the FP versus NFP World**

In the NFP World, **personal income taxes** (which typically bear the brunt of taxation[733]) **are lower across all levels of earning**, and the overall amount of state revenue from personal income tax has decreased. While greater levels of employment mean more taxpayers[734], and average wages are higher relative to the wealth circulating in the economy, people are working less hours and there is less of a difference between top and bottom marginal tax rates (with progressive taxation systems remaining common), due to greater wage equality.

Although less necessary, there is a higher tax on private profit, capital gains, gifts and inheritance, as a safeguard against the re-emergence of an elite economy. A high frequency trading tax is almost universal and squeezes speculative activities out of financial markets[735].

While the environmental impacts of business and human consumption are greatly reduced, the effects of climate change and other ecological crises remain ever-present. Thus, environmentally-oriented taxes[736] (e.g., excise taxes on non-renewable resources), ensure production and consumption occurs with consideration for resource caps, and stimulates further innovation. Taxes on goods and services continue to keep downward pressure on overall consumption, and can be applied with particular emphasis on luxury or other items

733 In the U.S., personal income taxes constituted 47 percent of 2015 government revenue, compared to 11 percent for corporate taxes.
734 "...inequality costs the UK up to £33bn per year in productivity and lost taxes..." <REF>.
735
http://qed.econ.queensu.ca/pub/faculty/milne/322/IIROC_FeeChange_submission_KM_AP3.pdf.
736 Pigouvian taxes

that may have particularly large environmental impacts (and are not basic necessities). Environmentally-oriented taxes are often earmarked to fund sustainability education, training and innovation, and programs that reduce, reuse and recycle. Internationally, high customs duty taxes assist in incentivizing local trade and minimizing inefficient international exchanges.

**Public Funding**

There will be a range of government approaches to public funding; the NFP World allows for significant diversity in national economies. It's feasible, but not necessary, for countries in which democratic socialism is already working quite effectively and efficiently, to keep taxes high in order to maintain functional public services (e.g., health, transport, housing, education), but countries with less of a socialist culture can maintain a low tax structure, with NFP enterprises filling the gaps. It's a matter of what the citizens feel most comfortable with. It is feasible for governments to provide for a lot of their citizens' basic needs in the NFP World; but it's not necessary, as the NFP market can also take care of these needs very well.

Fiscal policy (i.e., what government does with its budget) becomes less important in the NFP World. This is another way that central governments' power declines in the transition to an NFP economy. In the for-profit system, the only reason deficit budgets are applauded is because they pump money into the for-profit market with the belief that wealth then trickles down via job creation, productivity gains and taxation. But in an NFP system, government deficits help drive the circulation of wealth, taking us beyond the debates of Hayekian-inspired austerity or Keynesian-inspired stimulus. When NFP banks and companies buy government bonds, this primes the wealth circulation pump, rather than the wealth extraction siphon. Yet, since the goal of economic growth has been replaced with the goal of having an economy that maximizes positive public health outcomes, governments have less of an appetite for debt, especially given they are no longer expecting growth to enable large debt repayments. Surpluses are still good because they can represent an efficient state, but there's not scarcity in the real economy's money supply or leaching to the wealthy via debt repayments anymore, so public debt is not a major issue.

Funding certain services fully is not only possible in the NFP World, it also makes a lot more sense. Healthcare means a widely contributing workforce that is contributing to the wellbeing of all (rather than inequality of the wealth extraction siphon). When all jobs are NFP, free tertiary education makes sense because it is the public investment in training people to serve public needs. This is in contrast to the current situation, where it's largely a subsidy for the wealth extraction siphon – training people to deliver private profit to owners. It just makes sense to fund public research when it is truly for the public good.

Social safety nets remain an integral part of the economy, so any individuals who are adversely affected by bankruptcies or other unfortunate events will have state support. Paid parental leave via social insurance is also much more possible.

While funding for certain services can come fully from government, the delivery can be a mix of public and private sector (NFP). Since a significant percentage of the education sector is composed of NFP enterprises, in addition to public schools, there is a lot of diversity in education styles, but educators can still be held to local, national and international standards.

### Pure philanthropy

While many nonprofit organizations rely fully on charitable support, some nonprofits self-generate a portion of their revenue. The reverse is also true – while many NFP businesses are fully self-sufficient, some receive a percentage of their revenue from philanthropic sources.

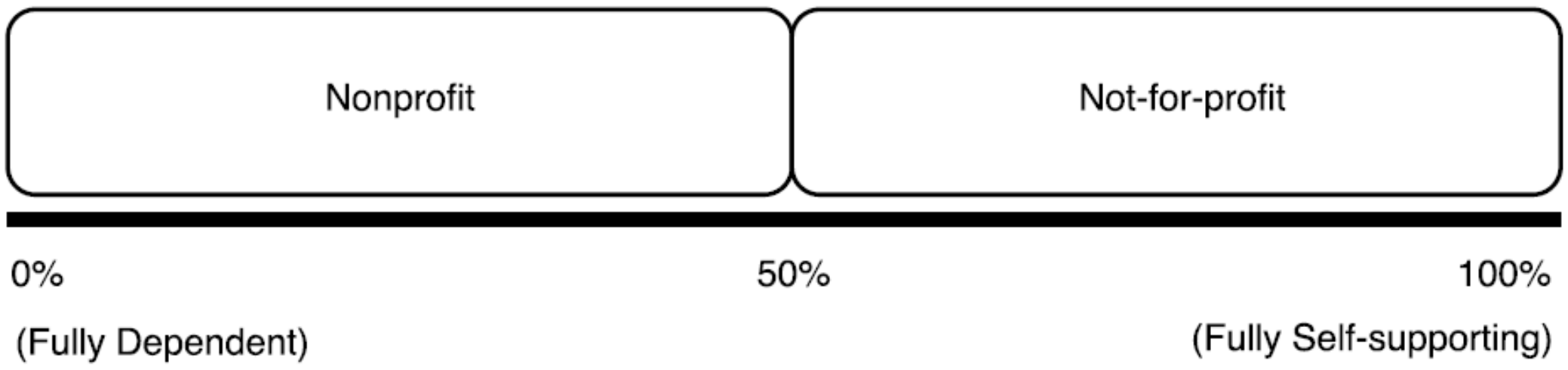

**The Spectrum of Revenue Sources for Nonprofits and NFPs**

Now that the elite economy has been absorbed by the real economy, private family foundations are less existent, and government grants are a minor contributor to philanthropic support. Corporate philanthropy has been maintained, with NFP firms seeking to strengthen the good work of others by sharing their profits. But most financial support comes directly from the public with adequate wages and the wealth circulation pump allowing employees and all community members to be more philanthropic. Tax deductions remain an incentive for donations, with the donor base more widespread and engaged. The benevolent are celebrated as heroes, given philanthropy has become purer, with less potential for conflicts of interest, now that profit-maximization doesn't underlie the economy.

With the story of international development focused on public health and wellbeing, rather than economic growth, and the wealth circulation pump having rectified many inequalities worldwide**, international aid is steered towards activities that build on existing strengths in order to achieve self-determined outcomes**. Support is commonly for small-scale entrepreneurs, NFP incubators, NFP enterprises and government programs, each exhibiting deepened forms of community accountability.

# Good corporations

Along with finance, the corporation has undergone a fundamental transformation.

In the for-profit system, corporations rule the world[737]. Their ability to influence policies, politicians, governments and the media, is compounded by abuse of the legal rights they are extended[738]. With 'limited liability' protection, corporate owners can behave recklessly, such as knowingly dumping harmful waste into the environment, and use bankruptcy as a business strategy rather than a last resort. Yet, how could we expect any different from a system focused on self-interested financial gain?

---

[737] Korten: When Corporations Rule the World; Klein, N: No Logo.
[738] Corporate personhood since 1886, for instance: https://en.wikipedia.org/wiki/Santa_Clara_County_v._Southern_Pacific_Railroad_Co. See, for example, The Corporation. And analyses on Citizen's United.

The scenario changes completely in the NFP World; the corporate form is restored to its original intent as an important driver of responsible innovation[739]. In addition to other proactive government measures[740], the limited liability of an incorporated NFP gives social entrepreneurs the protection and incentive they need in order to take risks in starting a business or scaling an existing sole proprietorship[741]. Given an NFP corporation can trade as its own legal entity, enter into contracts, buy and own assets, and sue others, 'corporate personhood' is now a force for good, rather than a means to control and dominate.

A successful company in the NFP World is not necessarily one that has the highest profit margins and produces the most goods, but one that produces the best social outcomes and has the most positive social impact. Inspiring trust and confidence in business, customer engagement has become truly genuine. Take banking, for example. Because an NFP bank's top priority is to meet their customers' needs, they are significantly more likely to do what is in the best interest of the majority of their customers, such as lending responsibly, charging fair rates on loans, being responsive to customers' feedback and changing needs, and investing financial and other resources in projects and businesses that benefit the community. The NFP ethic guides these banks to be sensitive, accountable and transparent, compared to the for-profit ethic which guides banks to be greedy, tricky (disguising loan terms and interest rates to look nicer than they actually are), and, in some cases, downright devious[742]. Or consider what happens when the media is entirely NFP. There is much less of a tendency for the concentration of media management in a few hands, so it has more space to be decentralized and diverse. In this system, the distributors of media are NFP too, so there is less editorial pressure from advertisers, which makes them more accountability-focused (accountable to their social missions which would have to do with the fair and honest dissemination of information for the benefit of society).

With wealth flowing through the economy, and the absence of the profit-maximization ethos, companies respond to client needs that were previously deemed 'too costly'. For example, goods and services are now commonly designed for use across a wide range of abilities. Businesses can afford to make their products and services truly simple and user-friendly, designed for practical use, thanks to the lean, open approach. Technologies can be taken apart, fixed and reworked, rather than being designed for the dump in order to sell more. Companies can use decentralized design and manufacturing capabilities that enable local creation and wide disbursement of technologies that serve the greater good. Because businesses can be more sensitive to local needs and cultural norms, designs are more easily customized for local needs (e.g., forking in software development). Through the participant-driven lean development model, innovators seek to incorporate multiple perspectives and multiple uses in their innovations. Manufactured needs have disappeared, enabling agreement on what constitutes 'social purpose' to be more participatory.

---

739 When 'corporations' were first formed in Britain and the US, they had to be granted a charters by the government, which required a statement of their benefit for society (Micklethwait & Wooldridge, 2003)

740 For example, tax breaks and grants.

741 For example, foregoing steady wages to start a new business, whether it be a childcare facility, a homeless shelter, or a bakery, as well as the risk of personal liability (including access to personal assets) for any debts incurred in the growth process.

742 For example, when Wells Fargo tricked thousands of their customers across the U.S. into signing up for bank and insurance accounts (and fees) they did not want and, in some cases, signed them up without their consent at all (Ref: https://www.npr.org/2017/08/02/541182948/who-snatched-my-car-wells-fargo-did).

With more widespread funding, what gets funded is also much more diverse, and based on community needs. The ecological impact of goods and services is more readily factored into corporate decision-making. Companies are more inclined and able to make sure their goods and services are benign or benevolent towards humans and the environment. Technological design is more systemic and interdisciplinary, ensuring innovators think about technologies not only in terms of direct effects, but also in terms of indirect effects and unintended consequences. Open innovation provides greater public scrutiny, and using the precautionary principle[743] is now much more possible due to relief from the pressure to innovate and sell at all costs. That is why not-for-profit enterprises create more durable products, and more innovation is geared towards helping reduce harm to the environment and aiding ecosystem recovery from existing damage. Remember, the NFP economy has stepped beyond the triple-bottom line, with profit now only serving outcomes for both people and the planet. Businesses are motivated to design products in a way that the parts can be re-used, re-purposed, or at least fully recycled. This is a purpose-maximizing business strategy, not a profit-maximizing one. Companies in the NFP World also look for ways to be closed-loop and zero-waste, by forming contracts with other companies, wherein one company's waste is used as a feedstock for another company's production[744]. And an expanded number of NFP projects actively drive environmental remediation, through efforts such as improving soils, protecting ecosystems, and planting trees.

But even NFPs without a direct environmental focus are more ecologically sensitive. Although sometimes NFPs don't take into account indirect damage they cause, very rarely do they intentionally try to profit from damage, as so many for-profit companies currently do. And with profit no longer a goal in and of itself, there is little incentive to dodge environmental standards. To the contrary, NFP business leaders are more likely to embrace environmentally-oriented policies, treaties, taxes and agreements, be they state-regulated or industry self-regulated. **The NFP World enables ecological stewardship because not-for-profit business and the environment are fundamentally allied, rather than at odds with each other.**

## Internal and market accountability

With enterprise models to support their social missions, NFPs have developed more accountability to both beneficiaries and customers (frequently one and the same). The market emphasis has brought a deeper level of professionalism to organizational decision-making, especially regarding business risks. Responding to market demands, both customers and beneficiaries more readily participate in the design and monitoring of goods and services. This blending extends to the high portion of the general public who now serve on NFP boards as a part of their civic contribution.

---

[743] The precautionary principle refers to the idea that we should choose not to do something if it might entail significant risks for people and/or planet.
[744] I.e., eco-industrial parks, industrial symbiosis, industrial ecology

Boards fulfill a governing role in most NFP companies[745], strengthened by the vesting of greater powers in the workers[746]. The shift to member models, involving stakeholders (not shareholders), and limited term policies for board members, adds a further layer of accountability, with the possibility of board member replacement and scrutiny at the company's annual general meeting. And with regular communications expected between organizations and their members, there is improved accountability, as well as increased avenues for representative feedback.

Systems for measuring impact prove more effective without the profit motive. They build on initiatives developed in the for-profit world, such as the Global Reporting Initiative, the Ecological Footprint Analysis for business, and the Common Good Balance Sheet. However, greater accountability in the NFP World is paradoxically accompanied by greater acceptance that not everything needs to be measured in the precise manner that was demanded by the for-profit world; good judgment is often relied upon for decisions regarding matters where outcomes may be intangible or hard to measure.

Given there are no business owners[747], the corporate structure is 'flatter' in the NFP World than experienced in the for-profit world. NFPs disperse power among operational units, rather than concentrating it within senior management and governance boards. Drawing on cooperative ideals, most firms in the NFP World are directed by the workers themselves, using new tools and practices to facilitate democratic decision-making[748]. Budgets are often created by the entire organization[749], typically include the listing of salaries, often include input from outside stakeholders, and are generally accessible to the entire company. Annual financial reports and board meeting minutes are shared with the public, as is already common practice by nonprofits in the for-profit economy, for purposes of transparency and accountability.

## Employment and working conditions

Most workers in the NFP World seek to contribute to the greater good, while having a high level of autonomy, self-expression, creativity, and skill development in their job. This is possible when work is created based on a combination of society's needs, and the strengths and needs of employees and organizations as a whole. And it is supported by a partial reversal in the division of labor – less hierarchically-structured companies employ multi-skilled workers, ensuring greater organizational resilience and enabling semi-autonomous teams to be constantly assembled and reassembled for project-specific work.

Labor unions and worker ownership have traditionally provided a crucial counterbalance to the destructive workplace tendencies of for-profit systems, fighting for and ensuring worker rights in a world in which workers are seen as a means to generating profits for owners.

---

745 Their independence relates more to the absence of a financial conflict of interest; they may well be recipients of the NFP's services. They may remain more hands-on/managerial in smaller start-ups.
746 Via, for example, provisions in a company's bylaws that provide for worker self-direction of the nonprofit <REF: SELC>.
747 In the sense of appropriation rights – there may be nominal or symbolic member-owners, but their ownership rights are limited to control rights (see Chapter 2 'Ownership and Assets' for more detail).
748 For example, sociocracy/Loomio.
749 Or representatives from all different departments of the organization.

However, in the NFP World, labor unions, collective bargaining and worker ownership are largely obsolete because organizations are more accountable, participatory, and considerate of worker wellbeing by nature. NFP enterprise is actually a major evolutionary step for worker rights.

Worker wellbeing is seen as inextricably linked to the greater social wellbeing NFPs seek to create. Leading the way is more equal pay[750]. In addition to wealth being more equally distributed in the NFP system, companies generally ascribe to voluntary salary ratios between 1:1 and 20:1[751], depending on the nature of each company[752]. Competition and worker self-governance, along with an abundance of highly skilled labor, maintains downward pressure on upper wages. Minimum wage laws remain, but prove less necessary in the NFP world. And in the absence of wage suppression (and lobbying against raising the minimum wage) by profit-hungry owners and managers, which is commonplace in the for-profit system, the minimum wage is always a living wage in the NFP World.

Worker hours are highly flexible, and many workers (with jobs that can be fulfilled remotely) work part of their hours from home or in local co-working environments. To avoid romanticizing things, many people still work overnight shifts, 'on call' jobs, overtime, and holidays to ensure the economy functions, despite certain processes and services having been automated, and companies regularly using workers across time zones to ensure around the clock service provision[753]. Physically demanding, and even unpleasant jobs remain in many sectors.

However, thanks to the cycle of wellbeing, people spend dramatically less time working. So much so that, for logistical reasons, annual leave policies in most countries remain between 2-4 weeks. However, globally, paid parental leave is much more generous. While it is often largely government subsidized, it is rarely government mandated. Indeed, it has become one of the most competitive aspects of the market, with many companies offering between 6 and 18 months of full time equivalent paid parental leave for each parent. Because the average workweek is only 20 hours in the NFP World, it's not as difficult to compensate for an employee who takes leave. Workers may even choose to absorb the hours for their absent colleagues. Or the company might hire a temporary replacement. Many new parents might even decide to keep working 10 – 20 hours per week because the workweek is much shorter and more enjoyable. If not, job protection is guaranteed throughout. And the shorter workweek and more flexible working hours means that full employment (or near full employment) is much easier to maintain on the aggregate level of the economy.

As mentioned before, people often choose to work longer into their lives, given work is more purposeful and people's physical health improved. This is completely unlike the for-profit economy, in which work is largely seen as nothing more than a means to earn a living (which is cashed in upon retirement[754]).

---

750 Gender inequities in pay have largely been resolved due to the shift in societal values and more attention paid to all kinds of inequity.
751 Good research on pay ratios here: http://b.3cdn.net/nefoundation/15c112d0bb14368496_ukm6ib653.pdf. Plato's Laws for the new republic suggested an income ratio of 4:1.
752 E.g., $10hr – shelf stacker/$200 – CEO of a large grocery store
753 e.g., Lifeline.
754 Roosevelt's covenant.

In the NFP World, automation is no longer seen as a threat to employment and people's livelihoods. This is because the robots and AI that take up work in the economy are owned and operated by NFPs, which use the surplus generated by automation to help the community through their social missions. Finally, the benefits of automation can be enjoyed by all.

## Efficiency and innovation

With a different approach to workers and work, **innovation in the NFP World proves both faster and more sensitive to human and ecological needs.**

Innovation that is primarily driven by purpose, rather than the prospect of financial gain, often comes with a more of a spirit of openness to collaborate and share, which leads to much more efficient innovation processes. Thanks to the 'open revolution'[755], knowledge (about logistics and supply chains, technological advances, and social innovations) is shared freely, along with the source code for most forms of software. 'Copyleft' has replaced copyright – meaning that most work is shared without restriction or with a creative commons or other form of peer production license[756]. This saves a lot of time, energy, and resources that might otherwise be spent reinventing the wheel. Thanks to the collective, overarching mission of greater social and ecological wellbeing, public research institutions and private NFP companies work collaboratively, ensuring that publicly-funded research translates into NFP market goods and services where appropriate, and society at large benefits.

Online platforms continue to facilitate the flow of information from citizens and service-users to companies and governments, ensuring efforts to improve social and ecological wellbeing are informed by wisdom from the public[757]. This is indicative of the now widespread lean development model, in which technological and social innovations begin as the simplest of prototypes testing their riskiest key assumptions through a process of ongoing user feedback.

In the for-profit economy, much social innovation from outside the market has gone unnoticed for a long time, as highlighted in Chapter 4. This kind of innovation flourishes and is better acknowledged in the NFP World. The formal market benefits from the innovation introduced via 'backyard' tinkering by individuals, like hackers and 'pro-ams'. And more people have the time to do such tinkering and experimenting, due to the shorter workweek.

In the for-profit world, a faster speed of technological innovation equates with less ethical and environmental scrutiny[758]. But in the NFP World there is greater sensitivity to human and ecological needs. When designing products or services, innovators ask themselves, "How does this benefit society?" not, "What can I create that will generate the greatest sales?". In a related way, marketers ask themselves, "How do we ensure greatest access to and knowledge about these innovations?", not, "How can I convince people they need these products and services?". Crucial here are decentralized control, participatory development, open innovation, and the associated 'purpose motive'. This motive is highly effective when it comes to accelerating innovation[759]. As witnessed with open source software in the for-profit world, ‘reputation’ can be a powerful currency. And in the NFP World, industry and public appreciation is greatest for those who innovate most openly and selflessly for the greater good.

---

755 Price, D. "Open".
756 E.g., https://wiki.p2pfoundation.net/Peer_Production_License
757 What social theorist, Clay Shirky, calls the cognitive surplus. See, Shirky, C , "Here Comes Everybody".
758 See, for example, 'beware the nanoethics divide'.
759 Dan Pink.

In a milieu of less desire or need to own and protect ideas, the securing of intellectual property (IP) rights, especially patents, is greatly reduced. Some people might still file patents, but the practice is considered socially unacceptable for private individuals. And, in a crucial development for IP law, innovators are unable to patent a design that has been released by the original creator on an open-access basis.

But the primary role of IP has also transformed from offering innovators a financial incentive to ensuring quality control in the marketplace. **In the NFP World, patents and trademarks play a valuable role for governments and NFP companies seeking to reinforce their reputation for delivering high quality goods and services.** In this sense, IP rights are considered valuable for their ability to enhance social outcomes, and to keep the community from being misled by imposters. An NFP may also patent an idea to recoup invested capital, via license fees. But, to safeguard against a company patenting to purposefully eliminate competition, most nations have reduced the standard twenty-year patent term to just a few years. Moreover, given patenting is generally motivated by better serving society rather than making a profit, the costs of licensing an NFP’s innovations prove affordable to most companies seeking to enter the NFP market.

For the most part, concerns about piracy and IP theft have faded, because piracy is essentially a for-profit problem. People are willing to pay for freely accessible products and services when they know their money benefits the community, not just a few business owners[760]. But more importantly, there is less drive to make as much money as possible from digital information because there’s a healthier respect for the collective value of the commons. As long as an NFP can cover its operating costs, greater free access to its products (digital or otherwise) is actually considered a very positive thing.

Reduced costs are the final factor driving the increased efficiency of innovation in the NFP World. For large projects that remain capital intensive[761], government bonds, social impact bonds, crowdfunding, and Community Public Offerings remain viable funding mechanisms. However, innovation generally proves less costly, due to:

a) *increased availability of free information and ideas;*
b) *participation from volunteer users in product development;*
c) *greater sharing and collaborative use of resources;*
d) *improved systemic efficiencies*[762]*;*
e) *decreased license costs;*
f) *lower material and energy costs*[763]*; and*

---

760 Think of any sort of pay-what-you-wish scheme, or even just donations to churches, public libraries, radio stations, and public parks, whose services are freely available.
761 For instance, despite more collaborative approaches to innovation, the development of new medicine remains extremely expensive due to long testing cycles and extensive regulatory requirements.
762 Increased lean product development approaches. Clear focus on social and environmental needs (no conflicts of interest between profits and purpose). Greater efficiency with logistics and supply chains.

g) *reduced material waste*[764].

# The healthy market

The Not-for-Profit World has a market economy[765]. But it's a market economy unlike any we've seen before. The primary motivation of trade in the NFP market is the desire to fulfill social and ecological needs, as compared to the for-profit market in which the guiding motivation for sellers and buyers is private gain. In this sense, the market still operates on 'supply and demand' principles, but both the supply and the demand are now entirely real rather than the result of marketing to manufacture demand and subsidies to promote production.

In the for-profit economy, if demand and consumption don't grow at a certain rate, then the economy can't grow fast enough to substitute for growing debt and financial extraction. This is often framed as the demand side of the supply and demand mechanism of the market.

Yet, when the economy is composed completely of NFP enterprises, levels of consumption go up and down, more closely correlating with the ebbs and flows of society's needs. By voting with their dollars, the public largely decides which social and ecological needs should be addressed through the market. And the information that people receive about the state of the world in order to make such decisions is not filtered through the political and financial desires of media magnates, as it is now. Different media outlets still present information according to the worldview they represent, but profit-motivated media and advertising are no longer constantly manipulating ordinary people in order to financially benefit their owners. In fact, **individual choice and freedom are more privileged than in the current economy.**

When there is an unmet need or one that is being inadequately addressed, NFP enterprises will enter the market to meet that need. The market is constantly adapting and morphing in order to find better ways to meet needs. This doesn't come from the unquenchable thirst for profit, but rather from the innate human desire to create, innovate, collaborate, and solve problems in order to contribute to a better world. Supply and demand continue to guide the level of business attention given to each sector and field, as well as employment. Oversupply leads to reduced consumer costs, profit, and wages, reducing business activity. Sometimes there might be too many companies in one sector, causing an oversupply of products or services, which means that the businesses in that sector that are least effective at delivering social outcomes will die out. The employees of the firms that died out will move on to other firms, just as they do in the current economy (except that full employment is a lot easier to maintain in the NFP World, where people only need to work 15 - 25 hours per week in the formal economy). The same is true when an NFP business has achieved its mission and is no longer needed by society. It dies out and the employees move on to greener pastures in other jobs.

---

763 Even with externalities included, energy costs decrease due to the widespread shift to renewables infrastructure. And as the companies that produce the equipment and materials used in innovation are NFP, they're not charging exorbitant prices.

764 And thus less cost for disposal of waste.

765 The Investopedia website gives a broad definition of a market economy as: "An economic system in which economic decisions and the pricing of goods and services are guided solely by the aggregate interactions of a country's citizens and businesses and there is little government intervention or central planning. This is the opposite of a centrally planned economy, in which government decisions drive most aspects of a country's economic activity". (Ref: Investopedia). Simply put, a market economy takes shape based on the dynamic interactions between producers, distributors, suppliers, retailers, traders and consumers in the market. In a market economy, prices, wages and the supply and demand of goods are all determined by marketplace interactions.

Due to the shift in what society values, as well as the widespread fulfillment of basic needs, the focus of the market moves away from the production of new goods based on raw material extraction, towards a service-based economy. In the NFP World, what companies are supplying has changed. The business sector in the NFP World focuses on providing *access* to goods, rather than ownership, and providing services, rather than consumer products (including product service systems, which refers to businesses maintaining and lending goods rather than selling them). The very role for primary producing companies is reduced by the dramatic rise in sharing of goods and services, the open-sourcing of designs, the shifting of production to the de-marketized community economy (e.g., material production via home-based 3D printing, repair through local makerspaces, food growing in community garden, and DIY and DIO[766]), and the takeover of certain fields by 'the commons'. In essence, thanks to better sharing and mapping of individual offers and needs, there is a state of hyper-accessibility to basic goods and services. This trend means that, in the NFP World, a lot of production happens at the periphery of the market or outside of it altogether. Whereas the for-profit economy is trying hard to figure out how to privatize the outcomes of peer-to-peer production through the market, the NFP ethic and goals of the NFP World celebrates this de-marketization.

Without the drive to continuously create jobs and profit margins, much of the production in the NFP World is actually the refurbishment, repair and recycling of secondary goods, and programs that seek to enhance energy and material conservation (reduce, reuse, recycle).

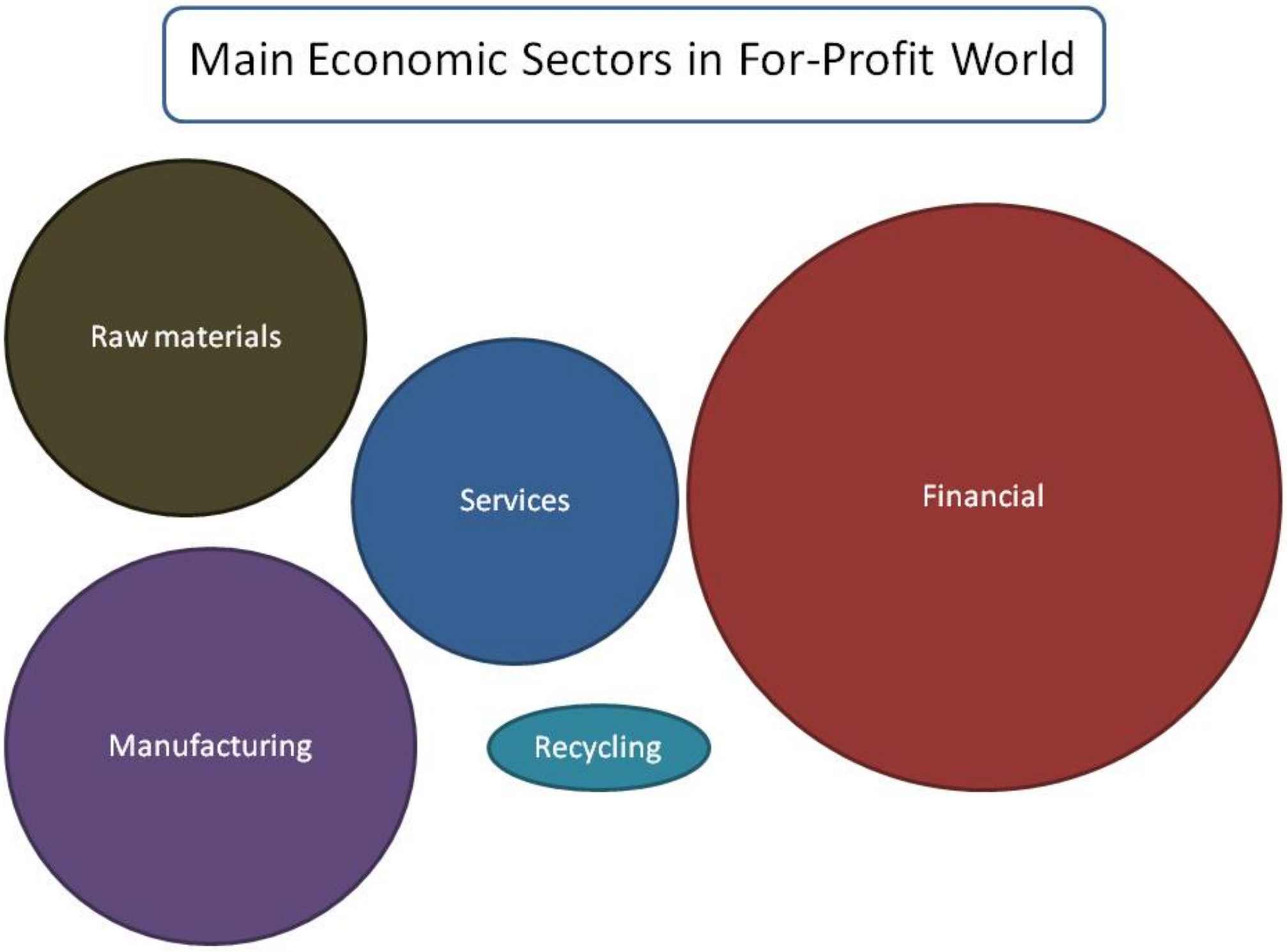

[766] Do-it-yourself and do-it-ourselves

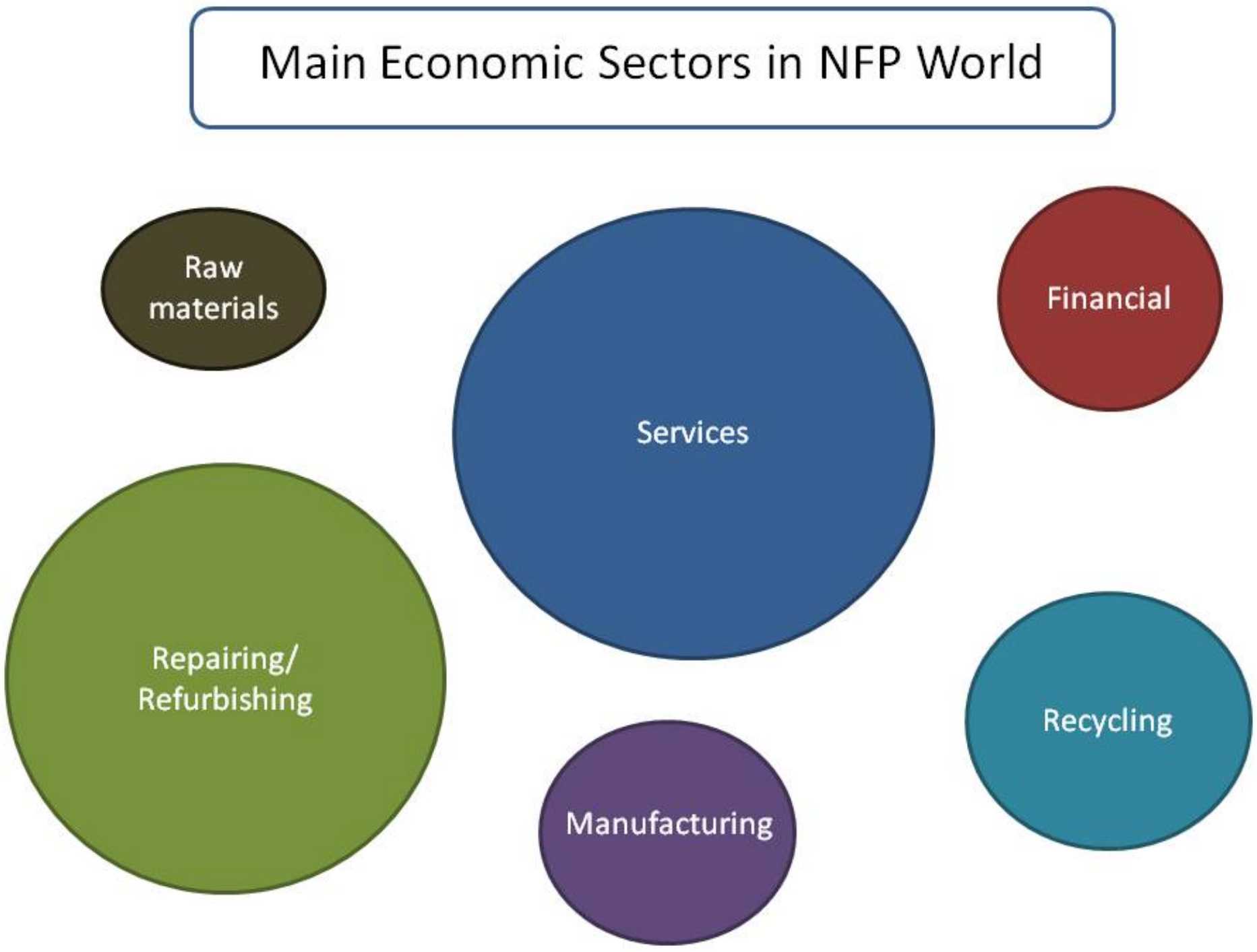

[767]

In general, the market dynamics of the NFP World lead to the efficient satisfaction of needs (at least those that can be met through the market).

**When needs are real, rather than manufactured, market value aligns with social values** (it's no coincidence that we use the word 'value' for both of these things). This shifts behavior of the economic actors and, thus, the outcomes of their aggregate behavior, making the prices of the things deemed 'valuable' for social-ecological health more affordable and things deemed 'not so valuable' or even 'destructive' for social-ecological health more expensive. In a world of reduced competition and demand-driven advertising, monetary values more easily align with our true values about what is worthwhile. The things society truly needs (the goods and services that maximize health) create purpose-driven competition that ensures greater affordability (supplemented by sliding-scale models), and less needed items and things that are explicitly destructive, like fossil fuels, become costlier, as they are not valued as conducive to wellbeing. When the goal of the economy and businesses change from financial gain to social purpose, activities like war-profiteering are much less enticing, far less socially acceptable, and a lot less common. Rare commodities remain expensive, due to their physical scarcity and the internalization, via taxation and fair trade practices, of previously unaccounted for costs.

**In a healthy market, competition is functional, not fierce.** It allows for the generation of the goods society wants (now more synonymous with 'needs'), in the quantities society desires, and at the prices society is able to pay. It remains a key driver of high quality outcomes, innovation, and price discovery, as well as employment stability (with the most successful businesses surviving), but it does so in a collaborative context. In the NFP World, competition occurs within a greater context of cooperation (in order to achieve positive outcomes for society). This contrasts greatly with the for-profit economy, in which collaboration only happens in service of beating out other competitors, and your collaborators might turn against you at any moment if they find a better partner to collude with.

[767] These illustrations are not meant to be accurate, but just to give an idea of the kind of changes that happen in the transition from the current economy to the NFP World.

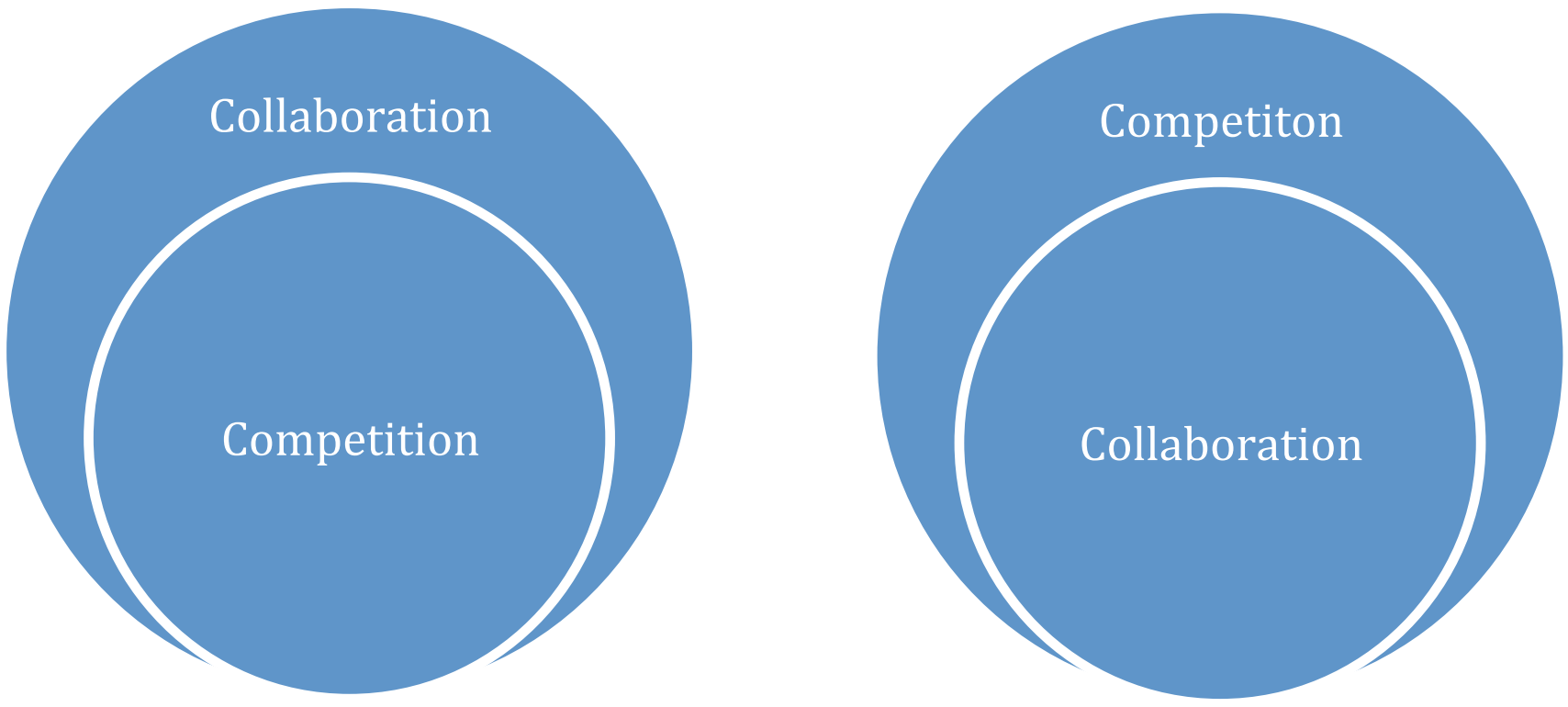

**Competition the NFP World** **Competition in the FP World**

As with ecosystems, competition in the NFP market is based on who can best serve the needs of the community, rather than who can accumulate the most for oneself. The NFP World encourages cooperation, because it is based on the purpose motive and the story of interconnectedness. We're all in this together and it's better for everyone if we collaborate. When the relationships of interdependence are based on who can accumulate as much wealth and grow as big as fast as possible, the whole system is unstable. But when interdependency is based on all actors working for the common good, it keeps everyone focused on their social mission and how their actions impact the wider system. Competition returns to its original etymological root 'competere' – to strive together.

For instance, the NFP tendency to share ideas and resources at all stages of a project's development creates a totally different dynamic in the market than the current for-profit tendency to privatize, protect, and hoard information and resources in order to effectively compete and survive in the dog-eat-dog market.

This process of cooperating and competing in the market to deliver positive social outcomes drives social innovation and, thus, diversity in the marketplace. Because different companies try a variety of strategies to meet needs more effectively, there are constantly new ways of managing and running NFPs. New NFP business models constantly emerge. We see new pricing strategies, as well as new and improved ways of handling supply chains, of tracking success and measuring progress, and of ensuring accountability and enhancing transparency in order to gain long-term trust and loyalty from customers so that they can achieve their social missions.

The price discovery mechanism remains central to determining what's fair for parties in an exchange. Pricing in the NFP World doesn't mean sacrificing one's own wellbeing for the wellbeing of others. It means figuring out what is fair for everyone involved, which isn't far from how price discovery works now (what people are willing to pay and be paid). It's just that Homo economicus encourages people and businesses to be dishonest (or at least not very transparent) about their needs in the current system (i.e., try to get as much as they can, not just enough or what's fair). Thus, there is often a lack of trust and a resulting sense of insecurity in the price discovery process in the for-profit economy. **Economic actors in the**

**NFP World enjoy more trust and security when negotiating prices, due to a widespread ethic of enough**. And the whole economy benefits from more accurate pricing as a result, again leading to better access to essential goods and services.

The digital revolution, combined with the NFP market ethic ensures better information in the geo-coded, reputation economy. Marketing is much less deceptive, because reputation is everything now that people can share and verify information faster, leading to faster blacklisting of fraudulent or misleading companies. As a result of the reduced pressure to make profit, you don't need to market as aggressively. This is a networked market, with the ability for appropriate information to reach people more effectively via online means, which helps demand to drive supply. Knowledge is much more available to the producer and consumer, alike.

NFPs can die (creative destruction remains), but it is seen as a valuable part of the business cycle. There's now a real composting, as should happen with any natural cycle. The end of businesses is an important part of the innovative economic life cycle. Businesses are allowed to fail, go bankrupt, and dissolve, just as for-profit businesses are in the current economy. Just as employees of a dissolving firm find work in other NFP businesses, so are assets reinvested in other NFPs, allowing them to be put to good use. And this helps to ensure that the NFP market is efficient and effective in meeting society's needs.

There are three main ways NFPs die in the NFP economy. Firstly, they fail as a business for the same reason that businesses fail in the present market - a poor business plan, poor timing for market entry, poor management, inefficient use of resources, a damaging lawsuit, etc.

Secondly, they collapse due to lost confidence, via mission drift and an associated lack of public support (e.g., they don't receive enough philanthropic support to plug the gap for the shortfall in their revenue generation activities.)

Finally, NFPs can voluntarily choose to dissolve, especially if their social objective has become obsolete (or another company has shown its ability to address the need much more effectively). Ideally, many NFP businesses have this end goal in mind when they commence their operations; how they will dissolve once they achieve their mission or what they will do next, if anything. They addressed the need they aimed to address. It could happen that a company has been putting its resources into producing high-quality washing machines, but now that market is saturated because the washing machines last so long that people don't need to buy a new one very often - a good thing, meaning there will be less waste and social and environmental damage as a result of producing more washing machines. Instead of seeing that as a threat to its business, the company can now pivot or dissolve. If it pivots, it turns its energy to something else that adds value to the community, perhaps repairing and maintaining those washing machines. The whole point of starting an NFP company is to meet society's needs. Some needs are constant, like the need for high quality food. But some NFPs seek to fulfill needs that we all hope will become history, like homelessness. When an NFP goes into business to resolve an issue like homelessness, it will have a plan for what it will do when it achieves that mission. Otherwise, the company's identity depends on the existence of the very problem it's trying to resolve. The notion of there being an endgame goes back to the ethic of enough that is embedded in NFP enterprise. This ethic allows for a point at which a company has done enough. And with less social ills, less wasted energy, and less consumption in the NFP economy, employment is not framed by the scarcity mentality; it's easier to find a job to compensate for 20-hours of work per week than 40-hours of work per week.

Mergers and acquisitions (M&As) still occur. By a financially stable NFP, they provide a route to avoiding bankruptcies while maintaining service provision. They're less about centralizing power, and more about a functional decision to ensure ongoing (or even expanded or enhanced) service delivery. A business that is failing, but still has a reliable workforce, a strong social mission, or a large amount of physical assets (like office space, a factory, or computers) might be seen as an ideal opportunity for another NFP business to diversify by acquiring it. When M&As are in the service of meeting social needs, they involve greater consideration than profit-motivated M&As. There is a need to consider the social mission and welfare of the workers of the NFP company being acquired. One present-day example of this is when Baylor, Scott and White Healthcare in 2013 acquired another NFP healthcare company, "motivated by a number of factors, including economics of scale to reduce cost, access to capital, access to covered lives to control risk in population health management and to maintain industry influence."[768] There is no shareholder to benefit in this kind of acquisition. This means that there is a much higher likelihood for NFP companies to take into account the ideas and feelings of the managers and staff of the NFP that they are acquiring or with whom they are merging. In contrast, for-profit mergers and acquisitions can often be ruthless and very dismissive of the needs and concerns of the staff of the acquired company[769]. This is another area in which the difference in purpose and goals has an enormous impact on how business is carried out.

Companies that can't stay financially viable, whether due to generating insufficient income, using resources poorly, drifting from their missions, overleveraging, or engaging in corrupt behavior, may go bankrupt. Liquidation processes in which debts are amortized with company assets. The asset-lock in NFP business models also means that nobody can privately gain from an NFP's assets upon its winding up. When an NPF business closes down, all of its assets must be donated to another NFP, so assets are also cycled back into the real economy in order to meet society's needs. Depending on the terms of the loan, individuals and NFP businesses that had lent to the NFP can be compensated with assets in the liquidation process.

With all financial surplus being constantly circulated back into the real economy, any unpaid debts can be absorbed by the market more easily than in the current economy. Cooperation and mutual support between NFPs helps lenders take any financial hits that result from bankruptcies. (And there's a bigger pool of wealth and resources available for retraining of the workforce, as well as a more viable safety net for those out of employment).

In nature, two trees grow in competition and one dies, because it is weaker and there is not enough water or sunlight for both. The one that dies becomes compost to feed the other. Similarly, NFP companies that cannot effectively serve the community die out, just like the weakest organisms in an ecosystem. The closure of a community enterprise after a number of years may not necessarily be deemed by all stakeholders as a 'failure'[770]. It will usually leave behind an important legacy of social benefit as well as more publicly accessible materials. For example, members may go on to develop other enterprises and contribute to community building in other ways.

---

768 <REF>
769 Micklethwait & Wooldridge, 2003
770 Cameron, J. and Gibson, K. (2005).

'Dirty' industries, like fossil fuels, tobacco, and weapons were taken over by NFPs with the mission of eradicating their use. In the case of fossil fuels, for instance, coal mines and oil companies were bought up by NFP businesses who used the profits from the sell of the fossil fuels in order to build up renewable energy infrastructure and train their employees in this new field. In this way they paved the path for a smooth transition from fossil fuels to renewable energy. Likewise, NFP businesses and state enterprises bought out big tobacco and weapons manufacturing companies with the purpose of minimizing their use in the first place. To this end, they advertise the negative impacts of their products and tobacco and alcohol producers and retailers use their profits to offer rehabilitation for addicts. Their aim is to end addiction altogether. **Social problems like addiction, alcoholism, violence, and war have not been eradicated, but they have been shrinking in prevalence and, at the very least, no one is financially profiting from them any longer.**

# The partner state

The state has transformed and is significantly smaller in many places. The rise of NFP enterprise, the wealth circulation pump, and the resulting positive effects on ecological and public health translate into less need for the welfare state, taxation and the associated bureaucracy. Incentives exist to make oneself redundant in order to move on and do more impactful things. There is greater automation and productivity. Many of the state's functions have been automated (e.g., census, voting, taxation, referenda). But government is also more decentralized. There is widespread support for the principle of subsidiarity, strengthened by NFP enterprises, which are generally local. In accordance, greater power has been disbursed to local, municipal, and provincial governments. Bioregions, areas defined by their natural biophysical traits, now play a key role in defining governance structures.

**In this context, the state's role is to support the fair advancement of social-ecological health, in partnership with the market and civil society.** In the NFP World, the state is fully divested from driving economic growth, and is instead fully invested in helping maximize public health, in all of its dimensions. It continues to provide local, regional, state, and national services and represent its people in the international arena. It manages key physical infrastructure that is best centrally governed. And there are still different national approaches to what should and should not be centrally governed. In Australia, for example, there are very clear distinctions in roles - the federal government doesn't deliver health services, rather the states do. But federal government does manage health insurance. This kind of government offers legislation, fiscal policy and infrastructure that seeks to maintain the lean society, prime the wealth circulation pump, encourage the cycle of wellbeing, and reinforce the ethic of enough.

As mentioned in a previous section, the state is also more entrepreneurial. So it funds its services not only through taxes, but also through a greater amount of revenue earned through the sale of goods and services (which most people are much more able to afford, due to the wealth circulation pump). Throughout, the state seeks to empower communities and create the conditions for the non-marketized fulfillment of needs. The government recognizes and incorporates the immense contribution of the community and family to satisfying needs in non-monetary ways. States see their role as providing a base of support for NFPs, communities, families, and 'the commons', more generally[771].

[771] Similar to what is depicted in 'The Partner State' by Bauwens

Without employment, it's still possible, although rare, to not be able to meet your basic needs without state intervention, so the state is also an important last resort form of support, providing essential safety nets for people who have nowhere else to turn[772].

**The state is generally less corrupt and there is less political capture.** When NFPs or peak professional bodies lobby government now, they do so with a healthier mandate. The profit-seeking trends of corporations lobbying against taxes and regulations, using tax havens to hide income, threatening to move offshore in order to get subsidies or tax breaks, and using revolving doors between regulation agencies and industry have all greatly reduced. Perverse, counter-productive subsidies and other forms of market manipulation are uncommon, and while there is wide variation between countries, all government subsidies are generally minimal, allowing healthy competition to run its course in the NFP economy. But of course, governments continue to provide tax advantages for certain NFPs, giving extra help to those that might have a harder time generating revenue (e.g., homeless shelters, ecological conservation, etc.).

The NFP World is, of course, not an entirely free market (no modern market has ever been). The notion of 'free' markets has been debunked and superseded by 'functional' markets. Even with the corporation having become a force for good, and although the state is less interventionist, there is a need for ongoing market oversight and regulation, especially when it comes to environmental issues. The government still serves to regulate and check market forces, as it does in the for-profit economy. In fact, using fines and legal measures to ensure the integrity of NFP businesses is a very important role that the government must play. Given the absence of the profit motive (and associated political capture and corruption), regulation proves highly effective. Regulators and state agencies work closely with the judicial system to make sure that NFPs and government enterprises alike adhere to basic environmental, health, and workers' rights laws. Without the profit motive influencing legislation, these laws are based on state-of-the-art knowledge and are sufficient to maintain basic levels of social-ecological health. Governments must enforce these legal standards and encourage transparency and accountability. The regulators (like the tax agency) also continue to play an important role in holding NFPs accountable to their stated missions and annual financial auditing has become a common, streamlined process. Testing agencies, like the Food and Drug Administration in the US, remain. There is also tight regulation of the debt markets to ensure minimal levels of speculation for the sake of economic stability. Credit rating agencies are NFP and independent.

One ongoing problem with the state is that it can use its revenues (or borrowing) to subsidize inefficient service delivery. But two factors minimize the significance of this problem: 1) more transparent budgets, and 2) NFPs provide a competitive benchmark for costs. If a government is spending a lot of taxpayer money on health and education, in an inefficient way (compared to NFP competitors), citizens can pressure governments for change, or increasingly support the (perhaps costlier) NFP option, where quality will likely be higher, given the expected correlation with flexibility and efficiency.

Although, anti-trust and market competition laws are less needed, given more decentralized local markets without monopolistic tendencies, such laws remain in an NFP World to restrict the formation of cartels and prohibit other collusive practices regarded as monopolistic. They restrict mergers and acquisitions that could substantially lessen competition. And they prohibit the creation of a monopoly and the abuse of power by large NFP businesses. But these laws have been revised to accommodate for greater collaboration between organizations, as long as it is clear how collaboration will benefit society and help the NFPs involved fulfill their purpose.

[772] Less likely with countries that have a Guaranteed Basic Income.

Another crucial role of the government is to continue to be a key steward of the commons, such as land, air, water, and space. This involves maintaining vast swathes of land and water for nature protection and conservation, as well as for public use (as is commonly done by government around the world in the for-profit economy).

Although NFP businesses will surely be more environmentally-friendly, in response to ongoing ecological crises, and the NFP World is a non-expansionary economic model that leaves space for overall material consumption to shrink significantly, there is a need for governmental agencies and other civil society organizations to work on keeping track of the big picture in terms of economic flows and ecological health. Environmental regulation agencies work with the Earth system scientists and ecological economists in order to monitor and set standards for ecological health, and governments work together to monitor domestic and global stocks and flows of energy and materials. Sustainable resource use is enforced via strict resource and emissions caps and thresholds. Use of resources and emissions of pollutants and greenhouse gases can also be discouraged using environmental taxes, although such financial incentives and disincentives are less necessary in the NFP World, where people no longer consume for consumption's sake. Unfortunately, this does not resolve all ecological issues, because there may be serious feedback loops from the legacy of the for-profit system, but it does allow global human society to eventually move back into the 'safe operating space' for life on Earth.

The rule of law is central to running the NFP economy and with greater equality, mental health and worker wellbeing, there are drastically less abuses by the state in terms of inappropriate police force. Indeed, much faith has been restored in the government, largely thanks to a deepening of democracy. The state is more independent and representative due to less political capture. But there are also more structured means for direct democratic participation, like participatory budgeting for spending[773], and more forums for public discussion. This has evolved because people now have the time and an increased interest in participating more fully in their political systems, due to a shorter workweek, lower levels of stress, and less time spent trying to assuage the stress (e.g., the work-watch-shop treadmill).

Perhaps the most exciting means by which the state is truly representative of the people is the different progress indicators that have been adopted. While some countries use set variations of more traditional indicators, such as the Genuine Progress Indicator[774], many are citizen-defined, as happened in the Bhutanese metric for Gross National Happiness and the Oregon Shines index.

# The global community

---

[773] E.g., . http://projectrobinhood.budgetallocator.com
[774] With a lot less weight on GDP (which is, afterall, still a useful measure for knowing the tax base).

The global economy is a network of highly integrated, yet localized NFP economies, in which people buy most of their products and services from local suppliers and the production (including manufacturing) of goods takes place much closer to where they are bought. Long-distance trade happens a lot less in the NFP World, but it is still important for essential items that are found only in certain geographies. For instance, if a certain raw material for building photovoltaic panels only exists in a few places in the world, it will of course be shipped long distances in order to build photovoltaic panels locally. **Yet, the overall volume of physical trade in the international market is significantly reduced.** There are five main reasons for this: reduced consumption, circular economy, dematerialization, relocalization, and full environmental cost accounting.

Firstly, there is decreased consumption and production. As mentioned before, this is partly due to a shift away from materialistic values and consumerism and partly due to more sharing and collaborative consumption.

Second, products are designed for the circular economy. That to say that goods and services are designed in a way that makes their components and basic materials easy to repair, refurbish, reuse, repurpose, and recycle. In the absence of the profit motive, planned obsolescence no longer makes sense. For the purpose of maximizing social-ecological health, frugality and leanness make sense. So, in the example of the photovoltaic panels above, the rare raw materials would not need to be extracted or shipped very often because those materials will be reused, repurposed, and recycled locally until it is no longer possible. Only then will there be more demand for more raw materials. With a circular economic system, the material economy can become quite circular as well[775].

Third, consumption has dematerialized and miniaturized. The economy has become even more digitally-based, with electronic exchanges replacing physical exchanges. And products can be made with fewer material and energy inputs, thanks to innovation. Although digitalization and miniaturization will never be enough to absolutely decouple economic activity from material flows, the trend of relative decoupling continues to lessen the material and energy requirements of economic exchanges.

Fourth, the market has re-localized. With the drive for a more democratic and sustainable system, there was also a push for local control of NFP businesses. Due to the crises of the for-profit system, people increasingly came to understand the importance of maintaining local jobs and production. Technological platforms enable distributed production (e.g., open source design and 3D printing), minimizing the benefits of economies of scale. With smaller global socio-economic disparities and greater wage parity, there is less 'offshoring', as well as less labor migration and 'brain drain'[776]. People don't feel they have to leave their families and homeland in order to have a high standard of living, which is also good for social-ecological wellbeing (coherence of families and local communities). As a result, it is lot easier and more natural for people to value their local traditions, culture, and landscapes than it used to be, without the constant undermining of the for-profit story, which told people from non-Western cultural traditions that they're 'backwards' or 'behind'. This also allows indigenous cultures to survive, less impacted by globalized Western lifestyles. There's still migration in the NFP World, but there's not the same pressure for people to leave their home countries to just make money to send back home, as remittances. Also, to ensure imports don't undercut local markets, there is a revival of state-driven trade protectionism for nascent markets in low wellbeing regions/countries. This includes import tariffs, capital controls (to guard against the flight of capital), and limitations on foreign ownership in order to avoid predatory competition from abroad, as well as tax credits for local business to give them a leg-up.

[775] Although there will always be limits to full circularity due to entropy – this is why we also need to reduce consumption.

[776] In the for-profit world, most international migration is economically motivated (Ref: Population and Society, pp. 198-199).

The final trend that contributes to lower levels of international trade is that energy and transportation costs have gone up. Environmental taxes on fossil fuels have made long-distance trade less feasible and desirable. This is on top of the crucial impact of high oil prices in a peak oil world. Eating bananas in Norway, for instance, is a lot more expensive in the NFP World than in the current economy[777]. This ensures that banana growers and all of the workers involved in harvesting and transporting those bananas are paid fairly, that no harsh chemicals or radiation are used to keep the bananas 'fresh', and that they were grown organically. But for the most part, food systems rely much more on the local and seasonal growing conditions.

As a result of all of the points above, business and tourism-related travel has decreased. In the tourism industry, slow travel (involving longer traveling times and stays) has replaced the jet-set lifestyle that had become the norm in the era of cheap fossil fuels and highly subsidized for-profit airlines.

The shift to more functional, domestic and local economies has been embraced, along with a simplification of the global value chain. As a result of less trade and shorter supply chains, trade imbalances no longer carry the importance they used to. This adds yet another element of stability and resilience to the economic system.

Reformed international institutions, whatever institutions (whether they continue to exist as the UN, WIPO, IMF, World Bank, or Asian Development Bank or not), no longer have a growth mandate. The free market, Washington consensus style approach to these institutions is long gone, and unjust free trade agreements have been rolled back. Their new mission is to maintain balance, reduce distortions and undue influence in the international market, and offer viable loan rates to states who need an external financial boost. For instance, the World Trade Organization rules have shifted from non-discriminatory to needs-based. Of course, such loans are less necessary now that businesses are primarily NFP and locally controlled for the benefit of the community. And there's a lot less foreign ownership to extract wealth from local economies in the Global South. In the for-profit world, protectionism in the Global South risks leading to reduced foreign investment and innovation. But in the NFP World, these two issues don't matter so much because wealth and knowledge are both circulating freely.

The evolution of international markets and international relations is largely guided by the developments in domestic economies. There is an open dialogue between national governments about their needs and how they might be able to help each other improve the wellbeing of their people, always taking into account the ecological health of the planet as well. They share knowledge about better ways to monitor public health and Earth system functioning and how to achieve better outcomes. Of course, there are still ideological clashes about how best to do this, but without the influence of the profit motive, discussions can be much more fruitful.

---

[777] There are some ways around this, as shown by Iceland, where bananas are currently grown in greenhouses, using geothermal heat. But it is hard to imagine this being done in a sustainable way for most imported produce in the world.

With the global economy and national economies focused on positive social outcomes, rather than dominating the global market, the competition to out-trade and out-wit each other (a current race to the bottom) no longer really exists. Along these lines, there has been an end to the belief that the maximization of trade based on comparative advantage is desirable, and so we no longer see competitive devaluation of currencies.

**In this economic context, international relations are vastly different.** There are more trading and political confederacies to add a layer of negotiation and knowledge-sharing that might be more focused on a certain region or characteristic of states (e.g., confederacies for small states or island states). Regional groupings like the EU work better in the NFP World. This is due to fewer regional disputes over trade, migration, currency stability[778], and power games[779]. A global reserve currency system remains, managed by a supranational agency[780].

The most prominent of changes is the application of the belief that collective prosperity ensures individual prosperity. Due to the shift in beliefs and values, the NFP World enables the kind of international cooperation of which the for-profit economy is incapable. There has been a major transformation of foreign policy (no longer driven by private, for-profit interests) from 'the national interest' to 'the international interest', and multilateralism (i.e., focused on interdependence[781]). This means less dominating political strategies in order to maintain control and more partnership strategies to create shared social-ecological prosperity on Earth[782]. Guided by the story of interconnectedness, politicians no longer see countries as disconnected, competing entities. The NFP Era politicians see an ecosystem of many different cultures and economies interacting with each other in order to deliver the highest levels of wellbeing for the greatest number of people, for the long-term, globally. As a result, there are less disruptive elements like wars and terrorism, and international peace-keeping forces have replaced some standing armies. This atmosphere of trust has allowed for countries that previously spent the most on defense to shrink their militaries. Greatly reduced opportunities for war profiteering have had a lot to do with this development.

The latest form of the UN has been significantly transformed and strengthened by its member states as an international arbiter, with widespread domestic translation of international law reinforcing this newfound trust. But it is only the sum of its national parts. The possibility for unilateral and multilateral embargoes and sanctions remain, but there is less unilateral action generally, because national leaders have a lot less reason to suspect that other nations are trying to dominate them. The hard lessons of pillaging other countries for resources and cheap labor only to find that the resulting damage to people and planet affect us all have, for the most part, been learned. Corruption, tax evasion, and other illegal activity still occurs and both domestic and international criminal courts still enforce the protection of human rights and provide safeguards against corruption. Additionally, they enforce the protection of the natural environment. The main difference is that there is less motivation to commit such

778 Less pressure on currency exchange, with localized economies, sharing, and barter now a greater share of trade.
779 Think of the economic motivations that Russia had for invading and occupying Crimea.
780 As proposed here: http://www.un.org/ga/econcrisissummit/docs/FinalReport_CoE.pdf.
781 Even now, politicians are increasingly coming to see that when one country is doing poorly, like Syria or Somalia, it affects the entire world and other countries cannot do well.
782 I.e., Rianne Eisler's partnership versus dominator cultures.

crimes in the NFP system, it's more difficult to get away with such crimes (due to increased transparency and higher expectations and accountability from society), and the wealthy cannot so easily buy their way out of being punished for crimes.

But just as the NFP World succeeds capitalism, so is it merely a stage in our ongoing economic evolution. While it may be necessary, it certainly is not sufficient. In any case, the picture of the NFP World painted here may seem so different from the current system that it is impossible to achieve. And it is true that the kind of transformation this kind of vision demands is enormous and it runs deep – it goes all the way to the core beliefs held by society about what it is to be human. However, we believe it is possible. The following chapter outlines just one way such a transformation might realistically unfold in the coming decades.

# 6. The Great Transition

## *A Not-for-Profit World is possible*

Known as the mother of the modern Civil Rights movement, Rosa Parks is a household name in the United States. On December 1, 1955, Parks was riding on a bus in Montgomery, Alabama, when the bus driver told her she had to give her seat to a white passenger. Under the discriminatory laws of the time and place, Parks was arrested by the police for her nonviolent refusal to surrender her seat, an action that gained considerable publicity and set off waves of protests across the South.

Parks is rightly remembered as a hero, but why did her simple action of defiance have such a lasting impact? Likewise, why did Mahatma Gandhi's nonviolent resistance lead to Indian independence from the British Empire? And why did Nelson Mandela's release from prison spur the end of Apartheid in South Africa four years later? It is because their actions occurred in contexts that were ready for change. When people can no longer tolerate the dysfunction of an antiquated system, when there's a compelling vision for a better way forward that draws on the best of human nature and contemporary trends, and when enough people become aware of and collectively act on a vision, change can occur surprisingly fast.

The present global context is rapidly moving toward an economic shift. A growing number of people can no longer countenance our dysfunctional capitalist system. The seeds of a post-capitalist future are being sown. And we now have a vision of how, in theory, an NFP economy could create material security for all while ensuring we flourish within ecological limits.

Because our economy is such a dynamic, complex system, we cannot accurately predict where it is heading. In any complex system, everything is in a state of flux, with all aspects of the system coevolving. Seeking to specify precisely the steps required to reach the NFP World is therefore futile.

Thus, what we offer in this chapter is a story of transition—a blueprint whose order of events matters less than its shared characteristics. We seek to inspire and influence, suggesting trends and possibilities worth strengthening as well as those worth weakening. We are sure to overestimate some outcomes, underestimate others and not foresee many potential consequences.

Indeed, given the vast number of factors involved in any economic transition, we will be forced to limit our focus to explaining how the key components of an NFP system might emerge. These primary aspects include: the rise of NFP enterprise within the economy; triple crises of inequality, well-being, and ecological devastation; and emergence of the *lean society model*, *wealth circulation pump*, *cycle of well-being*, and *paradox of enough*.

Since the first human societies began, economic paradigms have evolved rather than suddenly supplanting the status quo.[783] Thus, there won't be a discrete day on which we can say we shifted from capitalism to the Not-for-Profit World. Rather, as NFP businesses take center stage in the economy and the social ethic shifts, the economy will increasingly become less extractive and more generative, and there will be a complementary trend toward greater equality and ecological restoration. This shift will be gradual at first and then will speed up as feedback loops in the system strengthen.

[783] Ref: Harrari, Rifkin, The Third Industrial Revolution

Significant upheaval is a standard component of economic evolutions. The collapse of irrelevant tenets and institutions paired with the birthing of new ones from what remains is a necessarily chaotic process. Indeed, modern capitalism developed out of the chaos and horrors of the eighteenth century. Accordingly, we have resisted the temptation to idealize the transition process. Social change on such a grand scale inevitably faces unintentional inertia and intentional resistance.

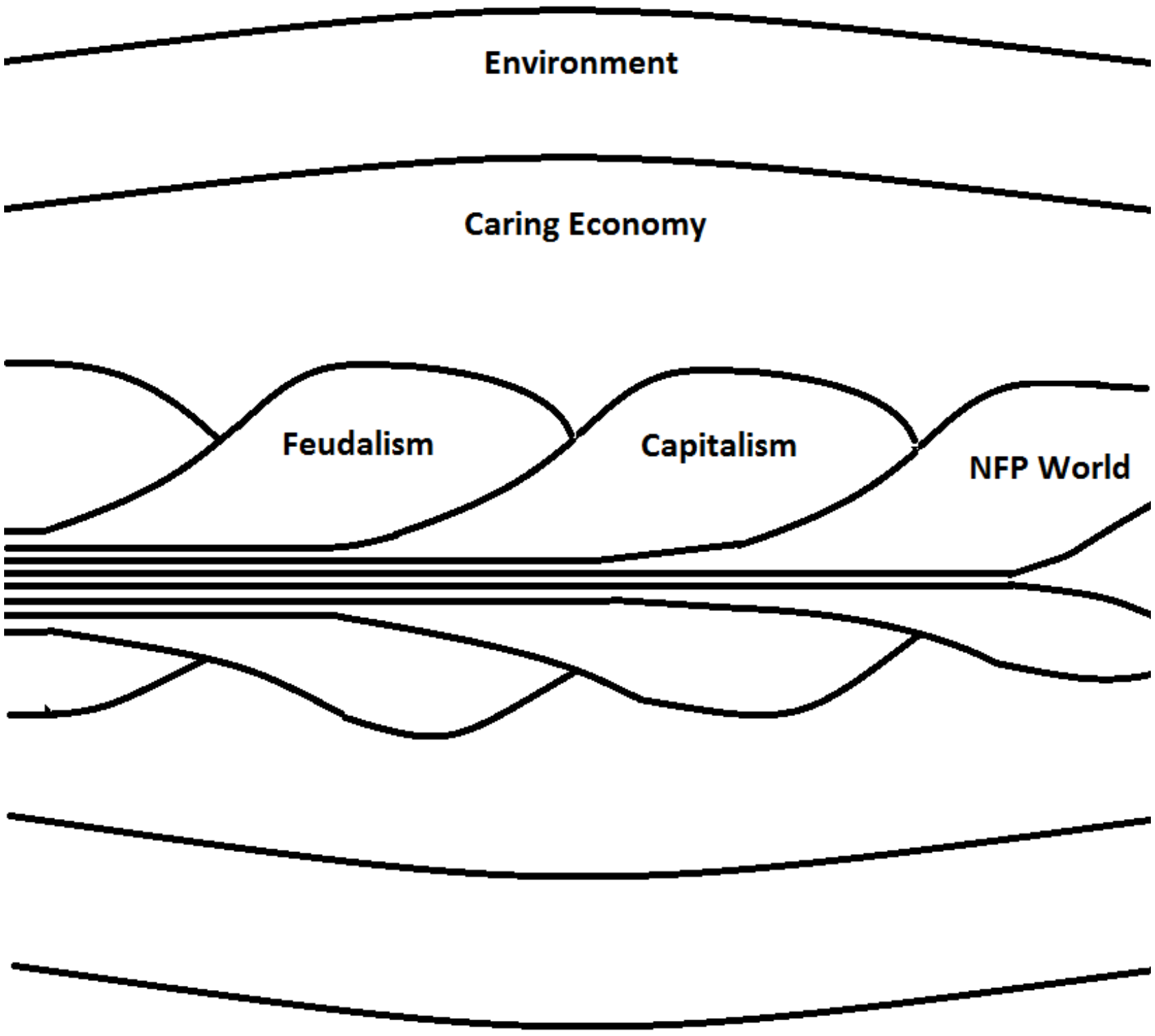

**The Evolution of Economic Systems**

The inherent nature of economic evolution involves two simultaneously occurring transitions. The evolution to an NFP World will require the simultaneous decline of the for-profit system and rise of the NFP system. The current for-profit story of isolation will fade as the story of interconnectedness emerges, just as happened when capitalism overtook feudalism in Europe. Whatever the economic system and its ideology, it is always embedded in the caring economy, which is always embedded in the planet's biosphere. The limitations of profit maximization are becoming more visible to those living to see how the capitalist system is bumping into economic, social and ecological limits to growth. At the same time, purpose maximization is becoming more appreciated for its ability to respond to the new needs, demands, and challenges of this century. As these consciousness grows about how dependent the economic system is on social systems of caring and on ecosystems, we will increasingly see the rejection of the ethic of "never enough" and the embracing of "enough".

At the outset of the Civil Rights movement, Dr. Martin Luther King Jr. said, "The arc of the moral universe is long, but it bends toward justice." We believe **the arc of the economic universe bends toward justice** as well and our economy is destined for greater fairness[784], but it is largely up to us to define the speed and nature of its emergence—and ensure it happens within our lifetimes, for all our children's sakes.

[784] This is due to the wealth extraction siphon of capitalism, which entails unacceptable social and ecological consequences. This will, at some point, lead to capitalism's superseding.

## The Coming Collapse

Although many of us may not feel it in our day-to-day lives, the for-profit system is in a state of rapid decline. This is transpiring due to the culmination of an array of global trends, including escalating inequality, thriftier consumers, the declining rate of corporate profit, asset bubbles, financial speculation, unsustainable levels of debt, environmental crises, resource shortages, and, as a result, the end of economic growth.

As we have shown, the for-profit system requires constant growth to compensate for the money being extracted into the elite economy. This doesn't simply mean the economy requires strong demand—it means ever-increasing demand. In a for-profit system, a lack of economic growth spells disaster; as theorist Serge Latouche argues, "There would be nothing worse than a growth economy without growth."[785]

There is an internal contradiction at the heart of the for-profit capitalist system: the economic growth it relies on ultimately ends up undermining its further growth as well as creating widespread social and ecological crises. This contradiction is the reason our economy is presently imploding.

### The End of Economic Growth

The evidence is stark. The global rate of economic growth has been steadily slowing for the last half-century.[786, 787] Even the economies of China and India are experiencing a "premature" decline of growth rates.[788]

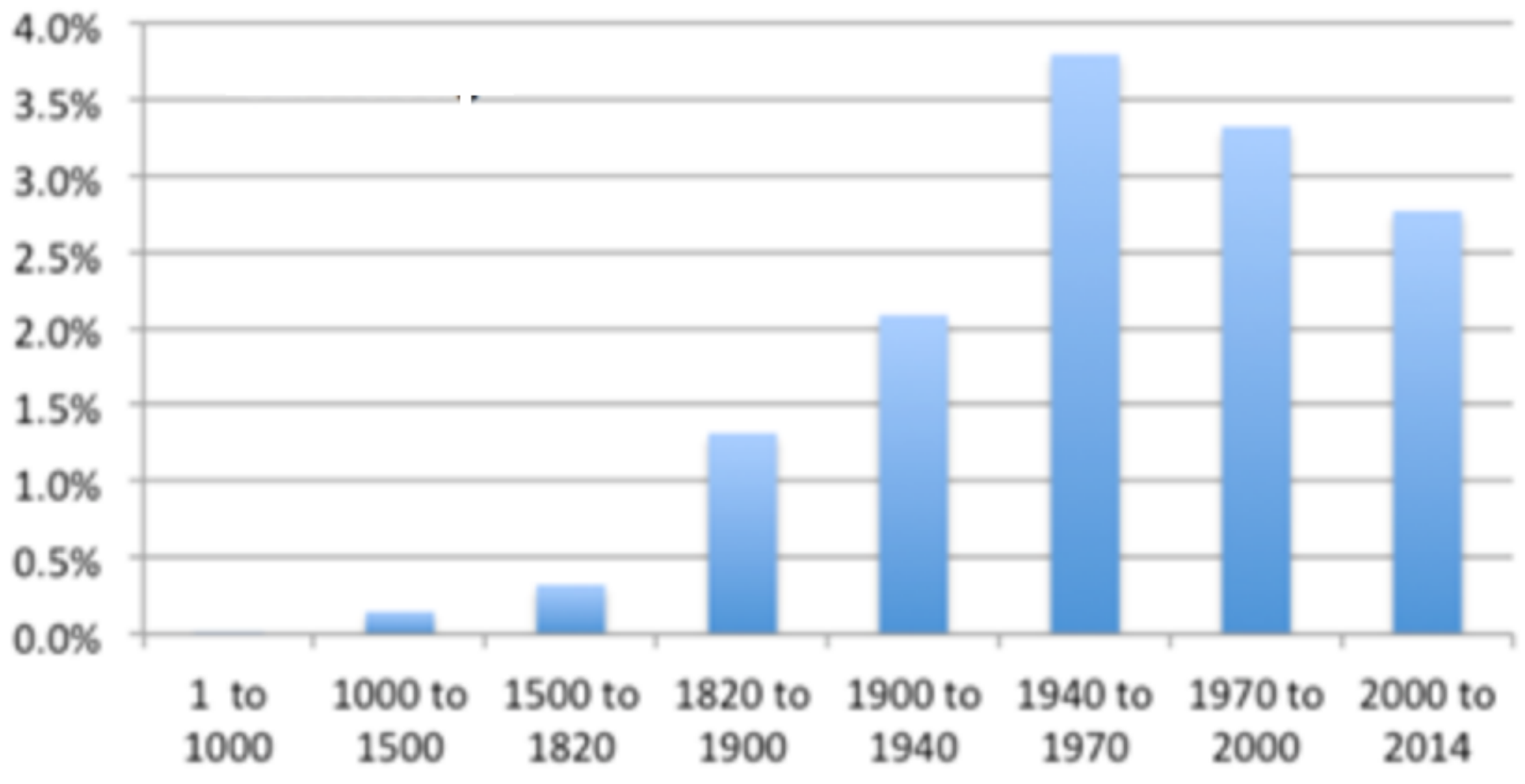

Average Annual World GDP Growth[789]

---

785 <REF 2003>.
786 Ref: IMF article
787 Ref: Haque, p. 4 (Ask Donnie which book if you have questions)
788 REF: ??http://www.wsj.com/articles/imf-cuts-2016-global-economic-growth-outlook-to-3-2-1460466006
789 Ref: Gail Tverberg, Charts Showing Long-Term GDP-Energy Tie

From 2010 to 2016—a post-crisis period during which the global market would have been expected to return to strong growth—the International Monetary Fund has been forced to downgrade its global growth forecasts six times.[790] Accompanying the fall in global market growth is the worldwide decline in corporate profits.[791, 792]

This pattern confirms Marx's claim that the rate of profit tends to fall[793], with an overall downward trend occurring since 1855[794].

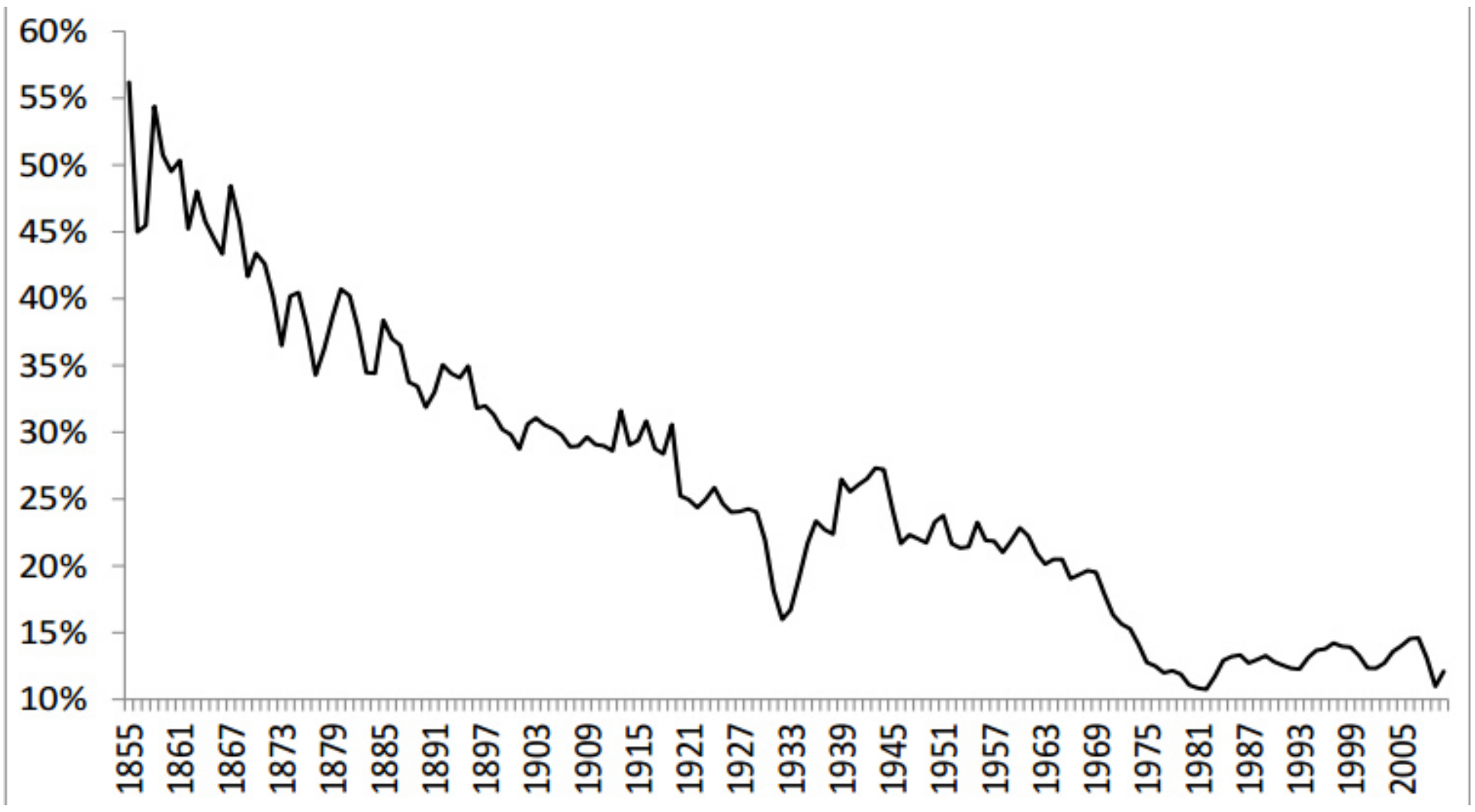

World Rate of Profit[795]

Marx foresaw internal contradictions of accumulation driving a series of worsening crises that would lead to political revolution.[796] While other economic philosophers weren't so extreme in their predictions, many also envisaged an end of sorts to the capitalist economy. Adam Smith, for example, thought the system would reach a plateau when the accumulation of wealth became "complete," bringing about a deep and long economic decline. John Stuart Mill believed a "stationary state" would evolve when accumulation ceased and capitalism gave way to a form of socialism. John Maynard Keynes postulated we would need a "somewhat comprehensive socialization of investment" in the future. And Joseph Schumpeter anticipated a kind of managerial socialism taking shape.

Each of these philosophers predicted some aspects of the presently emerging decline. Yet as we swiftly approach the end of economic growth[797], the full extent of the factors driving capitalism's demise is becoming clearer. We now understand social and ecological causes are synchronized.

---

790 REF:
791 Ref: IMF article
792 Ref: McKinsey
793 REF: Marx, chapter 13 of Das Kapital, Volume 3.
794 REF: Maito.
795 Ref: Roberts, M. (2015) "Revisiting a world rate of profit", Paper presentation: Conference of the Association of Heterodox Economists, Southampton Solent University, July 2015. (Data for the graph includes the strongest economies of the eras examined (e.g.; the G-6 countries after 1963)).
796 Ref for whole para: Heilbronner, R. L. (1985), The Nature and Logic of Capitalism (New York & London, W. W Norton), 143-4
797 REF: Paul Gilding.

### *The Chasm of Debt-Based Inequality*

Recall from Chapter 3 that the wealth extraction siphon is fundamental to the way the for-profit system seeks to create value and economic growth. Money introduced into the common economy mostly finds its way to the elite economy via for-profit business ownership, thereby reducing its general circulation. Given that all money is matched by equivalent debt in a fiat, fractional-reserve system, the concentration of money in the elite economy is matched by the expansion of debt—predominantly in the common economy.[798] Ongoing economic growth thus requires ongoing increases in debt.

Because the common economy has so little material security (assets) on which to fall back and leverage, it's not surprising household debt (along with all other forms of debt) increased in the wake of the 2008 crisis. That is to say, the crash did not "reset debt." Rather, since the Great Recession, global debt has increased by $57 trillion, outpacing world GDP growth[799], with the ratio of debt to GDP having increased in all advanced economies since 2007[800].

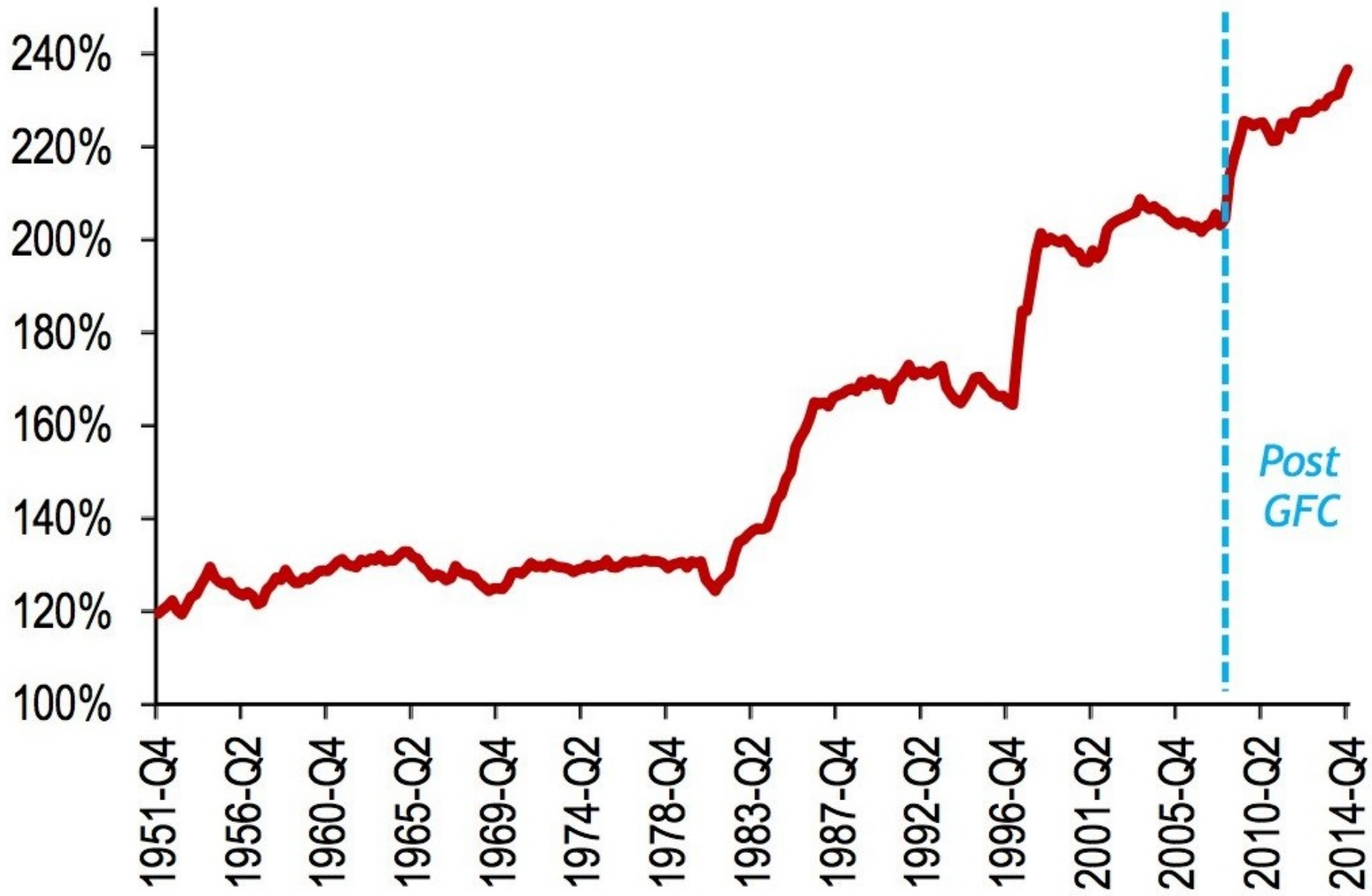

**Rising Global Debt[801]**

The lack of financial liquidity (money moving around) in the common economy creates deflationary pressure (falling demand for goods and services, even in the face of falling prices). We're not alone in believing this deflationary crisis in the common economy will continue to worsen in most parts of the world[802] as long as the system continues to require economic growth. However, we see the exact opposite trend happening in the elite economy, where the abundance of financial liquidity creates *inflationary* pressure (rising demand for assets along with rising prices). This means the small portion of people who live off passive income and other forms of extractive wealth creation are demanding more assets, causing asset prices to rise.

---

798 Although government, corporate and financial debt have all expanded significantly since 2000, as well <REF: McKinsey>
799 REF: [http://www.mckinsey.com/global-themes/employment-and-growth/debt-and-not-much-deleveraging](http://www.mckinsey.com/global-themes/employment-and-growth/debt-and-not-much-deleveraging)
800 Ibid.
801 Ref: http://www.businessinsider.com/baml-global-debt-has-rise-by-50-trillion-since-the-financial-crisis-2015-10?r=UK&IR=T
802 See, for instance: REF: HERE and HERE (SAME AUTHOR) and HERE>,

We call this phenomenon of deflation in the common economy and inflation in the elite economy *(de)inflation*. Rather than balancing each other out, (de)inflation reinforces the rising inequality caused by differences in income, reducing the key driver of economic growth: mass consumption. It does this by wiping out the disposable income of the consumer class—including in emerging economies[803]—through further stagnation of wage growth (justified by reduced consumption and profit margins), thereby ensuring asset ownership and material security are harder for people in the common economy to acquire.[804]

### *Shifts in Production and Consumption*

Technological innovation is also driving the end of economic growth.[805] Historically, labor-saving productivity gains (especially automation) have undermined middle-class wealth and purchasing power in the for-profit system by driving unemployment, wealth extraction, and deflation. This is because resources formerly invested in labor are shifted into new technological infrastructure.

The digital revolution is particularly relevant to the end of growth. We are heading toward what economic and social theorist Jeremy Rifkin calls a *near-zero marginal cost society*. Rifkin's hypothesis is based on evidence that, across many sectors, the price of producing an additional unit of a good or service (the marginal cost) has fallen dramatically, challenging industries in which revenue models are primarily built on individual sales. The music and publication industries are the most obvious examples: the cost of producing additional units of an album or book is close to zero because users can simply stream digital versions or download an electronic file onto their computers. Combined with the advent of online sharing, companies in these sectors can't sell enough of their traditional products (like music CDs, magazines, and newspapers) to cover their costs. Economic theories predicated on scarcity are becoming less relevant as it is obvious we are running a digital economy on an outmoded marginal cost operating system.[806]

The impact of the digital revolution extends to all sectors, with offline access to goods and services made increasingly possible through online platforms (e.g., Craigslist, Airbnb, Freecycle, and Uber). While tension arising from the privatization of such platforms[807] is indicative of the larger tension between the for-profit and NFP ethics, the disruptive impact of such models on traditional, for-profit notions of control over market interactions is undeniable. As a result, hierarchical organizational models and "vertical" industries are being replaced by networked models and "lateral" industries.

Perhaps most importantly for the long-term decline of the for-profit system, falling costs of production combined with distributed manufacturing methods like 3D printing are enabling more distributed production and competition. In addition to sharing with and shopping from each other, people can increasingly make their own physical goods in cost-efficient ways[808], which enables more people to opt out of the market[809] and thus put downward pressure on corporate profits.

---

803 The growth of consumer markets in emerging economies cannot continue indefinitely; mainly because of the impending liquidity crises. After the financial crises in the US, Brazil and Europe, and a painful economic slowdown in China, many people are reluctant to choose to take on further debt to buy superfluous products or services. This is why the anticipated emergence of a broad middle class in places like China and Brazil has not happened and instead we see rapidly growing inequality (Ref: Bloomberg article, Unit of Mich article, Financial Times article).

804 Today, 235 million households live in substandard housing and it is estimated that by 2025, 106 million more households will face the challenge of finding affordable housing. Ref: McKinsey report: Tackling the World's Affordable Housing Challenge

805 REF: Rifkin.

806 REF: Rushkoff.

807 REF: platform cooperativism movement.

808 REF: Joshua Pearce's research on the cost of producing household items using a RepRap.

Shifts in production have coevolved with shifts in consumption. A counterculture of conscious and ethical consumption has rapidly emerged, largely driven by the failings of the growth-obsessed economy. Collaborative forms of consumption focused on access rather than ownership are accelerating the drop in mass consumption. These factors combined with the powerful effects of customers seeking suppliers less obsessed with growth at all costs mean the for-profit system is taking a hit.

### *The End of Cheap Energy*

In addition to falling profits, astronomical levels of debt, financial speculation, and a declining consumerist culture, sustained economic growth is threatened by environmental factors. In economic terms, the most obvious of these factors is society is running out of cheap, easily accessible natural resources.

The growth system has relied on massive quantities of cheap energy. The era of cheap energy is now ending. With the low-hanging fruit having been picked, we are entering an age marked by shortages of resources vital to our present systems of industrial manufacturing.[810] Companies founded on the fact that fossil fuels are cheap and easy to extract are discovering the opposite will be true in the twenty-first century.[811]

Extraction of fossil fuels like oil and natural gas is already more costly due to the geological challenges presented by most remaining reserves.[812] The energy returned on energy invested (EROI) for the global production of oil and gas by publicly traded companies appears to be declining, with crude oil having peaked in 2006, according to the International Energy Agency.[813]

Alternatives to traditional fossil fuels—such as tar sands and oil shale—have contributed to a temporary glut[814] and lower prices[815], but they deliver a miniscule EROI[816]. And while new technology and production methods are said to be maintaining production, they are insufficient to compensate for the depletion of conventional oil.[817] Indeed, oil would need to be priced prohibitively high to justify drilling in the Arctic and ultra-deep-water reserves because those are such expensive operations. We know from the 2008 crash that the market has sensitivity to oil prices exceeding a certain point, so such drilling is not financially feasible.[818]

Oil companies are already struggling to distribute profits to shareholders.[819] In April 2016, Exxon Mobil, one of the biggest oil companies in the world, reported its lowest profits since

---

809 Ref for whole paragraph: Rifkin, Marginal Cost Society; [see www.fab.city and 'cosmo localization' – design global, fabricate local)

810 Ref: Geochemical Perspectives, Sverdrup & Ragnarsdottir: http://www.geochemicalperspectives.org/wp-content/uploads/2015/09/v3n2.pdf

811 Ref: Heinberg, The End of Growth

812 Ref: Heinberg, The End of Growth

813 In 2010, the International Energy Agency settled the matter. In its authoritative 2010 *World Energy Outlook*, the IEA announced that total annual global crude oil production will "never regain its all-time peak of 70 mb/d reached in 2006, p.6?" *World Energy Outlook 2010* (Paris: OECD/IEA,2010).

814 REF: http://www.postcarbon.org/the-peak-oil-dilemma/.

815 We refer here to the temporary fall of oil prices in 2016, to below $36 per barrel.

816 <REF>.

817 REF p.210.

818 <RIFKIN>.

819 Ref: Breaking Energy article

1999, and the company is in jeopardy because its overall debt is now greater than its annual profit.[820]

The data on coal is more limited, but researchers Charles Hall and Kent Klitgaard report that "the energy content of coal has been decreasing even though the total tonnage (mined) has increased."[821] Some of the largest coal-mining companies in the world recently declared bankruptcy.[822] And keep in mind much of the world's transport of coal, and therein its pricing, relies heavily on cheap oil.

An Oxford Economics study estimates that, by 2036, energy costs will be 166-percent higher than in 2005.[823] As companies across all sectors of the economy pay for energy—either directly or indirectly—the anticipated rise in energy costs will affect profit margins throughout the market. Moreover, the costs of manufacturing abroad are increasing, largely due to the rising costs of transportation and mineral extraction (compared to 2005 levels, the latter is expected to rise 35 percent by 2036[824]).

Surging consumer costs (particularly food prices, which, by 2036, are estimated to increase by 91 percent compared to 2005 levels[825]) combined with reduced disposable income in the common economy is a recipe for disaster. People either go deeper into debt or forgo necessities like lighting and heating.

Furthermore, with the outcome of the 2015 climate negotiations in Paris (COP21), carbon and other environmental taxes, fines, and trading schemes are set to rise.[826] This translates into even higher costs for companies, especially in resource-intensive fields.

### *The Costs of Ecological Damage*

The problem transcends the increasing costs of extracting resources to fuel economic growth. We have reached limits in terms of the amount of damage we can wreak on the planet's ecosystems without our economy being negatively impacted by it.

Given that our economy relies on a functioning biosphere, human impacts on the environment—including climate change, biodiversity loss, degraded land, and polluted water and air—put our economy (not to mention the existence of humans and all other species) in a perilous position.[827] These problems:

- threaten the amount of food we can grow and therein prices;
- wipe out material security, largely in the common economy, through natural disasters while also leaving some assets uninsurable; and
- exacerbate illness[828].

---

820 Ref: Exxon Mobil profits crash to lowest level since 1999
821 REF: C. Hall, K. Klitgaard, Energy and the Wealth of Nations: Understanding the Biophysical Economy, Springer Publishing Company, New York, USA (2012).
822 Ref: NPR article
823 Ref: Oxford Economics in http://www.pwc.com/us/en/corporate-sustainability-climate-change/assets/investors-and-sustainability.pdf
824 Ref: Oxford Economics in https://www.pwc.com/us/en/corporate-sustainability-climate-change/assets/investors-and-sustainability.pdf
825 Ref: Oxford Economics in http://www.pwc.com/us/en/corporate-sustainability-climate-change/assets/investors-and-sustainability.pdf
826 Ref: New York Times article: http://www.nytimes.com/2013/12/05/business/energy-environment/large-companies-prepared-to-pay-price-on-carbon.html?nl=todaysheadlines&emc=edit_th_20131205&_r=3&
827 Ref: Geochemical perspectives, Sverdrup & Ragnarsdottir; End of Growth, Heinberg.
828 Ref: http://www.forbes.com/sites/mikescott/2014/04/03/climate-change-threatens-economic-growth-un-report-how-should-investors-react/#5a1cc7783062.

### *The Millennial Difference*

These mounting environmental and socioeconomic problems aren't going unnoticed by the masses. Having grown up with the threat of ecological collapse, the Millennials have more ecologically conscious worldviews[829] and have been at the forefront of environmental activism, climate coalitions[830], and the fossil fuel divestment movement[831]. On many fronts, they are more open to questioning the status quo. Indeed, Millennials are the *x* factor threatening to magnify the aforementioned trends driving the end of economic growth.

The global youth were affected by the 2008 crisis and subsequent inequality more than the older generations. Many of them were just coming into adulthood and entering the workforce when the global economy cracked wide open, and they were met with record levels of job insecurity.

In this light, many young people are choosing to challenge the system, as witnessed by the massive rally of youth support for US Senator Bernie Sanders, a self-declared "democratic socialist," in the run-up to the 2016 US presidential election. A Harvard University survey conducted in 2016 found most members of the American Millennial generation reject capitalism.[832] This sentiment is manifesting in their increasingly selective spending patterns[833], how much they favor access over ownership[834], and their "opting out" of elements of what was previously considered mainstream culture, including the stock market[835]. In response, there is greater support for distributed production and the local economy (as witnessed in Greece and Spain following the 2008 crash).

Given the demographics of global population growth, most births in the coming decades will occur in the households most marginalized by the global economy, stretching the rubber band of inequality further as such families—and their respective governments—are forced to absorb debt even faster.

## A House of Cards

We are not alone in thinking a massive economic crash is coming.[836] We envisage this crash being deeper and more widespread than any the world has ever experienced due to the sheer size of global debt and the interdependencies woven throughout the contemporary economy. Combined, these factors form a house of cards ready to collapse as soon as the first few players go down.

---

829 REF? Steve Schein?
830 See, for example, the the International Youth Climate Movement.
831 REF.
832 Ref: Washington Post article (original study?)
833 See this study on 'fauxsumerism' for instance: http://www.dailylife.com.au/dl-fashion/fauxsumerism-is-the-latest-retail-trend-apparently-20140429-37ez9.html (primary source?)
834 REF
835 Millennials are already much less involved in the stock market than previous generations. They have much less of their retirement savings tied up in stocks and they have much lower trust in the stock market to handle their savings, as only 18 percent trust it. (Ref: Yahoo article referring to lots of other studies – primary sources?).
836 See the work of economists at the Royal Bank of Scotland, Ann Pettifor, Molly Scott Cato, Steve Keen, and Thomas Greco, to name a few.

The collapse will be swift. The lack of regulation in the global financial sector has enabled instruments such as derivatives and high-speed trading systems to create great reactivity and volatility in financial markets. Most worryingly, as of mid-2016, the total value of the unregulated derivative markets is US $492 trillion.[837] As avenues for high returns in productive sectors of the economy decrease, capital will continue to be steered to more speculative ends.

The coming collapse will also be longer not only due to the depth of the crash but also because many of the conditions leading to the end of growth will continue to be exacerbated.

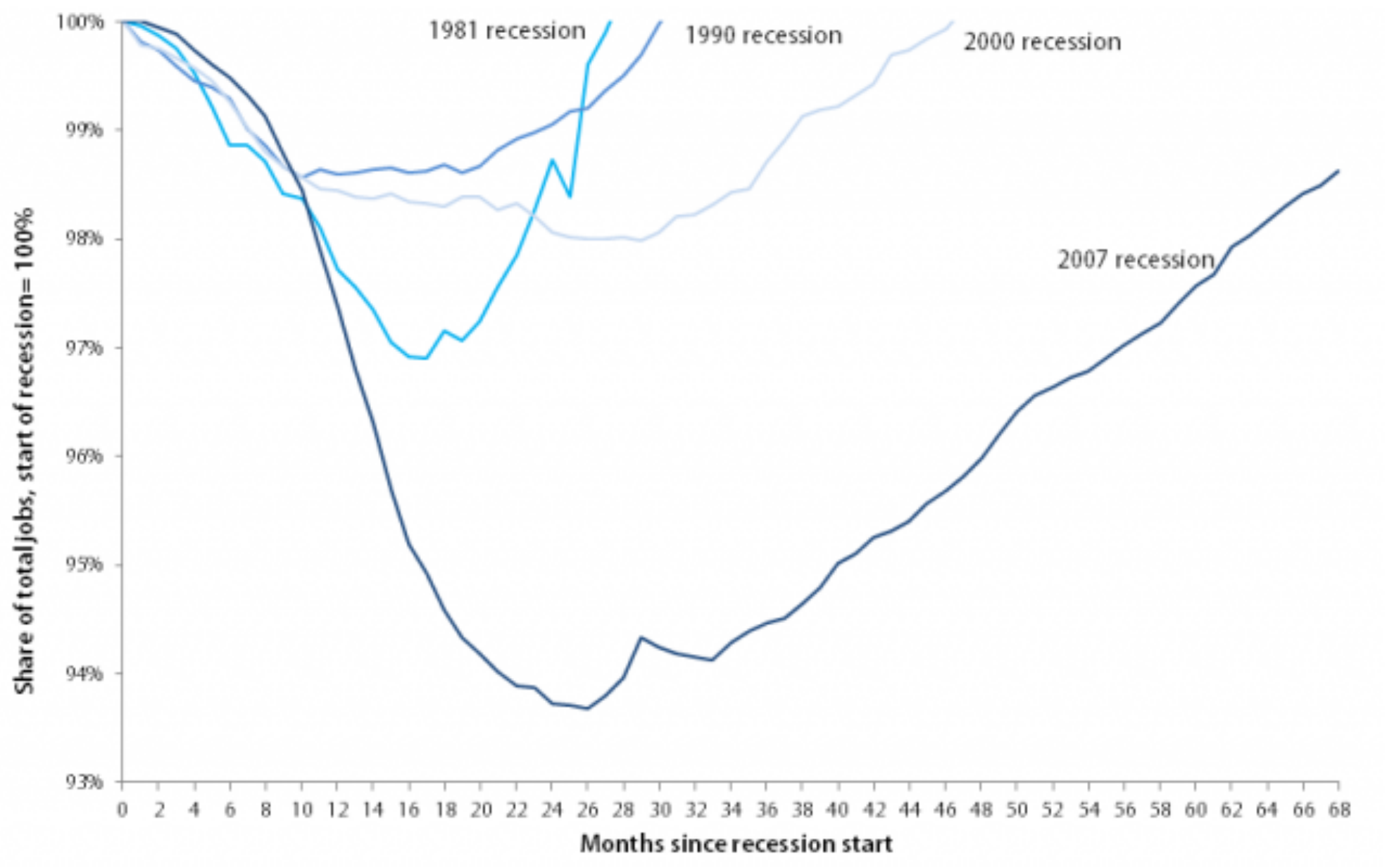

**Length and Depth of Recent Recessions on the US Jobs Market**

At the moment, last-gasp tactics such as quantitative easing and the 2015 rise in the US federal interest rate for the first time in seven years are creating unwarranted confidence in an economy loaded with debt and risk. Many for-profit companies have involved themselves in speculative forms of debt based on the questionable assumption of future growth. Likewise, governments have undertaken stimulus measures without a clear sense of how they will manage their debt obligations in the long run. More realistically, negative interest rates[838] signal the capitalist system is reaching its absolute limits.

The crash may begin with a large portion of corporate, household, and public debt coming due, triggering a huge wave of defaults and subsequent bankruptcies, as happened with housing debt in the US in 2007–2008. Indeed, speculative investment in housing is likely creating another global real estate bubble[839], while the steering of investment toward the speculative financial markets is driving more asset bubbles[840]. The crisis could also begin with a national debt default, causing what is known as a "run on the banks" and leading banks and large financial institutions to collapse or be bailed out again because they cannot stay in business without their loans being repaid.

---

837 Ref: http://www.bis.org/publ/qtrpdf/r_qt1606.htm.
838 An incredible example of this is Japan, where people are paid to borrow and charged to save (Ref: Bloomberg article and BBC article).
839 Ref: Ibid
840 Ref: Minsky?

In the next three sections, we present an entirely speculative account of a post-crash transition to an NFP system. We cover both the collapse of the for-profit story and the emergence of the not-for-profit story.

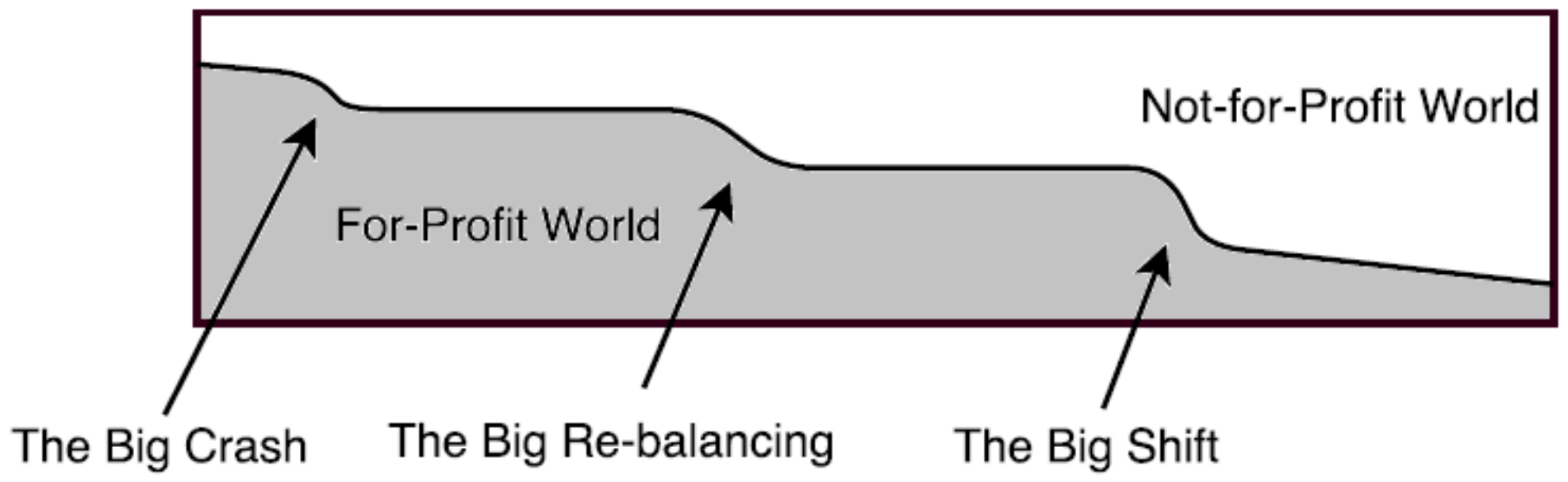

**The Three Waves of the Great Transition**

We have split the narrative into three sections outlining three waves of global experience we imagine might accompany the great transition beyond capitalism: the Big Crash, Big Rebalancing, and Big Shift. In our highly speculative account, we envision these waves rolling out over the period 2020 to 2050.

# The Big Crash

No matter the specific cause, the crash begins devastatingly, at around 2020. Stock prices, primary export markets, and currencies collapse. Efforts to suspend trading and foreign exchange do little to halt the hemorrhaging. In an act of desperation, people in the elite economy shift what investments they can to gold, cash deposits, cryptocurrencies, low-yield government bonds, and offshore bank accounts.

With the crash come widespread business bankruptcies. A number of multinational corporations fold due to collapsed demand for their goods and services or recalled loans that they can't service or renegotiate. It is discovered that many companies have been holding significant debt "off the books," unbeknownst to regulators, stockholders, and even some employees and company directors.

In the meantime, people in the common economy suffer the most. Small businesses are also forced to close their doors due to diminished demand and rising costs. Hundreds of millions of people lose their jobs and ability to pay their debts. Many of these people are unable to continue paying the mortgages on their homes, forcing them to try to sell their homes to avoid bankruptcy. With hardly anyone trying to buy a home, however, housing prices have dropped to the point that what the homeowner can secure through a sale is lower than the amount owed on the outstanding mortgage, meaning people lose their homes to the banks through foreclosure. This housing market collapse is as bad if not worse than the one that occurred in 2008. Vehicles and other valuable items are commonly repossessed as well, and additional savings are decimated with the bankruptcy of pension and retirement funds.[841]

[841] If this seems hard to imagine, just look at what happened in the US economic crisis of 2008, which left millions of people unemployed and homeless.

Prices of goods and services skyrocket due to a massive drop in global production.[842] Shortages of water, electricity, gas, fuel, and food create social devastation. Levels of famine rise along with suicide rates. Families are living in even more cramped conditions, sleeping in cars and on the streets, squatting[843], or swelling the world's slums. As extreme poverty surges, airborne and waterborne diseases spread. When natural disasters hit, the toll is even more horrific than usual.

There is an immediate rise in crime—especially assault, robbery, and looting—and the black market as well as exploitative lenders feed on the scarcity. While shock, despair, and survival instincts prevail, long-held rage also bursts through. Rioting, threats to wealthy landowners, and the forceful takeover of estates cause owners to flee their properties. Civil and international disputes turn into military escalations.

Given the extent of the collapse, it is no surprise many are claiming Armageddon has arrived. Governments immediately declare a state of emergency, calling on the military to assist in matters such as the rationing of food. Some leaders even temporarily declare martial law, leaving populations susceptible to fascist leadership and state brutality.

While the tax base has shrunk dramatically and publicly funded services are cut, governments rush through unopposed legislation to bail out and temporarily nationalize the "too big to fail" banks and other large companies deemed crucial to the economy. Extending methods from the 2008 crash, for-profit banks facing insolvency take depositors' money as their own[844] in a legal phenomenon known as "bail-ins"[845]. Governments follow suit, nationalizing retirement savings.[846] Corporate mergers occur with limited oversight, and many for-profit insurance companies rapidly mutualize.

Fear-based national protectionism sees the immediate introduction of strong capital controls (limits on how much money can be withdrawn from bank accounts or sent abroad), while countries overtly renege on free trade agreements, exacerbating pressure on prices. The collapse of the US dollar accelerates the shift to a basket of currencies acting as reserve currencies, which are ultimately replaced by a global reserve currency. Geopolitical power decentralizes as the wealthiest countries experience dramatic falls in GDP, requiring them to scale back military spending and overseas operations.

Communities and their leaders rise to the challenge. Of those that have survived, local businesses, nonprofits, and NFP enterprises play a significant role in providing stability amidst the chaos, responding to the greatly expanded need for essential services such as food, accommodation, and health care. With widespread collapse in government and donor support for nonprofits, the extension of essential services relies heavily on volunteered labor and donated goods. Even government services are propped up with volunteer teachers, nurses, doctors, police officers, firefighters, and emergency services personnel. In local communities around the world, people increasingly work together to meet each other's needs directly. They hire one another informally; barter goods and services; and share meals, seeds, tools, energy, fuel, vehicles, and land more readily than ever. Systems of mutual credit and alternative, community-based currencies swell as a means for brokering local exchange.

---

842 Interestingly, in times of crisis, (de)inflation largely flips – the prices of high-cost assets, except for rare materials like gold, drop and the prices of regular goods and services rise. However, this doesn't reduce the inequality gap because the common economy can no longer afford investments, as income declines in crises as well.
843 If this seems hard to imagine happening in a rich, Western country, look at what happened in Greece in 2010. Within a matter of months, rates of homelessness, crime and suicide skyrocketed as people lost their livelihoods (Ref: BBC article and Greece's Health Crisis article).
844 This is what happened in Cyprus in the financial crisis of the 2010s (Ref: New York Times article).
845 REF: Ellen Brown
846 As we've seen in Poland, Hungary, Portugal, and Argentina (REF).

## The Implications of Ongoing Recession

In contrast to previous recessions, the Big Crash is so severe there is a fall in the overall amount of debt in the system[847] (and therein the money supply). This is because throughout the crash, governments, corporations, and households are forced to default on debts and declare bankruptcy. When bankruptcy is declared, the unpayable debt is often canceled from the associated financial ledger. When this happens with a lot of large companies over a short period of time, the amount of overall debt (and money) in the economy decreases. Debt in the financial sector also shrinks as the financial markets contract severely. While governments seek desperately to expand the money supply, trust is so low that financial institutions (other than public banks) restrict their lending, and citizens and businesses try to avoid taking on more risk.

In the process of overall debt contracting, insolvencies ensure ownership of assets such as homes transfer from the common economy to the elite economy. With most debts involving payments to for-profit banks, global inequality actually rises further throughout the crash, and the gap between those with investment assets and the means for material security and those without widens. Indeed, many in the elite economy grow their equity investments by buying back stock at various stages of the market bottoming out, and some even make money by shorting the markets before and during the crash. Even the elite economy has shrunk during the crash, with many upper-middle–income households losing everything because they had accumulated substantial debt to support speculative activities or aspirational lifestyles before the crash and lacked sufficient assets to sustain the market's immediate losses.

Having already shifted their money offshore before heavy capital controls were enacted, many wealthy people from high-income nations migrate to middle-income nations to avoid harassment and maintain a lavish lifestyle amidst the strife.

The Big Crash leaves the economy in a protracted recession. Much more than a "market correction" that returns asset values to "reasonable" levels and clears out unproductive actors in the economy, the crash leaves few nations capable of producing even a single quarter of economic growth.

With mass unemployment, many are living in extreme poverty. Average consumer demand is close to flatlining. In India and China, where economists were counting on a large consumer class to develop and spur economic growth throughout the twenty-first century, consumption has dropped with devastating effects.

Reflecting the state of the economy, financial markets remain decimated. Investor confidence is at an all-time low, and there is minimal appetite for high-risk ventures, especially new businesses and startups. In isolated cases (and often as a result of direct government subsidies), rumors spread about booming assets. Small bubbles form but pop once the fervor evaporates and the market corrects to reflect the real value of the underlying assets. Rather, during the entire decade, capitalists shy away from equity-based investments since stocks in most sectors, businesses, and commodities offer little security (let alone returns), and those that do are rarely traded. This means few new businesses emerge, and a notable slowing of innovation occurs worldwide.

[847] However, the debt-to-GDP ratio is increasing, given GDP has dropped so significantly and government debt is rising so fast in many parts of the world.

Many for-profit business and political leaders intuit the deep connection between the for-profit system and the collapse being experienced, yet they know addressing these issues means overhauling the entire economy. They often present themselves as offering systemic solutions, but ultimately this is a façade for business as usual. There is still a lot of fear involved in contemplating those kinds of changes because the most powerful people in the world (who continue to bankroll candidates for political office) stand to lose much of their remaining fortunes and power in the transition away from a profit-oriented economy.

Most leaders cannot see any viable alternative. The vast majority of people remain stuck in thinking the economy can only model some form of capitalism or state socialism. A bitter debate continues to rage about the free market versus the regulatory state. Some even lay blame for the crisis on the for-profit system not being dominant enough, claiming the 2008 corporate bailouts didn't allow the market to operate freely and effectively, creating inefficiencies that led to widespread overleveraging.

Thus, in the longer term, official responses to the crisis largely involve denying its real causes, with efforts geared solely toward restarting the for-profit system. Throughout, there is little public acknowledgment that the global economy is in a state of protracted collapse. Governments try to keep people's faith by distorting statistics in ways that suggest growth is just around the corner.[848]

Indeed, growth at all costs is the mantra, and the mainstream argument is social and environmental concerns can only be addressed once growth is restored. This allows lowest-common-denominator corporate behavior to abound.

Following mass layoffs, for-profit companies in high-income nations automate what labor they can while outsourcing most of their service and production centers to low-income countries, where labor is cheap (except for the brightest talent, which remains lured by promises of stock options once the global economy returns to growth).

High unemployment and a massive labor surplus mean bonuses and benefits are curtailed, while working conditions and wages deteriorate. In a reinforcing loop, there is a loss of confidence and membership in workplace unions. Largely out of fear, governments do little to enforce good business practices.

With the prices of goods and services so volatile, major for-profit companies commonly collude in attempts to gain market share from smaller players, including controlling the flow of information through the media and even the Internet.[849]

---

[848] In our present world, governments are already distorting unemployment, efficiency and price index data (Ref: Shutt).
[849] This struggle has already started, with debates about net neutrality (Ref: ).

Corporate fraud—including that committed at nonprofits and some NFPs—has remained high ever since the Big Crash commenced. With governments and the courts scrambling to respond to the chaos, the managers, CEOs, and owners complicit in corporate fraud escape prosecution.

On the environmental front, standards have dropped due to inadequate budgets for enforcing regulatory measures as well as the focus on growth at all costs. Ironically, while pre-crash feedback loops continue to drive environmental problems, the global recession and massive drop in demand and global trade (along with the misery it has brought to people's lives) has driven a sizable drop in population growth, consumption of resources, waste, and greenhouse gas emissions (as happened during the 2008 crisis[850]). Overall demand for coal, oil, gas, and energy-intensive products such as meat dropped in a phenomenon that extended through the decade.

## Unsuccessful Attempts to Kick-Start Growth

Efforts to kick-start growth focus on stimulating market demand *and* incentivizing private sector investment and job growth. There are still many differences, however, in the ways countries approach fiscal policy as well as their attitudes toward the size of government debt.

Many national governments try to stimulate consumer demand through public investment. Measures include the creation of major infrastructure projects and direct payments to citizens (sometimes referred to as "helicopter money"). Although less severe in countries with sovereign wealth funds and public banks, recurring budget deficits are funded via low-yield government bonds, higher taxes on the wealthy, and the privatization of public assets (especially land) at below-market rates. Across all economies, there is public messaging such as "consume to get the economy going" and "have a child for the nation."

Other governments focus on maintaining minimal government debt by implementing austerity-based measures. They provide tax cuts for the rich to stimulate investment (thereby increasing relative taxes for the poorer classes). They also privatize public goods (especially services) as part of a targeted plan to cut government services and create more space for private sector businesses in addition to contracting work traditionally performed by public agencies to for-profit companies.[851]

Across all governments, standard measures to drive business investment and employment include corporate tax breaks (fearing capital flight), returning to the private sector the businesses and industries that were nationalized after the Big Crash (having absorbed the debt that would have otherwise bankrupt such companies).

Although stimulus and austerity measures occasionally provide short-term success in reviving growth, by the mid-2020s, nothing has succeeded in getting the economy back on track. State-led responses have failed primarily because they have not been able to counter the rate at which money is siphoned out of the common economy, and the perpetual expansion of sovereign debt has undermined investor confidence.

[850] REF:

[851] We already see this happening in political campaigns like the Big Society, in the UK, which is gradually de-nationalizing many goods and services previously provided by the government, like care for the elderly (Ref: NESTA?). Other examples include austerity measures in Spain, Italy, and Greece.

Market-led responses have failed primarily because they have been unable to create the necessary material security in the common economy to allow people to consume at an ever-greater rate each year. On the contrary, most approaches have driven greater inequality as the wealth extraction siphon—although weakened by the anemic economy—remains the dominant mechanism through which money flows within the economy. The official story is an extended recession is a necessary part of creating a new and better form of capitalism and such an economy lies waiting on the horizon.

## Demands for an Alternative

A considerable segment of the global population has lost faith entirely in capitalism by this time. For two decades, the global economy has been steadily worsening with no real signs of recovery, despite numerous efforts by governments to show otherwise. Millennials, now in their 40s and 50s, face a future without a secure retirement, even in high-income countries. This generation is the first in high-income countries to have less material security than its parents. In the rest of the world, the majority are struggling to simply survive.

The result is widespread participation in anti-establishment demonstrations, protests, and organized social movements. Through the strengthened community, bonds have been formed between people of all political persuasions. Protesters focus their energy on the wealthy and corporate capitalism, demanding that the elite economy and corporations pay more taxes, unfair subsidies and limited accountability end, and money in politics and widespread corruption be addressed. But there remains a struggle (and even contest) to clearly articulate an overarching alternative.

Among other proposals[852], word spreads about the not-for-profit market economy as an alternative. Yet many misinterpret the descriptions they hear, dismissing the model as state socialism or communism. They confuse NFP enterprise with the traditional charity-dependent nonprofit model, which conjures up thoughts of inefficient bureaucratic organizations as well as lobbying groups and partisan think tanks. Ignorance, combined with attachment to the old for-profit story, ensures most of the social entrepreneurs seeking to start businesses during this period still choose for-profit legal structures. Some others, who understand more about NFP business, believe it is corrupt for certain nonprofit organizations to make money, especially in light of tax exemptions.

What constrains the emergence of the NFP World even more is that, despite shifts, the profit motive and for-profit story remain deeply embedded in the collective psyche, even as capitalism is increasingly challenged. For many, the thought of an economic system based on NFP business is too great a conceptual leap. Many find it hard to imagine a market based on purpose or the idea that human nature is anything more than greedy and competitive. Economists and business schools reinforce this belief. Although there is now greater openness to social enterprise and cooperative models operating with a profit motive, economists and business leaders perpetuate the myth that business is always a for-profit activity and a for-profit market is the only way to ensure appropriate incentives, competition, and the economic diversity needed for innovation and kick-starting growth.

[852] Proposals like the Pluralist Commonwealth (Alperovitz); reinvigorating the Commons (Bollier); Resource-Based Economy (Fresco), and the Circular Economy (Ellen MacArthur Foundation).

People have made the connection between capitalism; the Big Crash; and the crises of inequality, well-being, and environmental degradation, but the pivotal role capitalism's profit-maximizing logic played in triggering these crises is not universally understood. The majority of people still don't comprehend the financial system and how money is created as debt. And most—including many environmentalists—are still unaware they live in a for-profit system that requires an ever-growing throughput of resources incompatible with life on a finite planet. Acknowledging it is our *economy* causing the sixth mass extinction—and not just some isolated bad behavior—relies on the kind of nuanced understanding of our economic system most people do not yet possess.

## Practical Shifts Toward the NFP Ethic

Nonetheless, global movements result in action that steers the economy and social values in an NFP direction. "Conscious capitalism," "social enterprise," and "purpose-driven business" make headlines, and underneath these concepts lies a powerful longing for distributed power and ownership.

The rise of community economies—especially the expansion of the digital and social commons as well as informal barter networks—gives people a chance to realize what is possible without shareholder ownership. Technologies such as the blockchain (supporting decentralized innovations such as cryptocurrencies) continue to show how interconnected economic activities can be governed by collective authentication systems that don't privilege any one individual. Community purchasing and the management of failing businesses such as cafés and pubs—often under cooperative structures—has proven one of the more dependable avenues for viable local business. Indeed, the movement to divest from the largest corporations is growing, particularly the momentum behind people moving their money out of big for-profit banks.

Nowhere has the divestment movement been stronger than in the energy sector, where people, organizations, and governments are encouraged to remove their investments from companies involved in extracting fossil fuels. Despite the ongoing recession, there has been a faster shift to renewable resources than believed possible before the Big Crash. As nations, regions, and communities recognize the desperate need for local energy security amidst shortages and rising prices, there is a strong impetus for the development of a renewable energy infrastructure, especially in countries in which the nationalization of foreign energy companies was not an option. In addition to government initiatives funded by green bonds, subsidies assist communities with forming local NFP energy cooperatives. With the renewal of mass environmental concern in the second half of the 2020s—driven by resource shortages, escalating climate impacts, and conflicts—communities are eager to reduce their ecological footprints, with localization seen as a powerful means by which to achieve strong outcomes.

More broadly, there are shifts to systems demanding less energy, such as mass transportation and organic agriculture, with the transition somewhat paralleling what happened when Cuba lost more than half its oil imports following the collapse of the Soviet Union in the early 1990s[853].

But it's not just energy systems that have been transformed. The Big Crash, and the associated drop in philanthropy and government support, have forced traditional nonprofits and those beginning their social enterprise journey to become even more innovative about generating revenue through the sale of goods and services. This means NFP businesses now form an ever-greater percentage of the overall not-for-profit mix, reinforcing the story of what is possible when NFPs run as businesses.

While most people still prize employment of any form, there is movement in the jobs market. Not-for-profit businesses are now able to hire previously inaccessible talent, thanks to the labor surplus. Many of those pushed out of large corporations seek work with purpose-driven companies, like NFP businesses, that are increasingly showing their capacity to provide greater job security.

Yet many individuals working in for-profit enterprises genuinely believe in the social value of their work and think for-profit business remains the best structure by which to deliver social value. In trying to maintain market share, some continue to use terms such as "sustainable," "sharing," "ethical," and "eco-friendly" as well as seeking to create markets for activities that commonly occur outside the market. The need for businesses to demonstrate social impact is so great most for-profit companies start claiming to be some form of ethical business, social enterprise, or cooperative with a social mission. The for-profit business community, entrepreneurs, and certain consumers and citizens rally behind triple bottom line models like B Corp certification, obfuscating the importance of the NFP distinction.

# The Big Rebalancing

As the 2020s roll on, inequality continues to rise. Unemployment, inadequate wages, and the ongoing extraction of wealth from the common economy to the elite economy mean the majority of the world's population remains enslaved to debt. While debt defaults are still widespread, groups are encouraging people to stop paying down their debt to big banks and corporations. There is growing talk of a global debt jubilee in which a range of debts around the world might be simultaneously forgiven or central banks print and give citizens money on the condition it is used to pay down debt.

In a repeat of the Jubilee 2000 actions[854], multilateral institutions like the International Monetary Fund have already developed a schedule for canceling large amounts of government debt in parts of Africa, Asia, Latin America, and Southern Europe without associated political prescriptions[855]. Such debt cancellations have proven especially beneficial for people in countries where governments had accumulated substantial unjust (odious) debt dating from as far back as the energy crisis of the 1970s through to the profit-oriented bailout schemes following the 2008 financial crisis. Although debt forgiveness has allowed for an increased focus on well-being–related investments by governments in these countries, it has not been enough to restart economic growth and business worldwide.

[853] REF: Power of Community documentary?
[854] The Jubilee 2000 was the cancellation of $34 billion in debts owed by 'heavily-indebted poor countries' to high-income countries. REF: https://www.imf.org/external/np/exr/ib/2001/071001.htm.
[855] What we are describing here is different from the debt-forgiveness of Structural Adjustment Programs, which offered loans and debt-forgiveness to economies in crisis, but the support was given on the condition that the debtor countries made certain economic reforms that often involved privatizing national services, private foreign investment, and reorienting to export-led growth.

Moreover, corporate leaders are worried ubiquitous debt forgiveness will undermine the capitalist system, threatening the viability of lending companies and damaging confidence in future lending. And central banks worry about the inflationary potential of a massive increase in the money supply.

Sensing the rubber band of inequality is about to snap and afraid "the pitchforks are coming," the world's wealthiest decide on a different form of debt jubilee that actually enables them to be cast as heroes. They use some of their cash reserves to voluntarily create a global relief and stimulus fund for scheduled debt alleviation. The fund purchases and forgives junk debt[856], helps people and nations by restructuring various forms of debt, and—with the assistance of governments—distributes the remaining money in direct one-off payments to citizens.

The debt jubilee provides temporary breathing room for the common economy as well as respite from riots and attacks on governments and the elite economy. Because the financing comes from existing money held by the wealthiest—not governments—global inequality drops, and mass inflation is evaded. With a swift injection of liquidity, consumption jumps, giving companies a sales boost. Nevertheless, these efforts are insufficient to address long-term job security, given far more is needed to provide households with the material security that would encourage a return to strong levels of borrowing and business investment. Moreover, the market has fundamentally shifted over the previous decade. Not only have people been forced to develop frugal lifestyles, but they are increasingly choosing an ethic of "enough" as they see the impact of a stronger community on their daily well-being. For many whose debts are cleared, the jubilee presents an opportunity to further opt-out of the for-profit economy.

The philanthropy from the wealthiest segment of society is largely cycled back into the elite economy via rents as well as company profits from increased consumption and, in the case of for-profit banks, debt repayments with interest. With ongoing job shortages and insecurity, debt soon re-enslaves people in the common economy.

Fearing anarchy, many governments respond by placing strong constraints on predatory lending, instituting taxes on high-frequency trading, and cracking down on tax havens. As a reflection of the rising social unacceptability of extreme wealth, citizens pressure governments to instate higher taxes on privately owned land and inheritance[857] to redistribute wealth and keep a few families from hoarding obscene amounts of assets.

The biggest move many governments make is to increasingly take on ownership in the financial sector. In addition to breaking up the biggest banks (a process that includes separating banks from investment firms), governments establish public banks—sometimes as iterations of previously nationalized institutions—at the local, regional, and national levels in countries where they did not previously exist. With all state revenue flowing into public banks, governments can now lend to themselves.[858] The gradual nationalization of the financial sector builds on a more general trend among many governments to become more business-minded and entrepreneurial.

---

[856] Contemporary examples of this are movements like RIP Medical Debt and Rolling Jubilee.
[857] Japan currently charges a 55 percent death tax, so this is not as drastic as it might sound (REF).

Along with income from remaining sovereign wealth funds, the interest on loans to non-government entities provides governments with a new form of revenue[859], allowing many countries to introduce a universal basic income in which the government gives every citizen a base income regardless of employment or financial status. This helps in some ways to increase fundamental levels of well-being for the most vulnerable segments of society. In response to collapses in for-profit health care provision due to the inability of the majority of people to pay for it, most countries move to implement or expand universal health care provision[860], with many favoring NFP providers of services and goods over for-profit equivalents. It is increasingly seen as best practice to implement universal health care systems in a way that complements rather than replaces traditional community and cultural networks of care.

The debt jubilee and basic income guarantee have given people some freedom to reflect on the sobering previous decade. By now, the ways in which the general population thinks have changed significantly.

The unrelenting discomfort of the last decades has paved the way for more critical thinking and a newfound openness to different perspectives and alternative approaches to social challenges. Issues like emotional well-being, inequality, and the decline of ecosystems—once considered secondary to the all-important goal of economic growth—are now widely viewed as the foundation of a healthy economy. People are more open to considering ideas for new systems to resolve these issues. Even proposals once considered utopian or out-of-touch with reality at the turn of the century—when it felt like we didn't really need systemic change—are now accepted as viable.

Accordingly, formal education shifts have also transpired, playing a critical role in reshaping ideas, attitudes and perspectives. When funding to public education was cut at the beginning of the Big Crash, teaching became more community-based, hands-on, and, in many cases, informal. In rural areas especially, local volunteers replaced the teaching of standard curricula with trades and skills like plumbing, carpentry, sewing, building, farming, and foraging. Of course, there was still a great need for software developers, electricians, chemists, biologists, and engineers, and such skills continued to be taught. There was a new appreciation for skills like counseling, facilitating dialogues, and creating a space for group decision-making.[861] Education moved into unconventional places like peoples' homes, makeshift community buildings, and even the outdoors.

---

858 Governments that lend to themselves can develop infrastructure for as much as 50 percent of what it would normally cost (REF: Ellen Brown).
859 Note that public banks typically create more revenue than is lost via the absence of for-profit private banks <REF>.
860 Most high-income countries already have universal health care systems, but many of these countries are currently cutting back on their provision due to austerity and privatization measures. We envision these countries expanding their provision of universal health care provision.
861 For more examples of the kinds of subjects receiving greater focus, see: http://postgrowth.org/upskilling-for-post-growth-futures-together/

With some stability restored by the early 2030s, formal education is expanding once more, but with a generally altered purpose seen as complementary to traditional forms of knowledge and modes of learning. With increased humility, more people accept the limitations of Western science. They increasingly accept more traditional approaches to health and education. Learning is seen through an evolutionary lens—people understand we are always learning and developing, without an endpoint. Students are encouraged to cultivate wisdom, knowledge, and skills that align with their passions and interests while also serving practical needs. What is considered practical has taken on a new meaning beyond a skill set that neatly fits the capitalist division of labor. Out of the strengthened community, the arts, for example, are valued for their contribution to improving human well-being and ability to inspire insights and creativity, rather than the "unproductive" label they often carried in the for-profit story. It is far more common for people to learn about diverse fields of knowledge, so even though someone might focus on engineering, they may also study ecology and sociology to understand how these disciplines are connected to engineering.

The altered purpose of education enables a more holistic approach to lifelong learning. The experience of strengthened communities has reinforced the value of providing children with a relaxed, natural, and outdoor experience of play-based activities through their childhood. As a result, more children grow up with a deeper connection to nature than in previous generations—even in big cities—developing what has been termed an "ecological worldview."[862] As children progress into their pre-teen years, they are more frequently sheltered from violence and emotional themes inappropriate for their level of development, especially in light of the ever-present challenges still characterizing people's daily lives. The forced community living of the 2020s also encouraged appreciation for different learning styles and needs. Through the economic collapse, people have come to see the systemic complexity that underlies everything within a larger, interconnected system.

With more breathing room in the economy, there is time to think in longer time horizons. Issues such as population pressure and reproductive health come to the forefront, supported by increased NFP services and a caring ethos that strongly values women's rights. Groups use entertainment-education soap operas on radio, television, and the Internet to change attitudes and behaviors toward family planning, reproductive rights, and gender equity.[863]

## A New Ethic Driving a New Path

Changes in education and thinking coevolve with a more holistic understanding of human nature that transcends the belief that the economy can only function on appeals to individual self-interest. Indeed, people are sharing and spreading ideas and information that flies in the face of the theoretical foundations and assumptions of capitalism.

The media reflects this shifting social consciousness. When advertising revenues collapsed at the beginning of the Big Crash, several mainstream media outlets were taken over by NFPs, governments, and nonprofit foundations. The depressed economy resulted in less advertising overall, and with different ownership structures, reporting has become more critical of what's not working (like the "consume for growth's sake" narrative) as well as providing hope by shining a spotlight on what's working.

[862] REF: Steve Schein.
[863] For a contemporary example of this, see the Population Media Center, which already runs such radio and T.V. programs.

A wider evolution of the public service develops, led by government-wide reviews undertaken by each country's auditor-general. Democracy is experiencing a revival, particularly in low-income nations, with the spread of fair and open elections, citizens' juries, people's parliaments, and online referenda.

Seeing the paradigm shift is forging an economy that works for social good, the younger generations of the elite economy want to be part of the positive change as well.[864] Having inherited wealth from their Baby Boomer grandparents[865], some of them are ashamed of the role their families played in creating inequality and ecological devastation. When they realize they can help remedy the primary challenges of the twenty-first century—becoming local heroes as respect and social status shift from wealth accumulation to social impact—they increasingly have an urge to:

- charge fair rent on their properties, rather than trying to maximize their personal gain;
- focus on the service aspects of property management;
- allow for the low-cost use of vacant housing;
- expand affordable housing opportunities;
- donate land to community land trusts and land conservancies;
- be more philanthropic as well as lending money to NFPs; and
- start NFP enterprises themselves.

While the wealthy play a crucial role in influencing the direction of the economic system, the NFP ethic has been most strongly pioneered by "poorer" communities. In the early stages of the crash, **necessity proved the mother of connection** for communities worldwide, causing people's compassion and empathy to deepen. Yet in poorer rural and urban communities, including slums, such deep connection already existed. Because people living in extreme poverty had come to depend on each other rather than state welfare, they could more quickly adapt to the devastating decline in global conditions. Building on their strengths, many combined traditional knowledge with the decentralized innovations of the twenty-first century (including those the Big Crash accelerated)[866] in fields such as energy, communications, agriculture, and manufacturing to create resilient local economies.

At the international level, low-income countries focused through the 2020s on substituting imports with local production because it was the only way to respond to the collapse of global trade. Now, thanks to the liberating effects of debt forgiveness, they are charting a new development path focused on well-being and the collaborative fulfillment of needs, not economic growth. Not-for-profit businesses (especially cooperatives) are taking off especially fast across Africa, Latin America, and Southeast Asia, where there is considerable openness to new approaches and cooperation happens more easily. Following the villages' lead, cities undertake retrofitting measures to ensure peoples' needs can be met while maintaining low ecological footprints.

---

864 This sort of sentiment can already be seen in the documentary that heir to Johnson and Johnson's fortune Jamie Johnson made, called *The One Percent*.
865 Baby boomers were born in the years following the Second World War.
866 These are innovations such as mobile Internet, mesh wireless networks, solar power and LED lighting, open source hardware, and permaculture.

## Rising NFP Advantages

Globally, there is increasing acknowledgment of the alignment between the NFP business model and the direction in which popular values and behavior are heading. The not-for-profit ethic is seen as representative of the next step in the moral evolution of humanity, with NFP enterprise progressively viewed as a functional business model on which the entire economy could be centered, efficiently delivering the goods and services society truly needs.

More aware of their potential to transform the economy for the better while harnessing and maximizing their ethical, financial, and workforce advantages, NFP organizations are intentionally driven to outperform for-profit businesses in the market. Many NFP enterprises display their NFP status with pride to show they are 100-percent mission-driven. <pq>"Not-for-profit" has become a mantra for the new economy movement and those yearning for economic justice.<pq>"

The rigid for-profit mentality is unable to keep up with the flexibility demanded by the social and cultural shifts toward sharing and collaboration embraced by NFP businesses. For-profit corporations grasp for ways to keep knowledge and innovation private, but they are swimming upstream in the face of free-flowing information, rise of the digital commons, and evolution of the patent system, which rewards more open forms of innovation.

Continuing to emerge in traditionally for-profit fields such as law, real estate, marketing, manufacturing, and IT, NFP companies tend to offer concessional pricing to—and procure goods and services from—fellow NFP entities, while for-profits still contract for-profits at typically higher cost.

In part due to ongoing government subsidies, goods and services offered by companies that don't have an active commitment to social and ecological sustainability remain cheaper than those offered by socially- and environmentally-minded companies. This gap in prices is narrowing, however, as increasing demand for ethically produced local items allows NFP businesses to take advantage of economies of scale and environmental tax breaks while avoiding worldwide fine increases and the rising costs of long supply lines (favored by for-profit multinational corporations). Despite some places still wishing to attract foreign investment and business, cutting costs by offshoring operations to places with lax environmental regulations and fewer labor rights becomes less feasible due to global pressures—especially the expanded Fair Trade movement. Indeed, labor costs in low-income nations (especially China and India) have been rising[867] through greater equality, global campaigns for working conditions and wages, and the introduction of the universal basic income in many countries.

## The For-Profits Strike Back

By the mid-2030s, the rise of the NFP economy and corresponding decline of the for-profit economy are so evident that those with a major stake in maintaining the for-profit status quo start to wage a serious fight. Up to this point, polite condescension and casual mockery about the role of NFP business in the economy had formed the extent of the pushback. Now for-profit businesses launch a concerted campaign touting their social and economic credentials, along with warnings that a world with more NFP businesses (i.e., the absence of the profit motive) would be doomed. They run media campaigns spreading misinformation and fear about NFP agencies. The larger context involves individual, for-profit entities sabotaging NFP competitors by spreading rumors and seeking to ruin their reputations. Such smear campaigns damage trust in the NFP business model among people who don't possess a solid understanding of its fundamental importance to a sustainable future. In some cases, scandals and poor actions by NFP businesses and their leaders reinforce the damaging messages, making the NFP World appear less desirable.

[867] REF.

Further escalating matters, investors and for-profit banks squeeze NFP access to capital, and for-profit companies restrict their business-to-business dealings with NFPs. For-profit companies sue NFPs for patent and copyright infringement, with certain large corporations taking legal action against governments that actively favor NFP enterprise.

Many politicians fall in line. They declare their intentions to outlaw NFP discounts to other NFP businesses (under anti-competition/anti-monopoly laws); cease grants to not-for-profit organizations with business activities; remove tax exemptions[868]; and force all NFP companies to pay taxes on unrelated business income. In extreme cases, there is even talk of NFP businesses being forced to take on the for-profit legal status, as happened to the credit unions in Canada in the early 1970s.

The for-profit campaign to derail the rise of NFP business is largely in vain. Given NFPs have reached the point where they are seen as such a threat by the for-profit world, they have also achieved a critical mass of supporters[869] who, in the digitally networked era, rapidly mobilize to halt attempts to undermine the NFP sector and its rise.

The central role of the for-profit ethic in scandals is evident, and there is now a powerful social disdain for this behavior. In the end, the scandals even solidify the importance of the NFP ethic and drive greater transparency in business more generally.

Foundations and NFP banks continue to make loans to NFP businesses at below-market interest rates[870], and NFP's access to capital enjoys sustained growth thanks to an emerging class of NFP investors. As more people understand the prudent nature of NFP governance, like the reduced ability for vested interests to drive board oversight, they become more willing to contribute to NFP capital-raising campaigns, mainly through debt-based investment (such as social finance) and crowdfunded contributions. Of major importance, pension funds now regularly invest in NFPs, which are increasingly perceived as a safe and socially valuable avenue for investment.

Campaigns successfully defend the ability of NFPs to provide discounts to other NFPs and conduct unrelated business while receiving tax exemptions due to their social impact. Such campaigns are also quick to point out that taxes are still collected through such activity via taxes on wages and the sale of goods and services.

---

868 A current example of this can be seen in Governor Paul LePage of the US state of Maine stating that he believes that the state should impose property taxes on nonprofits, because they are "takers, not givers" Ref: Christian Science Monitor article

869 For a contemporary example, look at the campaign to keep tax exemptions for US credit unions or the campaign to maintain access to generic HIV/AIDS drugs in South Africa. There are numerous historic examples of mass protests and demonstrations when a moral line of this nature is crossed. Once something is allowed and gains enough citizen support, it's very difficult to reverse its allowance. One very clear example of this is how big a failure alcohol prohibition has been all over the world. Or in a more similar vein, imagine what would happen if the American government tried to do away with civil rights, women's suffrage, basic environmental protection laws, or the minimum wage. There would be enormous uprisings.

870 This already happens. Ref: Fruchterman article.

# The Big Shift

As a result of for-profit scare tactics, the struggle to secure an NFP system gains valuable exposure on the world stage. The more visible NFP enterprise becomes, the more people become aware of it—and start looking to buy from, work for, and even start NFP businesses.

By the late 2030s, a significant portion of the population has come to realize the NFP economy can deliver outcomes the for-profit economy simply cannot: greater well-being, economic stability, increased equality, and respect for the biosphere on which we depend. In places that have chosen to actively support for-profit activity, it is now clear staying profit-oriented has made the environmental and social crises worsen, while well-being, inequality, and ecological conditions have improved in the places that have chosen to go in an NFP direction. Many citizens now favor NFPs over for-profits in whatever consumption choices they can.

Under enormous public pressure, policymakers make a U-turn on their earlier plans to strengthen for-profit business at the expense of their NFP competitors. They intentionally move their support to the NFP sector, proactively favoring NFPs in state contracts, subsidies, and new policies. They increase the number of grants provided to NFP startups and offer additional training, mentoring, and physical spaces for business incubation. This U-turn is partly possible because a growing number of people have been voting for politicians focused on creating a purpose-driven economy, especially at the local level. With most people able to see the impact of the wealth extraction siphon more clearly now, a movement for financial reform assembles in the strongest of ways. In many countries, public demands carry greater weight due to corporate lobbying restrictions instituted by campaign finance reform, which keeps big money from taking over political campaigns. Indeed, politicians are now able to campaign on an NFP platform.

As 2040 approaches, the Big Shift occurs: NFP enterprise replaces for-profit enterprise as the central mode of business. Most remaining for-profit corporations fail as the majority of customers shifts their patronage to local companies, worker cooperatives, and NFPs. Governments nationalize certain failing businesses but let others collapse, a limited number of which are acquired by thriving NFPs. To avoid failure, some for-profit companies shift to NFP structures themselves[871] or create new NFP structures and gift over residual assets. But most medium to large for-profit businesses simply shut down as they are outcompeted by NFPs and react too slowly to changing conditions to survive. For solo entrepreneurs and small business owners seeking to scale up, the NFP route is now the most favored form of incorporated business.

The end of the for-profit model's dominance coincides with a larger evolution from stock markets to bond markets. This transition includes many intermediate developments, including experiments with capped returns and limited term stocks[872]; systems that reduce high-frequency trading[873]; and local, regional, and social stock exchanges[874]. Such developments greatly reduce speculative financial activity.

[871] In some parts of the world, this is as simple as filing an amendment to their articles of incorporation.
[872] See Enspiral's capped returns for examples of this in the current economy, or Professor Steve Keen's proposals for a seven-year cap on investment returns.

## A Glimpse of Human Flourishing

We have entered an NFP World. <pq>Rather than the invisible hand somehow delivering collective well-being via everyone working toward their own short-term self-interest, the visible hand of NFP enterprise delivers collective well-being through the reinvestment of company profits into further social outcomes.<pq> The wealth circulation pump is now stronger than the wealth extraction siphon. Differences in pay (including gender-based) are smaller than in the for-profit economy and not disproportionate enough to stagnate the circulation of money. With the NFP market's surplus largely circulating throughout the economy, the wealth that has been concentrating in the elite economy is working its way back into the common economy. Prices of everyday goods and services have stabilized, and housing—along with other factors influencing material security—is increasingly affordable, thanks to the free circulation of money and debt, which enables real wealth and value creation. Over the coming years, the elite and common economies merge, ushering in the first large-scale improvements in the human condition in decades. Around the globe, people are filled with hope as they see the bigger picture of change emerging.

As people gain material security through greater equality, their fear and belief in the need to dominate others and nature to survive (the dominator model) are replaced with calm and the understanding that when we nurture each other and nature, there is more than enough for everyone to thrive (the partnership model). Geared toward positive social outcomes, the NFP World benefits vulnerable groups (often minorities, women, and children) and creates more widespread socioeconomic equality while building on developments from the community level, all of which alleviates tensions between races, nations, cultures, genders, and socioeconomic classes.

The prevalent notion that we are all in this together creates more space to discuss and address social oppression. Local forms of truth-and-reconciliation commissions are ongoing, allowing the voices of those who suffered under the for-profit era to be heard and facilitating healing between the dominators and dominated. These commissions give people the chance to express the pain and residual anger that has accumulated over millennia of domination and suppression.

Religious extremism and the threat of international terrorism are also fading, and armed conflicts are waning. As British economist and *Small Is Beautiful* author E.F. Schumacher noted in the 1970s, "People who live in highly self-sufficient local communities are less likely to get involved in large-scale violence than people whose existence depends on world-wide systems of trade." Indeed, less consumption and trade mean less motivation for countries to extract natural resources from each other. And the absence of the for-profit ethic frees governments to focus on real security rather than military prowess, defense, and economically motivated control. A new generation committed to peace is emerging in a world that is dramatically fairer, encouraging extensive multilateralism in policy spheres. Youth are cognizant that moving beyond the for-profit system is a turning point for humanity.

---

873 <REF?>.

874 Canada, the UK, Singapore, South Africa, Brazil, and Kenya have already started their own social stock exchanges. <REF>, while in the US we see examples of local stock exchanges such as California Stock Exchange, Hatch Oregon, and NPOEx (an NFP stock exchange that offers virtual shares and a purely social return on investments (Ref: http://npoex.strikingly.com).

## An NFP World Is Necessary but Insufficient

Not everything is rosy, though. The shift to the NFP World is not a utopia, nor is it a silver bullet. International, regional, and subregional disparities in wealth remain, along with struggles for land and autonomy. Regional conflicts over resources continue as the pressures of climate change are further felt. A lot of residual and silenced suffering from the for-profit era is now bubbling to the surface. In the for-profit world, this pain had been dismissed, denied, repressed, or trivialized because acknowledging our economy systematically led to social oppression was equivalent to questioning the fundamental assumptions and very legitimacy of our existing system.

Strong differences of opinion on social issues still exist, as does sociopathic, ruthlessly competitive, corrupt, and greedy behavior. Individuals and communities face the ongoing challenge of learning how to get along better, with growing awareness of the need for everyone to work on healing their own internal struggles.

Crime, violence, mental illness, addiction, suicide, homelessness, and even a black market still exist. But all of these have decreased greatly from levels in the previous twenty years as numerous businesses and organizations have emerged to help people with these issues in every community, and most people lead more physically active, connected, and fulfilling lives.

The legacy of the for-profit era has combined with new issues emerging in the wake of the Big Crash to engender significant concerns. A series of unforeseen health crises has been exacerbated by superbug resistance to antibiotics. The former use of toxic profit-driven agricultural production methods and of dangerous materials in products and buildings are triggering new health problems amongst the global population.

Environmentally, matters remain precarious. The material and energy footprint of the digital revolution (and even the costs of the mass transition to renewable energy) were severely underestimated. Feedback loops remain problematic, ensuring ongoing weather extremes and impacts from the era of environmental degradation continue to be felt. Further planetary boundaries have been crossed, such as ocean acidification, which has led to the collapse of global fisheries as well as the sweeping extinction of much marine and land life.

In the process of goods and services becoming more accessible through greater economic equality, some people who have been liberated from true scarcity temporarily consume as much as they can. Unaccustomed to the availability and accessibility of goods and services—and after many generations of the for-profit story dominating society—they try to compensate for feelings of emptiness, scarcity, and inadequacy through material consumption. We call this phenomenon the *liberation effect*.

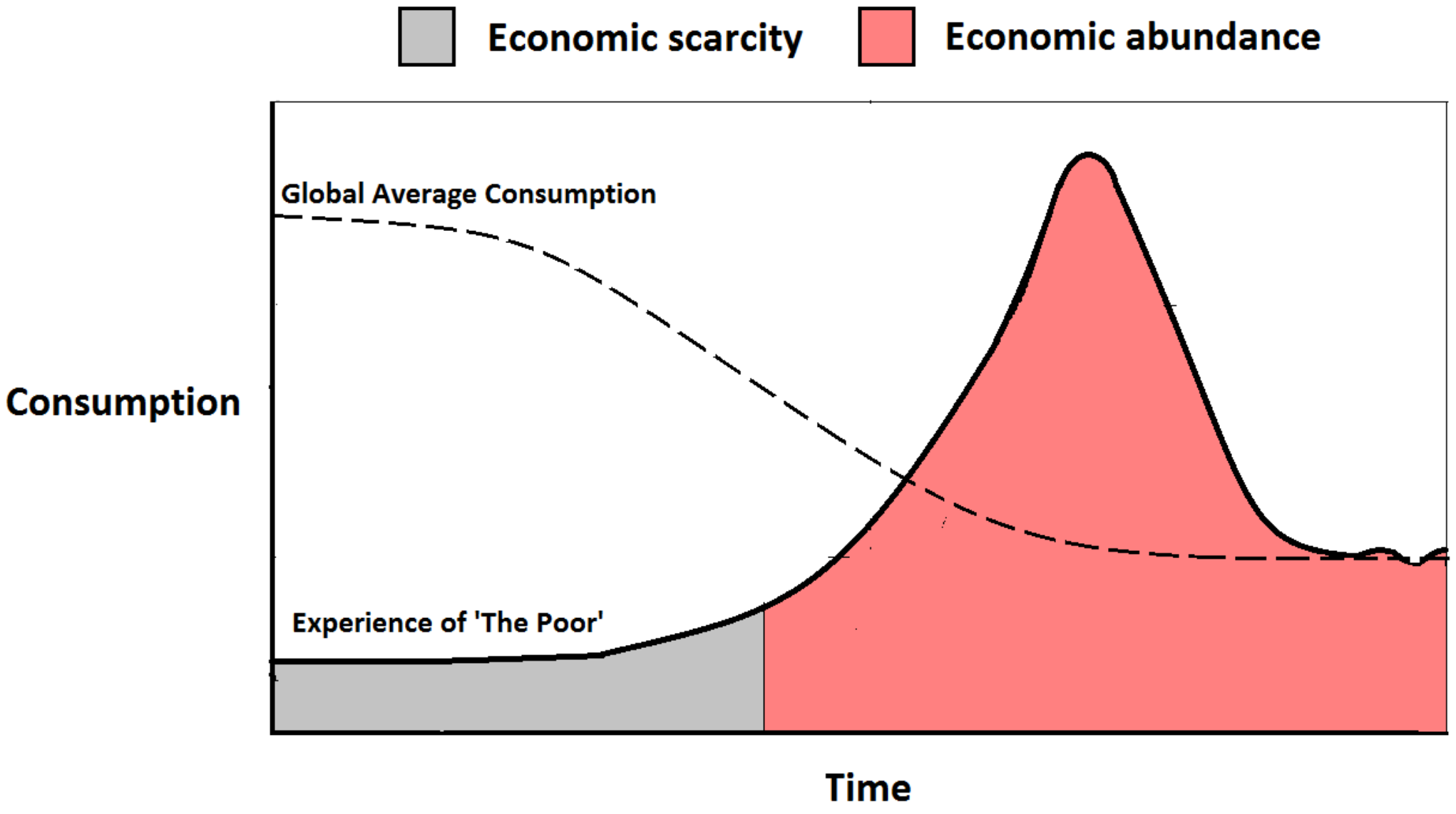

**The Liberation Effect**[875]

Yet the liberation effect is likely short-lived, if it happens at all, thanks to the evolving NFP ethic. Consumption falls rapidly when people realize that not only is there no longer a need to accumulate material goods to show one's status but such behavior is increasingly considered socially unacceptable. People who consume more than they need are seen as heralding another destructive for-profit era, which many are eager to ward off.

By 2040, the lean society model has emerged. Many needs are met most efficiently outside the market, with phenomena like the gift economy, time banking, community gardening, "upcycling," peer-to-peer production, and sharing now widespread. It is common for neighbors, friends, and family members to "share the care," helping each other with activities that used to be paid for, like caring for children, elderly or sick family members, and each other; cooking meals; cleaning homes; growing food; and making repairs. In the Big Shift, many for-profit companies died out completely rather than being replaced by NFPs because people started doing so much more for themselves and for each other—often simply by matching what they could offer with what others needed through offers and needs markets[876].

For needs met through the market, NFP businesses (including government enterprises, various forms of cooperatives, and industrial foundations) comprise the primary providers along with sole traders and nonprofits. Residual for-profit social enterprises are fading through social pressures. Charities now receive corporate philanthropy from businesses that benefit the community instead of profit-maximizing businesses that exacerbate the very conditions many charities seek to ameliorate.

---

875 Note, that while the end average consumption of 'the poor' is higher than to begin with, the overall footprint may be lower due to relative decoupling, over time.
876 See: http://postgrowth.org/running-an-offers-and-needs-market/, or for a current example of an offers and needs market, see online Freecycle groups that are spread around the world.

Financially motivated political capture has all but ended. Elections, policies, subsidies, and tax breaks can no longer be "bought" by for-profit corporations, leading to the associated end of the military industrial complex[877].

But there is carryover from the for-profit world. While there is a welcome influx of for-profit expertise, capital, and connections in the NFP sector, for-profit managers, owners, and executives also bring aspects of the for-profit ethic and culture into their NFP businesses, such as the focus on corporatization, efficiency, hierarchy, and obedience to authority—to the point of it becoming rigid dogma.

Likewise, some legacy nonprofit bureaucracy remains, but it is counteracted by the decentralization of power within the market. With a large proportion of most communities employed by NFPs, not-for-profit companies become even more embedded in community contexts. Local stakeholders influence programs, policies, and budgeting, leading to effective and efficient outcomes thanks to smaller and more appropriate economic flows.

### *How Life and the World Are Changing*

Most people in paid employment are hired to work twenty to thirty-five hours a week, an amount that is fast dropping. A much larger portion of society derives a sense of purpose from their paid work than at the turn of the century, and many no longer see Mondays as the worst day of the week. With purpose at the heart of business, even the mundane tasks that haven't been automated are experienced with greater joy.

In the case of unpaid caregivers, the universal basic income has reduced the need for additional paid work to survive. The reduced number of hours in paid employment permits a more equal distribution of household tasks within families.

Overall, most people have more time for doing non-work–related activities, like playing sports, games, and music; enjoying entertainment; working on new ideas and inventions; creating art; spending time with loved ones and in nature; gardening; making fresh food; and savoring all the other beautiful experiences life has to offer.

The end of the era in which business had a mandate to maximize profit no matter the social and environmental costs combined with the merging of the elite and common economies extinguishes the culture of consumerism and the imperative for economies to grow. By the mid-2040s, the ethic of enough is firmly embedded in society. The anxiety that drives people to buy things they don't need or want has faded. When people do still buy unnecessary goods and services, it is rarely as a form of "therapy," compulsion, or addiction but instead primarily involves purchasing recycled, refurbished, or secondhand goods. Upcycling, reusing, repairing, and recycling manufactured products is more common than buying completely new items. Even products that seem new often contain recycled components, with significantly reduced and different forms of packaging. Products are made according to the principle of built-in-resilience—they are designed to last as long as possible, with the ability to be recycled or reused if they break.

---

[877] National militaries still exist, as do weapons manufacturers, but they are all government-owned. And in the absence of for-profit weapons manufacturers, war-profiteering has all but faded away.

Levels of consumption ebb and flow, staying mostly within the planet's ecological limits. Indeed, we are close to achieving a steady state economy. The innovative efforts of NFP enterprises combined with the decrease and localization of trade, production, and consumption mean greenhouse gas emissions are close to peaking. Renewable resources provide for the bulk of the world's energy needs, and the global human population is close to stabilizing in number. Together, these factors provide the earth with breathing room to heal from the destruction caused by the relentless growth of the for-profit system.

To address the ongoing legacy of environmental feedback loops, businesses, nonprofit organizations, and communities are working to restore and regenerate ecosystems through processes such as mass reforestation and soil remediation.

While there was no single technological innovation that magically addressed all of the world's environmental and social ills[878], there have been amazing advances since the Big Crash began, especially in the fields of communications, IT, material science, energy storage, and energy generation. Importantly, open-source science and the hacker culture in the digital commons make amazing technological innovations widely available.

We have not colonized other planets by 2050, but we have something better than that: a thriving human civilization that can continue to coevolve with other species on this magnificent planet we call home.

Looking back from 2050, it is hard to reconcile what transpired during the latter stages of the for-profit era. How did businesses get away with damaging nature and hurting people for a few individuals to profit? Why did charities take money from companies that were doing more damage to society than they could ever compensate for? And why were citizens and consumers ever okay with those kinds of businesses? The Millennials tell their grandchildren that was just the way things were back then—in the absence of a better alternative.

## A Not-for-Profit World Is Necessary but Not Inevitable

As we have explored throughout this book, the circulation of money (and therein debt) is the lifeblood of a healthy, modern economy. But as we have discovered, any system that privileges profit-maximization will ultimately result in economic stagnation due to the concentration of money and reciprocal accumulation of debt.

When businesses are driven primarily by purpose—not profit—money and debt more readily circulate in the economy, allowing broad wealth creation, material security, and well-being.

---

[878] Examples of these technological panaceas include the singularity, or a 'free' energy source being discovered that defies the laws of thermodynamics. If such a source was discovered, this could break down the market system altogether, because people would be able to do almost everything for themselves with free energy. However, this still wouldn't address the issue of environmental degradation. To the contrary, without a change in mentality and values away from Homo economicus, free energy would only exacerbate the ecological crises, as it would enable people to consume even more resources, more quickly, creating, accumulating and throwing out as much as they want. Resource shortages would soon ensue and result in an economic crisis and we'd be suffering major health crises due to environmental pollution. A world with free energy would still have to figure out ways of making sure we take care of the natural environment.

This also allows for reduced levels of consumption. Thus, moving from the for-profit world to the NFP World is absolutely necessary if we desire a healthier, more equal society that can flourish within ecological limits.

Yet an NFP system is not inevitable. Capitalism may be coming to an end, but what follows it could be disastrous. If a ruthless form of corporate capitalism undermines community-building efforts and the NFP sector in the ensuing years, we could enter the Big Crash facing even deeper ecological and social distress. Our environment could be on an unstoppable path to full systems collapse[879], likely leading to our mass extinction. And if things head in a more for-profit direction—with NFP activities being usurped or outcompeted by for-profit businesses[880]—we could lack the cushion of collective resilience (reinforced by low levels of trust) needed to buffer the worst impacts of the Crash.

Or perhaps the Crash will be so devastating that we will enter a "fortress world" scenario, as described by the Global Scenario Group[881], in which authoritarian figures take over amidst the chaos of social collapse and survivalism, enforcing prolonged martial law and a culture of war and terror.

Even the path to an NFP World can take many different routes, some better for society than others. Keeping these weaknesses and challenges in mind is crucial as we advance with the intention of creating a flourishing Not-for-Profit World economy. How we reach the NFP World depends, to a large extent, on how quickly social norms change in the not-for-profit direction. Will most businesses continue to seek profits above (or even alongside) all else? Or will they learn to embody an ethic of enough?

Will the richest people in society continue to be publicly lauded for their accumulation of wealth and "charity"? Or will they become local heroes for widely sharing their wealth to create social impact, establishing not-for-profit businesses and community land trusts to steward the land they aren't directly using? Will feeding off exploitative rent and interest payments from poorer classes remain socially acceptable? Or will popular culture shift toward deeper definitions of success and prosperity, which praise contribution rather than accumulation?

Will governments see the rise of NFP enterprise as a threat to for-profit market competition and economic stability? Or will they see the value of having a higher moral standard in business and, as a result, promote NFP enterprise? Will for-profit businesses sabotage NFP competitors and create misinformation campaigns? Or will for-profits continue to move in the NFP direction? Will citizens and consumers support NFP businesses and favor them over for-profit competitors? Will a significant number of people become aware of the potential of NFP business to create better outcomes for everyone before for-profits begin to perceive NFPs as a threat?

---

879 We have assumed throughout this chapter that, through the coming great transition, the feedback loops driving ecological devestation are not strong enough to lead to a full ecosystems collapse in the next few decades.

880 The Skoll Foundation, for example, advocates that nonprofit organizations should transition to for-profit structures. This is an attitude also found in an article published by Forbes called 'The Nonprofit Sector Should Not Exist'. (Ref: Forbes article; Skoll Foundation article). Despite the rise of NFP business, the nonprofit sector is still largely made up of traditional nonprofits that might tend to operate inefficiently. As a result, for-profit businesses might move into traditionally NFP sectors, rather than the other way around. We can already see some of this happening, with the advent of for-profit charter schools, for-profit nursing homes, and for-profit daycare services <Ref: Hold the Fort report>. It is possible that all of the trends we've mentioned will not strengthen fast enough to counter for-profits outcompeting inefficient nonprofits. Consumers could continue to see things as a choice between business and charity-dependent nonprofit programs. If most people aren't aware of NFP enterprise as an option in the near future, and if NFP enterprise doesn't rise fast enough to secure a solid place in the economy, the for-profit economy could take over completely.

881 Ref: Global Scenario Group

It is impossible to say precisely what will happen, and there are an infinite number of pathways forward. We all play a role in the way this story unfolds. Nonetheless, all kinds of extraordinary changes are happening over the globe. The transition to NFP enterprise is just one of them. Together, they form the basis for a story of hope.

What, then, can we actually *do* to create the more beautiful world our hearts know is possible?

# 7. Our Shared Story

## *Let's create a Not-for-Profit World together*

The day Kristen Christian heard her for-profit bank was introducing fees on checking accounts with a balance below $20,000, she decided enough was enough. She moved her money to a not-for-profit credit union and set up a Facebook event page encouraging 500 of her friends to do the same. The twenty-seven-year-old art gallery owner from Los Angeles had no idea what she was about to inspire.

The ensuing five weeks involved one of the largest consumer migrations in US history. Between September 29, 2011—the day Bank of America announced its (now-defunct) monthly fee—and November 5, which Christian dubbed Bank Transfer Day, credit unions received $4.5 billion in funds and 440,000 new customers, equating to a 50-percent increase in new accounts.[882] An estimated six million Americans changed the way they bank in the year following the first Bank Transfer Day,[883] with credit unions experiencing exponential membership growth[884]. In America alone, credit union membership has passed 106 million people,[885] who are saving a combined $6.3 billion a year in lower fees and better rates thanks to their banks being purpose-driven and not-for-profit[886].

Christian's action galvanized everyday citizens into choosing the NFP way when it came to banking, but it also encouraged change at the upper end of town. Fed up with not mattering to the big banks, one Seattle businessman moved $3 million out of Chase and Bank of America into his local credit union.[887] Following that first Bank Transfer Day, half a dozen additional "move your money" movements emerged across the globe, with many of these movements continuing to this day to promote community-driven finance.[888]

## We Are the Leaders

While capitalism is devolving toward collapse, we all have meaningful choices to make. What sort of post-capitalist world do we want to help create? What roles do we want to play in defining this new world? The decisions we make and the ways we choose to direct our energy can profoundly influence the formation of a future that would truly sustain us. How we will shape the post-capitalist world is up to all of us.

For too long, we have placed considerable faith in decisions made by experts in government, business, and science. In reality, as Kristen Christian's example illustrates, evolution to a new economic system will involve everyone's input. This requires us to thoughtfully consider innovative possibilities and actively participate in the transformation process. It is inadequate to simply reject the status quo and blame those who hold power.

---

882 King, B. (2012), "Bank 3.0: Why Banking Is No Longer Somewhere You Go But Something You Do", John Wiley & Sons.
883 http://www.thenews.coop/38943/news/banking-and-insurance/capitalism-no-business-too-big-fail-kristen-christian/.
884 http://www.huffingtonpost.com/bill-cheney/bank-transfer-day_b_2056292.html.
885 http://www.cuna.org/Research-And-Strategy/DownLoads/mcue/.
886 http://www.huffingtonpost.com/bill-cheney/bank-transfer-day_b_2056292.html.
887 https://www.youtube.com/watch?v=cTzFdworUI0&feature=youtu.be.
888 For example: http://moveyourmoney.org.uk/, http://www.huffingtonpost.com/news/move-your-money/.

**We are *all* the leaders we have been waiting for**, and our leadership role requires us to intervene in different ways and on different levels throughout our entire system. Here are some of the actions you can take to bring forth the NFP World.

## Citizens, Advocates, and Workers

As a citizen, you have incredible power. You change the world every day through your behavior, purchasing, communication with others, and interactions with the wider community.

You can support the emergence of the NFP World by strengthening your participation in the non-market community economy—caring for family and community members, sharing items with friends, renting rather than owning equipment, banding with others to undertake home and neighborhood renewal projects, and partaking in local groups and activities. By meeting your needs and the needs of others outside the market, you will be reducing your dependence on the (for-profit) economy. By consuming less and creating space in your own life, you will become a living example to those around you of the NFP ethic of enough.

Shifting your purchasing patterns is also powerful. You can start[889] by:

- moving your money to an NFP bank such as a credit union;
- changing your health, life, automotive, and home insurance to a company with the word "mutual" or "cooperative" in its name;
- sourcing your utilities (energy, water, Internet, phone, cable television) through your town or city's service or a community-owned company;
- shopping for groceries at a store with the word "co-op" in its name;
- purchasing clothing and household items from local thrift stores[890];
- purchasing books from an NFP provider online[891], a cooperative bookstore, or sales at your local library or charity sale[892];
- shifting your gym membership to an NFP fitness center like your local YMCA;
- choosing NFP institutions for education;
- staying in NFP accommodations when you travel, like Hostelling International rooms or YWCA hostels[893]; and
- moving to an NFP Internet browser like Mozilla Firefox as well as considering the numerous other NFP software options available (e.g., Linux).

Locally, NFPs are apt to offer an even more comprehensive range of services. Most likely, there are NFP hospitals, health care clinics, social services, childcare centers, museums, arts and entertainment organizations, radio and television broadcasting services, primary and high schools, public transport, sporting clubs, and zoos in your area. Less obvious but more common than you might imagine are NFP options for car-sharing services, bicycle repair stores, flying clubs, pubs and bars, restaurants, cafés, catering services, bakeries, agricultural providers (e.g., CSAs), timber companies, pharmacies, cinemas, bookstores, construction companies, waste management businesses, manufacturing companies, (eco)tourism companies, book publishers, and funeral services. Not-for-profits also commonly offer venue rentals at competitive rates.

---

889 These are just some suggestions from a bigger list of ways we've documented for you to shift your support to NFP activities, which you can fine at: http://howonearth.us/NFPbusinesses.
890 E.g., Goodwill, Red Cross, St Vincent de Paul, Smith Family.
891 E.g., GoGoodBooks, or for a for-profit cooperative model, see: http://fairmondo.de.
892 E.g., Lions and Rotarian clubs.
893 https://www.hihostels.com/; http://www.worldywca.org/Member-Associations/Find-a-YWCA-hotel.

You may need to ask around to discover which of your local businesses are NFP (typing "not-for-profit" or "nonprofit" and your location into the Internet can also prove a helpful start). If you don't know whether a company is for-profit or not-for-profit, ask, "Do you have private owners, or are you not-for-profit?" This has the secondary benefit of raising the importance of this distinction amongst business owners and giving NFP employees even more pride in their work and the company's choice of legal structure.

If you can't replace for-profit goods and services with NFP equivalents, try to avoid the most money-hungry for-profit suppliers, such as those traded on the stock market and ones that prioritize shareholder value over almost everything else. This is where social enterprise directories can save you considerable time. Cooperatively run-and-owned platforms like Fairmondo[894] also offer important alternatives to giants such as eBay and Alibaba.

<s/b> Social Enterprise Directories Around the World

| Country | URL |
|---|---|
| **Australia** | http://socialenterprisefinder.com.au/ |
| **Canada** | http://www.socialenterprisecanada.ca/purchase/ |
| **Hong Kong** | http://socialenterprise.org.hk/en/sedirectory |
| **India** | http://khemkafoundation.in/directory/ |
| **Malaysia** | http://www.hati.my/category/social-enterprise/ |
| **Northern Ireland** | http://www.socialenterpriseni.org/find-social-enterprise |
| **Philippines** | https://www.choosesocial.ph/ |
| **Singapore** | https://www.raise.sg/directory/ |
| **United Kingdom** | http://buysocialdirectory.org.uk/<br>http://supportsocialenterprise.org.uk/find/<br>http://www.socialenterprisescotland.org.uk/our-story/directory/ |
| **United States** | http://www.givetogetjobs.com/social-enterprise.php |

Sometimes purchasing from NFPs can prove an inconvenient or costlier option, but it is worth considering the way NFP purchasing keeps money in your community and ends the constant drain of money from the real economy. Likewise, purchasing from a smaller, local, or cooperative company (even if for-profit) draws power from the giant for-profit corporations listed on the stock exchange since it is consumers who keep them in business. Bear this in mind when it comes to administrative costs for NFPs[895]; their employees need adequate compensation, and good salaries attract talent.

[894] Although yet to expand beyond an online market in German, by the end of 2015 Fairmondo was offering over 2 million products, including a wide variety of high quality, fair-trade, and sustainably produced items throughout most categories. See: https://www.fairmondo.de.

Another great starting point is moving your money to a credit union or other NFP financial institution.[896] Most people are pleasantly surprised to discover the range of services credit unions offer, including widespread access to ATMs, which hasn't always been the case. In America and elsewhere, certain phone apps make it easier for you to find local options.[897] If there are no NFP banks in your community, consider switching to a smaller local institution or even working with other community members to form your own credit union[898]. You can also do something powerful with any credit card debt you currently hold with a for-profit bank. Open or extend a line of credit with your credit union and pay down your debt using this line of credit. Not only will you likely receive a much lower interest rate on any balance you carry forward, but you will know the interest you pay is going back into your community thanks to the NFP structure.

In addition to moving your money (and continuing to donate) to NFPs, if you have investments, you can begin to divest from the for-profit economy. If you have funds under management, see if your fund manager offers an ethical option for your investments.[899] Pressure your fund manager to divest from the most destructive industries and companies or sign yourself up to the growing ranks of shareholder activists[900].

While we can't offer financial advice, when it comes to self-managed funds, it is worth noting government bonds offer an established avenue of investment in an NFP activity. Other investments through debt-based crowdfunding and peer-to-peer lending are in an exciting stage of development. You may also wish to start with the broader field of impact investing, which often focuses on purpose-driven for-profit initiatives but can be a less risky investment.

For those who are accredited investors, keep an eye out for opportunities to loan money to younger social entrepreneurs. One UK study[901] showed younger people are more likely than the general population to want to start a social enterprise. Startup weekends and public pitch events offer great opportunities to make such connections.

If you are seeking to purchase a home, investigate your local options with respect to community land trusts, eco-villages, intentional communities, and housing co-ops (especially those that are non–equity-based). If none exist, consider starting one of these forms yourself, with the help of family and friends.[902] If you are selling your house, try to avoid transferring it to a property developer, unless the intention is to create affordable housing. If you are a landlord, think about adding affordable housing to your portfolio or selling your property to an NFP or into a housing cooperative. If you want to go even further, you can gift your property into a community land trust or bequeath it to an NFP in your will.

Some of these steps are bolder than others. It's not necessary for everyone to move into an eco-village for the NFP World to emerge. Taking any of the actions we have outlined above helps. As philosopher Charles Eisenstein says, when you act from a different story, you create a break in the status quo story for those around you, spurring them to question their assumptions, even if just for a few seconds. If you act in an unexpectedly generous or kind way, for instance, it weakens the story that human nature is mostly greedy, selfish, and competitive.

---

895 <REF>.
896 E.g., saving and loan associations (that are member governed), building societies, and trustee savings banks.
897
898 http://www.woccu.org/financialinclusion/bestpractices/startingacreditunion.
899 Preferably one that includes both positive and negative screening.
900
901 Young people in the UK, are more likely than the general population to want to start up a social enterprise (27% compared to 20%) <REF>.
902 http://www.shareable.net/blog/how-to-start-a-housing-co-op; http://www.shareable.net/blog/how-to-start-a-community-land-trust; https://valhallamovement.com/6-guidelines-start-successful-intentional-community/

This is why the work of millions of people around the world on issues of inequality, declining well-being, consumerism, social justice, and ecological devastation is so vital. These movements create the fabric for a new story, stirring us from our preconceptions of what is "normal."

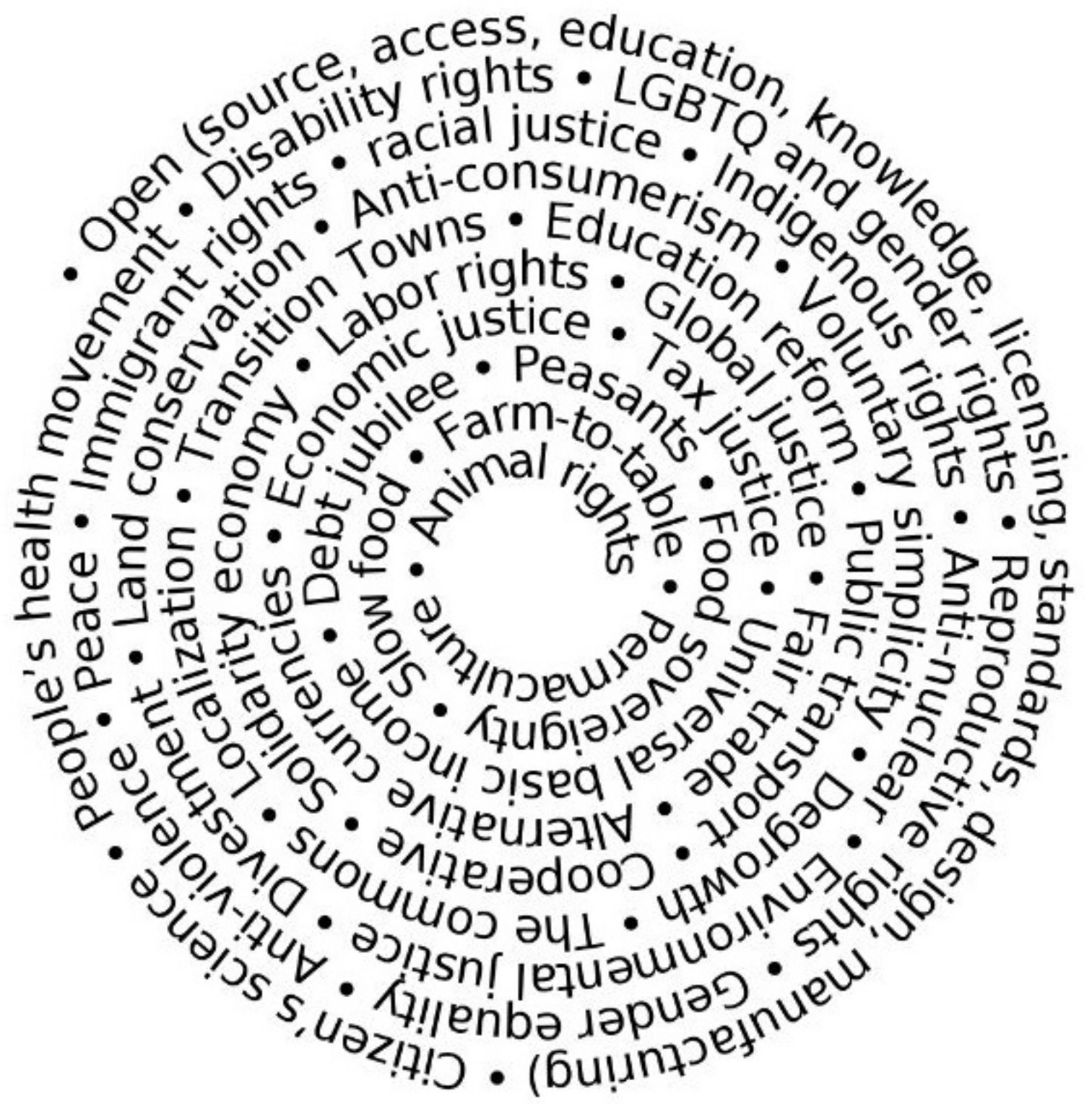

**Global Movements for a New Story <SHIFT TO CHAPTER 4>**

Moving beyond the for-profit economy to a flourishing NFP World strengthens all of these movements. We must evolve past an economic paradigm focused on private individual gain to address the root causes of the issues these movements seek to ameliorate. In a for-profit world, all we can do is apply temporary fixes that are wholly inadequate for the size and depth of the challenges we face. The NFP model, supported by the growing movement for a new economic story, provides the missing key for which we have been yearning. It offers a focal point for social and environmental movements in desperate need of a unified approach to their diverse thinking[903], speeding up the systemic change we so desperately need.

Simply speaking about NFP enterprise and the NFP World vision is of great value. The more people use the terms "for-profit" and "not-for-profit" (or "NFP") and discuss the NFP World, the more others will learn about this alternative and the quicker the movement will grow.

At the more specific end of advocacy, you can publicly support the ability for NFPs to conduct unrelated business and directly compete with for-profit companies, without threat to their business tax exemptions[904]. In this way, you can show how we do not so much need new legal structures for social enterprise as we need a re-evaluation and broadening of existing not-for-profit structures. You can also voice support for new forms of debt-based investment in NFPs, helping to expand the community's ability to finance new purpose-driven initiatives. To gain a better understanding of the NFP legislation in your part of the world, visit the International Center for Not-for-Profit Law's comprehensive online database[905].

[903] <REF: http://thenextsystem.org/wp-content/uploads/2015/10/GettingToTheNextSystem.pdf>.

Even if you are not an outspoken advocate for the NFP World, you can make a vital contribution by aligning your actions with purpose. If you currently work in an unsatisfying job at a for-profit company, consider working for an NFP enterprise that aligns more with your passions instead. In addition to performing more meaningful work, you are more likely to enjoy long-term job security as the economy shifts in the NFP direction. We recommend investigating your options closely. When applying for an NFP job, ask your potential employer, "What's your business model?" This will reinforce the growing social expectation that NFPs work toward financial independence rather than reliance on charity.

If you work at a for-profit cooperative, raise with your colleagues the idea of transitioning to an NFP model while retaining cooperative values and practices. This could involve shifting to a worker self-directed NFP or a nonprofit mutual corporation.[906]

Wherever you work, there is always the opportunity to volunteer for an NFP business. Even though you aren't being financially compensated, you will be increasing that organization's market competitiveness, creating a healthier community, and—given that it feels good to volunteer with an organization whose ideals you support—contributing to your own well-being.

Finally, you can consciously avoid the downward spiral of working longer hours and refuse to maximize productivity at the expense of your well-being. Rather, seek balance. Carrying the ethic of enough with you into your work and personal life can be empowering and can send an inspirational message to others.

## Nonprofit Employees, Managers, and Board Members

If you work at a traditional nonprofit or charity—whether you are a volunteer, paid worker, manager, CEO, or board member—you can help shift your organization to a more efficient, independent, and enterprising model.

At organizational meetings, raise the idea of becoming more financially self-sufficient. Use lean business concepts[907] as an alternative to time-intensive business plans. Test the innovative finance mechanisms we have outlined in this book. Create revenue generation strategies that build on your organization's existing strengths, integrating sliding scale or scholarship models where possible. Throughout, see if you can use NFP-friendly alternatives for status quo vocabulary to help spread awareness of the NFP distinction: for example, use "not-for-profit" instead of "private"; "patronage returns" instead of "dividends"; "stake" instead of "equity"; "members" instead of "owners"; and "stakeholders" instead of "shareholders."

---

904 For example: http://www.donttaxmycreditunion.org/. In fact, Canadians could even pressure their political leaders to reinstate tax exemptions for credit unions, returning them to their NFP roots.
905 http://www.icnl.org/research/library/ol/
906
907 Including the lean business canvas.

Use your NFP status to your competitive advantage. Seek procurement discounts, especially for software[908], and explore tax deductions for volunteers, pro bono consultants, or property owners who are willing to offer your organization reduced-cost rent. Make your NFP status an integral part of your marketing strategy—something that distinguishes you from competitors. For example, when you download Mozilla Firefox, a message appears saying, "Thanks for downloading Firefox! As a non-profit, we're free to innovate on your behalf without any pressure to compromise. You're going to love the difference." Indeed, as an enterprising NFP, you are at the leading edge of business, contributing to a better economic future for us all.

You can also make sure your NFP follows best practices such as:

- establishing thorough policies, procedures, and bylaws[909];
- establishing a compensation committee;
- undertaking voluntary financial audits;
- ensuring public transparency and accountability (e.g., publishing annual reports, financial statements, and possibly board meeting minutes);
- treating volunteers and staff with equal dignity;
- ensuring financial literacy across the organization (especially the board);
- maximizing participation in budgeting processes;
- using democratic decision-making processes and creating a culture where objections are embraced for their potential wisdom;
- sharing and collaborating with other NFPs whenever possible[910];
- providing thorough onboarding (induction) and exit procedures; and
- conducting periodic organizational reviews.

Investing in ongoing training and professional development is the most effective way to ensure high organizational standards. It also builds the pool of potential NFP consultants, who will be essential during the great transition ahead.

## Entrepreneurs and Business Owners

If you are an entrepreneur, consider starting your next venture as an NFP enterprise. Perhaps you have even become an NFP entrepreneur by accident because you didn't like what an FP business was doing in a certain sector (as was the case with the folks behind VTC Cab, an NFP alternative to Uber in France[911]). Whatever your motivation, there are numerous resources available to help you on your journey.[912]

If you're not yet sure whether you want to choose the NFP route, keep your options open by avoiding the distribution of company equity (i.e., stick with crowdfunding, loans, and other forms of debt-based financing). At the least, try to avoid the restrictions that come with venture capital.

---

908 even if your organization is not US-based, most US software companies will honor foreign organizations that can prove their legal NFP status.

909 Especially those relating to: conflicts of interest; corporate structure and the delegation of powers; whistleblower protection; discrimination and grievance; conflict resolution; document retention and destruction; financial controls; ethical investment; gift acceptance; maximum employee income differentials; joint ventures; and board term-limits and rotation.

910 Given your organization is part of a complex, interdependent ecosystem, seek to strengthen fellow NFPs in every way possible. Move into deeper, smarter models of partnership that are grounded in the strengths of each organization and save partners time and money. Join with other groups to database investment needs, and support the peak bodies representing NFPs in your industry.

911 http://www.theverge.com/2015/12/3/9841562/french-uber-drivers-launch-app-vtc-cab.

912 See, for example: https://www.councilofnonprofits.org/tools-resources; and our forthcoming book, The Not-for-Profit Handbook: http://notforprofithandbook.org.

If you are an owner of an existing for-profit business, consider making your company not-for-profit. Based on the trends we have outlined in this book, shifting to an NFP legal structure may be the smartest way to ensure your company exists well into the future, both financially and in terms of your vision (via the anti-takeover protection NFPs offer).[913]

The first route to taking a company not-for-profit is to buy back existing shares. Ironically, this strategy is one large companies have long employed to make their earnings look better.[914] If your profits or net capital reserves are strong, you may be able to do this without securing a loan. You may also choose to reduce the overall volume of stock via an equity-for-debt swap, either with existing shareholders becoming the lenders or by taking out a loan to finance the repurchasing of shares. Once the company has been divested of all its private equity ownership (i.e., it is "self-owning," either through the absence of stock or complete ownership of reacquired stock), it can be rechartered as an NFP. Nationwide Mutual Insurance, a Fortune 100 company, did something similar in 2008 when it purchased all the outstanding shares it did not already own in Nationwide Financial Services (NFS), making NFS a wholly owned subsidiary[915]—except in this case, the parent company, Nationwide, was already an NFP company.

This isn't just an idea for existing NFP companies. It is possible a passionate CEO, innovator, and investor like Elon Musk might see the economic and innovative benefits of having his companies (such as Tesla) repurchase their publicly traded stock (including his own) to allow for their transition to NFP companies. Given Tesla's industry-shaking decision in 2014 to make all its patents publicly available and free to use[916], there might even be a case for Tesla receiving corporate tax-exemption in the United States because its research-based production activities could be considered in the public interest.

An easier route to converting your for-profit company to an NFP structure might be to start an NFP foundation that receives 100 percent of the profits from your business, as in the case of multinational food company Newman's Own, which donates all after-tax profits and royalties to the Newman's Own Foundation.

<SHIFT TO CH. 2> Alternatively, full or partial ownership[917] of a company can be donated to an NFP trust or foundation, creating what is referred to as industrial or shareholder foundations[918]. In Northern European countries and elsewhere (less so in America[919]), such a structure has a long and successful history. Prominent companies that are solely or mostly owned by NFP foundations or trusts include:

- Trader Joe's (100 percent – Markus Foundation)
- Aldi Süd (100 percent – Siepmann Foundation)

---

913 http://link.springer.com/article/10.1023%2FA%3A1008605309347.
914 https://hbr.org/2014/09/profits-without-prosperity.
915 To raise capital, in 1997 Nationwide Mutual Insurance had publicly issued 20 percent of its stock in Nationwide Financial Services, a separate legal entity.
916 https://www.teslamotors.com/blog/all-our-patent-are-belong-you.
917 Can relate to voting rights. Minority outside ownership, including on publicly traded platforms, exists for many industrial foundations. 42 percent of companies are not 100% foundation owned, 13 percent have publicly listed shares, and 73 percent of the foundations have a general charitable purpose <REF>.
918 Industrial foundations are defined as "…independent legal entities without owners or members typically with the dual objective of preserving the company and using excess profits for charity" Børsting, Christa and Kuhn, Johan and Poulsen, Thomas and Thomsen, Steen, Long-Term Ownership by Industrial Foundations (January 31, 2016). Available at SSRN: http://ssrn.com/abstract=2725462 or http://dx.doi.org/10.2139/ssrn.2725462
919 "Industrial foundations were common in the US, prior to 1969 tax legislation that effectively prohibited private foundations from owning more than 20% of the voting shares in a business corporation" REF: Fleishman in Hansmen and Thomsen http://economics.mit.edu/files/8783.

- Carl Zeiss (100 percent – Carl Zeiss Foundation)
- IKEA (100 percent – INGKA Foundation)
- *The Guardian* (100 percent – Scott Trust Limited)
- Rolex (100 percent – Hans Wilsdorf Foundation)
- Lidl (99.9 percent – Dieter Schwarz Foundation)
- Bosch (92 percent – Robert Bosch Foundation)
- Velux (90 percent – Villem Foundation)
- Pierre Fabre (86 percent – Pierre Fabre Foundation)
- Aldi Nord (60 percent – Markus Foundation)
- Bertelsmann (77.4 percent – Bertelsmann Foundation)
- DNV GL (63.5 percent – Stiftelsen Det Norske Veritas)
- Tata Group (66 percent – numerous trusts, primarily Sir Dorabji Tata Trust and Ratan Tata Trust)
- William Demant (61 percent – Oticon Foundation)

Other companies—such as Carlsberg, Hershey's, Novo Nordisk, Maersk, and Trelleborg—are partially owned by trusts and foundations, which have less than half of the ownership rights but the majority of voting shares. What is more, the largest ever research project on industrial foundations (focused on Denmark, given its prominent concentration of such structures[920]) shows that compared to other types of companies, industrial foundations last five times longer, have less employee turnover (including at the management level), pay higher average salaries, and hold less debt.[921] Crucial to these advantages is the fact that they are required to have independent boards rather than directors who have a personal stake in the company's profits, as is common in large, for-profit companies.[922]

If converting to an NFP structure is too much of a stretch, you can still move your company in the NFP direction. Consider adding a distribution cap to your company bylaws and writing a social purpose into your corporate charter. Other legal structures (see Chapter 2)—such as employee-ownership, a worker cooperative, a community interest company limited by shares in the United Kingdom, or benefit corporation in certain US states—may serve such a shift to an NFP direction as well as providing economic advantages.

The for-profit business community has a great deal to offer emerging NFP enterprises. Business mentorship, board membership, and joint ventures are just a few of the ways business people (and NFP entrepreneurs themselves) can aid with the rise of NFP enterprise.

## Public Servants and Legislators

Accelerating the shift to the NFP World requires strengthening and expanding existing legislation. If you work in this field, you can help maintain existing NFP tax exemptions for eligible groups.[923] You can also help expand the types of NFP activity that are eligible for tax exemptions. For many countries, this likely involves clarifying and widening the legal meaning of "charitable purpose," as was done in Australia in 2013[924].

Given the decision to forgo private ownership can be viewed as a socially oriented act of goodwill, we see any NFP with a broadly defined social purpose as worthy of tax exemptions.

---

920 There are more than 1300 foundations in Denmark. REF: Thomsen, S. (2013) 'Industrial Foundations in the Danish Economy' Center for Corporate Governance. Department of International Economics and Management. Copenhagen Business School, Frederiksberg.
921 http://www.cbs.dk/en/research/departments-and-centres/department-of-international-economics-and-management/center-corporate-governance/news/industrial-foundations-live-forever.
922 REF:
923 E.g., credit unions in the US, and NFPs with unrelated business activities in Australia.
924 *Charities Act 2013 - https://www.legislation.gov.au/Details/C2013A00100*

Indeed, in Nordic countries, an NFP foundation "running a business for the good of society is considered an acceptable charitable aim."[925]

Expanding tax exemptions across the spectrum of enterprises by broadening the criteria for what is required to "serve society's needs" could provide incentive for social entrepreneurs considering the NFP route in typically for-profit industries such as hospitality, construction, energy, and manufacturing.

More fundamentally, you can support the right for NFP organizations to engage in trade. This means allowing NFPs to maintain surplus funds from year to year[926], enabling them to function as any other business would. You can lobby for the rights of NFPs to conduct unrelated business (to fund their primary mission) without jeopardizing their tax exemptions. You can also make the case that NFPs forming business partnerships with other NFPs to create greater market accessibility should be excluded from anti-trust legislation, as long as they are still in accordance with the law, using all of their resources for social benefit.

Since your government role confers a degree of public authority, you are well-positioned to raise awareness about the ability of NFPs to generate their own income. If you are working for a finance or business ministry, chamber of commerce, business bureau, or economic arm of government, consciously including NFP businesses (as well as for-profit co-ops and employee owned companies) when describing "the business community" in official communications helps change the story. The more you publicly acknowledge the contributions NFP businesses make to our economy and collective well-being, the more people will think of NFPs in a new light.

Even more proactively, you can establish public awareness campaigns encouraging people to start an NFP business. In most countries, for-profit startups receive significant government support. It is time to develop the equivalent for NFP startups. This can include providing financial backing; coworking spaces; equipment; startup guides; incubation programs; mentoring; legal and accounting assistance; training; platforms for NFPs to connect with customers and supporters; and opportunities to bid for government contracts. Listen closely to startup entrepreneurs, seasoned veterans, and representative groups to learn what is working and what is still needed. If you really want to get ahead of the curve, set up a ministry dedicated to NFP enterprise, just like the government of Indonesia has done for cooperatives.

In supporting the rise of NFP enterprise, you create co-evolutionary potential. Given the state's own drive toward entrepreneurialism[927], many of the resources it creates or makes available to NFPs—such as manuals, training programs, and software—may be equally useful to government agencies. Moreover, the insights gained while championing NFP innovation could prove relevant to governments as both realms share a mission-driven focus.

Beyond providing support, you can strengthen the integrity of the NFP legal status. If, in the shift to an NFP economy, there is a watering down of the three principles that differentiate NFPs from for-profits (a social mission-driven mandate, the inability to privatize company profits, and the absence of individual owners), the NFP World will suffer from many of the ills witnessed in the for-profit alternative. Reducing the rate of fraud across both nonprofit and NFP activities is essential. Tax agencies will need greater resources to expand auditing, and regulators will remain crucial for monitoring all aspects of business activities. A strengthening of regulatory requirements (e.g., financial/salary ratio reporting, mandatory executive compensation committees, labor laws) should parallel the associated rise in expected public accountability, especially for large NFP organizations.

---

925 http://www.tifp.dk/wp-content/uploads/2011/11/What-Do-We-Know-about-Industrial-Foundations.pdf.

926 Currently not allowed in countries such as Greece.

927 See: Osborne, David, and Ted Gaebler. 1992. *Reinventing Government: How the Entrepreneurial Spirit Is Transforming the Public Sector*. Reading, MA: Addison-Wesley; And: Reinventing State Capitalism: Leviathan in Business, Brazil and Beyond.

New scrutiny, legislation, and protocols will be needed to buttress emerging finance mechanisms, including crowdfunding, cryptocurrencies, peer-to-peer lending, community public offerings, and the bond market. Policymakers should pay particular attention to the rapid advances threatening centralized regulation, such as open-source software, the blockchain (currently providing the operational basis for cryptocurrencies), and Creative Commons licensing.

Increased NFP scrutiny builds on the greater accountability many legislators are already trying to bring to the for-profit world. These include efforts to limit tax evasion; strengthen progressive and environmental taxation systems; improve financial literacy among all citizens; and heighten transparency regarding the financial, social, and environmental impacts of business. All of these changes accelerate the shift to a new economy.

## Researchers, Educators, and Economists

As a researcher, you can provide pertinent data to promote the momentous transition we outlined in the previous chapter. Research in fields such as behavioral economics and other social sciences is crucial to challenging unhealthy orthodoxies and understanding their underlying assumptions and motivations. As researchers, you play a significant role in directing innovation toward real needs.

Exciting research opportunities also exist in the field of NFP enterprise itself. There is inadequate data on the rise of NFP enterprise and the challenges they face, especially data that distinguishes NFP enterprises from traditional nonprofits. There is a particular need for studies comparing FP and NFP businesses in terms of operational efficiencies; productivity; innovation; salary differentials; approaches to decision-making and risk; social impact; collaboration; competition; ecological impact; and market share by sector and country. As a researcher, you can assist by investigating NFP successes and failures in financial, social, and ecological terms. There is endless potential for case studies and comparative case studies around the world. Both specific and aggregate data is needed to influence policymakers, business leaders, and the public. If you can, publish in open-access journals or make your data publicly available in other ways.

If you have skills in economic analysis, we would appreciate your assessment of the NFP World proposal outlined here as well as analytically rigorous modeling of the abstractions, phenomena, and scenarios we have described in this book. You may agree it is time to rethink not just mainstream neoclassical theories but also many heterodox theories, insofar as the NFP model breaks the state-versus-market binary that sits at the heart of conventional studies in political economy. You can help disseminate such thinking through conferences; public and media engagements; journals; and textbooks and other pedagogical tools.

The most influential avenue may be integrating the NFP model into the teaching of economics and business, including consideration of the complex interplay between consumption choices and well-being. Whether your field is business, commerce, economics, sociology, anthropology, environmental science, history, or psychology—or whether you teach at a high school, university, or online MBA program—consider using the FP/NFP lens to analyze historical and present-day phenomena.

The teaching of economics should evolve to become more appropriate, accurate, and relevant for the contemporary contexts we collectively face. Both authors of economic textbooks,

Neva Goodwin and Jonathon Harris highlight the significant ideological bias present in economics education in the United States, claiming there is[928]:

- *very limited treatment of environmental and ecological problems;*
- *glossing over realities of class, race, and gender divisions and discrimination;*
- *very limited treatment of income and wealth inequality;*
- *the acceptance of current institutional structures as given;*
- *the misrepresentation of these institutional structures as being consistent with models of perfect competition, ignoring concentrations of economic power; and*
- *acceptance of increased consumption as the primary measure of wellbeing*

Fortunately, there is a growing body of literature on alternative approaches to teaching economics[929] as well as a strong movement in popular economics to simplify economic complexity through graphic animation and storytelling. Examples of this simplification include the RSA Animate series, *The Story of Stuff* documentary, and the work of groups such as Positive Money[930]. Additionally, students all over the world are increasingly demanding more heterodox economic ideas be taught, such as the International Student Initiative for Pluralism in Economics and Rethinking Economics[931].

As educators, you can ensure economics remains tied to the real world by involving your class in studies and projects requested by NFP enterprises. For example, Professor Vince Smith of Southern Oregon University had his class conduct a pilot survey of customer attitudes for the local NFP association that runs operations at the Mt. Ashland Ski Area. The data was of significant value to the association's business planning, influencing their Learn to Ski packages as well as their marketing and operational strategies.[932]

Some universities and other educational centers provide even more hands-on support for NFPs, like the University of Chicago, whose Community Programs Accelerator supports local nonprofits by facilitating business networking, free workshops, and connections with campus volunteers[933]. Encouraging students, especially those in business, to consider internships with NFPs is another great way to speed up the shift to an NFP economy.

## Journalists, Celebrities, and Online Influencers

If you are a journalist, blogger, or podcaster, you can help shine a spotlight on the NFP World. Platforms like *Positive News*, HuffPost Good News, The Extraenvironmentalist, and *Yes! Magazine* publish interviews with social entrepreneurs, exploring the motivations and experiences of innovators running or working for businesses they don't own. You could reinforce the concept of NFP enterprise in your reporting by presenting business and not-for-profit activities as mutually inclusive. While maintaining transparency about any conflicts of interest, you should also not hesitate to investigate NFPs in your journalistic work. This will help keep them accountable to their own standards.

If you are a celebrity or online influencer, you have a compelling ability to stir people's curiosity about new economic models. We advise doing research before talking about alternative economics. People like Naomi Klein and Pope Francis show that possessing a thorough understanding of the details behind a popular economic message inspires others to investigate the ideas more rigorously. Whenever possible, remind people of humanity's

928 http://ase.tufts.edu/gdae/publications/working_papers/principles.pdf.
929 http://link.springer.com/article/10.1007/s12143-009-9033-1.
930 Add relevant hyperlinks: http://positivemoney.org/videos/
931 http://www.isipe.net/ and http://www.rethinkecon.co.uk/.
932 Personal comms: Annette Batzer, May 24, 2016.
933 https://communityprograms.uchicago.edu/.

interconnectedness. This message carries even more weight when it comes from people frequently idolized by the media in unhealthy ways.

## The Six Layers of Engagement

As we transition to the NFP World, our different contributions will effect change on different levels. Citizens are changing *behaviors* relating to their spending and tapping into *feelings* such as empathy and unrest. Public servants and legislators are establishing the *conditions* for NFP tax exemptions. Workers, NFP managers, and business leaders are adopting new *frameworks* for participatory, decentralized business. Advocates are promoting *values* such as social justice and ecological stewardship. And researchers and journalists are helping to create and spread new *constructs*, such as the story of interconnectedness. We call these the *six layers*.

**The Six Layers of the Great Transition**

| Layer | Explanation | Examples |
|---|---|---|
| **Behaviors** | Our actions, both as individuals and groups | • Purchase from NFPs<br>• Work and volunteer for NFPs<br>• Move your money to a credit union |
| **Feelings** | Our emotional states | • Empathy<br>• Sense of purpose<br>• Unrest |
| **Conditions** | The visible, tangible or observable phenomena around us with which we may interact on a regular basis | • Social entrepreneurs<br>• NFP incubators<br>• Tax exemptions for NFPs |
| **Frameworks** | Mental structures we use to organize our thoughts and actions | • For-profit versus NFP structures<br>• Decentralized, participatory business models<br>• The circular economy |
| **Values** | Things we prioritize, both as individuals and groups | • Social justice<br>• Well-being<br>• Ecological stewardship |
| **Constructs** | Our individual and collective worldviews, assumptions, narratives, goals, and beliefs | • Human nature is complex<br>• The NFP World<br>• The story of interconnectedness |

All six layers constantly influence each other, and all layers must be actively engaged for there to be a high likelihood of whole system change. The electric car's failure to emerge in the 1990s[934] provides a perfect example of what can go wrong when some layers are missing.

In the 1990s, there were strong *feelings* of despair and frustration about the state of the planet, and conservation was emerging as a *value*, but people's *behaviors* as both individuals and groups were still heavily consumption-oriented. Transportation relied on *conditions* of cheap oil and policies that were strongly influenced by lobbying from the automotive and fossil fuel industries. Energy infrastructure was still largely conceived in terms of hierarchical and centralized *frameworks*. The overarching *construct* was the triumph of humanity's ability to harness nature's potential through ingenuity. The necessary ingredients for systems change were missing on multiple levels. What author Malcolm Gladwell terms "the tipping point" simply could not eventuate, and, as a result, the great potential of the electric car languished.

Reports suggest the electric car is finally approaching a global tipping point.[935] When we examine the layers where ingredients were previously missing, we see why: consumption has become a more conscious *behavior* (Chapter 4). Networked energy and ridesharing services represent the increasingly decentralized *frameworks* being used by designers, and environmental advocacy has changed the overarching *construct* to increasingly focus on the fragility of nature in relation to human impact. The missing ingredients remain the political will to develop legislative measures that would accelerate this shift as well as price parity (oil remains cheap). The stage is set for developments supporting these two remaining layers to catalyze a systemic shift in transportation.

Not only is the time right for people to embrace the NFP story since the key ingredients now exist across each layer, but it is also becoming obvious that no one's contribution is more important than any one else's in driving this systemic change. Because the layers are interconnected and complementary, *all* are necessary. We need one another to bring the NFP World into existence. Anything we do, across any level, reinforces all other actions in the system. When enough of us act in a certain way, a pattern is created. Each one of our actions is either reinforcing the for-profit story and patterns to some extent or telling a new story and helping to strengthen the new NFP World pattern.

As author Charles Eisenstein points out, at present, no one can live in this new story 100 percent of the time .[936] We live in both old and new stories simultaneously. Most of us cannot fully align our lifestyles with this new story yet, but there is a definite mass movement in that direction. And the momentum is accelerating.

Given your unique background and connections within the wider world, how will you use your talents to bring the NFP World story closer to fruition? What behaviors can you change that will encourage others to follow your lead? What are you feeling that might motivate you to act even more passionately for social change? What can you create that will inspire hope? What assumptions will you question? What values will motivate your choices? And what story of human nature will you seek to embrace?

# The Not-for-Profit Way

934 Brought to wider attention by the 2006 documentary 'Who Killed the Electric Car?'
935 http://www.businessinsider.com/the-2020s-could-be-the-decade-of-the-electric-car-2016-2.
936 REF.

A healthy economy thrives on circulation. In such an economy, money, resources, and value constantly flow to where they are needed, creating shared wealth. We can realize such an economy by shifting our central mode of businesses to NFP enterprise. In that way, money (including profit) is constantly cycled through the economy in service of purpose, ensuring human needs are adequately met while eradicating the conditions of false scarcity and culture of consumerism driving overconsumption.

These bold claims have required us to reconsider our understanding of profit and recognize its potential to be either generative or destructive, depending on the circumstances. This more realistic approach also acknowledges the massive potential inherent in an NFP World, together with the significance of clearly demarcating the differing wider consequences of the FP and NFP forms of business[937]. Let us take a moment to review the argument presented in this book.

We have discovered an NFP structure enables far more activities than charity, and successful businesses—across all types of industries—can be run as not-for-profit. We have realized co-ops and social enterprises are umbrella structures and that it is valuable to learn whether a cooperative or social enterprise activity is run as FP or NFP, given the former facilitates wealth extraction while the latter supports the principle of wealth circulation. When we see the world through the FP/NFP lens, we tend to assign less blame to "the market," "big business," "corporations," "private banks," and "sociopathic CEOs" for the state of our world. Instead, we realize the *essential* qualifier for each of these important critiques is "for-profit." Without this distinction, it has been difficult to develop comprehensive macroeconomic alternatives, given that it is the *for-profit* market economy, *for-profit* corporations, *for-profit* CEOs, *for-profit* banking, and, perhaps most importantly, the *for-profit* story that need changing. Until now, "for-profit" has been as invisible to us as the water is to a fish. Most of us were raised with the for-profit story taught as the entire story, with the underlying assumptions that all businesses must be run in an FP way and that we have to ruthlessly compete to get ahead. This isn't some huge conspiracy—it's merely the ideology and assumption of our time.

**<pq>**The NFP business model proves the profit motive is not necessary for business efficiency.[938]<pq> The existence of new financing mechanisms and falling startup costs combine to nullify the rationale that businesses must start out as for-profit enterprises. Not-for-profit business offers a model for corporate sustainability that creates real value through its purpose-driven focus. It also increases employee and community participation, promotes greater ecological sensitivity, and drives the circulation of money and resources throughout our economy. We have not romanticized NFP business but rather have highlighted its vulnerability to the shortfalls of traditional nonprofits (e.g., fraud, bureaucracy, inefficiency, and mission drift). We have, however, witnessed how NFP enterprises are less susceptible to these shortfalls than traditional nonprofits, thanks to their structure. Particularly in the context of the NFP World—in which the NFP ethic has become paramount—this potential problem would shrink even further.

We have explored the grave need for an economic alternative given that FP enterprise and the for-profit story are at the root of our economic, social, and environmental crises. We have seen how the for-profit system has a tendency to exacerbate economic inequality and environmental devastation, and any reformation of capitalism will ultimately prove inadequate since its very nature requires the extractive siphon.

---

[937] Cooperative federalism.
[938] Especially by basic agency theory.

We have discovered how the rising NFP ethic worldwide is setting the stage for an economic evolution. And we have seen the many advantages NFP enterprises hold in the market—including being more financially efficient and affordable, having an increased capacity for innovation, and better aligning with the more ethical market and more purpose-motivated workforce of the twenty-first century.

We have investigated how a global economy in which NFP companies constitute the primary business mode can function effectively. We illustrated how this one shift enables four underlying mechanisms to emerge that in turn create a system of abundance within ecological limits. Since NFP finance eliminates the necessity for debt to constantly expand, other seemingly intractable structures and institutions (such as compounding interest, money creation, limited liability, "the corporation," corporate personhood, and intellectual property) become irrelevant or even benevolent in an NFP economic framework. Shifting to an NFP paradigm feels like less of a leap when you realize how familiar such a system could be while at the same time being so fundamentally different.

We explored the seemingly inevitable fall of the for-profit system and journeyed through a scenario for transitioning to the NFP World. There are many ways we can each accelerate the evolution to an NFP World. Not only do we all have roles to play, but our very success relies on each of us playing these parts.

We have shared these ideas with you in the hope that you will become an ambassador for this new vision. We ask that you share the parts that most resonate for you as well as your insights regarding the parts that do not.

This vision has something for everyone. It can unite a diverse range of politics, ideologies, and worldviews, from the most idealistic to the most pragmatic.

The NFP World appeals to those who:

- yearn for the expansion of social services and safety nets because NFP enterprises prime the wealth circulation pump, ensuring people's needs are fully met;
- desire greater freedom and less government influence in society and their lives, or those who distrust big organizations (in the NFP World, wealth and power are decentralized, *and* the system operates with interconnected efficiency. State control and the need for taxation have decreased, and institutions are highly accountable to the wider community.);
- believe in the power of business and markets as a force for good, value profit, and feel individual choice and competition are needed to ensure the efficient allocation of resources;
- crave status and recognition as social entrepreneurs, and their financial supporters are celebrated as true heroes in the NFP World;
- wish to encourage technological innovation (the NFP model spurs progress faster than ever since competition occurs within a wider context of genuine collaboration);
- are concerned about the well-being of the planet and future generations, thanks to the inbuilt paradox of enough and associated reductions in per capita consumption; and
- are simply tired of the daily grind and eager to enjoy the path to rest, leisure, and greater well-being promised by the NFP World.

Beyond our rising discontent, disillusionment, and depression, beyond the misery conditioned by poverty, injustice, and enslavement, there is a way. Beyond our fixation with the market and state, beyond our aggressive attempts to force ideologies on others, there is a way. Beyond our belief that progress comes from domination of nature and our fellow human beings, beyond the theory that we are fundamentally greedy and will never change, there is a way.

There *is* more than enough for us all. If our economy builds circulation into its very heart, we can ensure a fair distribution of our common wealth. We can counter the argument that dominating each other and nature is inevitable. The best of human characteristics will emerge when we create the conditions for flourishing. Consuming less while having more, we can honor Mother Earth as custodians of her bountiful riches and immense diversity. We can celebrate our innovative genius when we know it has been inspired by love rather than fear. By standing together, we can cocreate a new story based on a life worth living and a world that works for the benefit of us all.

There is a way. And it is not-for-profit.

Let us discover this new way together. Join us on this revolutionary journey toward a sustainable economy, greater equality, and deeper interconnectedness.

---

[i] We use the terms 'capitalism' and 'for-profit economy' interchangeably, as the private ownership of business is a central component of capitalism (Ref: Shleifer, A. (1998) 'State versus Private Ownership', *Journal of Economic Perspectives*, 12(4): 133-150).

[ii] All dollar amounts in this book refer to U.S. dollars unless otherwise specified.

[iii] All of the NFP enterprises mentioned in this book generate at least 50% of their revenue through the sale of goods and services and many of them consistently generate a profit. They all operate very much like 'normal' businesses.

[iv] Black Friday is the day after Thanksgiving in the U.S., and most retailers lower their prices and extend their hours to encourage people to spend as much money as they can on consumer goods and gifts for the holidays.

[v] Not-for-profit enterprise is not a panacea or a silver bullet and we will discuss the weaknesses of this model in more depth later in the book, but the first chapters are an exploration of its strengths.

[vi] Regardless of whether the residual, cyclical or hybrid dividend method is used (Ref: Investopedia)

[vii] A few countries, such as Canada, allow for a loophole in this. They allow companies to pay dividends before taxes, as a business expense, to a legal body called an 'income trust', which is owned by the shareholders. (Ref: Investopedia)

[viii] Of course, much peer-to-peer lending happens within a for-profit context and investors are just looking for a financial return on investment. However, we are focusing on peer-to-peer lending in the NFP context.
[ix] All bonds are basically loans that investors make to companies and governments (Ref: Personal Finance book, Biedenweg, p. 37).
[x] Other ways to raise startup finance for NFPs include: revenue-based finance, direct public offerings, refundable membership shares (which involve no equity or means to make a capital gain); social investment funds; and community development funds.
[xi] To learn about other kinds of capital-raising strategies available to NFPs, see *How on Earth's* sister book, *The Not-for-Profit Handbook*.
[xii] The other 9% is labeled as 'other revenue' in their annual report.
[xiii] See for instance: Cooperatives and the Sustainable Development Goals, UK Cooperative economy , Coops Europe, Consumer Co-operatives Worldwide website
[xiv] In consumer co-ops, there are also generally provisions which prohibit trading member shares in the co-op (non-transferable shares) or prohibiting members from profiting from selling shares in the co-op (limited transferability of shares).

[xv] Credit unions can be legally categorized as for-profit or not-for-profit depending on local legal frameworks. In the U.S., for instance, all credit unions are legally not-for-profit, while in Canada all credit unions are legally categorized as for-profit (although they still act like NFPs). We consider all credit unions to be NFP, as no one makes a private gain from them.

[xvi] In addition to the NFP business models mentioned in this chapter, here's a fuller (but still not comprehensive) list of NFP models found around the world: non-equity housing co-ops, community land trusts, mutual benefit corporations, community benefit societies, community development corporations, community investment corporations, mission-controlled corporations, social enterprise cooperatives, restorative corporations, common good corporations, community advancement co-ops, non-distributing purchasing co-ops, non-distributing producer co-ops, NFP social franchises, NFP disability enterprises, international associations without lucrative cause, common profit organizations, gift-based business and barter. Communities and people living subsistence lifestyles, who focus on meeting their own needs and do not typically participate in the formal market, are also considered to be part of this NFP realm (referred to in the figure below as 'non-trading self-sufficiency').
[xvii] Freelancers and self-employed persons who embody the 'ethic of enough' should probably be in the Zone of NFP Enterprise, too, but there's no formal way of defining a 'non-acquisitive freelancer' at this point, so we have left it out of the depiction.

[xviii] We know that Harry Shutt, for instance, has also explored this and made important contributions in his book, "Beyond the Profits System". For better or worse, we weren't aware of his book until the final editing stage of our book, so we weren't able to incorporate much of the wisdoms put forth in his writings.
[xix] See, for example: Urban Institute, Salamon, Casey?, Shutt
[xx] We will explain this more fully in the 'Incredible Inequality' section of the chapter.
[xxi] Companies report their financial positions every 3 months, known as quarters.
[xxii] Some early observers, like Marx, noted many of the internal contradictions of the capitalist system, including its tendency to create inequality.
[xxiii] And it's likely to be more extreme now, as those numbers are from 2008.
[xxiv] This trend has spurred social responses, such as Go Home on Time Day, a campaign that invites employees to pledge to leave work on time on one day each November, as well as calculating how much unpaid overtime is being donated to the employer – see: www.gohomeontimeday.org.au
[xxv] Of course this is not limited to for-profit companies, but it is an effect of the for-profit way of organizing the economy.
[xxvi] Small, family-owned businesses
[xxvii] It is too large for us to show in this book, but we encourage readers to view it online at: http://www.motherjones.com/files/legacy/news/feature/2007/03/and_then_there_were_eight.pdf

[xxviii] Furthermore, many corporations are lobbying to tilt the Internet in their favor. They are lobbying for policies that will make companies pay fees in order to have more Internet reach (Ref: Guardian article). Along the same lines, big telecom companies in the U.S. for instance, lobby to keep hundreds of cities from using their high-speed fiber cables to provide Internet to their citizens at affordable prices (Ref: Motherboard article).
[xxix] Lobbying refers to the legal act of trying to influence decisions made by government officials.
[xxx] Companies on the stock market that seek to raise shareholder value in ethical ways are an exception to this general rule; however, when it comes to making decisions about tradeoffs between shareholder value and environmental or social concerns, the profit motive and the legal structure of these companies pushes them to prioritize shareholder value. And that's generally what they'll do.
[xxxi] And this ratio has been steadily increasing: they were worth 55% of GDP in 2006 and 17% of GDP in 1995 (Ref: Politifact calculations).
[xxxii] See, for example, economist Joseph Stiglitz's 2008 article in The Guardian.
[xxxiii] Take, for example, Sweden, in which state-owned enterprises contribute approximately 8 per cent to the national GDP. See: http://www.government.se/content/1/c6/24/81/93/30fcec38.pdf.
[xxxiv] See, for example, Transparency International's "Corruption Perceptions Index 2014", available at: http://www.transparency.org/whatwedo/publication/cpi2014.
[xxxv] First used by John Erhenfeld in <1999?>.
[xxxvi] In updating the draft of this book, we found that Gates went from having $80.1 billion in 2015, to $87.4 billion in 2016.
[xxxvii] A complex system is different from a complicated system. A complicated system is made up of many different parts and it's not very clear how it works, but it can be taken apart and put back together. A computer, for instance, is a complicated system. Whereas a complex system, like a living organism, cannot simply be taken apart and put back together, because all the parts are so interdependent.
[xxxviii] While it remains obvious that crime and violence disproportionately affect the poorest in society.
[xxxix] Often referred to as minimalism or voluntary simplicity, which are both about doing more with less.
[xl] See Human Development Index: http://hdr.undp.org/en/content/human-development-index-hdi; Genuine Progress Indicator: http://rprogress.org/sustainability_indicators/genuine_progress_indicator.htm; Happy Planet Index: http://www.happyplanetindex.org/; and Social Progress Index: http://www.socialprogressimperative.org/data/spi
[xli] As a cohort, the Millennial Generation was born between the early 1980s and the early 2000s. They are the first generation to begin their adult lives in the new millenium.
[xlii] MBA refers to the Master of Business Administration professional degree.
[xliii] We are certainly not the first to look at the evolution of business models over time, but we are probably the first to say that trends are leading in a NFP direction. We see the evolution of business lining up with the evolution of global society as described by Spiral Dynamics, a theory put forth by Clare Graves, which is why we've chosen this specific order of colors (see, for instance: *The Five Levels of Understanding*, by Natasha Todorovic and Chris Cowan).
[xliv] After adjusting for inflation.
[xlv] These countries are Australia, Belgium, Brazil, Canada, the Czech Republic, France, Israel, Japan, Kyrgyzstan, Mexico, Mozambique, New Zealand, Norway, Portugal, Thailand, and the United States.
[xlvi] The GDP contribution of the nonprofit sector in all 8 countries for which longitudinal data are available outpaced overall economic growth.
[xlvii] Not all NFP enterprises necessarily have all of the advantages we explore here, but they are unique to the not-for-profit realm. These advantages also vary from place to place.
[xlviii] At least in most 'developed' countries.
[xlix] This commercial aired in the U.S. in May 2013
[l] The Green Bay Packers is the exception, as an NFP member-'owned' team.
[li] The study indicates that the sectors with the highest concentration of purpose-driven employees are education, nonprofits, agriculture/forestry/fishing, entertainment, and

healthcare. As education and healthcare are largely made up of NFPs, we feel it's safe to conclude that the not-for-profit sector has more purpose-driven employees than the for-profit sector.
[lii] Value Added Tax free